%% file: Thesis_AOzaeta.tex
 \documentclass[oneside,12pt]{book}

\usepackage{setspace}
\usepackage{amsmath}
\usepackage{amssymb}
\usepackage{amsfonts}
\usepackage{mathptmx}
\usepackage{graphicx}
\usepackage{epstopdf}
\usepackage{dcolumn}
\usepackage{bm}
\usepackage{morefloats}
\usepackage{enumerate}

\DeclareMathOperator{\tr}{\mathop{\mathrm{Tr}}}

\DeclareMathOperator{\re}{\mathop{\mathrm{Re}}}
\DeclareMathOperator{\im}{\mathop{\mathrm{Im}}}
\DeclareMathOperator{\arctanh}{arctanh}

\newcommand{\Eq}[1]{Eq.~(\ref{#1})}
\newcommand{\Eqs}[1]{Eqs.~(\ref{#1})}

\newcommand{\ket}[1]{\begingroup\lvert#1\rangle\endgroup}

\begin{document}

%\language{english}

% A page with the abstract on including title and author etc may be
% required to be handed in separately. If this is not so, then comment
% the below 3 lines (between '\begin{abstractseparte}' and 
% 'end{abstractseparate}'), normally like a declaration ... needs some more
% work, mind as environment abstracts creates a new page!
% \begin{abstractseparate}
%   \input{Abstract/abstract}
% \end{abstractseparate}

% Using the watermark package which is in StyleFiles/
% and to remove DRAFT COPY ONLY appearing on the top of all pages comment out below line
%\watermark{DRAFT COPY ONLY}

\begin{titlepage}
    \begin{center}
        \vspace*{1cm}
        
        \Huge
        \textbf{Transport phenomena in superconducting hybrid nanostructures}
        
        \vspace{0.5cm}      
        
        \vspace{1.5cm}
        
        \textbf{Asier Ozaeta}
        
        \vfill
        
       Advisor: Dr. Sebastian Bergeret
        
        \vspace{0.8cm}
        
        \includegraphics[width=0.4\textwidth]{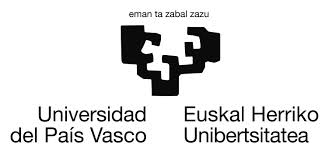}
        
        \Large
        2014
        
    \end{center}
\end{titlepage}
%\maketitle

%set the number of sectioning levels that get number and appear in the contents
\setcounter{secnumdepth}{3}
\setcounter{tocdepth}{3}

\frontmatter % book mode only
\pagenumbering{roman}
%\include{Dedication/dedication}
%\begin{dedication} %this creates the heading for the dedication page
%
%I would like to dedicate this thesis to my loving parents ...
\newpage\null\thispagestyle{empty}\newpage

%
%\end{dedication}
%\include{Acknowledgement/acknowledgement}
%\begin{acknowledgementslong} %uncommenting this line, gives a different acknowledgements heading

      %this creates the heading for the acknowlegments
\thispagestyle{empty}
\singlespacing

This thesis would not exist as it is without the help and support from a number of people.

First of all, I want to express my deep gratitude to my scientific advisor Sebastian Bergeret for permanent attention and help in my work. I have learned a lot from you with respect to both physics in general and mesoscopics in particular, together with some extra lessons about life.

My thanks to the members of the thesis committee: Andres Arnau, Nerea Zabala, Carlos Cuevas, Francesco Giazotto and Jacobo Santamaria.  

I am obliged also to my coauthors: Niladri Banerjee, Christopher Berg Smiet, Mark Blamire, Carlos Cuevas, Alexander Golubov, Tero Heikkil\"a, Frank Hekking, Shiro Kawabata, Jason Robinson, R.G.J. Smits, Andrey Vasenko and Pauli Virtanen for pleasant and fruitful collaboration. Also thanks to Teun Klapwijk for meaningful discussions specially at the early stages of my doctorate. In addition to those mentioned above I wish to thank the following people: Felix Casanova, Francesco Giazotto, Ion Lizuain, Guillermo Romero, Mihail Silaev, Enrique Solano and Estitxu Villamor.

Looking to the future, I wish to thank Michael Crommie, Nacho Pascual and Dimas G. de Oteyza for giving me such great opportunities and for their time and help. 

My thanks to the present and past members of our mesoscopic physics group, Alvise Verso and Vitaly Golovach, and to all my colleagues in the CFM. I am indebted to Eneko Malatsetxebarria for his help at the beginning of my scientific career.  

I am also grateful to Tero Heikkil\"a for hosting me in Espoo and to the people of the Low Temperature Laboratory for their hospitality. Specially to Pauli Virtanen and Teemu Ojanen for those nice lunch times.

I would also like to acknowledge the people who proof-read part of the manuscript: David Pickup and Vitaly Golovach. 
%specially Sebastian Bergeret who read the whole document (few times).

I would also like to mention my friends in Gasteiz, those in Donosti and Madrid, the friends I met in Leioa, the ones that I met during my Erasmus year and those in Helsinki and Hamburg. I do not want to forget about my BJJ teachers Andre Crispin and Iury Martins, and all my team-mates both in Donosti and Gasteiz.  

I want to thank also my parents and sister for support during all these years.

Special thanks to Tea for being the way she is. 

%I would also want to thank myself for being an awesome individual and for my hard work.

% 

%\end{acknowledgmentslong}

\let\cleardoublepage\clearpage
%\include{Abstract/abstract}

%\listoffigures
%\printnomenclature  %% Print the nomenclature
%\addcontentsline{toc}{chapter}{Nomenclature}
\newpage\null\thispagestyle{empty}\newpage

\tableofcontents

\newpage\null\thispagestyle{empty}\newpage

\addcontentsline{toc}{chapter}{List of Publications}
\thispagestyle{empty}
%The following document is an overview of those publications.

This thesis has resulted in the following peer-reviewed publications. Below we list them in chronological order.

\begin{enumerate}[I]
\item F.S. Bergeret, P. Virtanen, \textbf{A. Ozaeta}, T.T. Heikkil\"a and J.C. Cuevas, \textit{Supercurrent and Andreev bound state dynamics in superconducting quantum point contacts under microwave irradiation}, Physical Review B \textbf{84}, 054504 (2011).

\item \textbf{A. Ozaeta}, A.S. Vasenko, F.W.J. Hekking and F.S. Bergeret, \textit{Electron cooling in diffusive normal metal–superconductor tunnel junctions with a spin-valve ferromagnetic interlayer}, Physical Review B  \textbf{85}, 174518 (2012).

\item\textbf{A. Ozaeta}, A.S. Vasenko, F.W.J. Hekking and F.S. Bergeret, \textit{Andreev current enhancement and subgap conductance of superconducting SFN hybrid structures in the presence of a small spin-splitting magnetic field}, Physical Review B \textbf{86}, 060509 (2012).
\item A.S. Vasenko, \textbf{A. Ozaeta}, S. Kawabata, F.W.J. Hekking and F.S. Bergeret, \textit{Andreev current and subgap conductance of spin-valve SFF structures}, Journal of superconductivity and novel magnetism \textbf{26}, 1951-1956 (2013).

\item S. Kawabata, \textbf{A. Ozaeta}, A.S. Vasenko, F.W.J. Hekking and F.S. Bergeret, \textit{Efficient electron refrigeration using superconductor/spin-filter devices}, Applied Physics Letters \textbf{103}, 032602 (2013).

\item A.S. Vasenko, S. Kawabata, \textbf{A. Ozaeta}, A.A. Golubov, F.S. Bergeret and F.W.J. Hekking, \textit{Detection of small exchange fields in S/F structures}, Proceedings of the Vortex VIII conference, arXiv:1401.0646 (2013).

\item N. Banerjee, C.B. Smiet, R.G.J. Smits, \textbf{A. Ozaeta}, F.S. Bergeret, M.G. Blamire and J.W.A. Robinson, \textit{Evidence for spin selectivity of triplet pairs in superconducting spin valves}, Nature Communications \textbf{5}, 3048 (2014).

\item \textbf{A. Ozaeta}, P. Virtanen, F.S. Bergeret, T.T. Heikkil\"a, \textit{Predicted Very Large Thermoelectric Effect in Ferromagnet-Superconductor Junctions in the Presence of a Spin-Splitting Magnetic Field}, Physical Review Letters \textbf{112}, 057001 (2014).

\item S. Kawabata, A.S. Vasenko, \textbf{A. Ozaeta}, A.A. Golubov, F.S. Bergeret and F.W.J. Hekking, \textit{Heat transport and electron cooling in ballistic normal-metal/spin-filter/superconductor junctions}, proceeding of the MISM2014 conference (2014).
\end{enumerate}

\thispagestyle{empty}

%\newpage\null\thispagestyle{empty}\newpage
%
%\newpage\null\thispagestyle{empty}\newpage

%\include{Abstract/abstract}

%\listoffigures
%\printnomenclature  %% Print the nomenclature
%\addcontentsline{toc}{chapter}{Nomenclature}
%\section*{List of publications}
%\addcontentsline{toc}{subsection}{List of publications}

 % book mode only

\mainmatter
 
%\pagestyle{fancy}
%\fancyhf{}
%\fancyhead[RO]{\bfseries Chapter 1: Introduction }
%\fancyhead[LE]{\bfseries Chapter 1: Introduction}
%\fancyfoot[C]{\thepage}

%\renewcommand{\headrulewidth}{0.5pt}
%\renewcommand{\footrulewidth}{0pt}
%\addtolength{\headheight}{0.5pt}
%\fancypagestyle{plain}{
%  \fancyhead{}
%  \renewcommand{\headrulewidth}{0pt}
%} 
 
\chapter{Introduction}
\label{ch:1}

\include{intro2d}

%\newpage\null\thispagestyle{fancy}
%\fancyhead[RO]{\bfseries \rightmark }
%\fancyhead[LE]{\bfseries \rightmark}
%\fancyfoot[C]{\thepage} \renewcommand{\headrulewidth}{0pt}
%\newpage 

%\pagestyle{fancy}
%\fancyhf{}
%\fancyhead[RO]{\bfseries Chapter 2: Fundamentals of the theory of superconducting nanohybrids }
%\fancyhead[LE]{\bfseries Chapter 2: Fundamentals of the theory of superconducting nanohybrids}
%\fancyfoot[C]{\thepage}

%\renewcommand{\headrulewidth}{0.5pt}
%\renewcommand{\footrulewidth}{0pt}
%\addtolength{\headheight}{0.5pt}
%\fancypagestyle{plain}{
%  \fancyhead{}
%  \renewcommand{\headrulewidth}{0pt}
%}

\chapter{Fundamentals of the theory of superconducting nanohybrids}
\label{ch:2}

\include{chap2_SB}

\newpage\null\thispagestyle{empty}\newpage

%\pagestyle{fancy}
%\fancyhf{}
%\fancyhead[RO]{\bfseries Chapter 3: Subgap charge transport in superconductor-ferromagnet junctions }
%\fancyhead[LE]{\bfseries Chapter 3: Subgap charge transport in superconductor-ferromagnet junctions}
%\fancyfoot[C]{\thepage}

%\renewcommand{\headrulewidth}{0.5pt}
%\renewcommand{\footrulewidth}{0pt}
%\addtolength{\headheight}{0.5pt}
%\fancypagestyle{plain}{
%  \fancyhead{}
%  \renewcommand{\headrulewidth}{0pt}
%}

\chapter{Subgap charge transport in superconductor-ferromagnet junctions}
\label{ch:3}

\include{3112}

%\newpage\null\thispagestyle{empty}\newpage

%\pagestyle{fancy}
%\fancyhf{}
%\fancyhead[RO]{\bfseries Chapter 4: Electron Cooling and Thermoelectric effect in nanoscale superconducting devices }
%\fancyhead[LE]{\bfseries Chapter 4: Electron Cooling and Thermoelectric effect in nanoscale superconducting devices}
%\fancyfoot[C]{\thepage}

%\renewcommand{\headrulewidth}{0.5pt}
%\renewcommand{\footrulewidth}{0pt}
%\addtolength{\headheight}{0.5pt}
%\fancypagestyle{plain}{
%  \fancyhead{}
%  \renewcommand{\headrulewidth}{0pt}
%}

\chapter{Electron Cooling and Thermoelectric effect in nanoscale superconducting devices}
\label{ch:4}

\include{312}

\include{thermo2}

%\newpage\null\thispagestyle{empty}\newpage

%\pagestyle{fancy}
%\fancyhf{}
%\fancyhead[RO]{\bfseries Chapter 5: Superconducting quantum point contact in a microwave field }
%\fancyhead[LE]{\bfseries Chapter 5: Superconducting quantum point contact in a microwave field}
%\fancyfoot[C]{\thepage}

%\renewcommand{\headrulewidth}{0.5pt}
%\renewcommand{\footrulewidth}{0pt}
%\addtolength{\headheight}{0.5pt}
%\fancypagestyle{plain}{
%  \fancyhead{}
%  \renewcommand{\headrulewidth}{0pt}
%}

\chapter{Superconducting quantum point contact in a microwave field}
\label{ch:5}

\include{ch43}

\newpage\null\thispagestyle{empty}\newpage

%\pagestyle{fancy}
%\fancyhf{}
%\fancyhead[RO]{\bfseries Chapter 6: Conclusions and future directions }
%\fancyhead[LE]{\bfseries Chapter 6: Conclusions and future directions}
%\fancyfoot[C]{\thepage}

%\renewcommand{\headrulewidth}{0.5pt}
%\renewcommand{\footrulewidth}{0pt}
%\addtolength{\headheight}{0.5pt}
%\fancypagestyle{plain}{
%  \fancyhead{}
%  \renewcommand{\headrulewidth}{0pt}
%}

\chapter{Conclusions and future directions}
\label{ch:6}

\include{conc_SB}

\newpage\null\thispagestyle{empty}\newpage

\appendix

%\pagestyle{fancy}
%\fancyhf{}
%\fancyhead[RO]{\bfseries Appendix A : The Keldysh quasiclassical Green functions technique }
%\fancyhead[LE]{\bfseries Appendix A : The Keldysh quasiclassical Green functions technique}
%\fancyfoot[C]{\thepage}

%\renewcommand{\headrulewidth}{0.5pt}
%\renewcommand{\footrulewidth}{0pt}
%\addtolength{\headheight}{0.5pt}
%\fancypagestyle{plain}{
%  \fancyhead{}
%  \renewcommand{\headrulewidth}{0pt}
%}

\chapter{The Keldysh quasiclassical Green functions technique}

\include{app2}

\backmatter % book mode only

%\include{Appendix1/appendix1}
%\include{Appendix2/appendix2}

%\bibliographystyle{plainnat}
%\bibliographystyle{Classes/CUEDbiblio}
%\bibliographystyle{Classes/jmb}
%\bibliographystyle{Classes/jmb} % bibliography style
%\renewcommand{\bibname}{References} % changes default name Bibliography to References
%\bibliography{References/references} % References file

\end{document}

%% file: intro2d.tex
Superconductivity was discovered by H. Kamerlingh Onnes (Leiden) in 1911~\cite{super}. It is a macroscopic quantum phenomenon\cite{macros} and although it has been widely investigated over the last century, the interest is far from declining\cite{reviewsuper}. Partly, because of the search for superconductors with high critical temperatures ($T_C$)\cite{highsuper} and also because superconductors are the basis for future emerging technologies as quantum computation and quantum information. 

Superconductors have been used for a wide range of practical purposes. Since the early days, they have been considered as zero-resistance conductors and ideal diamagnets. Among all the applications nowadays, the most notable still remain the use of superconductors as zero-resistance elements to produce strong magnetic fields (\textit{e.g.} Large hadron collider at CERN) and as ideal diamagnets to levitate objects (\textit{e.g.} Japan´s Maglev trains). Superconductors are also used as  precision detectors, due to the presence of a well defined superconducting gap, either to measure current~\cite{BTK,KBT} or magnetic fields with SQUIDs~\cite{squid} (superconducting quantum interference device). The latter being a sensitive magnetometer, used for measuring extremely small magnetic fields and based on superconducting loops.

Conventional superconductors, with low $T_C$, are easy to manipulate and to use for the fabrication of structures with sizes smaller than the characteristic coherence length, which is on the order of a micron. In recent decades, the great achievement in making high-quality contacts between superconductors and normal metals, ferromagnets, and insulators has allowed the building of nanostructures large enough to be implemented in a circuit but small enough to show quantum phenomena. 

In such hybrid structures, interesting physics takes place due to the leakage of superconducting correlations into non-superconducting materials. This phenomena is called \textit{the proximity effect}. In samples of a small size it leads to a wide range of interesting phase-coherence effects. The proximity effect underpins several of the effects discussed in the present thesis. Its study started back in 1932 in a work by R. Holm and W. Meissner~\cite{Holm}. They observed a zero resistance state in a junction of two superconductors separated by a normal metal layer. This phenomenon was later studied in the 1950s and 1960s~\cite{71,15,16} in thin layer systems. Related experiments in the 1970s and 1980s studied the effects of bias voltage\cite{181}, microwave irradiation~\cite{191}, and magnetic fields~\cite{20} in SNS junctions. Here S is a superconductor and N a normal metal.  The achievement of building high-quality contacts between superconductors and normal metals at the nano scale in the 90s enabled better understanding of the proximity effect~\cite{8,9}. Subsequent work on this topic broadened the numbers of materials in which the proximity effect could be studied to include: 2D electron gases in semiconductors~\cite{21}, novel materials such as carbon nanotubes~\cite{101,111} and graphene~\cite{121}, and \textit{ferromagnets}~\cite{22}. In particular, superconductor-ferromagnet (SF) structures have attracted the attention of several research groups in the last decade\cite{buzdinrev}. 

In conventional superconductors, the ground state is described by pairs of electrons (Cooper pairs) with opposite spins. It is well known that electrons with different spins belong to different energy bands. The energy shift of the two bands can be considered as an effective exchange field acting on the spin of the electrons. Therefore, at first glance, ferromagnetism and conventional superconductivity cannot coexist in bulk systems. However, in SF hybrids the interplay between superconductivity and ferromagnetism leads to interesting physics. For example the exponential decay of the condensate into the ferromagnet is accompanied by oscillations in space. This phenomenon leads, for example, to oscillations of the critical temperature $T_c$ as a function of the thickness\cite{buzkup,radovic,jiang}. Furthermore, due to the oscillatory behaviour of the superconducting condensate in the ferromagnetic region, the critical Josephson current changes its sign in a $SFS$ junction\cite{bulaevskii,bauer,blum, kontos2001,ryazanov,sellier}. Under certain conditions, it is also possible that the presence of a ferromagnet leads to a \textit{long-range triplet superconducting pair correlation}\cite{berg2001,kadigrobov}. Such pairs are not affected by the ferromagnetic exchange field and, therefore, can propagate in the ferromagnet over long distances. 

According to the theory, the triplet component of the superconducting condensate can create highly polarized supercurrents, \textit{i.e.} currents without dissipation, in Josephson SFS junctions\cite{eschrigphyto}. Such currents can be exploited for spintronics devices, an emerging technology based on the manipulation and control of the spin currents in electronic devices\cite{prinz}. One of the bottlenecks in the development of nanoscale spintronic devices are the large currents needed to control the spin states and the heat losses associated with them. Spin-polarized supercurrents can help to overcome this problem with the availability of fully polarized triplet supercurrents.

New technologies focused on miniaturization of electronic solid-state circuits face the same problem as spintronics. Decreasing the size and increasing the transistor speed leads to large ohmic dissipation and the associated heating is a significant obstacle. Therefore, there is an increasing interest in the study of heat management and control of heat at the nanoscale. The branch of electronics that studies the coupling between charge and heat currents is called \textit{caloritronics}. If one adds the spin degree of freedom, one talks about \textit{spin caloritronics}. Examples of effects studied in spin caloritronics are the spin dependence of thermal conductance, the Seebeck and Peltier effects, heat current effects on spin transfer torque, thermal spin, and anomalous Hall effects.

In the present thesis, we extend the field of caloritronics and spin caloritronics by adding superconductors as building blocks. On the one hand, this allows to reduce heat losses and, on the other hand, to exploit phase coherent effects~\cite{natgia,giaheat1,giaheat2,giaheat3,giaheat4,giaheat5}. Two types of heat-related topics are address in this thesis. First, we study superconducting hybrids for cooling applications. The flow of charge current in normal metal/insulator/ superconductor (NIS) tunnel junctions at a bias voltage $V$ is accompanied by a heat transfer from N into S. This phenomenon arises due to the presence of  the superconducting energy-gap which allows for a selective tunnelling of high-energy "hot" quasiparticles out of N. Such a heat transfer through NIS junctions can be used for the realization of microcoolers~\cite{Nahum,Leivon,Giazotto06,Muhonen}. We extend these studies with the aim of increasing the cooling power by considering superconductor-magnetic hybrids. The use of magnetic materials can reduce the Joule heating and, therefore, lead to an increase of the cooling efficiency. 

The study of the interplay between spin dependent fields and superconducting correlations is also important for the realization of structures supporting Majorana bound states, which are proposed as the basis for topological quantum computation~\cite{Fukane}. Recent experiments have suggested that Majorana modes are supported in a semiconductor with Rashba spin-orbit coupling in the presence of a Zeeman field\cite{kouwen}. Experimental results are, however, non conclusive and alternative explanations for the zero-bias anomalous peak observed in the experiment are being considered. Parts of the present thesis focus on the transport properties of nanostructures with induced superconducting correlations in the presence of spin-splitting fields and hence they contribute also to this active research field. 

Superconductors are also the building blocks for solid-state quantum bits (qubits). Realization of superconducting qubits encompasses charge and flux qubits\cite{revqubit}.  It has been also theoretically proposed that a superconducting small constriction with a few numbers of bounds states (Andreev bound states) could be used as a qubit (the Andreev qubit\cite{shumeiko1}). The states of this qubit can be manipulated, for example, by an external rf-field and read out by measuring the Josephson current through the junction. Part of this thesis is devoted to the study of the electronic dynamics of quantum point contacts in the presence of a microwave field. 

% In summary, the study of superconducting hybrid structures is important for both the development of new emerging technologies and a better understanding of fundamental properties of matter. 

% The present work is a contribution to this broad field. It is mainly focused on charge and heat transport in different nanohybrids consisting of superconductors, normal metals, ferromagnets and insulators. In the next section the outline of the thesis is presented. 

\section{Outline of the thesis}

In chapter 1 we introduce basic superconducting phenomena. Such as, the BCS theory, the Andreev reflection and the proximity effect, and the charge current transport in superconducting tunnel junctions. 

In chapter 2 we present the Keldysh nonequilibrium Green function formalism used to obtain the results of this thesis, together with clarifying examples corresponding to simple junctions. This chapter also includes the results of the critical temperature calculations in a superconducting nanohybrid junction. It consist in a $FSF$ spin valve with a spin mixer at the $SF$ interface. Here, the ferromagnetic layers surrounding the thin superconducting layer generates the long-range triplet component. The feasibility of superconducting spintronics depends on the spin sensitivity of ferromagnets to the spin of the equal spin triplet Cooper pairs. This structure provides evidence of a spin selectivity of the ferromagnet to the spin of the triplet Cooper pairs. As a second example, we describe the Hanle effect in a spin valve geometry. With the help of the quasiclassical Keldysh formalism we describe spin imbalance and spin injection in a normal metal. Furthermore, we compare this results with the macroscopic theories available to date.

%The main results of the thesis are presented in three different chapters. In chapter~\ref{sec:} shows the main results of the study of charge and spin transport in superconducting nanohybrid structures. In Sec.~\ref{sec:} we show the results corresponding to spin imbalance, using our microscopic theory we studied the spin injection in superconductors with an intrinsic exchange field. Phenomena such as the Hanle effect and spin imbalance effects were studied in this systems. We also managed to merge together our results with the macroscopic theories available to the date.

The main results of the thesis are presented in three chapters. In chapter 3, the subgap transport properties of a $SIF$ structure is studied. We compute the differential conductance and show that its measurement can be used as an accurate way of determining the strength of a spin-splitting field smaller than the superconducting gap. It is also shown that for an $SIFIF^\prime$ system with arbitrary magnetization direction, one can measure precisely the value of the effective exchange field. This is the averaged field acting on the Cooper pairs in the multi-domain ferromagnetic region. For exchange fields of the order of few $\Delta$, the density of states of the FS bilayer at the outer border of the ferromagnet shows a peak at the value of the field~\cite{golubov}. Thus, we propose a series of accurate ways for determining the exchange field.

We also show that, contrary to what it could be expected, the Andreev current at zero temperature can be enhanced by a spin-splitting field smaller than the superconducting gap. There is a critical value of the bias voltage above which the Andreev current is enhanced by the spin-splitting field. This unexpected behaviour can be explained as the competition between two-particle tunnelling processes and decoherence mechanisms originating from the temperature, voltage, and exchange field.

We devote chapter 4 to the study of thermal transport in superconducting nanohybrid structures. The first part of the chapter focuses on cooling (\text{i.e.} the heat flow out of the normal metal reservoir), where we introduce two new cooling devices based on spin filters and non collinear ferromagnets. The first contribution, consists of a cooling device based on a $SIFIF^\prime$ structure with arbitrary magnetization. Here, we study the role of the triplet superconducting component in the cooling phenomenon. We demonstrate that the cooling efficiency depends on the strength of the ferromagnetic exchange field and the angle between the magnetizations of the two F layers. Contrary to what we expected, for exchange fields lower than the superconducting gap, the cooling power has a non-monotonic behaviour versus the exchange field. We also study the dependence of the cooling power on the lengths of the ferromagnetic layers, the bias voltage, the temperature, the transmission of the tunnelling barrier, and the magnetization misalignment angle.

For the second cooling device, the spin-filtering effect leads to values of the cooling power much higher than in conventional $NIS$ coolers. The device, consisting of a superconductor and normal metal separated by a spin filter, $SI_{SF}N$,  shows a highly efficient cooling in both ballistic and diffusive multi-channel junctions.

In the second part of chapter 4, we study the thermoelectric effects in hybrid superconducting structures. In thermoelectric devices a temperature gradient can generate an electric potential ("Seebeck effect") and viceversa ("Peltier effect"). In electronic conductors a major contribution to thermoelectricity is given by the electron-hole asymmetry in the system. This is the reason why semiconductors with their chemical potential tuned to the gap edge are used for this purpose. The chemical potential in superconductors is not tunable, as charge neutrality dictates electron-hole symmetry, thus, thermoelectric effects are weaker than those of normal metals. However, we propose to inject spin-polarized current in a superconductor with a spin splitting field. This generates a huge thermoelectric effect: the resulting thermoelectric figure of merit can far exceed unity, leading to heat engine efficiencies close to the Carnot limit. We also show that spin-polarized currents can be generated in the superconductor by applying a temperature bias. This results look promising for a detector that measures precisely small temperature changes. Conversion of waste energy is not energetically favourable, due to the need of cooling and keeping the superconductor at low temperatures.

In chapter 5, we develop a general theory for the microwave-irradiated high-transmittance superconducting quantum point contact (SQPC), which consists of a thin constriction of superconducting material in which the Andreev states can be observed. We proposed using the Andreev bound states of a SQPC for quantum computing applications as qubits.

%This idea was later developed in detail in Ref.~\cite{romero}, where a general frame for studying the quantum dynamics of a superconducting point contact galvanically coupled to a single-mode resonator was presented. This theory allows the proposed setup to be considered as an alternative quantum device for circuit quantum electrodynamics (QED) technology.

We show that a microwave field is an ideal tool to make a direct spectroscopy of the Andreev bound states in a superconducting junction. This theoretical study allows to understand the influence of a microwave field on the supercurrent through superconducting weak links such as Dayem bridges or SNS junctions. Furthermore, depending on the values of the different parameters of the system, one may observe different physical phenomena, described in chapter 5. We predicted that for weak fields and low temperatures, the microwaves can induce transitions between the Andreev states leading to a large suppression of the supercurrent at certain values of the phase. In contrast, at strong fields, the current-phase relation is strongly distorted and the corresponding critical current does not follow a simple Bessel-function-like behaviour. More importantly, at finite temperature, the microwave field can enhance the critical current by means of transitions connecting the continuum of states outside the gap region and the Andreev states inside the gap.

The thesis concludes with a summary of the obtained results in chapter 6. The detailed derivation of the quasiclassical equations is presented in the appendix. Throughout this thesis we set the Planck constant, $\hbar=1$ and the Boltzmann constant, $k_B=1$. Consequently, energies, frequencies and temperatures have the same units.

\newpage

\section{Basic properties of conventional superconductors}
\label{sec:1.2}
%%
%\begin{figure}[b]
%  \centering
%  \includegraphics[width=0.5\columnwidth]{I-V}
%  \caption{Current voltage characteristic of a superconductor.}
%\label{fig:IV}
%\end{figure}
%%

The defining property of a superconductor is that below a certain temperature, $T_C$ (the superconducting transition temperature), the electrical resistance vanishes. This perfect conductivity of metals was discovered in 1911 by H. Kamerlingh Onnes. Another hallmark of superconductivity in bulk metals is the Meissner effect\cite{meissner} (1933): for an applied field smaller than the critical field $H<H_C$, the superconductor expels the magnetic field. This means that the superconductor behaves like a perfect diamagnet.

In 1935, the brothers F. and H. London\cite{london} proposed two phenomenological equations, the London equations, to describe these two properties of superconductors. If $\textbf{E}$ is the electric field and $\textbf{H}$ the magnetic flux density, the London equations read, 
\begin{equation}
\textbf{E}=\frac{\partial}{\partial_t} (\Lambda \textbf{J}_S) \; ,
\label{eq:lon1}
\end{equation}
\begin{equation}
\textbf{H}=-c \text{curl}\frac{\partial}{\partial_t} (\Lambda \textbf{J}_S)\; ,
\label{eq:lon2}
\end{equation}
where we define
\begin{equation}
\Lambda=\frac{4 \pi \lambda^2}{c^2}= \frac{m}{n_s e^2} \; .
\end{equation}
In these expressions $n_s$ is the density of superconducting electrons. These are the electrons in a metal that can transport current without dissipation. In contrast to the "normal" electrons with density $n$.

Equation \ref{eq:lon1} shows that any electric field accelerates the superconducting electrons rather than simply sustaining their velocity against resistance as described in Ohm´s law in a normal conductor. Thus, it describes perfect conductivity in superconductors. The second,  eq.\ref{eq:lon2}, combined with the Maxwell equation $\text{curl} \textbf{H}=4 \pi \textbf{J}/c$, leads to,
\begin{equation}
\nabla^2\textbf{H}=\frac{\textbf{H}}{\lambda^2} \; .
\end{equation}
This implies that the magnetic field penetrates the superconductivity over the length $\lambda$.

%Which implies that the magnetic field is exponentially screened from the interior of a superconductor with penetration length $\lambda$, the Meissner effect. 

%Later, it was established the existence of an energy gap $\Delta$, of the order $k_B T_c$, between the ground state and the quasiparticle excitations of the system.   

The next theoretical step was given by V.L. Ginzburg and L.D. Landau\cite{GL} in 1950, when they proposed a theory of superconductivity to describe the superconducting electrodes. They introduced a complex pseudowave function $\psi$. This is the order parameter within Landau´s general theory of second-order phase transitions. The local density of superconducting electrodes of the London equations \ref{eq:lon1} and \ref{eq:lon2}, is then given by,
\begin{equation}
n_s=|\psi(x)|^2 \; .
\end{equation} 
Ginzburg and Landau wrote an expression for free energy near $T_C$, the so-called Ginzburg Landau (GL) free energy. By minimizing the free energy with respect to the order parameter one arrives to the G-L equations. These are valid for temperatures close to $T_C$ and read,
\begin{equation}
\alpha \Psi + \beta |\Psi|^2 \Psi+\frac{1}{2m^*}(-i \nabla- \frac{e^*}{c}\textbf{A})^2 \Psi=0 \; ,
\end{equation}
\begin{equation}
\textbf{J}=\frac{c}{4 \pi} \text{curl}\textbf{H}=\frac{e^* }{2m^*i} (\Psi^* \nabla \Psi-\Psi \nabla \Psi^* ) - \frac{{e^*}^2}{m^* c} \Psi^* \Psi \textbf{A} \; .
\label{eq:glj}
\end{equation}
The first equation determines the order parameter $\Psi$ and the second provides the superconducting current $\textbf{J}$. Here $\alpha$ and $\beta$ are treated as phenomenological parameters. This theory allows the study of non-linear effects of fields strong enough to change $n_s$ and its spatial variation. It also can handle the coexistence of superconductivity and normal metal in the so-called intermediate state in a magnetic field comparable to $H_c$. Here the field is strong enough to destroy superconductivity rather than simply inducing screening currents to keep the field out of the interior of the sample.

Thus, according to the GL theory the superconducting state can be described by a many-particle wave function, $\Psi$, with a macroscopic phase. According to Eq.\ref{eq:glj}, a finite (equilibrium) current can be generated  by a gradient in the superconducting phase.

%\subsubsection{Properties of conventional superconductors}
Despite the good description of superconductivity given by the GL equations, there was still no microscopic explanation for superconductivity. First, in 1957, Bardeen, Cooper and Schrieffer\cite{BCS} (BCS) presented their pioneering microscopic pairing theory of superconductivity. The \textit{BCS theory} is based on (Cooper) pairing between electrons. At low temperatures electron-phonon interaction can lead to an effective weak attraction that can bind pairs of electrons. These bound pairs of electrons occupy states with equal and opposite momentum and spin. These are the so-called "\textit{Cooper pairs}".

The concept that even a weak attraction can bind pairs of electrons into a bound state was presented by Cooper~\cite{Coop} in 1956. He showed that the Fermi sea of electrons is unstable against the formation of at least one bound pair, regardless of how weak the interaction is, so long as it is attractive. The solution has spherical symmetry; hence, it is an s state as well as a singlet spin state. This means that a Cooper pair in a conventional superconductor is formed by an electron with spin up and momentum $\vec{k}$, and another with spin down and momentum $-\vec{k}$. Notice that unconventional pairs, for example those in a triple state, can also appear and will be discussed later in section\ref{sec:proximityeffect}. The size of the Cooper pair state is much larger than that of the interparticle distance and thus, the pairs are highly overlapping. 

The physical picture behind the idea of an effective attractive interaction is the following: By interacting with the crystal lattice, conducting electrons generate a deformation in the lattice, due to coulomb attraction.  The deformed lattice generates a higher density of positive charge that attracts a second electron, resulting in an effective attractive interaction between two electrons. If this attraction is strong enough, to override the repulsive screened Coulomb interaction, it gives rise to a net attractive interaction, which in turn leads to superconductivity. Historically, the importance of the electron-lattice interaction in explaining superconductivity was first suggested by Fr$\ddot{o}$hlich~\cite{frohlich} in 1950 and confirmed experimentally by the discovery of the "isotope effect"~\cite{isotope}, \textit{i.e.}, the proportionality of $T_C$ to $M^{-1/2}$ for isotopes of the same element.

\begin{figure}[h]
  \centering
  \includegraphics[width=0.5\columnwidth]{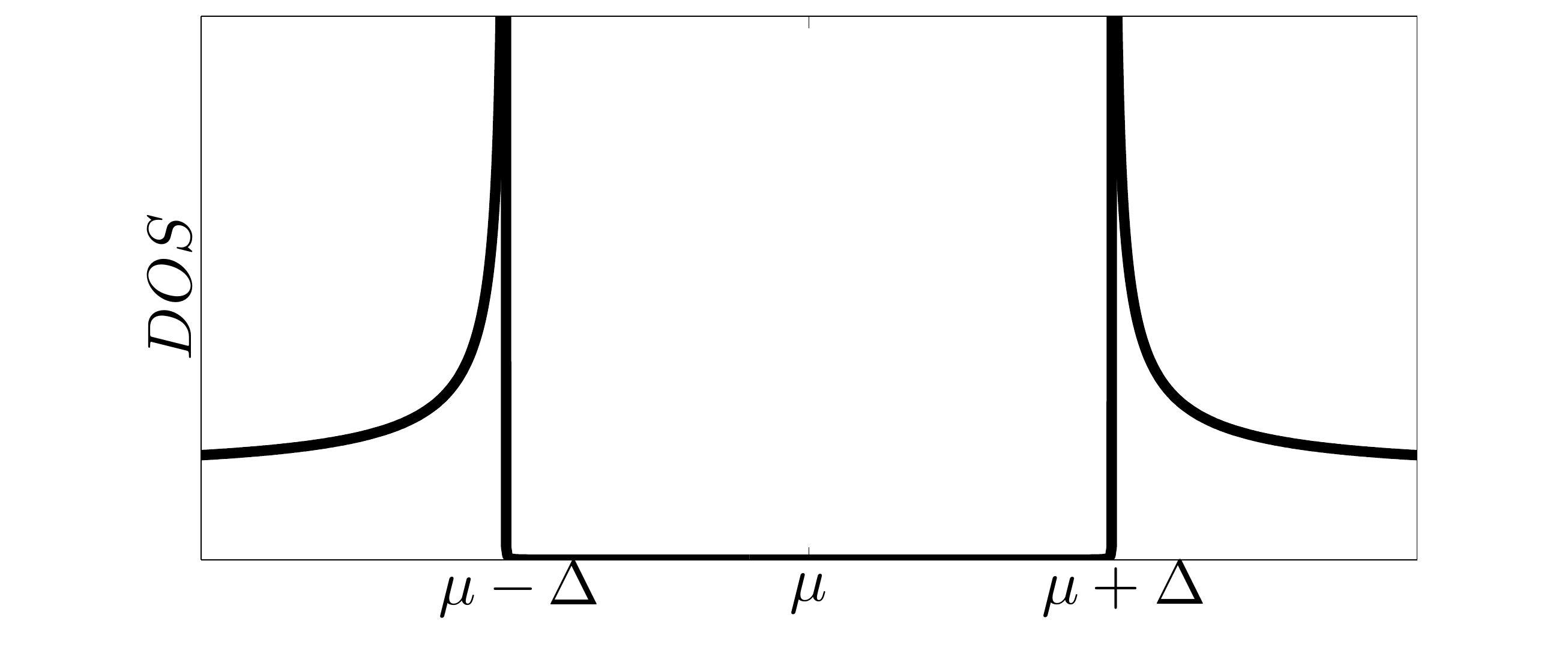}
  \caption{The density of states of a BCS bulk superconductor The chemical potential is located in the middle of the gap.}
\label{fig:BCSdos}
\end{figure}

The BCS theory predicts a gapped density of states described by
\begin{equation}
N_S(E)=\frac{E}{\sqrt{E^2 - |\Delta|^2}} \theta(|E|-|\Delta|) \; .
\label{eq:bcs1}
\end{equation}
The energy (E) is measured with respect to the Fermi level $E_F$. In fig.~\ref{fig:BCSdos} we show the DoS described by Eq.\ref{eq:bcs1}. There are two main features characteristic of the BCS DoS: The \textit{energy gap} $\Delta$, which leaves states unavailable for quasiparticles with energy $|E|<\Delta$ and the square-root divergence of the density of states at $E=\Delta$.

The energy gap is temperature dependent and the \textit{critical temperature} $T_C$ is the temperature at which $\Delta(T_C)=0$. So for $T>T_C$ the excitation spectrum becomes the same as in the normal state. According to the BCS theory the energy gap at zero temperature is related to $T_C$ in the following way,
\begin{equation}
\Delta(0)=\frac{\pi}{e^\gamma}~  T_C \; .
\label{eq:tcdelta}
\end{equation}
Here $\gamma$ is the Euler constant. This exact value is usually approximated as 1.764. This value has been tested in many experiments and found to be reasonably good.

The spatial extension of Cooper pairs is of the order of the \textit{coherence length}, $\xi_0$. The coherence length is the second characteristic length of a superconductor together with the penetration depth $\lambda$.  In a pure superconductor, \textit{i.e.} when $\xi_0$ is much smaller than the elastic scattering length $l$, it is given by
\begin{equation}
\xi_0^{clean}=\frac{ v_F}{\pi |\Delta|} \; ,
\end{equation}
and in the dirty case ($\xi_0 \gg l$) by
\begin{equation}
\xi_0^{dirty}=\sqrt{\frac{ \mathcal{D}}{2 \Delta}} \; .
\end{equation}
Here $v_F$ is the Fermi velocity and $\mathcal{D}$ the diffusion constant. Typical values for the elastic mean free path range from $l \sim 3$ to $ 60 \ nm$ . Thus, for the Fermi velocity of Al, where $v_F= 2.03 \times 10^6 m/s$, the diffusion coefficient $\mathcal{D}$ would be between $10^{-2}-10^{-4} \ m^2/s$. If, for example, we consider Al with $\Delta=180 \ \mu eV$, then the coherence length would be $\xi_0=190 \ nm$. While for Nb with $\Delta=1.8 \  meV$, the coherence length is found to be $\xi_0=38 \ nm$. 

In 1959, a couple of years after the BCS theory was first presented, Gor´kov\cite{gorkov1959} showed that the GL theory is a limiting form of the BCS theory, valid near $T_C$. He showed that $\psi$ is proportional to the gap $\Delta$. The formulation of the BCS theory in terms of the Gorkov Green functions will be presented in the next chapter.

\begin{figure}[h]
  \centering
  \includegraphics[scale=0.4]{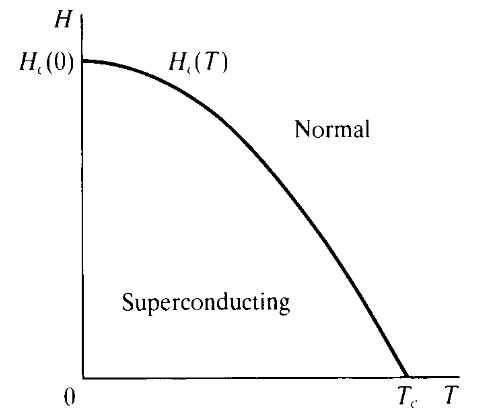}
  \caption{Temperature dependence of the critical field. (From ref.~\cite{tinkham})}
\label{fig:hcvst}
\end{figure}

As mentioned above another important property of the superconducting state has been introduced, the Meissner effect. For a certain value of the field, known as the critical field, the superconductivity is suppressed (orbital effect). This critical field is reduced as we increase the temperature, as depicted in fig.\ref{fig:hcvst}. This phenomena can be derived from the BCS theory. In the case of a thin superconducting film, if the field is applied in-plane, the critical field can reach values much larger than $H_c$,
\begin{equation}
H_{c||}=2\sqrt{6}\frac{H_c \lambda}{d} \; .
\end{equation}
Here $\lambda$ is the London penetration length and $d$ the film thickness. This means that, if $d/\lambda$ is small enough,  it can exceed the critical field by a large factor. 

\begin{figure}[h]
\begin{centering}
\includegraphics[width=0.65\columnwidth]{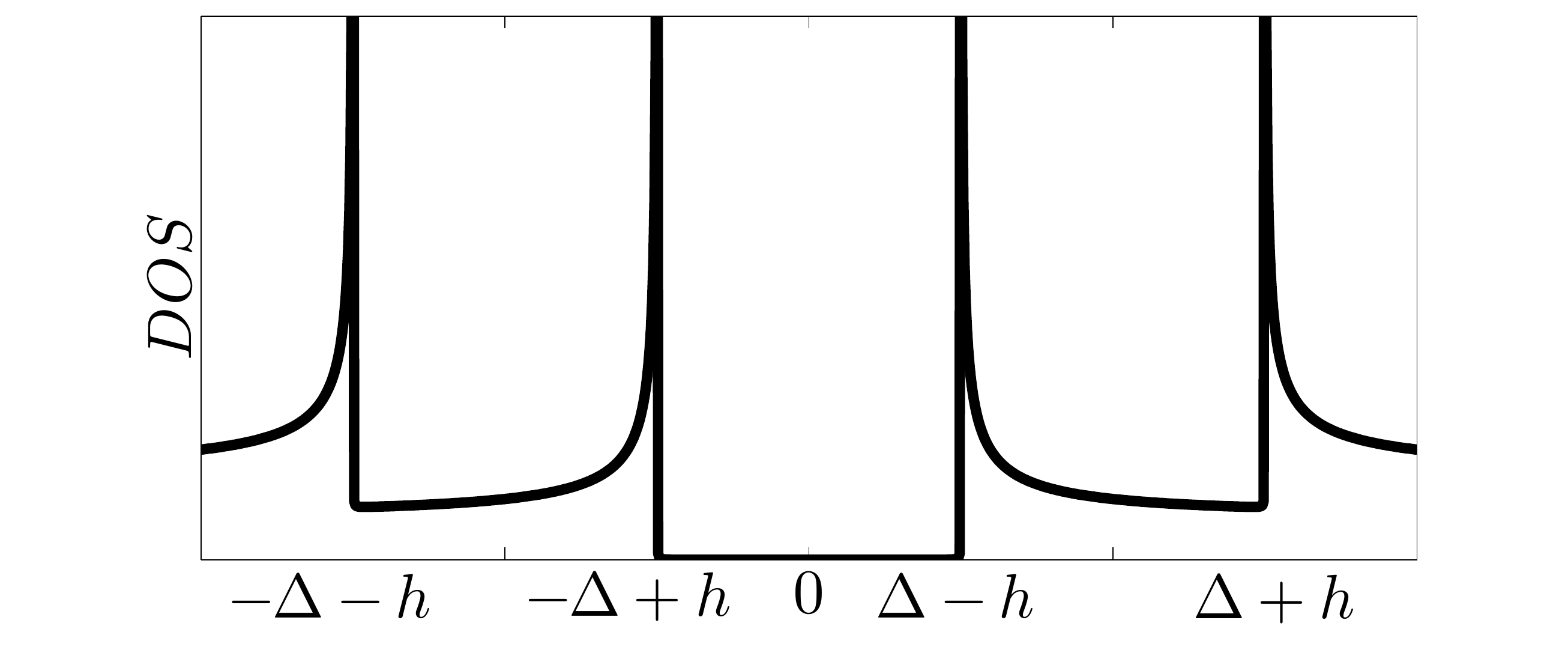}
\par\end{centering}

\caption{Density of states of a BCS superconducting thin film in a Zeeman field, where $h$ is the value of the Zeeman field. }
\label{fig:sdos}
\end{figure}

Besides the orbital effect, a magnetic field can destroy superconductivity by means of the paramagnetic effect. The magnetic field tends to align spins of Cooper pair in the same direction, preventing pairing. For a pure paramagnetic effect, the critical field of a superconductor $H_p$ at $T=0$ is given by the Chandrasekhar-Clogston limit\cite{clogston1962,chandrasekhar1962},
\begin{equation}
H_p(0)=\frac{\Delta}{\sqrt{2} \mu_B} \; .
\end{equation}
Furthermore, this paramagnetic effect causes a splitting in the BCS density of states (DoS). As shown in fig.~\ref{fig:sdos}, four peaks are observed, corresponding to the summation of the BCS DoS of spin up and down particles, now shifted in energy by $\pm h$, respectively. The density of states now reads,
\begin{equation}
N_{SF}(E)=\frac{1}{2} \left[ \frac{E+h}{\sqrt{(E+h)^2 - |\Delta|^2}} \theta(|E+h|-|\Delta|)+\frac{E-h}{\sqrt{(E-h)^2 - |\Delta|^2}} \theta(|E-h|-|\Delta|) \right] \; .
\end{equation}
Here $h$ is the spin splitting that is related to the external magnetic field by $h=g \mu_B H$. Note that in the absence of a magnetic field, there is no shifting and the BCS peaks for spin up and down are located at the same energies. The summation then leads back to the two peak DoS of fig.\ref{fig:BCSdos}. Following this brief summary of the main properties of homogeneous superconductors, we now introduce the superconductor-normal metal hybrid structures and focus on the Andreev reflections and the proximity effect.   

%When applying a magnetic field to the superconductor, it shifts the momentum of electrons with spin up and down forming the cooper pair in opposite directions, breaking the symmetry. Thus, reducing the gap and smearing the BCS peaks, eventually, for large enough fields destroying superconductivity. Y luego diria que debido a este ultimo efecto, el Hparalelo causa el splitting. 

%

\pagebreak

\newpage

\section{Andreev Reflection and proximity effect}
\label{sec:andreev}

In this section we introduce the concept of \textit{Andreev reflection} at the SN boundary. It is a key phenomena in nanohybrid superconducting systems.

\begin{figure}[h]
  \centering
  \includegraphics[width=0.5\columnwidth]{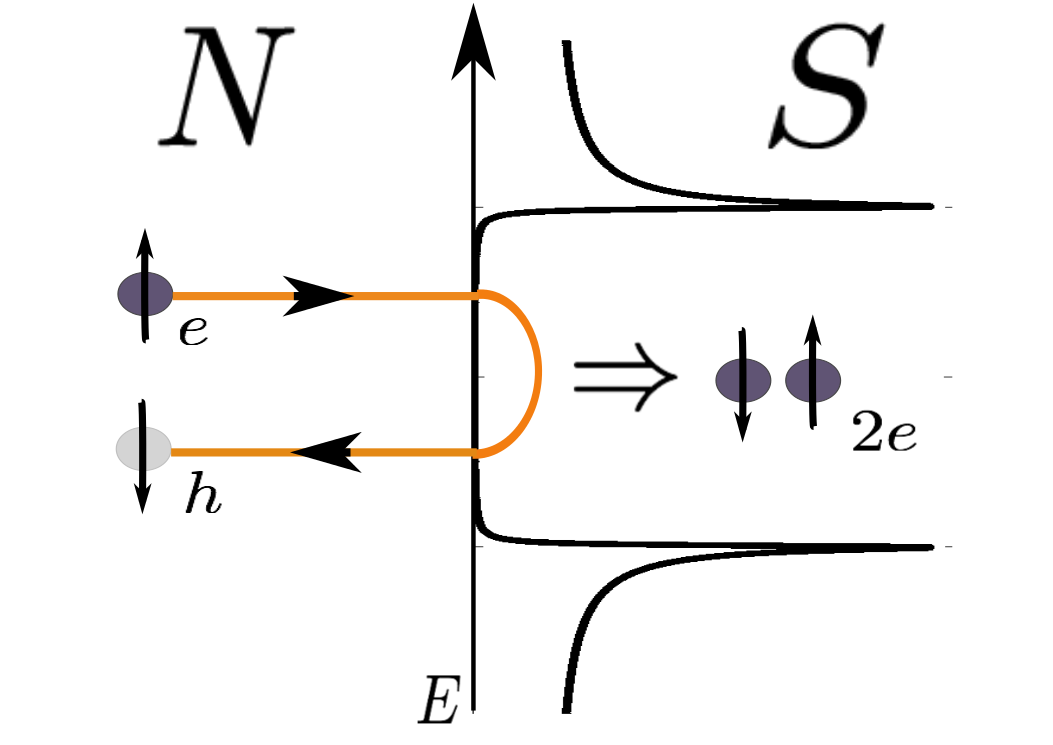}
  \caption{Sketch of the Andreev reflection process. The black circles represent electrons and the grey ones holes. Black arrows represent the spin direction. The orange lines represent the displacement direction. }
\label{fig:andRef}
\end{figure}
Let us consider an interface between a superconductor and a normal metal. In principle, electrons with energies $E<\Delta$ in the normal metal cannot cross the interface to the superconductor due to the gapped DoS. However, there is an additional reflection process, which was first identified by Andreev~\cite{Andreev} and subsequently treated by Artemenko et. al.~\cite{Artemenko} and by Zaitsev~\cite{Zaitsev}. Upon reaching the interface, electrons cannot be transferred as quasi-particles because there are no quasi-particle states in the gap. Instead, a Cooper pair is created on the right side of the interface. In order for the charge to be conserved, a hole (excitations below the Fermi energy) is reflected back into the normal metal. This process transfers $2e$ across the interface to the superconducting Cooper pair condensate (see fig.~\ref{fig:andRef}). In other words, Andreev reflection carries a charge current.

 The electron-hole pair remain phase-coherent over distances of the order of  the coherence length. For $ T \rightarrow 0$ and $eV<\Delta$, in a perfectly transmitting barrier, essentially all incident electrons are Andreev reflected. Thus, each reflection transfers a double charge, giving a differential conductance twice that of the normal state. As $T$ is raised this value falls continuously, reaching unity at $T_C$.
\begin{figure}[h]
\begin{centering}
\includegraphics[scale=0.75]{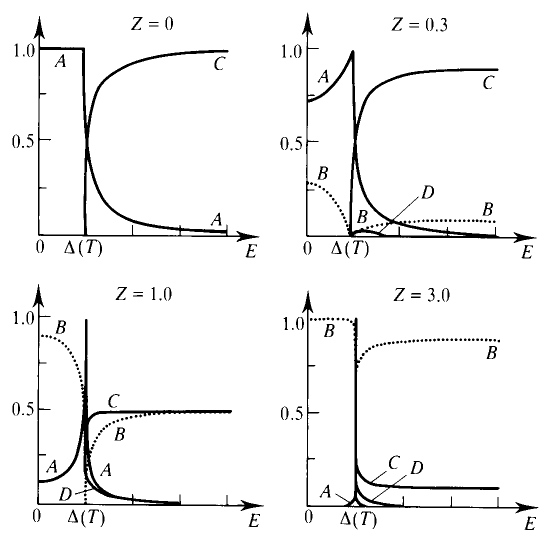}
\par\end{centering}

\caption{Plots of transmission and reflection coefficients at an NS interface. A gives the probability of Andreev
reflection, B of ordinary reflection, C of transmission without branch crossing, and D of transmission
with branch crossing. The parameter Z is a measure of
the barrier strength at the interface. (After Blonder et al.\cite{BTK} ) }
\label{fig:btk}
\end{figure}

More generally, there is normally some sort of barrier causing normal reflection at the NS interface, e.g. due to some oxide layer there, or because of the different Fermi velocities associated with the different metals. To allow a simple comprehensive treatment for the continuum of possibilities between no barrier and a strong tunnel barrier, Blonder, Tinkham and Klapwijk~\cite{BTK} (BTK) introduced a $\delta$-function potential barrier of strength $Z$ at the interface. They solved the Bogoliubov-de Gennes microscopic equations\cite{bogo} to find the probability for the various outcomes for an electron of energy $E$ incident on the interface, as a function of Z. The probabilities of the different possibilities are shown in fig.~\ref{fig:btk} for representative values of Z. \textit{A} is the probability of Andreev reflection as a hole on the other side of the Fermi surface. \textit{B} is the probability of ordinary reflection. \textit{C} is the probability of transmission through the interface with a wave vector on the same side of the Fermi surface, while \textit{D} gives the probability of transmission with crossing through the Fermi surface.

%\subsubsection{Andreev Bound States}

%
\begin{figure}[h]
\begin{centering}
\includegraphics[width=0.5\columnwidth]{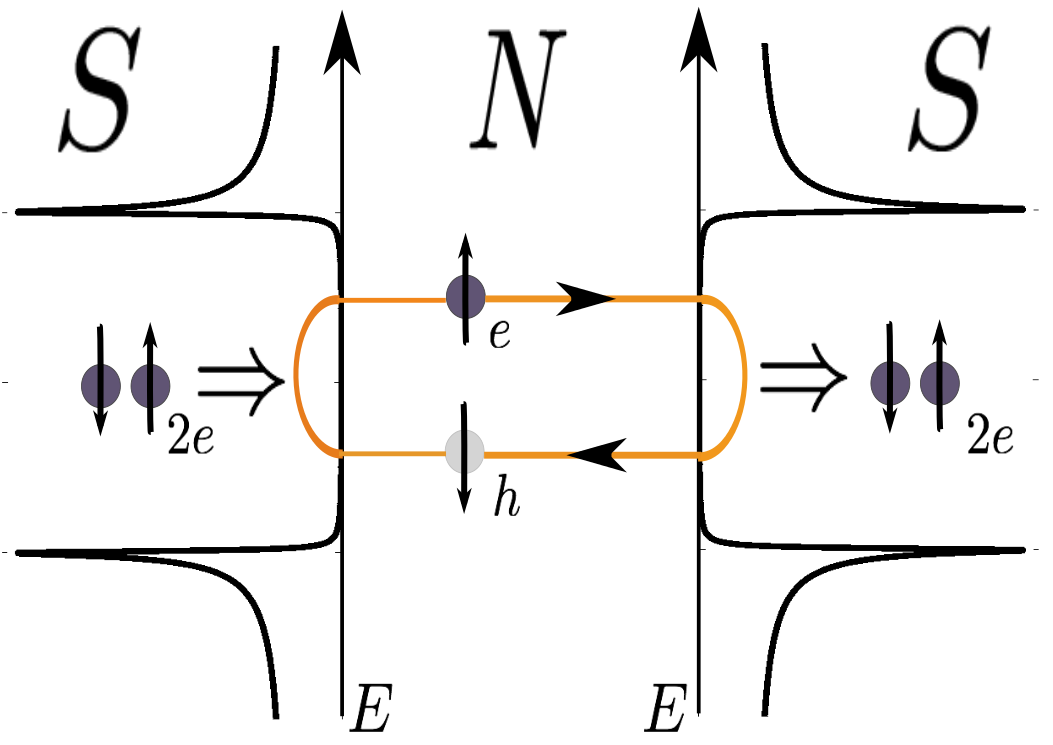}
\par\end{centering}

\caption{Sketch of the process that generates the Andreev bound states.}
\label{fig:abs}
\end{figure}
\begin{figure}[h]
\begin{centering}
\includegraphics[width=0.5\columnwidth]{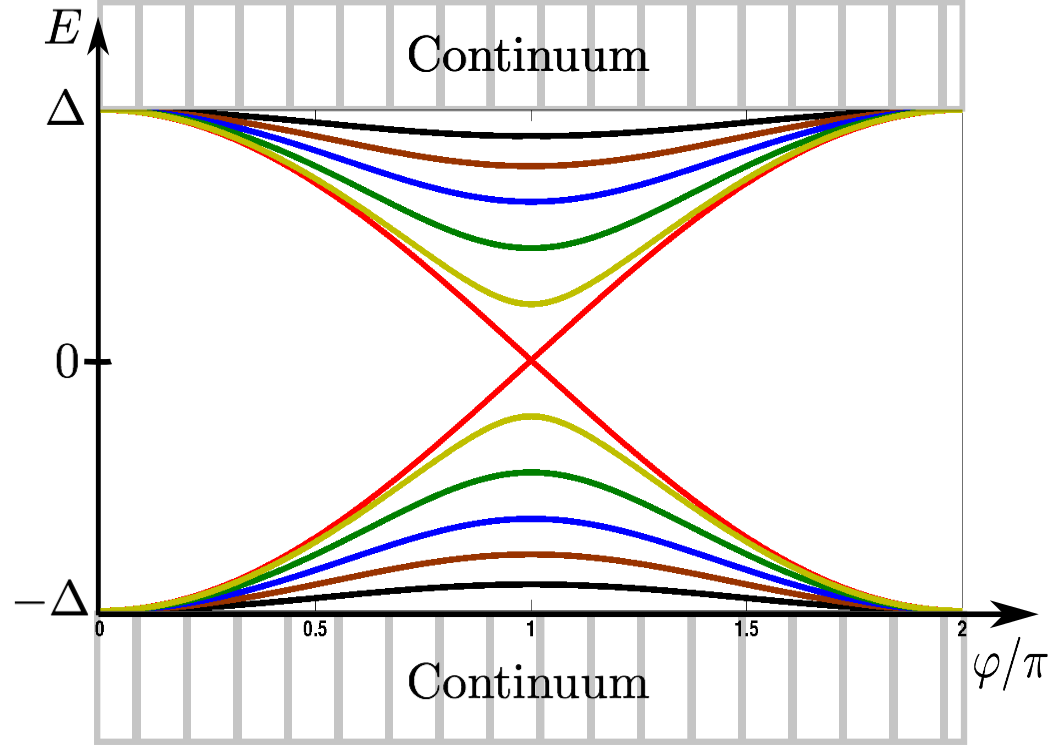}
\par\end{centering}

\caption{The energies of Andreev bound states versus phase difference between
the superconductors for different values of $T$. For $T$: 1(red), 0.95(yellow), 0.9 (green),
0.7(blue), 0.5 (brown) and 0.3 (black).}

\label{fig:ABS}

\end{figure}

We now consider a normal metal placed between two identical superconductors with different phases, as shown in fig.~\ref{fig:abs}. An electron incident from the left in the normal metal at energies smaller than $\Delta$, experiences an Andreev reflection. The same happens to the resulting hole in the left electrode, starting the process again. As it is Andreev reflected at both sides, discrete energy levels arise in the system. We can conclude that, a coherent conductor between two superconducting reservoirs rise to discrete bound energy states for quasiparticles, known as the \textit{Andreev bound states} (ABS). 

\begin{figure}[h]
\begin{centering}
\includegraphics[width=0.5\columnwidth]{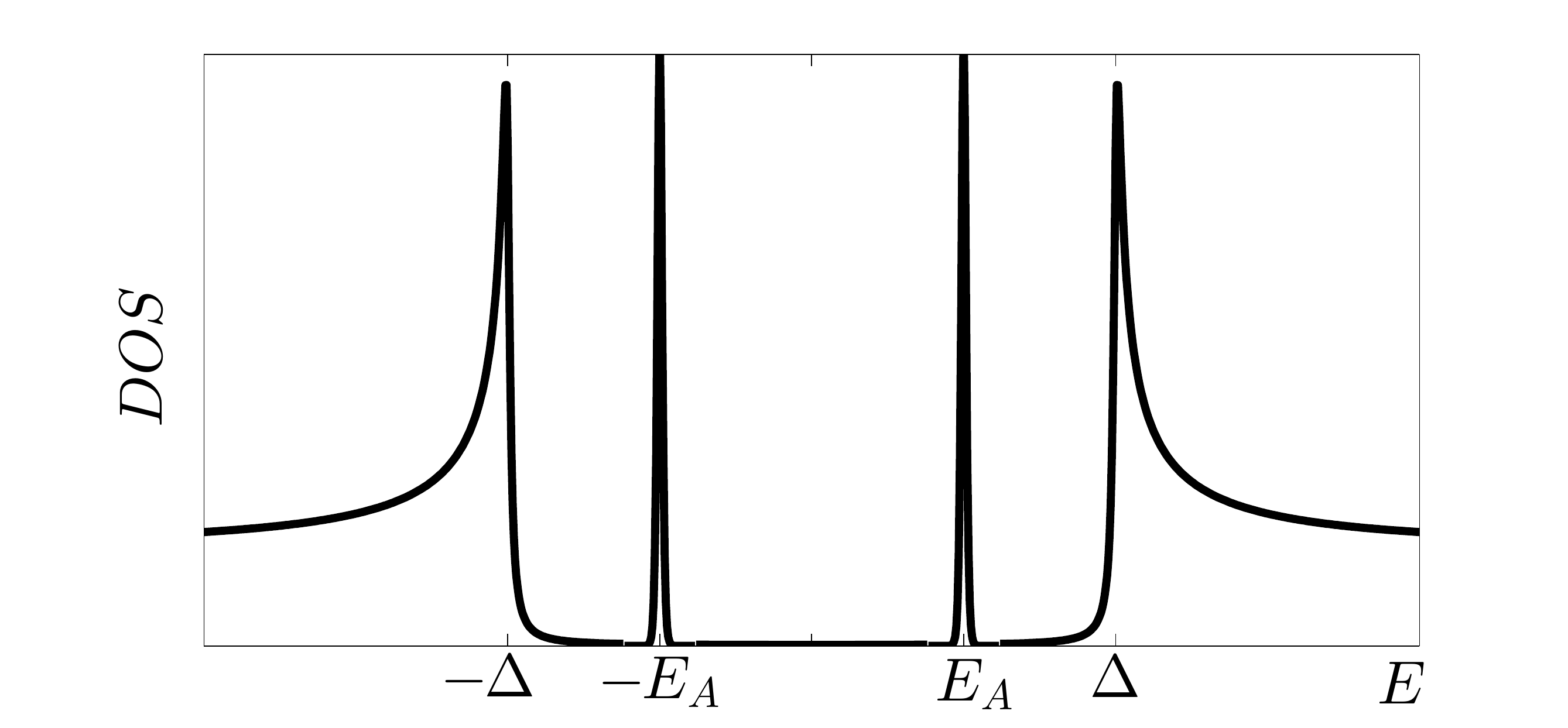}
\par\end{centering}

\caption{Density of states corresponding to a system of two BCS superconductors with different phases
and a nanostructure between them. The ABS can be observed inside the superconducting gap.}

\label{fig:dosABS}

\end{figure}

The energies of the Andreev bound states envisioned separately by Saint James\cite{stjam} and Andreev\cite{abs}, are described by to the expression,
\begin{equation}\label{eq:ABS}
E_A=\pm\Delta \sqrt{1- \tau \sin^{2}(\varphi/2)} \; .
\end{equation}
This is the expression for each transmission channel. Here $\tau$ is the transmission coefficient and $\varphi=\varphi_L-\varphi_R$ is the phase difference between the two superconductors~\cite{Beenaker}. The $\pm$ term corresponds to the existence of two Andreev bound states for quasiparticles, one with positive energy and another with negative. In fig.~\ref{fig:ABS} we plot the energies of the ABS as we vary the phase difference. Here the "continuum" represents the states outside the gap. It is shown that in order to observe the ABS deep inside the gap, high transmission values are required and they only touch each-other for perfect transmission values. On the other hand, for low transmission values the ABS are very close to the continuum. The DoS of this junction is plotted in fig.~\ref{fig:dosABS}. In the case of a nanostructure with multiple channels, each channel has a pair of ABS and depending on the transmission values the energies vary. For a sufficient number of channels, there are many pairs of ABS located in different energy positions inside the gap. In the case of low transmission values, they form a continuum at the edges of the gap.

\begin{figure}[h]
\begin{centering}
\includegraphics[scale=0.75]{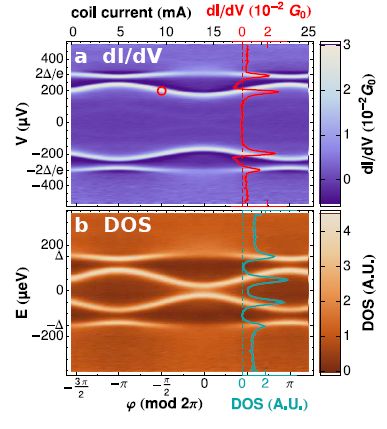}
\par\end{centering}

\caption{Conductance and DoS versus phase difference obtained experimentally, corresponding to the measurement of the ABS. (After ref.~\cite{absexp})}
\label{fig:absexp}
\end{figure}

%There are two cases in which these states are so close to the continuum that they are no longer considered as independent bound states. For phase differences between superconductors close to $\varphi=2 n \pi$, where $n$ is an integer, and for very small transmission values $\tau<<1$, the ABS take the value $E_A \sim \pm \Delta$. Thus, channels with high transmission values are required for ABS to be distinguishable from the continuum. 

We must bear in mind that the chemical potential is always located in the middle of the gap. This is an intrinsic property of superconductors. For zero temperature, only energies lower than the chemical potential are occupied, namely, the Andreev level and continuum spectrum with negative energies (see Fig.\ref{fig:dosABS}). Transitions of quasiparticles from the negative ABS to the positive ABS or to the continuum can occur by microwave irradiation. As we increase the temperature, the occupation of the negative energy ABS decreases, whilst the occupation of the positive energy one rises. A proof of the existence of the Andreev states has been provided in several experiments\cite{absexp2,absexp3}. For example, a very strong proof for the existence of these states in carbon nanotube is shown in fig.~\ref{fig:absexp}~\cite{absexp}. The measurement of the DoS shows that the height of the peaks of the ABS reduce as we increase the distance from the interface. The above description of the junction in terms of the ABS is valid in clean and small constriction.

%\subsubsection{The proximity effect}
 
%The above description of the junction in terms of ABS is valid in clean and small constriction. However, in realistic SNS junctions the situation is different. For example, the conductance of a SN junction in general cannot be described by the BTK model, as it cannot explain the zero bias anomaly~\cite{zba,zba1,zba2}. This anomaly consist of a peak in the conductance at zero bias voltage, that appears in SIN junctions in the low temperature limit. The interference between the electron and hole is what strongly enhances the subgap conductance. For this reason we now switch to the diffusive formalism. Using this formalism we show that the Andreev reflection is the key mechanism for the superconducting proximity effect. In a diffusive metal this effect leads to a loss of interferences beyond the energy dependent coherence length. That characterizes how far the two electrons from a Cooper pair leaking from the superconductor diffuse in phase into a normal metal. This means that the normal metal in contact with the superconductor exhibits superconducting properties near the interface \textit{i.e.} the proximity effect. The proximity effect and the Andreev reflection are just two faces of the same coin.

For planar diffusive $SN$ junctions the Andreev reflection leads to coherent electron-hole pairs in the normal metal over the so called normal coherent length $\xi_N=\sqrt{ \mathcal{D}/(2 \pi T)}$, where $\mathcal{D}$ is the diffusion constant. For low temperatures this may be much larger than $\xi_0$. This is the so-called proximity effect. For typical metal samples, at $T \sim 100 mK$, it is of the order of few microns and it is restricted by decoherence processes, namely inelastic or spin-flip scattering. The penetration of the condensate into the normal metal over large distances, allows current with no dissipation in SNS structures. The proximity effect leads to a change of the DOS of the normal metal, which in turn leads to an increase of the local conductivity. On the other hand, the proximity of the normal metal to the superconductor also has an effect on the properties of the latter. The superconductivity is suppressed over the correlation length $\xi_0$, meaning that $\Delta$ is reduced at the interface in comparison with the bulk value. This phenomena is called the inverse proximity effect.

%At low temperatures the characteristic length over which these decoherence processes occur may be quite long (few microns).
If the normal metal in a SN junction is substituted by a ferromagnet(F), the Andreev reflection and hence the proximity effect is suppressed. In a F metal electrons with different spins belong to different energy bands. The energy shift of the two bands can be considered as an effective exchange field acting on the spin of the electron and the reflected hole. The suppression of the Andreev reflection is due to the fact that all the electrons with spin-up do not find a "partner" with spin-down. This effect reduces the coherence length and the superconducting condensate decays fast in the ferromagnetic region for $SF$ interfaces. The estimation of the ratio of the condensate penetration length in ferromagnets to that in non-magnetic metals with a high impurity concentration, is of the order of $\sqrt{ T_C/h}$, where h is the exchange field. The highest possible value of the critical field is given by the Chandrasekhar-Clongston limit\cite{clogston1962,chandrasekhar1962} $h_c \sim 0.7 \Delta$. Using eq.\ref{eq:tcdelta} we can write $T_C$ as a function of $\Delta$. This leads to a value of the ratio of the order of $0.9$.  Making the penetration depth in ferromagnets much smaller than the one corresponding to normal metals, that we define as  $\xi_F \sim \sqrt{ D/(2 \pi h)}$.

In the next section we discuss transport phenomena in planar junctions involving superconductors.

\newpage

\section{Charge current through superconducting tunnel junctions}
\label{sec:tunnel}

We now proceed to describe \textit{transport phenomena} of simple junctions with superconducting elements. These correspond to tunnel junctions between a superconductor and a normal metal (SIN), where $I$ is an insulator, and between two superconductors (SIS), in which we impose a bias voltage. The most interesting phenomena is related to the nontrivial and energy dependent density of states of the superconductor. From quantum mechanics it is known, that there is a nonzero probability of charge transfer by tunnelling of electrons between two conductors separated by a thin insulating barrier. This probability falls exponentially with the width of the insulating barrier and depends on the properties of the insulating material.
\begin{figure}[h]
\begin{centering}
\includegraphics[width=0.6\columnwidth]{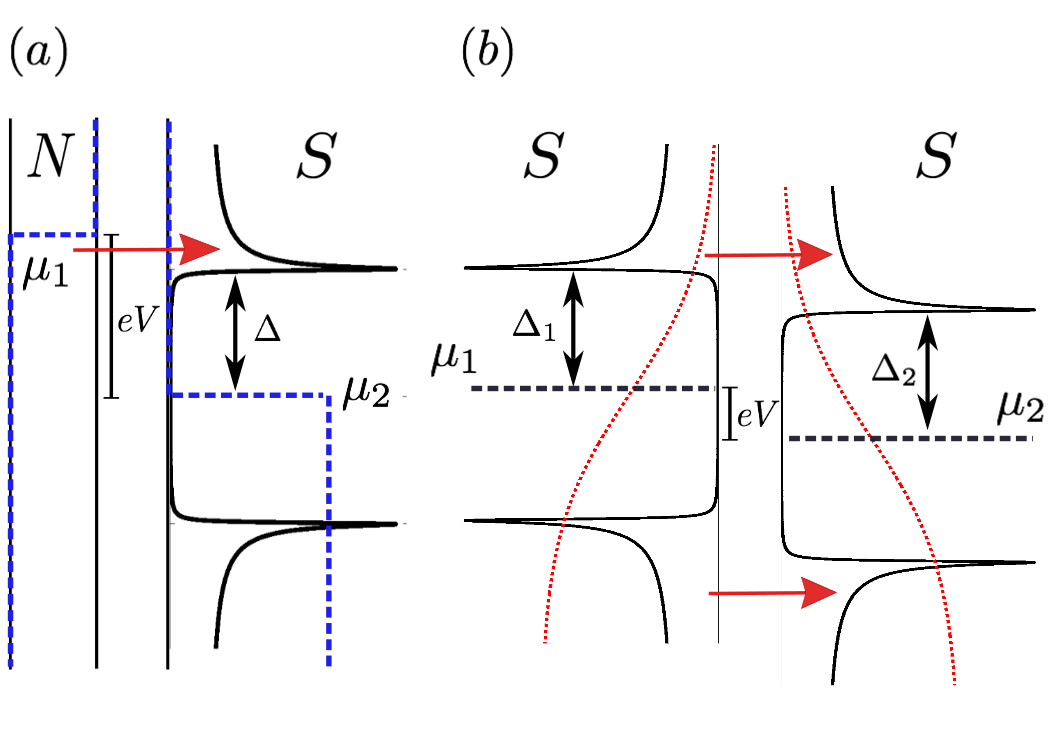}
\par\end{centering}

\caption{ Density of states (horizontal) versus energy (vertical) as an example of the semiconductor model to describe electron tunnelling. Here the dashed lines stand for the occupation of the states. We present two figures: (a) N-S tunnelling at T = 0, with bias
voltage that slightly exceeds the energy gap.
Horizontal arrow depicts electrons from the
left tunnelling into empty states on the right.
(b) S-S tunnelling at $T > 0$, with bias voltage
$eV < \Delta_1 + \Delta_2$. Horizontal arrows
depict tunnelling involving thermally excited
electrons or holes. }
\label{fig:semicon_mod}
\end{figure}
%

%\subsubsection{Semiconductor model}

In order to understand the physical picture of the transport at these tunnelling interfaces we use the so-called semiconductor model. In this method, illustrated in fig.~\ref{fig:semicon_mod}, the normal metal is represented in an elementary way as a continuous distribution of energy independent particle states. The superconductor is represented like an ordinary semiconductor with the BCS gap $\Delta$. It reduces to the normal-metal density of states as $\Delta \rightarrow 0$. The occupation of the states is described by the Fermi distribution function. For $T=0$, all states up to the chemical potential are filled, while for $T>0$ the occupation numbers are given by the Fermi function.

Within this model, tunnelling transitions are all elastic, \textit{i.e.} they occur at constant energy after adjusting the relative levels of the chemical potential in the two metals to account for the applied potential difference $eV$. This property facilitates the understanding of all the contributions  to the current.  As a simple example, in the next section we determine first the current through a $NIN$ junction.

%For further reading we recommend the book by Tinkham~\cite{tinkham}.

%Although our previous argument for simply adding the currents from the two degenerate channels is valid for the usual case, there can be an interference effect between them which causes an oscillatory variation of the tunnel current with voltage or sample thickness known as the Tomasch effect \cite{Tomasch}.

\subsection{NIN junction}

We consider here a NIN junction and calculate the current flowing from the left to the right normal metal when the junction is voltage-biased. The Fermi golden rule is applied in order to calculate the transition rate,
\begin{equation}
I_{L \rightarrow R}=\frac{1}{e R_T |T|^2 N^0_L N^0_R} \int_{-\infty}^{\infty}  T^2 N_L(E) f_L(E,\mu,T) N_R(E+eV)[1-f_R(E+eV,\mu,T)] dE \; .
\end{equation}
Here $eV$ is the applied voltage that results in a difference in the chemical potential across the junction. $N(E)$ is the normal density of states. For a finite temperature, one has to include the Fermi distribution functions of the electrons, $f_i(E,\mu,T)$, in each electrode. It reads $f_i(E,\mu,T)=\{ \exp[(E- \mu)/( T)] \}^{-1}$ and $\mu_R= \mu_L - e V$. The factors $N_L f_L$ and $N_R (1-f_R)$ give the numbers of occupied initial states and of empty final states in unit energy interval. If the initial states are empty or the final states fully occupied, the spectral current is zero. This expression assumes a constant tunnelling factor $T$. The prefactor of this expression is often written in terms of the resulting resistance $R_T$ and $N^0_{L,R}$, which is the normalization value of the corresponding density of states. Subtracting the reverse current (current from the left to right minus that from right to left), gives the total current
\begin{equation}\label{eq:semmod}
I=I_{L \rightarrow R}-I_{R \rightarrow L}=\frac{1}{e R_T |T|^2 N^0_L N^0_R} T^2 \int_{-\infty}^{\infty}  N_L(E) N_R(E+eV)[f(E)-f(E+eV)] dE \; .
\end{equation}
%
%The prefactor of this expression is often written in terms of the resulting resistance, $R_T=1/(e C \lfloor T \rfloor^2 N^0_L N^0_R)$. Here $N^0_{L,R}$ is the normalization value of the corresponding density of states. 

From this expression the current for a NIN system is obtained by replacing the densities of states by energy independent ones,
\begin{equation}
I_{NN}=\frac{1}{e R_T}\int_{-\infty}^{\infty} [f(E)-f(E+eV)] dE=\frac{e V}{e R_T} \equiv G_{T} V.
\label{eq:Inn}
\end{equation}
As expected it results in the ohmic behaviour. The conductance value reads $G_T=1/R_T$, independent of temperature and voltage.

%Thus, $E<0$ corresponds to those energies that are below the Fermi energy. 

%To help reduce any lingering confusion about the relation of this semiconductor, or independent-particles, scheme to the elementary excitation scheme, we will later show in the text how this simple case would have been treated in the other framework. 

\subsection{NIS junction}

The next step is to introduce a $NIS$ junction. The DoS of the superconductor is energy dependent, corresponding to a BCS DoS, while for the normal metal $N_N(E)=N^0_N$, which is energy independent. Eq.\ref{eq:semmod} thus converts to the following,
\begin{equation}
I_{NIS}=\frac{1}{e R_T N^0_S} \int_{-\infty}^{\infty} N_{S}(E) [f(E,T_R)-f(E+eV,T_L)] dE \; .
\label{eq:Ins}
\end{equation}
In general, this expression cannot be analytically integrated for an arbitrary voltage and temperature but numerical integration is straightforward. 

In the limit $T \rightarrow 0$ the expression reduces to,
\begin{equation}\label{eq:curnist}
I_{NIS}\lfloor_{T \rightarrow 0}=\frac{2}{e R_T N^0_S} \int_{0}^{eV} N_{S}(E) dE \; .
\end{equation}
There is no tunnelling current for $e  |V| \le \Delta$, as the voltage has to overcome the gap in order for tunnelling to be allowed. The magnitude of the current is independent of the sign of V because hole and excitations have equal energies. For $T>0$, the energy of excitations already present allows tunnelling at lower voltages, resulting in an exponential tail of the current in the region below $eV= \Delta$. 
\begin{figure}[h]
\begin{centering}
\includegraphics[width=0.6\columnwidth]{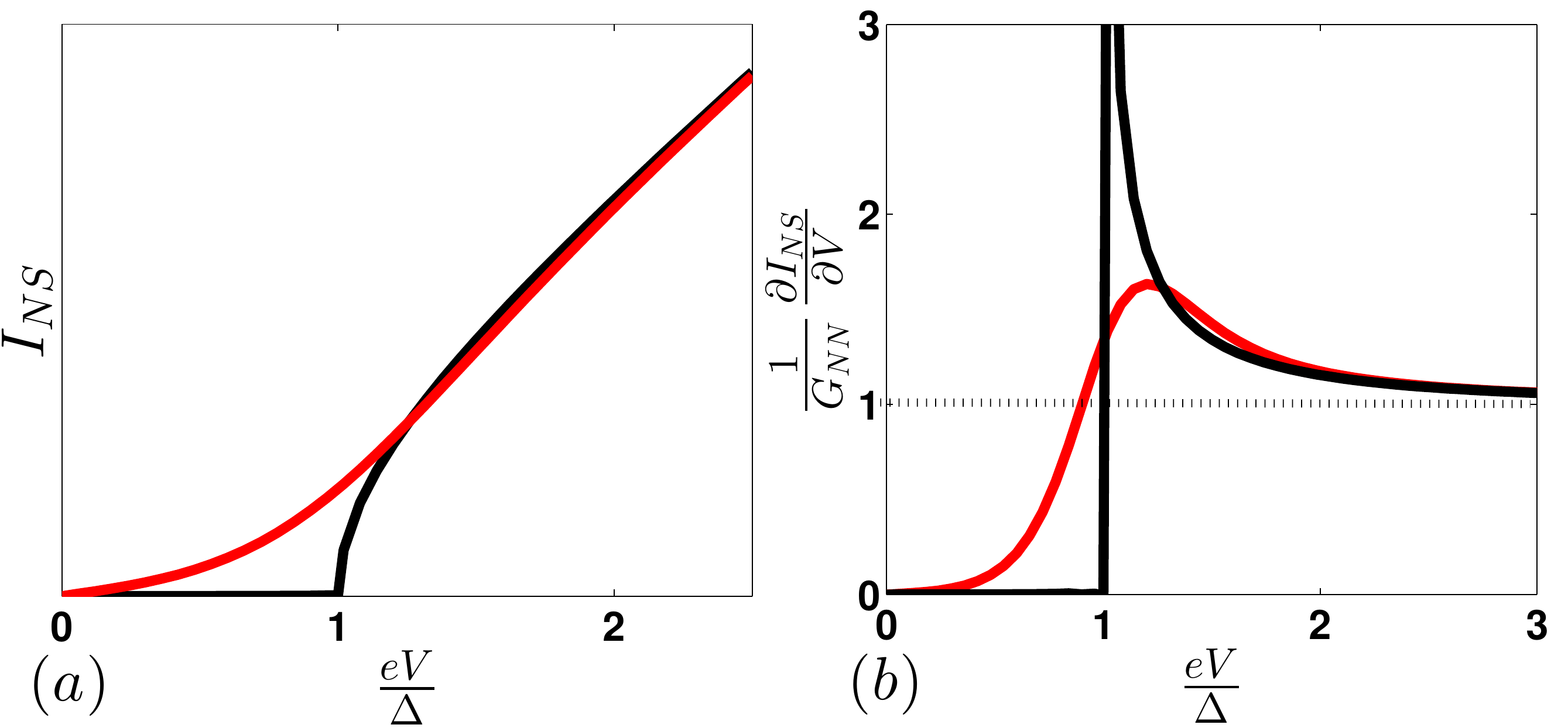}
\par\end{centering}

\caption{ Current and differential conductance versus voltage for a normal metal- superconductor tunnel junction. Black curves represent the $T=0$ case while red ones correspond to a finite temperature.}
\label{fig:cha}
\end{figure}

We calculate the differential conductance $G_{NIS}=dI_{NIS}/dV$ from eq.~\ref{eq:Ins},
\begin{equation}\label{eq:semmodS}
G_{NS}=\frac{dI_{NS}}{dV}= \frac{1}{R_T} \int_{-\infty}^{\infty} \frac{N_{2S}(E)}{N^0_S} [-\frac{\partial f(E+eV)}{\partial(eV)}] dE.
\end{equation}
It is important to notice that $-\partial f(E+eV)/\partial(eV)=[4 T \cosh^2((E+eV)/(2 T))]^{-1}$ is a bell-shaped weighting function peaked at $E=eV$, with width $\sim 4 T$. Therefore, as $T \rightarrow 0$ this function reads as a delta function, leading to 
\begin{equation}\label{eq:semmodG}
G_{NS}\lfloor_{T \rightarrow 0}=\frac{dI_{NS}}{dV}\lfloor_{T=0}= \frac{N_{S}(|eV|)}{N^0_S R_T}. 
\end{equation}
Thus, in the low-temperature limit, measurements of the differential conductance reveal information about the density of states. In fig.\ref{fig:cha}(b) the result corresponding to the conductance in this limiting case is plotted, resulting in the BCS DoS. Electron tunnelling was pioneered by Giaever~\cite{Giaever}, who used it to confirm the density of states and temperature dependence of the energy gap predicted in the BCS theory.

At finite temperatures, as shown in  fig.~\ref{fig:cha} (b), the conductance is smeared by $\sim \pm 2 T$ in energy, as a result of the width of the derivative of the distribution function. Due to the exponential tails, it turns out that the differential conductance at $V=0$ is related exponentially to the width of the gap. In the limit $ T \ll \Delta$, this relation reduces to
\begin{equation}
\frac{G_{NS}}{G_{NN}}\lfloor_{V=0}=  \left( \frac{2 \pi \Delta}{ T} \right)^{1/2} e^{-\Delta/ T} \; .
\end{equation}
The calculations of the tunnelling current do not take into account all the phenomena related to SIN interfaces, such as the Andreev reflection. As we shall see in the next chapter a large correction to the conductance is given by this subgap contribution. In the low temperature limit this subgap conductance tends to a finite value.

\subsection{SIS junction}

In this section we study a junction of two superconductors with different superconducting gaps ($\Delta_1$,$\Delta_2$) connected by an insulating barrier. In this case both DoS are energy dependent. Here we only consider the flow of quasiparticles and neglect the supercurrent that can flow through the junction (Josephson effect) In this case eq.~\ref{eq:semmod} reads,
\begin{equation}\label{eq:semmodS}
I_{SIS}= \frac{1}{e R_T} \int_{-\infty}^{\infty} \frac{N_{S}(E)}{N^0_{S(1)}} \frac{N_{S}(E+eV)}{N^0_{S(2)}} [f(E)-f(E+eV)] dE 
\nonumber
\end{equation}
\begin{equation}
= \frac{1}{e R_T N^0_{S(1)} N^0_{S(2)} } \int_{-\infty}^{\infty} \frac{|E|}{|E^2-\Delta_1^2|^{1/2}} \frac{|E+eV|}{|(E+eV)^2-\Delta_2^2|^{1/2}} [f(E)-f(E+eV)] dE \; .
\end{equation}
This integral excludes energy values that are inside the gaps of either superconductor, $|E|<|\Delta_1|$ and $|E+eV|<|\Delta_2|$. Numerical integration is required to compute complete I-V curves.

\begin{figure}[h]
\begin{centering}
\includegraphics[width=0.35\columnwidth]{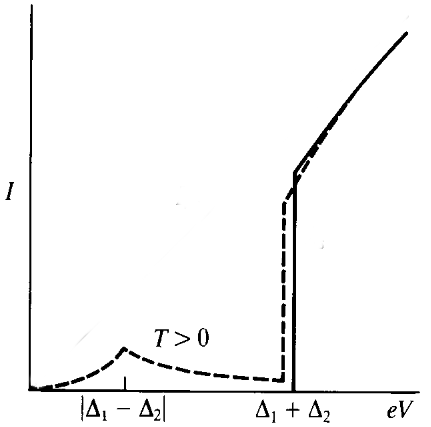}
\par\end{centering}

\caption{Characteristic I-V curve corresponding to the Superconductor-superconductor tunnelling case. Note that for finite temperatures $T>0$ there is a sharp peak at voltages 
corresponding to the difference of
the two gap values. (From ref.\cite{tinkham}) }
\label{fig:sscha}
\end{figure}

In fig.~\ref{fig:sscha} we show the qualitative features of the I-V characteristic. From the semiconductor model, depicted in fig.~\ref{fig:semicon_mod}(b), we can see that at low temperatures no current can flow until the voltage exceeds the value $\Delta_1 + \Delta_2$. At this point, the potential difference supplies enough energy for particles to tunnel. Since the density of states is infinite at the gap edges, it turns out that there is a discontinuous jump in $I_{SIS}$ at $eV=\Delta_1+\Delta_2$ even at finite temperatures.

 For $T>0$, current also flows at lower voltages because of the availability of thermally excited quasi-particles. This current rises sharply to a peak when $eV=|\Delta_1-\Delta_2|$ because this voltage provides just enough energy to allow thermally excited quasi-particles with energy $\Delta_1$, to tunnel into the peaked density of available states at $\Delta_2$. The existence of this peak leads to a negative resistance region $[(dI/dV)<0]$ for $|\Delta_1-\Delta_2 \le eV \le \Delta_1 +\Delta_2$. This region cannot be observed with the usual current-source arrangements since there are three possible values of V for a given I and the one where $dI/dV<0$ is unstable. 
 
 The existence of sharp features at both $|\Delta_1 - \Delta_2|$ and $\Delta_1+\Delta_2$ allows easy determination of $\Delta_1$ and $\Delta_2$ values from the tunnelling curves. The SIS tunnelling method is more accurate than the NIS tunnelling method in this regard, due to the existence of very sharply peaked densities of states at the gap edges of both materials, which helps to counteract the effects of thermal smearing\cite{giaeSS,ambaSS}.

\subsubsection{The Josephson effect}
\label{sec:josep}

Previously, we have focused on the quasiparticle contribution to the current. However, a finite current can flow thought the $SIS$ junction in the absence of a voltage drop. This is the so-called \textit{Josephson effect}. It results in a current without dissipation(supercurrent) generated by a gradient in the superconducting phase. In 1962, Josephson~\cite{josephson} made the remarkable prediction that a zero voltage, the supercurrent
\begin{equation}\label{eq:Is}
I_s=I_C \sin(\Delta \varphi)
\end{equation}
should flow between two superconducting electrodes separated by a thin insulating barrier. Here $\Delta \varphi$ is the phase difference between the wavefunctionsof the two electrodes. The critical current $I_C$ is the maximum supercurrent that the junction can support. This phenomena was confirmed experimentally shortly afterwards by Anderson and Rowell~\cite{AmRo}. 

\begin{figure}[h]
  \centering
  \includegraphics[width=0.5\columnwidth]{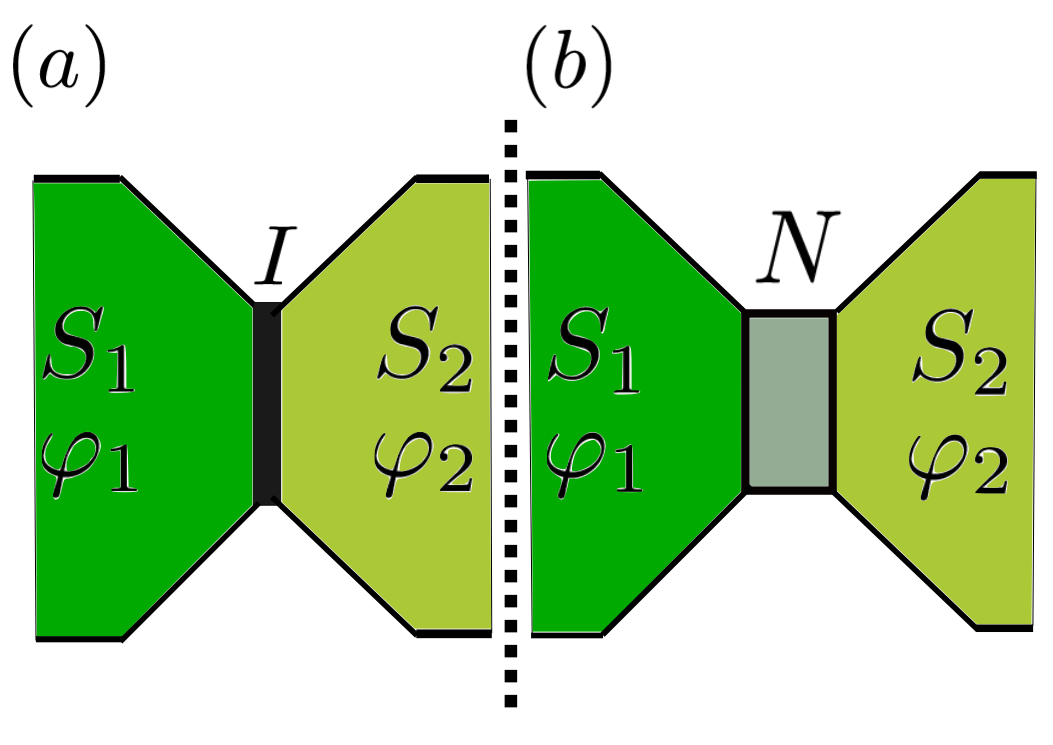}
  \caption{Two possible geometries for a Josephson junction: (a) two superconductors in contact by a tunnel barrier and (b) superconductors separated by a normal metal.}
\label{fig:josep}
\end{figure}

Josephson further predicted that if a voltage difference $V$ is maintained across the junction, the phase difference $\Delta \varphi$ would evolve according to 
\begin{equation}\label{eq:josepRe}
\frac{d(\Delta \varphi)}{dt}= \frac{2eV}{\hslash} \; ,
\end{equation}
so that the current would be an alternating current of amplitude $I_c$ and frequency $\nu=2eV/h$. Thus, the quantum energy $h \nu$ equals the energy change of a Cooper pair transferred across the junction. These predicted effects are known as the dc and ac Josephson effects.

Although Josephson$^\prime$s prediction was made for SIS junctions, it is now clear that the effects are much more general, and occur whenever two strongly superconducting electrodes are connected by a "weak link". This "weak links" can be an insulating layer as Josephson originally proposed, or a normal metal layer made weakly superconductive by the proximity effect. It can also be a short narrow constriction in otherwise continuous superconducting material. These three typical cases are often referred to as SIS (fig.~\ref{fig:josep} (a)), SNS (fig.~\ref{fig:josep} (b)) or ScS junctions (see also section\ref{sec:sqpc}), respectively. 

Given the two relations \Eq{eq:Is} and \Eq{eq:josepRe}, one can derive the coupling free energy stored in the junction by integrating the electrical work $\int I_S V dt=\int I_S (\hslash/2e) d(\Delta \varphi)$ done by a current source in changing the phase. In this way, we obtain,
\begin{equation}
F=const.- E_J \cos(\Delta \varphi) \quad where \quad E_J \equiv (\hslash I_C/2e) \; .
\end{equation} 
Clearly, the energy has a minimum when the two phases are equal, so that $\Delta \varphi=0$. This corresponds to the energy minimum in the absence of a phase gradients in a bulk superconductor. The critical current is a measure of how strongly the phases of the two superconducting electrodes are coupled through the weak link. This depends on size and material of the barrier. In the case of constriction weak links, it depends on the cross-sectional area and length of the neck. In most applications $I_C$ lies in the range of a microampere to a few miliamperes.

$I_C$ scales with dimensions of the bridge exactly as the inverse of its resistance $R_N$ in the normal state. Thus, $I_C R_N$ has an invariant value, which depends only on the material and the temperature, and not on bridge dimensions. The $I_C R_N$ product is frequently used as a measure of how closely real Josephson junctions approach the theoretical limit.

\begin{figure}[h]
  \centering
  \includegraphics[width=0.6\columnwidth]{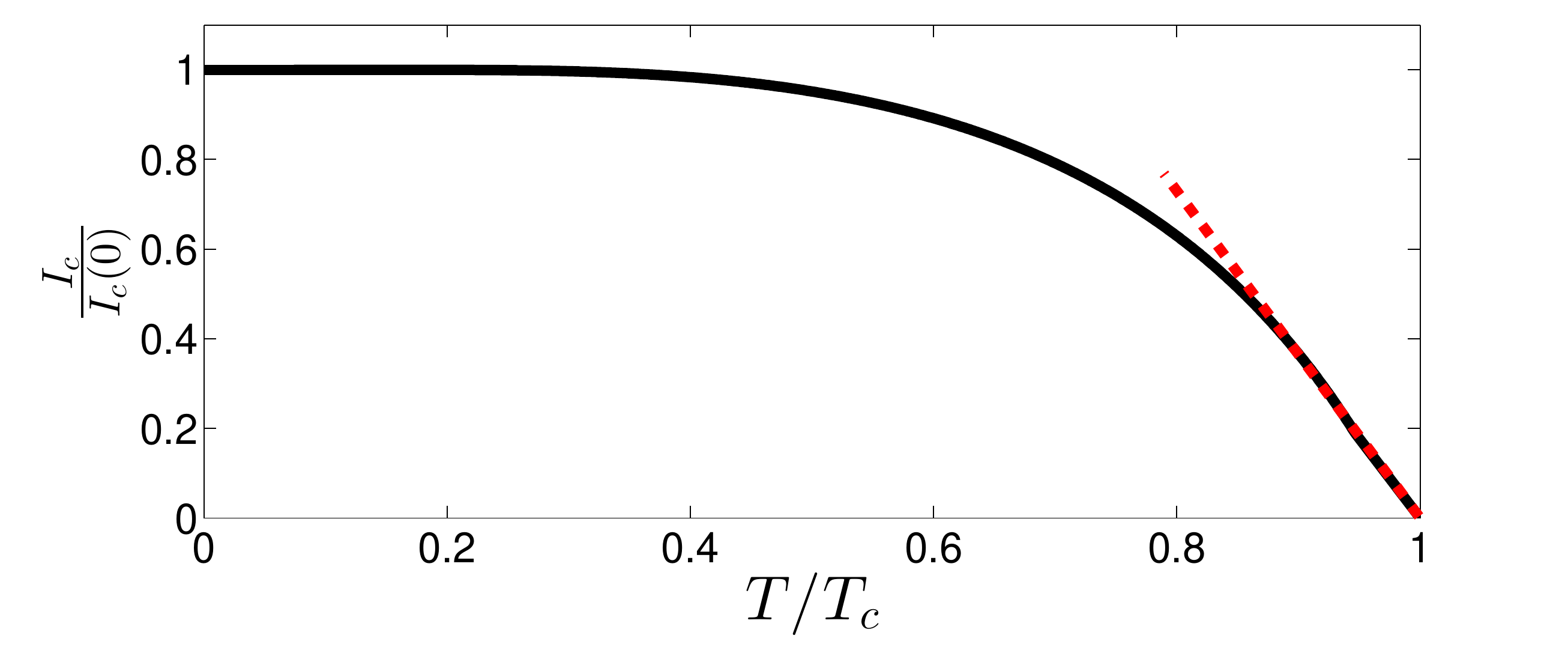}
  \caption{Ambegaokar-Baratoff dependence of the critical current with temperature. The linear dependence valid for temperatures near $T_C$ is added as a red dashed line.}
\label{fig:amba}
\end{figure}

\paragraph{Ambegaokar-Baratoff formula for tunnel junctions}
 Ambegaokar and Baratoff~\cite{ambaSS}, worked out an exact result for the full temperature dependence of $I_C$ in this system. They applied a microscopic theory to a tunnel junction geometry, as had Josephson in his original derivation. It reads,
\begin{equation}\label{eq:ambaSS}
I_C R_N=(\pi \Delta/2e) \tanh(\Delta/2 T) \; .
\end{equation}
This is an important general result. In the $T=0$ limit, $I_C R_N= \pi \Delta(0)/2e$. It is also convenient to note that by using the BCS $\Delta(T)$, we see that eq.~\ref{eq:ambaSS} varies linearly with $T$ near $T_C$ and can be approximated by
\begin{equation}
I_C R_N=(2.34 \pi /e)(T_C-T)\sim (T_C-T) \times 635 \mu V/K \; .
\end{equation}
Eq.~\ref{eq:ambaSS} holds for tunnel junctions. The temperature dependence of eq.~\ref{eq:ambaSS} is shown in fig.~\ref{fig:amba}, where we plot the linear dependence corresponding to temperatures near $T_C$.

\newpage

\section{Summary}

In summary, in this chapter the basic properties of conventional superconductors have been introduced. Furthermore, the properties that arise in junctions between superconducting and non-superconducting material have been studied. These include the Andreev reflection and the proximity effect, that are two faces of the same coin. Finally, the current through $NIN$, $NIS$ and $SIS$, planar tunnelling junctions has been described using a simple semiconductor approach of the superconductor.

\bibliographystyle{unsrtnat}
\renewcommand{\bibname}{Bibliography of Chapter 1} % changes default name Bibliography to References

%\end{document}

%% file: chap2_SB.tex
 As it has been discussed in the first chapter, the BCS theory  introduced in section\ref{sec:1.2} gives a microscopic description of superconductivity improving the phenomenological Ginzburg-Landau theory. In 1959, L.P. Gorkov\cite{GFGorkov} used the Green function (GF) method developed in the context of quantum field theory to show that the BCS theory reduced to the Ginzburg-Landau theory for temperatures close to the critical temperatures.

 The BCS theory formulated in terms of the Green function technique is a powerful tool for description of superconductivity, in particular for the study of  bulk superconductors. However, calculations involving microscopic Green functions are very cumbersome when dealing with inhomogeneous (hybrid) systems and in several cases the solutions are impossible to be found. A step forward in the method of the Green function technique was provided by  by G. Eilenberger~\cite{eilenberger} and in parallel by A. I. Larkin and Yu. N. Ovchinnikov \cite{laov}, who introduced the so-called quasiclassical method in the theory of superconductivity.

 This method is based on the fact that the energies and lengths  involved in the superconducting phenomena are usually much smaller that the Fermi energy and larger than the Fermi wave length respectively.  The quasiclassical approximation reduces the Gorkov equations for  the Green functions to  a kinetic-like equations. By this means, the two-coordinates dependent  GFs $G(r,r^\prime)$ transform to the quasiclassical GFs with dependence (\textbf{r},\textbf{p}$_F$), where \textbf{p}$_F$ is the momentum direction at the Fermi level. In the so called dirty limit, \textit{i.e.} when the mean free path of the electrons is much smaller than other characteristic lengths, the Eilenberger equation can be simplified and transformed to a diffusion-like equation, as shown in 1970 by K. Usadel~\cite{Usadel}. The resulting equation is an extremely useful tool that is used in several chapter of this thesis.

The purpose of this chapter is to show the most important steps in the derivation of the quasiclassical equations, whereas, most of the technical details are given in the Appendix. This complete formalism allows to study the electronic transport in a wide range of hybrid structures and phenomena related to the proximity effect.

\section{The quasiclassical Green function technique}
\label{sec:quasi}

Our starting point is a general Hamiltonian describing hybrid systems: 
\begin{equation}
\hat{H}=\hat{H}_{BCS}+\hat{H}_{imp}+\hat{H}_{so}+\hat{H}_{sf}+\hat{H}_{ex} \; .
\label{eq:hamil}
\end{equation}
%
%+\hat{H}_{pot}

The $\hat{H}_{BCS}$ term is the BCS Hamiltonian for the description of superconducting materials in terms of the order parameter $\Delta$  and reads, 
\begin{equation}
\hat{H}_{BCS}= \Sigma_{\{ p,s \}} a_{sp}^\dagger \left( \xi_p \delta_{pp^\prime} + eV  \right) \delta_{ss^\prime} a_{s^\prime p^\prime}- \Delta \left( a^\dagger_{\bar{sp}} a^\dagger_{s^\prime p^\prime} + c.c. \right) \; .
\end{equation} 
For $\Delta=0$,  $\hat{H}_{BCS}$ describes the  normal state. The summation is carried out over all momenta $(p,p^\prime)$ and spins $(s,s^\prime)$ (the notation $\bar{s}$, $\bar{p}$ means inversion of both spin and momentum), $\xi_p=p^2/2m-E_F$ is the kinetic energy counted from the Fermi energy $E_F$, $V$ is a smoothly varying electric potential. The superconducting order parameter $\Delta$ must be determined self-consistently. 
  
 In Eq.\ref{eq:hamil}, the term $\hat{H}_{imp}$ describes the interaction of the electrons with nonmagnetic impurities,  $\hat{H}_{so}$ the spin-orbit coupling due to impurities (extrinsic coupling), and $\hat{H}_{sf}$ the spin flip interaction at magnetic impurities. In-plane magnetic fields or the presence of ferromagnets (F) in the hybrid system are described by the Zeeman-like term $\hat{H}_{ex}$. Along the thesis we use the symbols $\breve .$  for $8\times 8$, $\check .$  for $4\times 4$ and $\hat .$  for $2\times 2$ matrices. The explicit expressions for all these terms can be found in the Appendix, section \ref{sec:quasiap}.
 
 %The term $H_{pot}$ describes eventual potentials at the interfaces between different materials in the hybrid structure.

 We introduce the time ordered matrix Green functions (in the particle-hole$\times$spin space) in the Keldysh representation,
\begin{equation}
\check{\textbf{G}}(t_i,t_k^\prime)=\frac{1}{i} \langle T_C \left( c_{ns}(t_i) c_{n^\prime s^\prime}^\dagger(t^\prime_k)\right)\rangle \; .
\label{eq:green}
\end{equation}
In the Keldysh representation the time coordinates have subindices ($k,j$) that take the values $+$ and $-$. These correspond to the upper and lower branches of the contour C, running from $-\infty$ to $\infty$ and back to $-\infty$. The new $c_{ns}^\dagger$ and $c_{ns}$ operators are related to the creation and annihilation operators $a_s^\dagger$ and $a_s$ by the relation
\begin{equation}
(n=1) \quad c_{1s}=a_s \quad (n=2) \quad c_{2s}=a^\dagger_{\bar{s}} \; .
\end{equation}
These operators (for $s=1$) were introduced by Nambu\cite{nambuop}. The new operators allow one to express the anomalous averages $\langle a_\uparrow a_\uparrow \rangle$ introduced by Gorkov as the conventional averages $\langle c_1 c_2^\dagger \rangle$ and therefore the theory of superconductivity can be constructed by analogy with a theory of normal systems.

%
%Here $\Psi$ are pseudo-spinors,
%%
%\begin{equation}
%\check{\Psi}(r,t_k)=  \begin{pmatrix} \Psi_\uparrow  \\ \Psi^\dagger_\downarrow \end{pmatrix} (r,t_k) \quad \check{\Psi}^\dagger(r,t_j)=\left( \Psi^\dagger_\uparrow, \Psi_\downarrow \right)(r,t_j) \; .
%\end{equation}
%%
%The coordinates $r$ and $t$ correspond to space and time. In the Keldysh representation the time coordinates have subindices ($k,j$) that take the values $+$ and $-$. These correspond to the upper and lower branches of the contour C, running from $-\infty$ to $\infty$ and back to $-\infty$.
%} HASTA AQUI : COPIAR DERIVACION DE NUESTRO ARTICULO

In the Keldysh space $G$ is a $2\times 2$ matrix. The four elements of this matrix are not independent and can be reduced to three. Thus, in the Keldysh space the matrix  Green functions read,
\begin{equation}\label{kelmat}
\breve{\textbf{G}} = \begin{pmatrix} \check{\textbf{G}}^R & \check{\textbf{G}}^K \\
0 & \check{\textbf{G}}^A
\end{pmatrix} \; .
\end{equation}
The upper scripts stand for Retarded, Advanced and Keldysh components. The Retarded and Advanced components contain information about the spectral properties of the system, such as the density of states (DoS). Whereas the Keldysh component describes how the states are occupied, \textit{i.e.} it contains the distribution function.

The above introduced Green functions satisfy the so-called \textit{Gorkov equation}\cite{gorkoveq}, which in coordinate space reads:
\begin{equation}
(\check{\textbf{G}}_0^{-1}-\tilde{\hat{\Delta}}\otimes \sigma_3+\textbf{h}\tau_3\hat{\textbf{S}}-\check{\Sigma}_{\textit{imp}}-\check{\Sigma}_{\textit{so}}-\check{\Sigma}_{\textit{sf}}) (r_1,t_1,r_2,t_2) \otimes \check{\textbf{G}}(r_2,t_2,r_1^\prime,t_1^\prime)=\delta(r_1,t_1,r_1^\prime,t_1^\prime)
 \label{eq:gorkov}
\end{equation}
Here $\otimes$ represents convolution over coordinates, while $\check{\Sigma}_{\textit{imp}}$, $\check{\Sigma}_{\textit{s.o.}}$ and $\check{\Sigma}_{\textit{s.f.}}$ are the self-energies. In this thesis the Pauli matrices $\tau_i$ and $\sigma_i$, correspond to Nambu and spin space respectively. In principle, this equation is valid for any self-energy but in the present case these are given in the Born approximation, as shown in the Appendix. The matrix order parameter equals $\tilde{\hat{\Delta}}=\tau_1 \text{Re} \Delta-\tau_2 \text{Im} \Delta$.  The spin vector takes the form $\hat{\textbf{S}}=(\sigma_1, \sigma_2, \sigma_3)$ and the exchange field vector reads $\textbf{ h}=(h_x,h_y,h_z)$. Here $h_i$ is the value of the exchange field in the "i" direction. The Green function of a non-superconducting bulk material reads,
\begin{equation}
\check{\textbf{G}}_0^{-1}(r_1,t_1,r_1^\prime,t_1^\prime)=\delta(r_1-r_1^\prime) \delta(t_1-t_1^\prime)  \left[ i  \partial t_1 + \frac{1}{2m} (\nabla_{r1}-i \textbf{A})^2 \tau_3 + E_F \right] \; .
\end{equation}
\begin{figure}[h]
  \centering
  \includegraphics[width=\columnwidth]{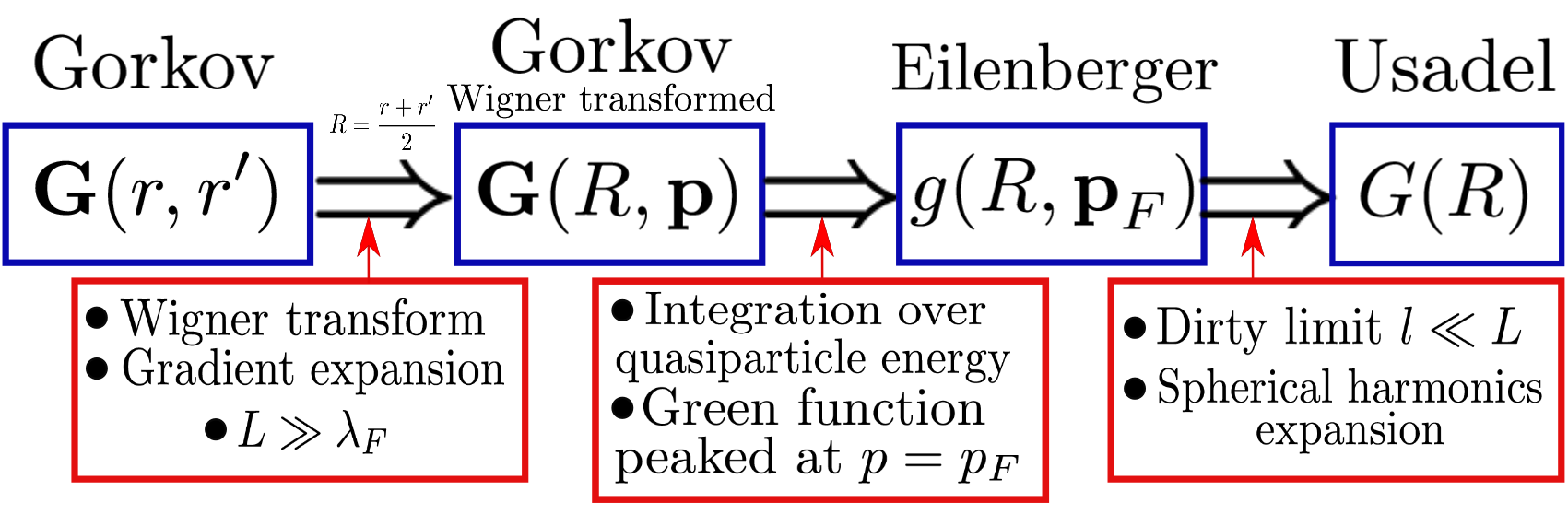}
  \caption{Main steps for the derivation of the quasiclassical Green functions. The time dependence of the GFs $(t,t^\prime)$ is omitted.}
\label{fig:gfs}
\end{figure}

For example, the Retarded and Advanced GFs are given from Eq.\ref{eq:gorkov} of a homogeneous superconductor after Fourier transform reads:
\begin{equation}
\check{\textbf{G}}^{R(A)}(\omega,\textbf{p})= \left[ (\omega \pm i \eta) \tau_3-\tilde{\hat{\Delta}}\otimes \sigma_3- E_F \right]^{-1} \; .
\label{eq:gorkovhomo}
\end{equation}
$\eta$  here  is the Dynes parameter\cite{eta1,eta2,eta} that  describes the inelastic scattering energy rate within the relaxation time approximation. In principle, the Gorkov equation can be used to describe hybrids structures consisting of different materials and interfaces. However,  dealing with full double coordinate Green functions  becomes very cumbersome and in several cases the solutions are impossible to be found. For this reason it is convenient to introduce the quasiclassical GFs that satisfy more simple kinetic equations. Its derivation can be found in several papers\cite{quasi1,quasi2,quasi3,quasi4} and in the appendix.  Fig.\ref{fig:gfs} shows schematically the main steps in obtaining from the Gorkov GFs $\textbf{G}(t,t^\prime;r,r^\prime)$, the quasiclassical GFs (Eilenberger)  $g(t,t^\prime;R,\textbf{p}_F)$ and Usadel GFs. 

In the most general case (arbitrary impurity concentration ($\tau$), spin orbit ($\tau_{so}$) and spin flip ($\tau_{sf}$)  impurity spin relaxation times, intrinsic spin splitting field ($h$) and superconducting gap ($\Delta$), the equation describing the quasiclassical GF $\check{g}(t,t^\prime;R,\textbf{p}_F)$ is the so-called \textit{ Eilenberger equation}~\cite{eilenberger} (we omit the coefficient for simplicity):
\begin{equation}
-(\tau_3 \partial_t \check{g} + \partial_{t\prime} \check{g} \tau_3 + \textbf{v}_F \hat{\nabla} \otimes \check{g})=[-i \check{\Delta}-i\textbf{h} \hat{\textbf{S}} + \frac{1}{2\tau} \langle \check{g} \rangle_{} \textbf{v}_F+ \frac{1}{2\tau_{so}}  \langle \check{g} \rangle_{} \textbf{v}_F+ \frac{1}{2\tau_{sf}} \tau_3 \langle \check{g} \rangle_{} \tau_3 v_F, \check{g}]_\otimes \; .
\label{eq:eilenberger}
\end{equation}
Here $\check{\Delta}=\tau_3 \tilde{\hat{\Delta}}$. The term $\hat{\nabla} \otimes C=\nabla_R C-i[\textbf{A} \tau_3,C]_\otimes$ is the gauge-invariant gradient involving the vector potential $\textbf{A}(t,t\prime)=\delta(t-t\prime) \textbf{A}(t)$. Here ${\textbf{v}_f}$ is the direction of the momentum at the Fermi surface and $|v_f|=k_f/m$ the Fermi velocity. This equation has to be complemented by the \textit{normalization condition} $\check{g}(t,t^\prime;R,\textbf{p}_F) \otimes \check{g}(t,t^\prime;R,\textbf{p}_F)=\check{1} \delta(t-t^\prime)$\cite{norm1,norm2}.

A further simplification can be done if the system is in the dirty limit, \textit{i.e.} when the mean free path of the electrons is smaller than other characteristic lengths. This transforms the Eilenberger equation to a diffusion equation for the momentum-averaged (\textit{i.e.} s-wave) Green function $\check{G}(t,t^\prime;R)=\langle \check{g}(t,t^\prime;R,\textbf{p}_F) \rangle_{\check{\textbf{p}}}$. These obey the \textit{ Usadel equation}~\cite{Usadel} (see Appendix):
\begin{equation}
i \mathcal{D} \hat{\nabla} \otimes \left( \check{G} \otimes \hat{\nabla} \check{G}\right)-i (\tau_3 \partial_t \check{G}+\partial_{t\prime} \check{G} \tau_3)=[\check{\Delta}+\textbf{h} \hat{\textbf{S}}+ \frac{i}{2\tau_{so}} \check{G} v_F+ \frac{i}{2\tau_{sf}} \tau_3  \check{G}  \tau_3 v_F, \check{G}]_\otimes \; .
\end{equation}
Here $\mathcal{D}=v_f l^2/3$ is the 3D diffusion constant corresponding to the elastic scattering length $l$. Retarded, Advanced and Keldysh components are not independent from each other. The first two fulfil the relation
\begin{equation}
\check{G}^A=-\tau_3 \check{G}^{R \dagger} \tau_3 \; .
\label{eq:ra}
\end{equation}
In addition, the Keldysh component relates to the Retarded and Advanced ones via the distribution function $\check{f}$ as
\begin{equation}
\check{G}^K = \check{G}^R \otimes \check{f}- \check{f} \otimes \check{G}^A \; .
\label{eq:keldysh}
\end{equation}
The normalization condition $\check{G}(t,t^\prime;R) \otimes \check{G}(t,t^\prime;R)= \check{1} \delta(t-t^\prime)$ in terms of the components implies $\check{G}^R \otimes \check{G}^R=\check{G}^A \otimes \check{G}^A=\check{1} \delta(t-t^\prime)$ and  $\check{G}^R \otimes \check{G}^K + \check{G}^K \otimes \check{G}^A=0$. 

The matrix distribution function that appears in eq.~\ref{eq:keldysh} has the general form,
\begin{equation}\label{eq:distrfun}
\check{f}= \tau_0 (f_L+f_{Ti} \sigma_i) + \tau_3 (f_T +f_{Li} \sigma_i),
\end{equation}
where $i=1,2,3$. The functions with an "L" subscript are odd in energy, while those with "T" subscripts are even. For example, $f_T$ is related to the charge imbalance and $f_{Ti}$ to the spin imbalance in the i direction. On the other hand, $f_L$ is related to the heat imbalance and $f_{Li}$ to the ”heat spin” imbalance (in the i direction). Here $i=1,2,3$ correspond to $x,y,z$ directions respectively.

In a stationary case, the GFs depend only on the time difference and therefore it is convenient to introduce the Fourier transformed  GF $\check{g}(E;R,p_F)$. The corresponding Eilenberger equation reads 
\begin{equation}
-\textbf{v}_F \hat{\nabla} \check{g}=[i E \tau_3 \sigma_0-i \check{\Delta}-i\textbf{h} \hat{\textbf{S}} + \frac{1}{2\tau} \langle \check{g} \rangle_{}v_F+ \frac{1}{2\tau_{so}} \langle \check{g} \rangle_{} v_F+ \frac{1}{2\tau_{sf}} \tau_3 \langle \check{g} \rangle_{} \tau_3 v_F, \check{g}] \; ,
\label{eq:eilenberger}
\end{equation}
while the  \textit{ Usadel equation} now reads, 
\begin{equation}\label{eq:FullUsadel}
\mathcal{D} \nabla (\breve{G} \nabla \breve{G}) +  i E \left[ \tau_3 \sigma_0, \breve{G} \right] + i \left[ \check{\Delta}, \breve{G} \right]+  \left[ (\textbf{ h} \hat{\textbf{S}}) \tau_3, \breve{G} \right] -\frac{i}{\tau_{sf}}\left[ \hat{\textbf{S}} \tau_3 \breve{G} \tau_3 \hat{\textbf{S}}, \breve{G} \right]-\frac{i}{\tau_{so}}\left[ \hat{\textbf{S}} \breve{G} \hat{\textbf{S}}, \breve{G} \right]=0 \; .
\end{equation}
Here $x$ is the transport direction. All equations, including the Usadel equation,  must be complemented by the corresponding boundary conditions, that we introduce later in section\ref{sec:kupri}. We define the matrix current, $\breve{J}$, together with the normalization condition that the solutions for the Usadel equation must obey.
\begin{equation}\label{Usadel3Din}
\check{J} = \check{G}\partial_x \check{G}, \quad \check{G}(\textbf{p}_F,\textbf{r};E)\check{G}(\textbf{p}_F,\textbf{r};E)= \check{1} \; .
\end{equation}
The superconducting gap $\Delta$ is determined by the so-called self-consistency equation (that we introduce in section \ref{sec:self}), which reflects the dependence of the order parameter on temperature, exchange field or proximity effect. Combining this equation with Usadel (or Eilenberger) and Maxwell equations, the set of equations is closed. 

%\paragraph{Kinetic Equation}
When studying out-of-equilibrium properties of a diffusive system, the Usadel equation (eq.\ref{eq:FullUsadel}) has to be solved. In such a case one  first  solves the equations for $G^R$ and $G^A$ (spectral part). The distribution function obey an equation which can be derived from the Keldyh component of the Usadel equation (eq.\ref{eq:FullUsadel}) and reads:
\begin{equation}
\mathcal{D}[ \hat{G}^R \nabla  \hat{G}^R \nabla \check{f} -  \nabla \hat{G}^R  \nabla \check{f} \hat{G}^A - \hat{G}^R \nabla \check{f}  \nabla  \hat{G}^A -   \nabla \check{f} \hat{G}^A \nabla  \hat{G}^A] +\mathcal{D}[\nabla^2 \check{f} - \hat{G}^R \nabla^2 \check{f} \hat{G}^A ]
 \nonumber
 \end{equation}
 \begin{equation}
 + \hat{G}^R(\check{\Lambda}^R \check{f} - \check{\Lambda}^K - \check{f} \check{\Lambda}^A)-(\check{\Lambda}^R \check{f} - \check{\Lambda}^K - \check{f}  \check{\Lambda}^A)\hat{G}^A=0 \; ,
 \label{eq:kinetic}
 \end{equation}
where
\begin{equation}
\check{\Lambda}^{R(A)}= i E \tau_3 - i \textbf{ h} \hat{\textbf{S}} -\check{\Delta}-\frac{1}{\tau_{sf}} \hat{\textbf{S}} \tau_3 \hat{G}^{R(A)} \tau_3 \hat{\textbf{S}}-\frac{1}{\tau_{so}} \hat{\textbf{S}}  \hat{G}^{R(A)} \hat{\textbf{S}}
\end{equation}
and
\begin{equation}
\check{\Lambda}^{K}= -\frac{1}{\tau_{sf}} \hat{\textbf{S}} \tau_3 \hat{G}^{K} \tau_3 \hat{\textbf{S}} -\frac{1}{\tau_{so}} \hat{\textbf{S}} \hat{G}^{K} \hat{\textbf{S}}.
\end{equation}

 \subsection{Quasiclassical Green Functions for bulk electrodes}

A typical hybrid structure that consider in this thesis  consist of a mesoscopic region connected to electrodes. We restrict the analysis to realistic  diffusive systems  and  solve Eqs.\ref{eq:FullUsadel}-\ref{Usadel3Din}  in the mesoscopic region.   The interfaces with the  electrodes will be described by suitable boundary conditions. In this section we introduce the GFs for the electrodes, which are bulk GFs and discuss their analytical properties. 
 
\subsubsection{Spectral functions}
\label{sec:spectral}

%\subsection{Green Functions}

The Retarded and Advanced components of the Green function read,
\begin{equation}
\check{G}^{R(A)}_S(E)= \sigma_0 (\tau_3 g_S^{R(A)}(E) + \tau_1  f^{R(A)}_{S0}(E)).
\label{eq:bulkgfS}
\end{equation}
Here we make a distinction between normal Greens functions $g^{R(A)}$ related to quasiparticles and \textit{anomalous} ones $f^{R(A)}_0$ that are related to superconducting properties or pair correlation. For a conventional superconducting electrode, $f^{R(A)}_0$ describes singlet superconducting correlations. Unconventional components of the condensate are introduced in section \ref{sec:proximityeffect}. They read,
\begin{equation}
(g^{R(A)}_S, f^{R(A)}_{S}) = (\pm E \pm i \eta, \; \pm \Delta)/\epsilon, \quad \epsilon =
\sqrt{(E \pm i\eta)^2 - \Delta^2}. 
\label{eq:gsfs}
\end{equation}
Here $\eta$ or the Dynes parameter is introduced in eq.\ref{eq:gorkovhomo}. It corresponds to a small parameter in the calculations ($\Delta \gg \eta \rightarrow 0$). Note that the sign of $\eta$ is different for the Retarded and Advanced component. 

\begin{figure}[h]
  \centering
  \includegraphics[width=0.4\columnwidth]{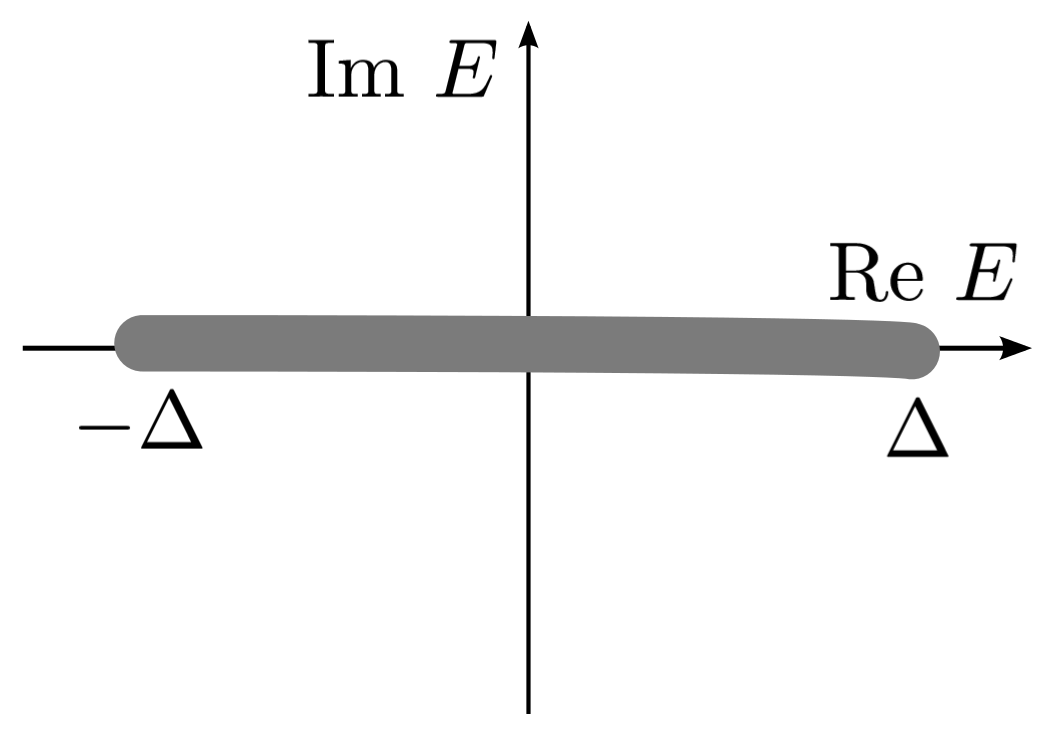}
  \caption{Plane of complex energy with the branch cut defining  the Retarded and Advanced components.}
\label{fig:compla}
\end{figure}

We can define $\sqrt{(E\pm i \eta)^2 - \Delta^2}$ as an analytical function on the plane of complex energy with the cut connecting the points $\pm \Delta \mp i \eta$, as depicted in fig.~\ref{fig:compla}. Applying this definition we can study the values of the Greens functions for each energy range. For $|E|>\Delta$ they read,
\begin{equation}
g^R_S=\frac{E+i\eta}{\sqrt{(E+i\eta)^2 - \Delta^2}} \quad g^A_S=-\frac{E-i\eta}{\sqrt{(E-i\eta)^2 - \Delta^2}}  \; .
\label{eq:gs1}
\end{equation}
In the region $|E|<\Delta$, for the Retarded component we take the value $\sqrt{(E+i\eta)^2 - \Delta^2}$ of the upper edge of the cut, that corresponds to $i\sqrt{\Delta^2-(E+i\eta)^2}$. Instead, for the Advanced component we take the lower edge of the cut, $-i\sqrt{\Delta^2-(E-i\eta)^2}$. Therefore, for this range of energies we obtain,
\begin{equation}
g^R_S=\frac{-i(E+i\eta)}{\sqrt{\Delta^2-(E+i\eta)^2}} \quad g^A_S=\frac{-i(E-i\eta)}{\sqrt{\Delta^2-(E-i\eta)^2}} \; .
\label{eq:gs2}
\end{equation}
In the limit $\Delta \rightarrow 0$, that correspond to the normal metal case, this definition result in
\begin{equation}
g^R_S=-g^A_S=1 \; .
\end{equation}

Similarly, for the anomalous component of the Greens functions we obtain for the range $|E|>\Delta$,
\begin{equation}
f^R_S=\frac{\Delta}{\sqrt{(E+i\eta)^2 - \Delta^2}} \quad f^A_S=-\frac{\Delta}{\sqrt{(E-i\eta)^2 - \Delta^2}} \; .
\end{equation}
On the other hand for $|E|<\Delta$,
\begin{equation}
f^R_S=\frac{-i\Delta}{\sqrt{\Delta^2-(E+i\eta)^2}} \quad f^A_S=\frac{-i\Delta}{\sqrt{\Delta^2-(E-i\eta)^2}}  \; .
\end{equation}
 In this case, the limit $\Delta \rightarrow 0$ corresponds to, as we could expect,
\begin{equation}
f^R_S=f^A_S=0 \; .
\end{equation}
Finite values of the anomalous Greens functions are related to superconducting properties. In normal metals, this properties are absent, making this anomalous components equal to zero. 
\begin{figure}[h]
  \centering
  \includegraphics[width=0.7\columnwidth]{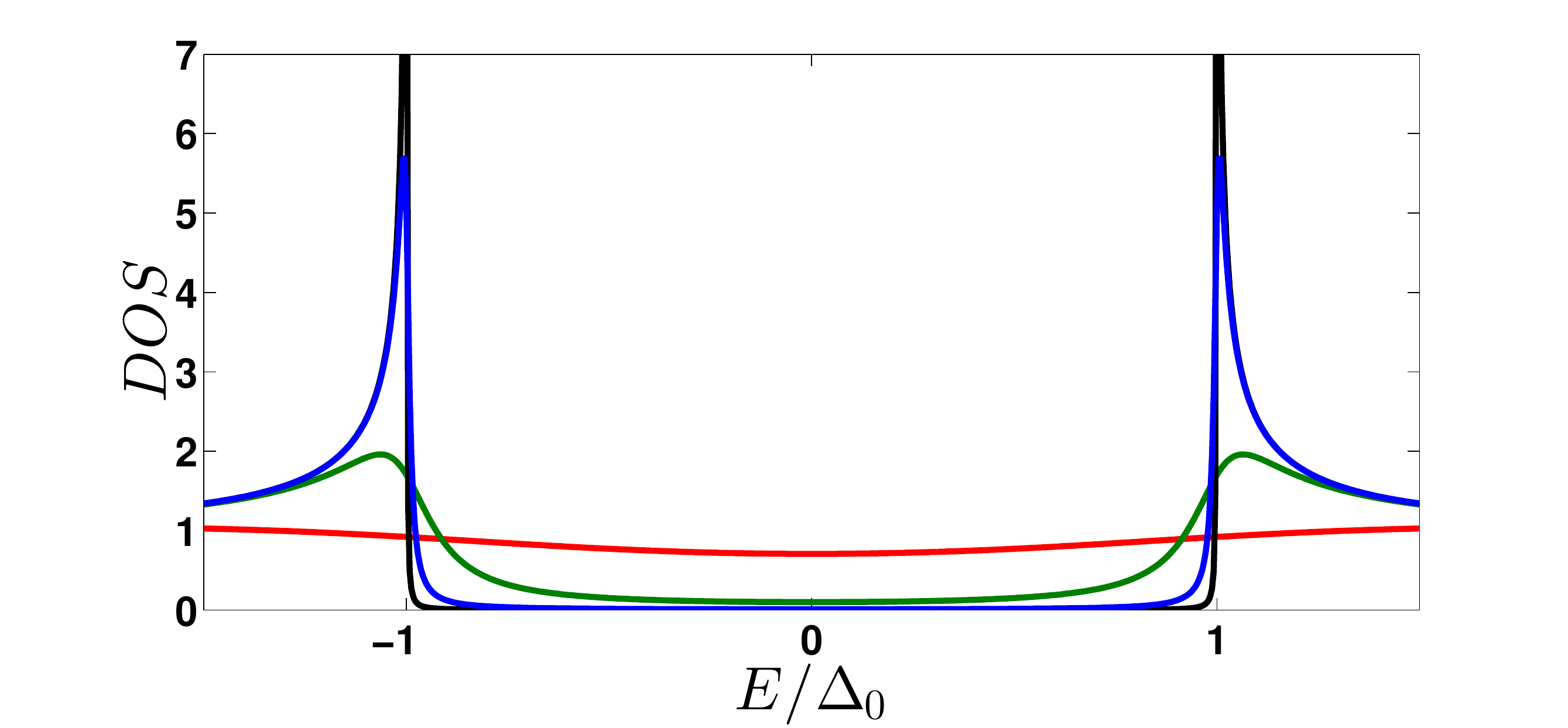}
  \caption{Dependence of the BCS DoS on the inelastic scattering parameter. For values of $\eta$: $10^{-3}$ (black), $10^{-2}$ (blue), $10^{-1}$ (green) and 1 (red).}
\label{fig:gapeta}
\end{figure}

The density of states (DoS) $N_S$ is defined in the following way,
\begin{equation}
N_S=(g_S^R-g_S^A)/2 \; .
\end{equation}
In fig.~\ref{fig:gapeta} we plot this quantity using the analytical expressions, eqs.\ref{eq:gs1} and \ref{eq:gs2}.  For very low $\eta$ values, the peak at $E=\pm \Delta$ diverges, increasing it we cut off this divergence and obtain a finite DoS inside the gap. For high values of $\eta$, the density of states takes the shape of that corresponding to a normal metal, almost energy independent and without a gap. Previous energy range dependent definitions, eq.\ref{eq:gs1} and \ref{eq:gs2} show that for the $\eta=0$ case the DoS is only finite for $|E|>\Delta$. 

The normalization condition for the bulk BCS superconductor Greens functions read,
\begin{equation}
(g_S^{R(A)})^2-(f_S^{R(A)})^2=1 \; .
\end{equation}

The Greens functions that correspond to a bulk normal metal can be derived from the expressions eqs.\ref{eq:gsfs} by taking, $\Delta\rightarrow 0$. The Retarded, Advanced and Keldysh components read,
\begin{equation}
\check{G}^{R(A)}_N= \pm \tau_3 \sigma_0, \quad \check{G}^K_N= 2 \tau_3 \sigma_0 \check{f}\; .
\label{eq:bulkgfN}
\end{equation}
The DoS of the normal metal is energy independent unlike that of a BCS superconductor. The distribution functions of the Keldysh component is that of the bulk electrodes shown below.

For a ferromagnet the quasiclassical GFs are the same as the ones for a normal metal reservoir.  This is a limitation of  this formalism, since the density of states for up- and down electrons is the same. In other words, polarization of the electrons at the Fermi level in a normal metal cannot be described.  In order to overcome this limitation we will introduce new boundary conditions  in section\ref{sec:spinfilters}. The idea is to model ferromagnets as normal metals with spin polarized interfaces at both ends.

The bulk Greens functions correspond to reservoirs or electrodes, big clusters of materials (normal, ferromagnetic or superconducting), that due to their size when placed in contact with other materials are not affected by the proximity effect. On the other hand, for small size samples the GFs may be corrected due to the proximity of other materials. Such corrections are described by the Usadel equation. 

\subsubsection{The Keldysh component of the quasiclassical Green Functions for bulk electrodes}

Per definition electrodes are in local thermal equilibrium. The distribution function in eq.~\ref{eq:keldysh} describing the electrodes is given by:
\begin{subequations}
\begin{equation}
\check{f}= \sigma_0 ( n_+ \tau_0 +  n_- \tau_3),
\label{eq:bulkgfdist}
\end{equation}
\begin{equation}
n_\pm = \frac{1}{2}\left( \tanh{\frac{E + eV}{2 T}} \pm
\tanh{\frac{E - eV}{2 T}} \right).
\label{eq:npm}
\end{equation}
\end{subequations}
Here $V$ is the voltage bias and $T$ the temperature of the electrode, we set $E_F=0$. It is useful for practical purposes to split the distribution matrix into even ($n_-$) and odd ($n_+$) in energy components with respect to the Fermi surface. Deviations of $n_+$ from equilibrium are related to a difference in energy parametrized by a different effective temperature between layers. Whereas, $n_-$ is related to a difference in particle number parameter by a shift of the effective voltage. This relation becomes more clear when the transport expressions for charge and heat current are introduced. In most of the cases $\tau_0$ and $\sigma_0$ terms are not explicitly expressed in the equations, nevertheless, we have added them for clarity in this section. 

In the absence of voltage bias ($V=0$), $n_-(V=0)=0$ and $n=n_+(V=0)=\tanh (E/2 T)$, the distribution function is proportional to $\tau_0$. In this case, we can write 
\begin{equation}
\check{G}^K = \tanh (E/2 T)(\check{G}^R - \check{G}^A) \; .
\end{equation}
That results in a trivial structure of the Keldysh component of the Green functions. 

For clarity we can write the previously introduced quantities $n_{\pm}$, eq.\ref{eq:npm}, in terms of the well known Fermi distribution functions  $f_F(E)=[1+\exp(E/T_N)]^{-1}$,
\begin{subequations}
\begin{align}
n_-(E,V,T)= f_F(E-eV)-f_F(E) \; ,
\\
n_+(E,V,T) = f_F(E-eV)+f_F(E) 
\end{align}
\end{subequations}
and $n=(1+f_F(E))/2$. 

\newpage

\subsection{Some experimental values for the main parameters of the theory}

The most common conventional superconductors used in experiments are Nb and Al, with critical temperatures, $T_C$, of 10K and 1K respectively. Although a very thin layer of Al can show higher $T_C$. The characteristic strong spin orbit coupling of Nb it is a concern if we are interested in the study of certain properties. Such as spin imbalance in the system. Generally, the use of Al is preferred in experiments if the corresponding $T_C$ is reachable. Although in order to reach this temperatures electronic cooling techniques are required.

Using the formula that relates critical temperature and superconducting gap, $\Delta_0=1.764~ T_C$, we can obtain the energy value of the superconducting gap for zero temperature and no applied magnetic field ($\Delta_0$). For example, the value of the order parameter for 10K Nb, $\Delta_{0(Nb)}$, and 1K Al, $\Delta_{0(Al)}$,
\begin{equation}
\Delta_{0(Nb)}=2.43 \times 10^{-22}~ J= 1.52 \times 10^{-3}~eV= 10~\Delta_{0(Al)} \; .
\end{equation}

In the numerical calculation plots we usually normalize the energies by $\Delta_0$. In order to compare then with experimental results it is useful to keep in mind its values. 

For energies as the exchange field, which is related to the external magnetic field, we normalize it using the energy $\Delta_0$. Remember that this field is the Zeeman field as the orbital effect can be neglected. Here we describe the coupling between the field and the spins of the cooper pairs. This gives us values of the external magnetic field,
\begin{equation}
B_{(Nb)}=\frac{\Delta_{0(Nb)}}{g \mu_B}=13.12~ T \; .
\end{equation}
This coupling allows to reach much higher magnetic field values without destroying superconductivity.

\newpage

\subsection{Self consistency calculation of the order parameter, $\Delta$}
\label{sec:self}

As mentioned above, the superconducting order parameter entering the quasiclassical equations has to be calculated self-consistently. It is  related to the condensate function (or anomalous component) via the self-consistency equation. That in the Matsubara representation (section \ref{sec:matsubaraap} and refs.\cite{w75,w76,w86}) has the form 
\begin{equation}
\Delta=\pi T \lambda \Sigma_n \text{Tr} \check{f} \; .
\label{eq:selfcon}
\end{equation}
Here $\lambda=N_0 V$ is the electron-electron coupling constant leading to the formation of the superconducting condensate. Here V is the interaction parameter in the BCS Hamiltonian and $N_0$ is the density of states at the Fermi level. $\check{f}$ is the anomalous GF and the trace should be taken over the spin variables. For a bulk superconductor in Matsubara formalism we obtain $\text{Tr} \check{f}=f_0=\Delta/\sqrt{\Delta^2+\omega_n^2}$. Here $\omega_n=\pi T (2n+1)$ are the Matsubara frequencies. The cut off for the Matsubara frequencies is the Debye frequency $\omega_D \gg \Delta$. In this case is easy to check that
\begin{equation}
\pi  T \Sigma_n \frac{1}{\omega_n}=\log(\frac{1.13 \omega_D}{T}) \; .
\label{eq:debfreq}
\end{equation}
For a bulk superconductor and $T=T_C$ eq.~\ref{eq:selfcon} reduces to
\begin{equation}
1=\pi T_C \lambda \Sigma_n \frac{1}{|\omega_n|} \; .
\end{equation}
Thus, using eq.~\ref{eq:debfreq} we obtain
\begin{equation}
\lambda^{-1}=\log(\frac{1.13 \omega_D}{T}) \; .
\end{equation}
Now, dividing eq.~\ref{eq:selfcon} by $\lambda$, subtract and add $\Delta/|\omega_n|$ and using the previous results eq.~\ref{eq:debfreq}, we obtain, 
\begin{equation}
\Delta \log\frac{T}{T_C}+\pi T \Sigma_n \frac{\Delta}{|\omega_n|}-f_0=0 \; .
\end{equation}
This expression allows us to obtain the $T_C$ of a superconductor if the anomalous singlet component is known. 

Other useful relations between the critical temperature and the zero temperature order parameter with the interaction parameter $\lambda$ are
\begin{equation}
T_C=1.13 \omega_D e^{-\lambda^{-1}} \; ,
\end{equation}
\begin{equation}
\Delta_0=2 \omega_D e^{-\lambda^{-1}} \; .
\end{equation}
This lead to the well known relation $\Delta_0 \approx 1.764 T_C$. In this thesis we define $\Delta_0$ as the order parameter for zero temperature and exchange field.
\begin{figure}[h!]\begin{center}
\includegraphics[width=0.5\columnwidth]{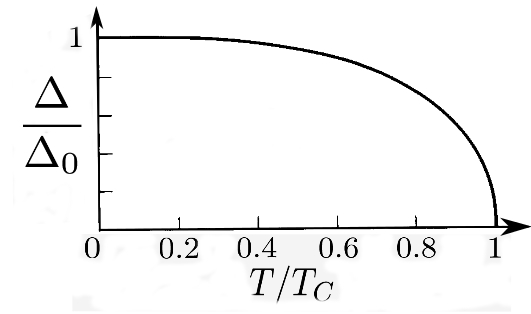}
\caption{Temperature dependence of the self-consistent superconducting gap $\Delta$.}
\label{fig:deltaT} 
\end{center}
\end{figure}

For weak coupling superconductors, those with a sufficiently weak attraction between the electrons, in which $ \omega_C / T_C \gg 1$, $\Delta(T)/ \Delta_0$ is a universal function of $T/T_C$. This decreases monotonically from 1 at $T=0$ to zero at $T_C$, as displayed in fig.~\ref{fig:deltaT}. Near $T=0$, the temperature variation is exponentially slow since $e^{-\Delta/ T} \approx 0$, so that the hyperbolic tangent is very nearly unity and insensitive to temperature change. Physically speaking, $\Delta$ is nearly constant until a significant number of quasiparticles are thermally excited. On the other hand, near $T_C$, $\Delta(T)$ drops to zero with a vertical tangent, approximately as
\begin{equation}
\frac{\Delta(T)}{\Delta(0)} \approx 1.74 \left(1-\frac{T}{T_C} \right)^{1/2} \quad T\approx T_C\;.
\end{equation}

\subsubsection{Dependence on the spin splitting field of the order parameter}

The spin splitting field affects greatly singlet pairing in conventional superconductors because electrons with different spins belong to different energy bands. Consequently, the critical temperature of the superconductor $T_C$ is considerably reduced in SF structures \textit{i.e.} a superconductor ferromagnet junction with high interface transparency. In this section we study, in particular in SF junctions with a spin splitting field, the superconducting order parameter and thus, the critical temperature.

\begin{figure}[h]\begin{center}
\includegraphics[scale=0.5]{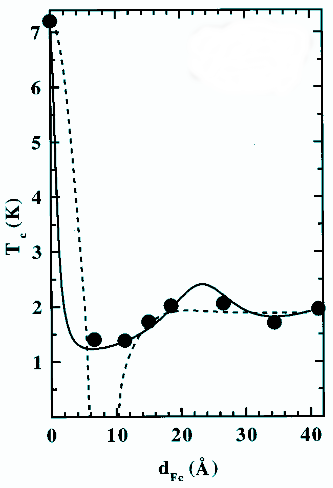}
\caption{Dependence of the critical temperature with the thickness of the F, as determined by resistivity measurements. The dashed line is a fit assuming a perfect interface transparency while the solid line corresponds to a non-perfect interface (From ref.~\cite{Lazar}).}
\label{fig:Tcvsd} 
\end{center}
\end{figure}

The critical temperature of the SF bilayers and multilayered structures was calculated in several works~\cite{Bagrets,BaladieBuzdin,Baladie,BuzdinKupriyanov,Demler,Fominov2002,Fominov2003,KhusainovProshin,
ProshinKhusainov,Proshin,Radovic,Tagirov,Tollis,You}. Experimental studies of the $T_C$ were also reported in many publications~\cite{Aarts1997,Guu,Jiang,Lazar,Muhge}. Fig.~\ref{fig:Tcvsd} is an example of the good agreement between theory and experiment that has been achieved in some cases. This shows the critical temperature dependence on the thickness of the F layer. Despite the many papers published, the problem of the transition temperature $T_C$ in this structures is not completely clear. For example ref.~\cite{jiang1995} and ref.~\cite{ogrin2000} claimed that the nonmonotonic dependence of $T_C$ on the thickness of the ferromagnet observed on Gd/Nb samples was due to the oscillatory behaviour of the condensate function in the ferromagnet. However, ref.~\cite{aarts1997} showed that the interface transparency plays a crucial role in the interpretation of the experimental data. That showed both, nonmonotonic and monotonic dependence of $T_C$ on the length of the ferromagnet. In other experiments (ref.~\cite{bourgeois}) the critical temperature of the bilayer Pb/Ni decreases with increasing length of the ferromagnet in a monotonic way. 

From the theoretical point of view, the $T_C$ problem in a general case cannot be resolved exactly. In most papers it is assumed that the transition to the superconducting state is second order, i.e. the order parameter $\Delta$ varies continuously from zero to a finite value with decreasing temperature $T$. However, generally, this is not true. In the case of a first order phase transition from the superconducting to the normal state, the order parameter $\Delta$ drops from a finite value to zero.

In order to illustrates these different behaviours, let us consider a SF junction with $d_F<\xi$, $d_S<\xi$, where $d_{F,S}$ are the lengths of the Ferromagnet and Superconductor respectively. In this case we can apply the short limit and the Usadel equation can be averaged over the length of the mesoscopic structure. The Greens functions are uniform in space and correspond to a superconductor with an intrinsic exchange field $\tilde{h}=r_F h$ and superconducting gap $\tilde{\Delta}=r_s \Delta$, where $r_F=1-r_S=N_F d_F/(N_F d_F+N_S d_S)$. Here $N_S$ and $N_F$ are the densities of states in the superconductor and ferromagnet respectively. The only difference between the bilayer and the superconductor with intrinsic exchange field cases is that for the former the quantities depend on the lengths of the layers, otherwise they are equivalent. The critical temperature of this bilayer was studied in ref.~\cite{Bergeret2001} . It was established that both first and second order phase transitions may occur in these systems if $\tilde{h} < \tilde{\Delta}$. 

Here we determine the critical temperature for thin SF bilayers, such that the GFs are position independent (see section\ref{sec:deter}). In the range of parameters for which a second order phase transition occurs, one can linearise the Usadel equation for $\Delta$ and use the Ginzburg-Landau expression for the free energy, assuming that the temperature is close to $T_C$. The equation is obtained from the self-consistency condition and in the Matsubara representation it reads,
\begin{equation}
\log{\frac{T_C}{T^*_C}}=(\pi T^*_C) \Sigma_\omega (\frac{1}{|\omega_n|}-i f_\omega/\Delta)\;.
\end{equation}
Here $T_C$ is the critical temperature in the absence of the proximity effect and $T_C^*$ is the same quantity affected by the ferromagnet. We introduce this equation to calculate the order parameter dependence on the exchange field $h$. The solutions for a superconductor with a homogeneous exchange field (as  described in section\ref{sec:superferro}) satisfying the normalization condition can be expressed as
\begin{equation}
f_\pm= \mp \frac{i \Delta}{\xi_\pm} \quad g_\pm=\frac{\omega_n \pm i h}{\xi_\pm} \; ,
\end{equation}
where $\xi_\pm=\sqrt{(\omega_n \pm ih)^2 + \Delta^2}$ and $f_0=f_++f_-$. This leads to a self-consistency equation,
\begin{equation}
1=\frac{1}{2} \pi T \mu_0 V \Sigma_n \frac{1}{\xi_+}+\frac{1}{\xi_-}\; .
\end{equation}
For numerical computation it is convenient to subtract the equation for the case $h=0$, $\xi_0$, obtaining
\begin{equation}
0=\frac{1}{2} \pi T \mu_0 V \Sigma_n \frac{1}{2} \left(  \frac{1}{\xi_+}+ \frac{1}{\xi_-} \right)  - \frac{1}{\xi_0} \; .
\end{equation}
\begin{figure}[h]\begin{center}
\includegraphics[width=0.7\columnwidth]{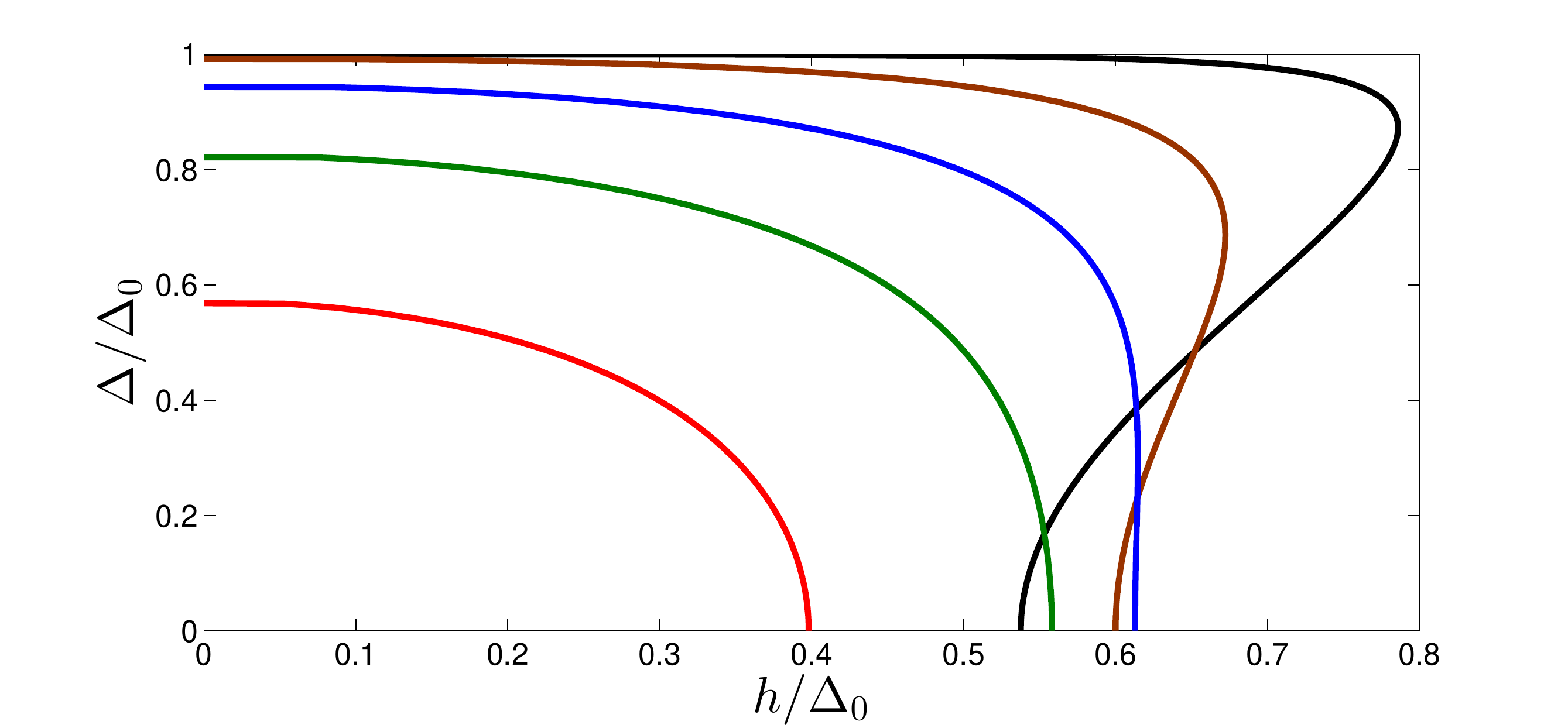}
\caption{Normalized self-consistent superconducting gap $\Delta/\Delta_0$ versus exchange field $h/\Delta_0$ for different temperatures: black ($ T= 0.1 \Delta_0$), brown ($ T= 0.2 \Delta_0$), blue ($ T= 0.3 \Delta_0$), green ($ T= 0.4 \Delta_0$) and red ($ T= 0.5 \Delta_0$).}
\label{fig:Scongap} 
\end{center}
\end{figure}

In fig.~\ref{fig:Scongap} we show how the self-consistent order parameter $\Delta$ varies as we increase the exchange field $h$, so when reaching high enough fields the superconductivity is destroyed ($\Delta=0$). As expected, lowering the temperature the overall self-consistent gap (and also its value at $h=0$) increases. For high temperatures ($ T=0.5, 0.4 \ \text{and} \ 0.3 \Delta_0$) the  self-consistent gap decreases monotonically by increasing the  exchange field. In this case the transition is of second order. 

In contrast, for low temperatures ($ T=0.2 \ \text{and} \ 0.1 \Delta_0$) and high exchange values, $\Delta$ is not monoevaluated. The upper branch (the highest $\Delta$ values of the two possible ones) correspond to a equilibrium state while the lower one corresponds to metastable states. For this kind of curves, the maximum value of the exchange field or the crossing point from the lower to the upper branch, is the critical exchange field in which the gap closes. A first order transition occurs at this point from superconducting to normal state.

The self-consistency equation alone is not enough to determine the order parameter. The nonzero solution for $\Delta$ implies only the local minimum of the free energy, this may correspond to a metastable rather than to an equilibrium state. In order to find out whether the transition to the superconducting state really occurs, we must find the free-energy difference between the normal and superconducting states. If the free-energy of the superconducting state is higher than that of the normal state, the superconducting transition does not occur. From the condition that both free-energies must be equal we find the Chandrasekhar-Clogston (1962)~\cite{clogston1962,chandrasekhar1962} limit, or paramagnetic limit at $T=0$, that the exchange field has to always fulfil,
\begin{equation}
h < \frac{\Delta_0}{\sqrt{2}}=0.707 \Delta_0 \; .
\label{eq:hlimit}
\end{equation}  
This means that in the previous figure\ref{fig:Scongap} as h reaches this point a first-order transition occurs form the superconducting phase with $\Delta$ to the normal phase with $\Delta=0$. 

A more detailed investigation shows that the real situation is more complicated. In the preceding treatment we were concerned only with the possibility of the appearance of superconductivity with $\Delta=\text{const}$. It turns out that at values of h larger than the critical value may generate a superconducting phase with an inhomogeneous order parameter $\Delta(r)$ by way of a second-order phase transition. This was found by Larkin and Ovchinnikov\cite{fflo1} and almost simultaneously but independently by Fulde and Ferrel\cite{fflo2} in 1964. This new phase is called the LOFF phase and at $T=0$ fulfils $h^{FFLO}_C=0.755 \Delta_0$. This FFLO state only appears in the temperature interval $0<T<0.56 T_C$. In the dirty limit it is suppressed and the first-order transition into the uniform superconducting state occurs instead. 

\begin{figure}[h]\begin{center}
\includegraphics[width=0.6\columnwidth]{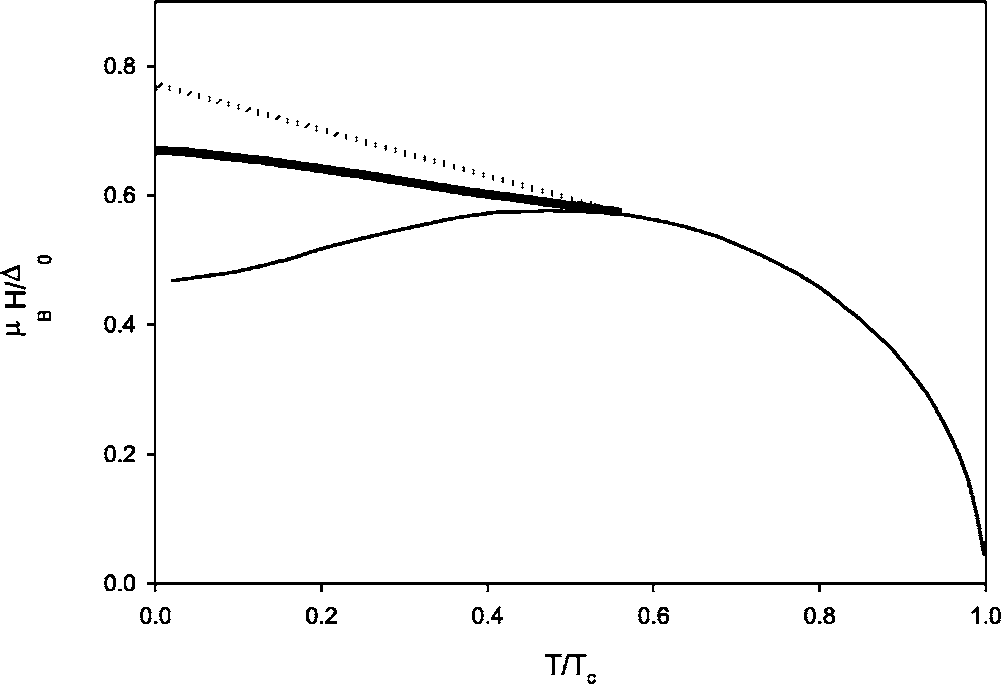}
\caption{Phase diagram of a superconductor for T and $h=\mu_B H$. At temperatures below T*=0.56Tc the second-order transition occurs from the normal to the nonuniform superconducting FFLO phase. The bold line corresponds to the first-order transition into the uniform superconducting state, and the dotted line represents the second-order transition into the nonuniform superconducting state. (From ref.~\cite{stjames})}
\label{fig:Hvst} 
\end{center}
\end{figure}

In fig.~\ref{fig:Hvst} we show the phase diagram of the superconductor for temperature and magnetic field (where $h=g\mu_B H$). As we could expect, values in the upper part correspond to normal states and lower ones to superconducting states. The bold line is the first-order transition corresponding to stable states, the ones limiting the self-consistency equation. Here we can see how it varies with temperature reaching the maximum value for $T=0$, the Chandrasekhar-Clogston limit\cite{clogston1962,chandrasekhar1962}. The critical exchange field follows approximately\cite{stjames},
\begin{equation}
\frac{h_C}{\Delta_0} \sim 0.707-0.1964 \frac{T}{T_C} \; ,
\end{equation}
valid only for $T<0.56 T_C$. The dotted line represents a second-order transition into the non homogeneous superconducting state (LOFF phase). This allows higher exchange fields in the superconductor than those of a homogeneous superconducting state. Finally, the simple solid line represents the previously introduced metastable states, those corresponding to the lower branches of fig.~\ref{fig:Scongap} in the non monoevaluated case.

\newpage

\section{Some properties of hybrid structures}
\label{sec:properties}

In this section charge and heat transport in simple hybrid structures is analysed. First, we provide a description of a planar junction within quasiclassics and  study junctions involving a single superconductor. In the second part of the section transport through Josephson junctions and superconducting quantum point contacts is discussed.

\subsection{Description of Planar Junctions}
\label{se:2.3}

We first consider junctions between  a superconductor (S) and a normal metal (N) or ferromagnetic (F)  with  thin insulating interlayer. For that purpose, we introduce boundary conditions (b.c.) that allow s to model interfaces complementing the Usadel equation. The transport equations presented in this chapter take into account the proximity effect, unlike those presented in section\ref{sec:tunnel}.

\subsubsection{The Kupriyanov-Lukichev boundary conditions}
\label{sec:kupri}

The Usadel equation needs to be complemented by boundary conditions that describe the interfaces.   Here we introduce the one presented by \textit{Kupriyanov and Lukichev} (1988)~\cite{KL} derived from the  Zaitsevs boundary conditions~\cite{Zaitsev1984}.  Assuming that the interface is located at $x=0$, the boundary condition corresponds to the value of the spectral matrix current at that point, 
\begin{equation}\label{eq:kl}
\breve{J}_{LR} \bigl |_{x=0}= \breve{G}_L\partial_x \breve{G}_L \bigl |_{x=0}= \kappa_t \left[
\breve{G}_L, \breve{G}_R \right]_{x=0} \; .
\end{equation}
Here the subscripts stand for the left and right sides of the interface, where $\breve{J}_{LR}$ is the matrix current, that leaves the Left side to the Right one. $\breve{G}_{R/L}$ correspond to the Green functions of the right and left electrode. Calculating the matrix current at the opposite side of the interface gives,
\begin{equation}
\breve{J}_{RL} \bigl |_{x=0}= \breve{G}_R\partial_x \breve{G}_R \bigl |_{x=0}= \kappa_t \left[
\breve{G}_R, \breve{G}_L \right]_{x=0} = -\breve{J}_{LR}\bigl |_{x=0} \; .
\end{equation}
This relation represents the current conservation at the interface and the subscripts of the matrix current are omitted from now on. Here we define $\kappa_t=1/(\sigma R_I A)$, where $\sigma$ is the conductivity of the electrode in the normal state, $R_I$ is the interface resistance and $A$ is the area of the junction.  The range of applicability of this b.c. corresponds to high resistance (tunnelling) interfaces, which limits the calculations to values of $\kappa_t \ll 1$. In order to model higher transmission interfaces we introduce Nazarov b.c.~\cite{nazarovbc} in section\ref{sec:nazarov}, these conditions reduce to the K-L case in the low transmission limit. 

%Even thought they are usually assumed to be equal, the conductance of the right and left electrodes can differ. We account for this phenomena by a prefactor, introduced in the equation the same way as $\kappa_t$. Boundary conditions for the Eilenberger equation were also obtained in ref.~\cite{eilenbc1,eilenbc2}.

The charge current $I$,  and heat current $Q$, at the interface  are given by~\cite{LOnoneq, Belzig, Vinokur, Golubov}
\begin{subequations}
\begin{align}\label{eq:I}
I(V,T) &= \frac{\sigma}{e} \int_{-\infty}^{\infty} I_- \; dE, \quad Q(V,T) =
\frac{\sigma}{e^2} \int_{-\infty}^{\infty} E I_+ \; dE \; ,
\\
\label{Ipm2} I_- &\equiv (1/16) \tr \tau_3 \check{J}^K, \quad I_+
\equiv (1/16) \tr \tau_0 \check{J}^K \; ,
\\
\check{J}^K &\equiv \left( \breve{G}\partial_x \breve{G} \right)^K
= \check{G}^R \partial_x \check{G}^K + \check{G}^K \partial_x \check{G}^A \; .
\end{align}
\end{subequations}
 In order to calculate the expressions for the current, we are interested in the Keldysh component of the spectral matrix current. For the Kupriyanov-Lukichev case, this term is proportional to the Keldysh component of the commutator, which reads,
\begin{equation}
\left[ \check{G}_L,\check{G}_R \right]^K=\check{G}_L^R \check{G}_R^K + \check{G}_L^K \check{G}_R^A- \check{G}_R^R \check{G}_L^K - \check{G}_R^K \check{G}_L^A \; .
\end{equation}
The Green functions in eqs.\ref{eq:I} and \ref{Ipm2} have to be determined by solving the Usadel equation eq.\ref{eq:FullUsadel} with the boundary condition eq.\ref{eq:kl}.

%Note that depending on the b.c. we use, the matrix current is different and, thus, the expressions for the currents. The most frequent in this manuscript are the K-L boundary condition.

\subsubsection{NIN and NIS tunnel junctions}\label{ss:reservoir}

As a first simple example we consider two electrodes tunnel connected and voltage biased. If the superconducting state the Green function of the electrode is given by eq.\ref{eq:bulkgfS} and in the normal state by eq.\ref{eq:bulkgfN}. The distribution functions are the equilibrium ones, and given by eq.\ref{eq:bulkgfdist}. Applying the voltage to the left electrode the distribution functions read: $\check{f}_L = \sigma_0 (\tau_0 n_+(E,V,T_L) + \tau_3 n_-(E,V,T_L))$ and $\check{f}_R = \sigma_0 \tau_0 n(E,T_R)$. Due to the low transmission interface we use the  K-L boundary conditions. 

\begin{figure}[h]
  \centering
  \includegraphics[width=0.5\columnwidth]{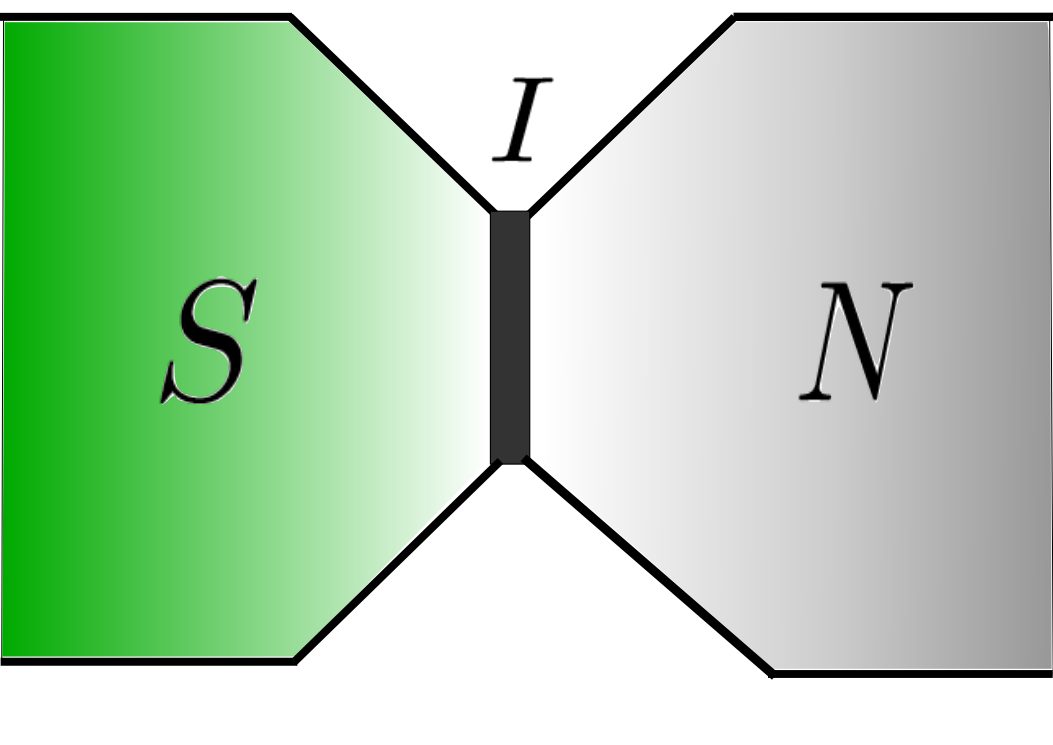}
  \caption{The SIN tunnel junction. The superconductor (S) and normal metal (N) are electrodes.}
\label{fig:snreservoirs}
\end{figure}

First, consider the simple case of two normal metal reservoirs, a $NIN$ junction, where $I$ is the insulating layer. Here we insert the Green functions and distribution functions into the expression for the matrix current corresponding to the K-L boundary conditions,  eqs.~\ref{eq:kl},\ref{eq:I} and \ref{Ipm2}. We obtain the following expression, 
\begin{equation}
I=\frac{\kappa_t \sigma}{2 e} \int_{-\infty}^{\infty} dE n_-(E,V,T) = \frac{\kappa_t \sigma}{e} \int_{0}^{V} dE=\frac{V}{R} \; .
\end{equation}
This is nothing but the Ohm´s law.

Let us now introduce a superconducting reservoir to the junction and study a $SIN$ system, as shown in fig.~\ref{fig:snreservoirs}. Here the superconductor is grounded while the normal metal is voltage biased. Superconductors have an intrinsic superconducting phase, in this case, as we have a single superconductor, the phase is set to zero. If a voltage is applied to the superconductor, the Josephson relation means that its phase is time dependent. In order to avoid this complication, we always consider that the normal metal is voltage biased and the superconductor grounded. 

%As explained in the introduction this kind of interfaces give rise to interesting phenomena as the Andreev reflection. 

The heat and charge currents using the K-L b.c. take the form,
\begin{equation}
I=\frac{1}{e R} \int_0^{\infty} dE n_-(E,V,T_N) \re g_S(E,\Delta,\eta),
\label{eq:Ibulk}
\end{equation}
\begin{equation}
Q=\frac{1}{e^2 R} \int_0^{\infty} dE E (n_+(E,V,T_N) - n(E,T_S)) \re g_S(E,\Delta,\eta).
\label{eq:Qbulk}
\end{equation}
Here $\re g_S=N_S(E) = |E| \Theta(|E| - \Delta)/\sqrt{E^2 - \Delta^2}$ is the BCS normalized density of states (DOS) and $\Theta(x)$ is the Heaviside step function. $T_{N(S)}$ stand for the temperature of the normal metal (Superconductor) and $eV$ is the voltage bias. Note that in the absence of bias voltage and with both reservoirs set at the same temperature there is no current of any kind. By applying $V=0$ but setting the reservoirs at different temperatures, we obtain a finite heat current.

All the expressions of the charge current of this section correspond to those of the tunnelling current, eq.\ref{eq:Inn} and eq.\ref{eq:Ins}. The phenomena corresponding to the Andreev reflections is not present in these results, since we have not included the proximity effect. In the next section we show how the proximity effect changes the transport properties of mesoscopic structures.

\subsubsection{ Subgap transport in a SIN junction}\label{sec:proximityeffect}

Due to the proximity effect superconducting correlations are generated in a normal metal. Mathematically this means that the anomalous component of the GFs is finite  in the non-superconducting region. Here, $g$ are normal components of the Greens functions and $f_i$ anomalous ones that correspond to superconducting properties. We define $f_0$ as the singlet component. In the case of a low transparency barrier ($\kappa_t \ll 1$) or for temperatures near the superconducting critical temperature, $T_C$, the proximity effect is weak. As a consequence, the anomalous Greens functions are small for all energies ($|\check{f}| \ll |\check{g}|$). Then the normalization condition reads,
\begin{equation}
(g^{R(A)})^2-(f^{R(A)})^2 \simeq (g^{R(A)})^2=1 \; .
\end{equation}
This allows us to \textit{linearise} the Green functions in the following way,
\begin{equation}
\check{G}^{R(A)}_N \simeq \check{g}^{R(A)} + \delta \check{g}^{R(A)} = \tau_3 \sigma_0 + \tau_1 f^{R(A)}_0 \sigma_0 \; .
\end{equation} 
Here the correction to the Green function $ \delta \check{g}^{R(A)}$ corresponds to the anomalous components. Using this approximation we only have to determine the anomalous component from the Usadel equation. Even though linearisation can be a useful tool as it simplifies the calculations, we should bear in mind that some part of the physics may be lost in the process, as shown in section\ref{sec:deter}.

We now determine heat and charge currents for the $SIN$ system shown in fig.~\ref{fig:snproximity}. Here S is a superconducting reservoir and N a thin normal metal under the non-linearised proximity effect. The charge and heat currents read,
\begin{subequations}\label{IQ}
\begin{align}
I &= \frac{1}{eR} \int_0^\infty n_-(E,V,T_N) \left( \re g(x=0) \re g_S + \re f_{S0} \re f_0(x=0) \right) \; dE,\label{II}
\\
Q &= \frac{1}{e^2 R} \int_0^\infty E (n_+(E,V,T_N) - n(E,T_S)) \left( \re g(x=0) \re g_S - \im f_{S0} \im f_0(x=0) \right) \; dE.\label{QQ}
\end{align}
\end{subequations}
The main difference between these equations and eq.\ref{eq:Ibulk} and \ref{eq:Qbulk}, is that here we have extra components corresponding to the pair correlation contributions. In other words, there is a new subgap contribution related to the anomalous components in the superconductor and normal metal. Here the Green functions without subscript correspond to the N.  As the normal metal is affected by the proximity effect, its Green functions depend on the position and are evaluated at the interface, where we apply the b.c. In order to obtain the expression of the current corresponding to the linearised case, the normal component of the normal metal Green function is set close to unity.

Depending on the energy range, there are two different contributions to the current, the quasiparticle contribution ($|E|>\Delta$) and the subgap (Andreev) contribution ($|E|<\Delta$). The former corresponds to single particles tunnelling through the barrier to the energy states outside the gap. Theses are the only processes that we have studied up to now and the only possible ones in the absence of the proximity effect. The subgap (Andreev) contribution corresponds to particles that can not tunnel due to the presence of the superconducting gap i.e. the absence of states to tunnel. This inability causes the particle to reflect as a hole generating a cooper pair (charge neutrality) on the opposite side, this is known as Andreev reflection, sec.\ref{sec:andreev}. It is important to notice that Andreev reflections only transfer charge and never energy (heat), as no particle is transferred.

Let us focus first on the single particle contributions that we denote as $I_1$ and $Q_1$. For energies larger than the superconducting gap ($E > \Delta$) only the terms proportional to $\re g_S$ are non-zero, as can be seen from eq.\ref{eq:gsfs}. Thus, the  single-particle contribution to the  electric current is given as follows,
\begin{equation}\label{I12}
I_1 = \frac{1}{eR} \int_\Delta^\infty \re g(E,x=0) N_S(E) n_-(E,V,T_N) \; dE.
\end{equation}
Here $N_S(E) = |E| \Theta(|E| - \Delta)/\sqrt{E^2 - \Delta^2}$ is the BCS normalized density of states (DOS) and $\Theta(x)$ is the Heaviside step function. Now the DoS of the normal metal depends on both energy and position, unlike in eq.\ref{eq:Ibulk}.

Let us study the limiting case of very low temperatures, $T_N \ll 1$. Now both (S and N) DoS are energy dependent, making the integration more complex.  We still require voltages higher than the superconducting gap $\Delta$ in order to observe a finite single particle current, so the expression reads,
\begin{equation}\label{I1T}
I_1(T_N \ll 1) = \frac{1}{eR} \int_\Delta^V \re g(E,x=0) N_S(E) \; dE.
\end{equation}

The single-particle contribution to the heat current can be obtained from eq.~\ref{QQ},
\begin{equation}\label{Q1}
Q_1 = \frac{1}{e^2 R} \int_\Delta^\infty E(n_+(E,V,T_N) - n(E,T_S)) \left[ N_S(E) \re g(E,x=0) - M_S^+(E) \im f_0(E,x=0) \right] \; dE,
\end{equation}
where $M_S^+(E) = \Delta \Theta(|E| - \Delta)/\sqrt{E^2 - \Delta^2}$. Unlike the expression of eq.\ref{eq:Qbulk} with no proximity effect, we now have an extra term given by the product of the two anomalous functions of N and S. 

For energies $E < \Delta$, the electric charge $I_A$ (as $Q_A=0$) is transferred by means of the Andreev reflection, the subgap contribution to the current  mentioned earlier. The new anomalous component present in the N makes the Andreev current finite and it reads,
\begin{equation}\label{IA}
I_A = \frac{1}{eR} \int_0^\Delta n_-(E,V,T_N) M_S^-(E) \re f_0(E,x=0) \; dE.
\end{equation}
where $M_S^-(E) = \Delta \Theta(\Delta - |E|)/\sqrt{\Delta^2 - E^2}$. This expression is similar in shape to eq.~\ref{I12} but for an "anomalous DoS". We neglect the contribution to the Andreev current due to the partial Andreev reflection at energies above the superconducting gap. In the case of strong enough tunnel barriers, this contribution leads to a small correction and can be neglected. 

In the opposite case, when we have a normal reservoir and a thin superconductor. The inverse proximity effect is maximized and the proximity effect is neglected. Green functions of the superconductor are perturbed and the superconducting gap starts to close taking the form of a "minigap".

\begin{figure}[h]
  \centering
  \includegraphics[width=0.5\columnwidth]{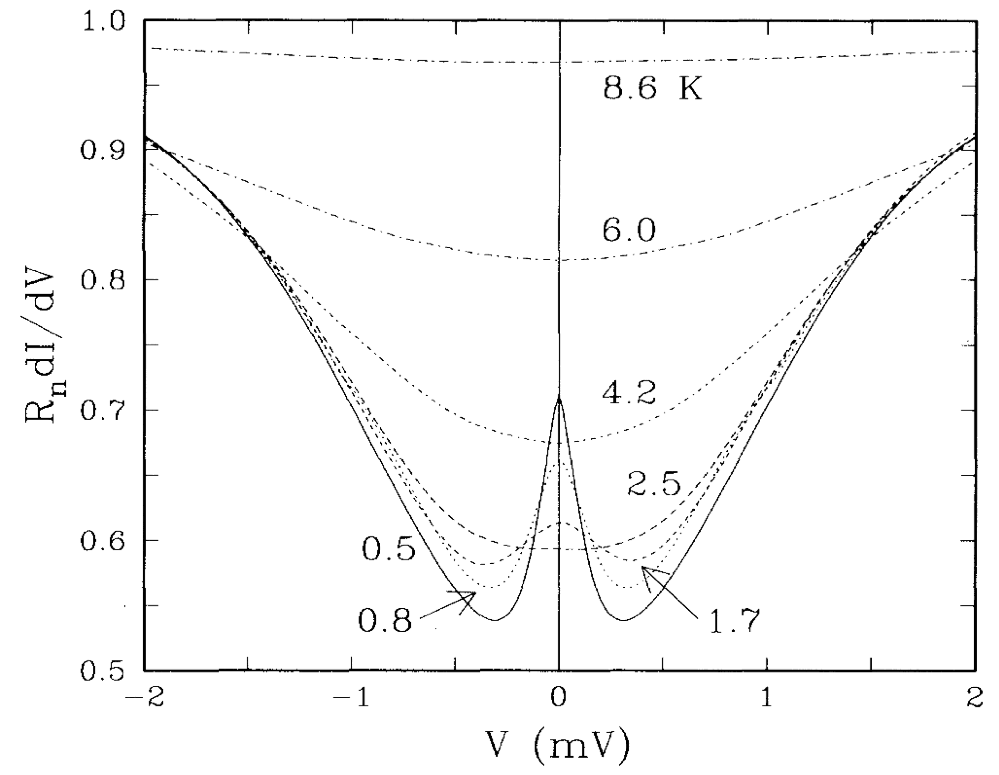}
  \caption{The measured normalized conductance-voltage characteristics at different temperatures and zero magnetic field. At low temperatures it is clear to see the zero bias anomaly (peak).(From Ref.\cite{kastalskii})}
\label{fig:zba}
\end{figure}

Let us now calculate the subgap differential conductance,
\begin{equation}
G_A=\frac{\partial I_A}{\partial V} = \frac{1}{eR} \int_0^\Delta \frac{\partial n_-}{\partial V}(E,V,T_N) M_S^-(E) \re f_0(E,x=0) \; dE.
\end{equation}
In the low temperature limit, the diffusive $SIN$ junction shows a \textit{zero bias anomaly} (ZBA) peak due to the impurity confinement and the electron-hole interference at the Fermi level. It occurs at zero bias since the electron is perfectly retro-reflected as a hole during the Andreev reflection process. Thus the electron and the reflected hole interfere along the same trajectory and the interference effect strongly enhances the subgap conductance at zero bias. First experimental evidences for such subgap conductance have been reported in 1991\cite{petrashov,kastalskii} (see Fig.\ref{fig:zba}). 

\subsubsection{Determination of the Green functions}
\label{sec:deter}

\begin{figure}[h]
  \centering
  \includegraphics[width=0.5\columnwidth]{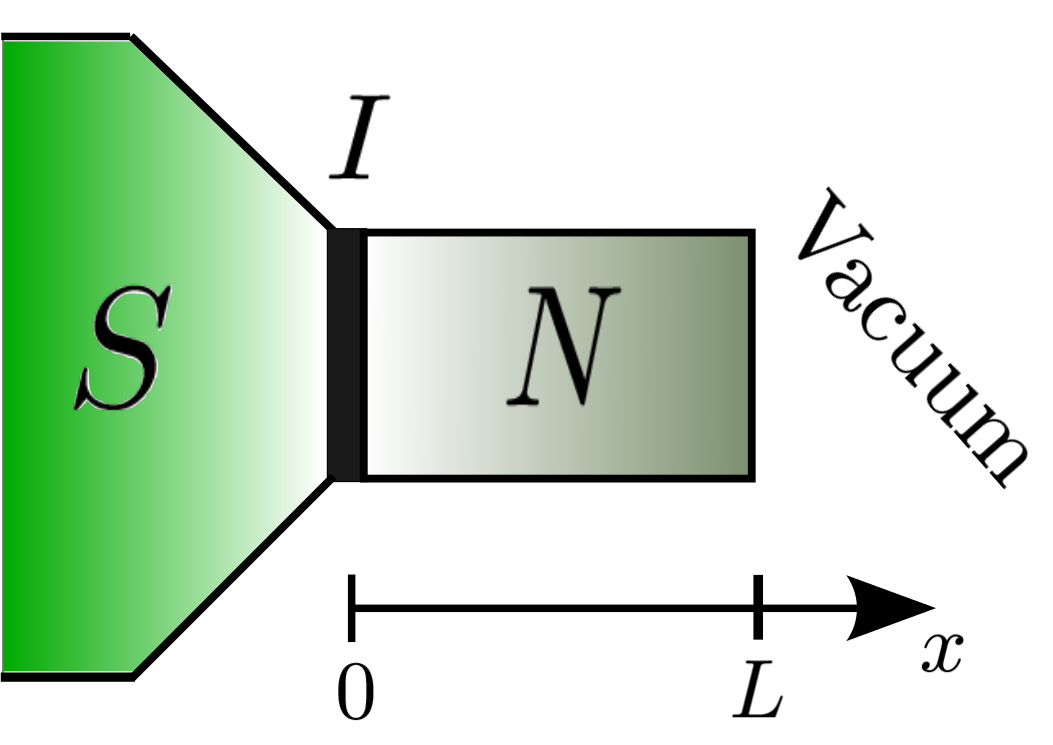}
  \caption{The SIN junction to illustrate the proximity effect.}
\label{fig:snproximity}
\end{figure}

The Green functions of the layers under the proximity effect depend on the characteristic properties of the junction. They must be determined by the Usadel equation (Eq.~\ref{eq:FullUsadel}), complemented with the boundary conditions at the interfaces and the normalization condition. Let us continue working with the SIN structure, depicted in fig.\ref{fig:snproximity} where we assume that the right side of N is in contact with vacuum. In order to model it we assume that there is no current flowing in that end, which leads to, 
\begin{equation}
\hat{J}^{R} \bigl |_{x=L} = \hat{G}^{R} \partial_x \hat{G}^{R}\bigl |_{x=L}=0 \; .
\label{eq:vac} 
\end{equation}
Here $L$ is the length of the N layer, so that the position of the insulating barrier is $x=0$. Our purpose is to determine the GFs of the normal metal, where the Retarded components of the Usadel equation read,
\begin{equation}\label{eq:usadelSIN}
\mathcal{D} \partial_x \hat{J}^{R} = \left[E \tau_3,
\hat{G}_N^{R} \right], \quad \hat{J}^{R} = \hat{G}_N^{R} \partial_x \hat{G}_N^{R}.
\end{equation}
We are only left with the meaningful components corresponding to the N from of the general equation~\ref{eq:FullUsadel}. Due to the absence of an exchange field, the spin space is trivial and the Retarded and Advanced GFs can be depicted as $2 \times 2$ in particle-hole space. 

%

%%%%%%%%%%%%%%%%%%%%%%%%%%%%%%%%%%%%%%%%%%%%%%%%%%%%%%%%%%%%%%%%%%%%%%
%
\begin{figure}[h]\begin{center}
\includegraphics[width=\columnwidth]{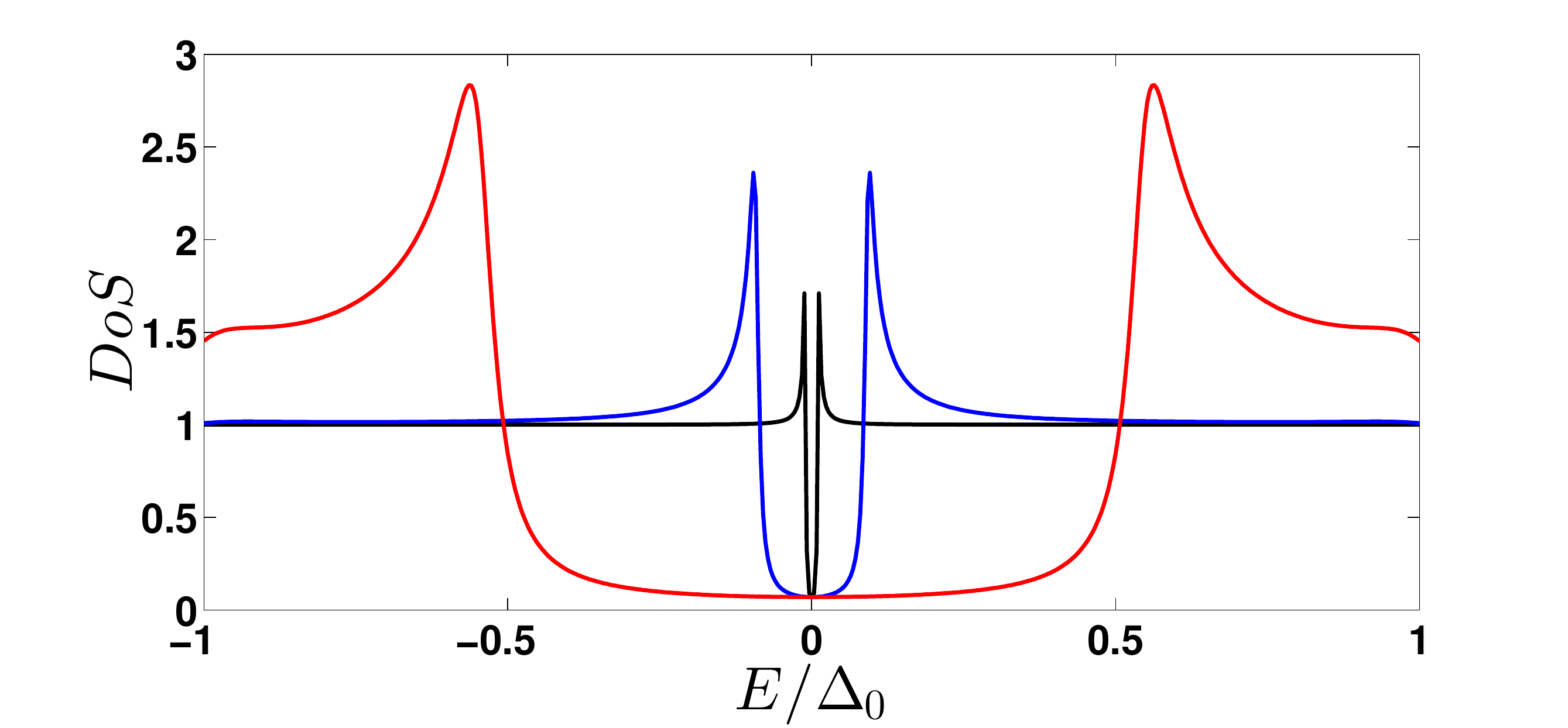}
\caption{Density of States of a short normal metal in contact with a superconducting electrode for different $\kappa_t$ values of the barrier: 0.1(red), 0.01 (blue) and 0.001(black). Here $\eta=0.07$ and $L=0.1$.}
\label{fig:minigap} 
\end{center}
\end{figure}
%
%%%%%%%%%%%%%%%%%%%%%%%%%%%%%%%%%%%%%%%%%%%%%%%%%%%%%%%%%%%%%%%%%%%%%%

Let us limit ourselves to the case of a \textit{short} normal metal, due to the small length of the layer, the Green functions inside this normal metal do not vary with position. In general, all GFs depend on position, unlike the electrodes discussed in section \ref{sec:spectral} and these short layers. This approximation allows us to integrate eq.~\ref{eq:usadelSIN} over the length of the N layer,
\begin{equation}
\mathcal{D} ( \hat{J} \bigl|_{x=L} - \hat{J} \bigl|_{x=0} ) = L \left[ E \tau_3,
\hat{G}_N \right] \; .
\end{equation}
Note that in the above equation we have omitted the R subscripts. The terms $\breve{J}|_{x=L/0}$ are given by the  boundary conditions at both sides of the normal metal. In this case they read,
\begin{equation}
\hat{J} \bigl|_{x=0}= \kappa_t \left[\hat{G}_L, \hat{G}_R \right]_{x=0},
\end{equation}
\begin{equation}
\breve{J} \bigl|_{x=L}= 0 \; .
\end{equation}
That correspond to the K-L and vacuum boundary conditions from eq.\ref{eq:kl} and eq.\ref{eq:vac} respectively. Substituting the Retarded component of the Usadel equation gives,
\begin{equation}
f^R_S g^R_N = f^R_N (\frac{L E}{\mathcal{D} \kappa_t } + g^R_S).
\end{equation}
Here $S(N)$ subscripts stand for the superconducting (normal metal) GFs, with those of the superconductor corresponding to the electrode values in eq.\ref{eq:gsfs}. We still lack one equation, the normalization condition of the GFs shown here,
\begin{equation}
(g^R_N)^2-(f^R_N)^2=1.
\end{equation}
These are the results which we obtain for the anomalous and normal GFs of the normal metal,
\begin{equation}
(f^R_N)^2=\frac{(f^R_S)^2}{(f^R_S)^2+ (\frac{L E}{i \kappa_t \mathcal{D}} + g^R_S )^2},
\end{equation}
\begin{equation}
(g^R_N)^2=\frac{(\frac{L E}{i \kappa_t \mathcal{D}} + g^R_S )^2}{(f^R_S)^2+ (\frac{L E}{i \kappa_t \mathcal{D}} + g^R_S )^2}.
\end{equation}
In the limiting cases of a very high interface resistance $ \kappa_t \rightarrow 0$ then $g^R_N \rightarrow 1$ and  $f^R_N \rightarrow 0$ , the GFs return to the normal metal reservoir values. By making the barrier very transparent $ \kappa_t \rightarrow \infty$, then $g^R_N \rightarrow g^R_S$ and  $f^R_N \rightarrow f^R_S$, the normal metal takes the values of the superconducting reservoir. As we decrease the resistance of the barrier a "minigap", generated by the proximity effect of the superconducting reservoir, is formed inside the normal metal as seen in fig.~\ref{fig:minigap}. The peaks on either side of the gap also decrease their height with increasing resistance. If we increase the length of the N layer, it has the opposite effect and reduces the proximity effect.
%%%%%%%%%%%%%%%%%%%%%%%%%%%%%%%%%%%%%%%%%%%%%%%%%%%%%%%%%%%%%%%%%%%%%%
%
\begin{figure}[h]\begin{center}
\includegraphics[width=0.5\columnwidth]{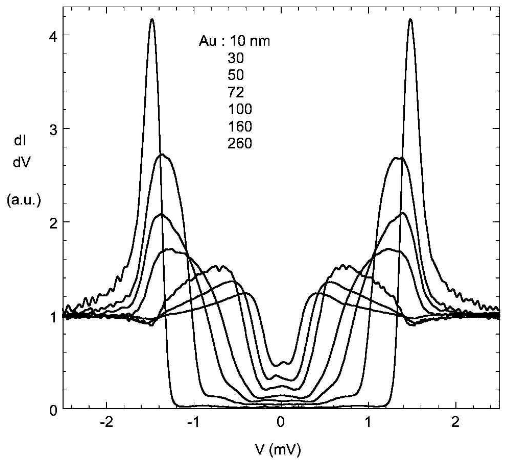}
\caption{(From ref.~\cite{Gupta}) Tunnelling density of states measured at 60 mK at the Au surface of Nb-Au bilayer samples with a varying Au thickness.}
\label{fig:minigapex} 
\end{center}
\end{figure}
%
%%%%%%%%%%%%%%%%%%%%%%%%%%%%%%%%%%%%%%%%%%%%%%%%%%%%%%%%%%%%%%%%%%%%%%

Experiments determining the density of states of SN bilayers with the help of tunnelling spectroscopy were performed many years ago~\cite{AdkinsKington,ToplicarFinnemore}. While spatially resolved density of states were later measured in refs.~\cite{Anthore,Gueron}. An especially interesting experiment is the one depicted in fig.~\ref{fig:minigapex}. Here we observe in a detail way the variation of the minigap with the length of the normal metal. With decreasing length, the minigap becomes smaller, whilst the DoS inside the minigap is finite and the peaks at both sides of the minigap smear.

%\paragraph{Derivation of a BCS superconducting reservoir Greens functions}
%
%As an example we derive the BCS superconductor bulk Greens functions, that has been defined in section\ref{sec:spectral} and widely used in the text. Let us assume that we have a superconductor with vacuum at both ends and the corresponding GFs are independent of position (equivalent to short limit). The Usadel equation (\Eq{eq:FullUsadel}) already integrated over length reads,
%%
%\begin{equation}
%0 = L_S \left( \left[ E \tau_3,\hat{G}_S \right] + \left[ \hat{\Delta},\hat{G}_S \right] \right) \; .
%\end{equation}
%%
%By assuming $\hat{G}_S=\tau_3 g_S + \tau_1 f_S$ where $g_S$ and $f_S$ are unknown. We obtain the relation,
%%
%\begin{equation}
%E f_S = \Delta g_s \; .
%\end{equation}
%%
%Imposing the normalization condition $f_s^2-g_S^2=1$ we obtain,
%%
%\begin{equation}
%g_S^2=\frac{E^2}{E^2-\Delta^2} \; , \quad f_S^2=\frac{\Delta^2}{E^2-\Delta^2} \; .
%\end{equation}
%%

\subsubsection{Ferromagnets in superconducting hybrid structures}\label{sec:ferro}

Following the study of superconducting nanohybrids we now introduce the effect of ferromagnets and how to model them in this notation. These ferromagnets, under the proximity effect, give rise to new and interesting phenomena, which are the core of the superconducting hybrids field. In this notation with respect to ferromagnets, all normal metal layers have an internal exchange field:  normal metals with externally applied magnetic field and in contact with ferromagnetic insulator. The finite exchange field term in the Usadel equation is characteristic of these kind of structures. 

\begin{figure}[h]
  \centering
  \includegraphics[width=0.5\columnwidth]{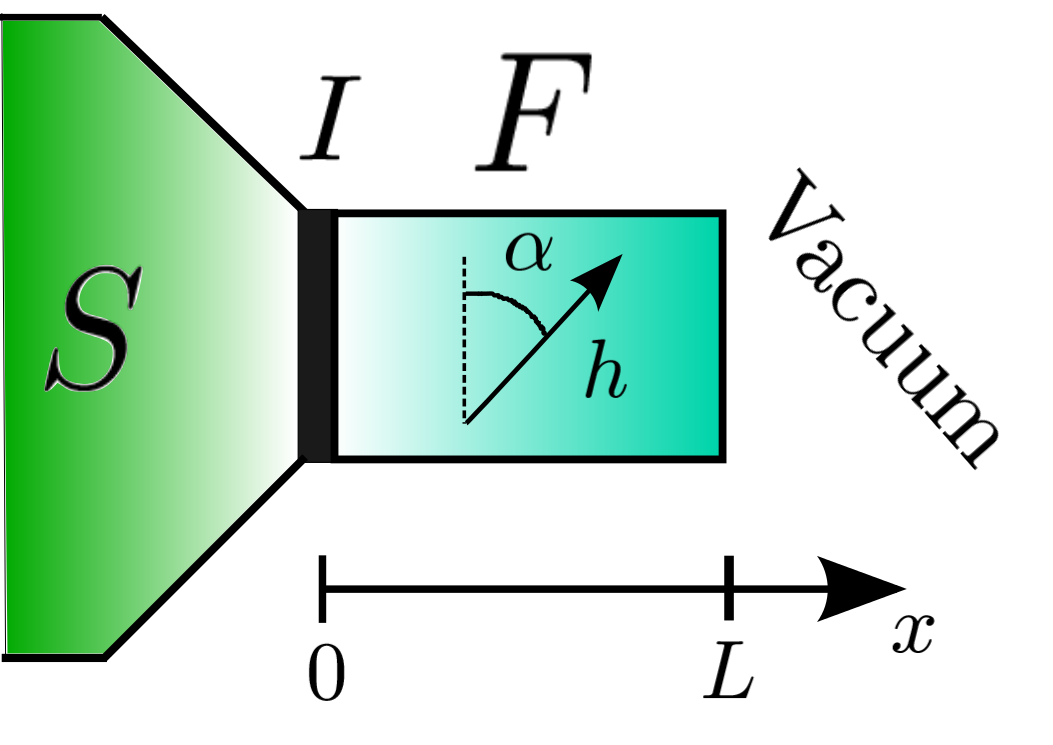}
  \caption{The SIF junction to illustrate the proximity effect.}
\label{fig:sfprox1}
\end{figure}

The GFs of a ferromagnet under the proximity effect have a non-trivial spin structure. They read,
\begin{equation}
\check{G}^{R(A)}_F= \tau_3  (g^{R(A)} \sigma_0 + g^{R(A)}_2 \sigma_2 + g^{R(A)}_3 \sigma_3) + \tau_1 (f^{R(A)}_0 \sigma_0 + f^{R(A)}_2 \sigma_2 + f^{R(A)}_3 \sigma_3) \; .
\end{equation}
Here $g_i$, $i=\emptyset,2,3$, are the normal components of the GFs. In general there are three anomalous components, $f_0$ is the singlet, whilst $f_3$, $f_2$ are the triplet components with zero and $\pm 1$ projections on the spin quantization axis respectively. In this case, the z axis. This means that in order for the $f_2$ and $f_3$ components to be finite, ferromagnets, spin polarisers or externally applied magnetic fields must be present in the system (the exact conditions are explained in sections \ref{sec:rotate}, \ref{sec:spinfilters}, and the current section). In the absence of exchange fields in the system, the only finite anomalous component is $f_0$. This is determined by means of the Usadel equation, complemented by the boundary conditions. 

Determining whether the $f_2$ or $f_3$ triplet term is the short range component depends on the definition of the quantization axis. That with zero projection on the axis is the short range component. The long range triplet component describes Cooper pairs with parallel spins  which survive the strong exchange splitting and can diffuse into ferromagnets over larger distances compared to the singlet component.~\cite{Bergeret, Klapwijk, Blamire, Birge}. It is shown in section \ref{sec:ferro} that while $f_0(E,x)$ and $f_3(E,x)$ decay into the ferromagnet over the  magnetic length $\sqrt{\mathcal{D}/2h}$ the long-range component $f_2(E,x) $ decays over the  length given by $\sqrt{\mathcal{D}/2E}$.  
 
In section\ref{sec:proximityeffect}, we introduced the concept of linearised GFs and in this section we continue to focus on it in more detail. For that purpose we present the linearised Usadel equation, that contains only the anomalous component of the GFs, $\check{f}$. The linearised GFs of a ferromagnet reads, 
\begin{equation}
\check{G}^{R(A)}_F \simeq \check{g}^{R(A)} + \delta \check{g}^{R(A)} = \tau_3 \sigma_0 + \tau_1 ( f^{R(A)}_0 \sigma_0 +f^{R(A)}_2 \sigma_2 + f^{R(A)}_3 \sigma_3 )\; .
\end{equation} 
Let us now study a SIF system where F is a ferromagnet under the influence of the proximity effect, as depicted in fig.~\ref{fig:sfprox1}. After manipulating the linearised Usadel equation for the Retarded component reads (we omit the R subscript),
\begin{equation}
\mathcal{D} \partial^2_x \check{f}= 2 E \check{f} - \left\lbrace \textbf{ h} \hat{\textbf{ S}}, \check{f}  \right\rbrace \; , \quad \check{J}= \partial_x \check{f} \; .
\end{equation}
Here $\left\lbrace .,. \right\rbrace$ stands for the anticommutator. We also show the non linearised Usadel equation obtained from eq.~\ref{eq:FullUsadel},
\begin{equation}\label{eq:Usadel3D2}
\mathcal{D} \partial_x \hat{J}^{R(A)} = \left[ \tau_3 (E- \textbf{h} \hat{\textbf{S}}),
\hat{G}_F^{R(A)} \right] \; .
\end{equation}
The result strongly depends on the direction of the exchange field. For example, in a ferromagnet with an exchange field in the $z$ direction, $\textbf{ h} \hat{\textbf{S}} = h \sigma_z$, the Usadel equation~\ref{eq:Usadel3D2} has the form,
\begin{equation}\label{eq:Usadel2}
\mathcal{D} \partial_x \hat{J} = \left[ \tau_3 (E- h \sigma_z),
\hat{G}_F \right].
\end{equation}
This is the simplest case for the appearance of an exchange field in the system. The direction of the exchange field is arbitrary but by rotating the GFs it can be reduced to this case. Assume that the exchange field takes the form, 
\begin{equation}\label{eq:noncollinear}
\textbf{ h} \boldsymbol{\sigma} = h \sigma_z \exp(-i \sigma_x \alpha)=h \sigma_z( \cos{\alpha} \sigma_0 - i \sin{\alpha} \sigma_x)=h (\sin{\alpha} \sigma_y + \cos{\alpha} \sigma_z ) \; . 
\end{equation}
The above represents an exchange field rotated from the \textbf{ z} axis to the \textbf{ y} axis, spun through $\alpha$ degrees. In this case it is convenient to introduce the Green functions rotated in spin-space~\cite{Bergeret2002},
\begin{equation}\label{eq:gauge}
\widetilde{\breve{G}} = R^\dagger \breve{G} R, \quad R =
\exp\left( i \sigma_x \alpha/2 \right) \; .
\end{equation}
In this new basis the exchange field is now located in the \textbf{ z} direction and the  rotated function $\widetilde{\breve{G}}$ can be determined by \Eq{eq:Usadel2}. Once the result is obtained, by applying the inverse rotation we can obtain the solution for the original GF. This method is especially useful when the direction of the exchange fields varies with the position, which is discussed in section\ref{sec:rotate}.

Let us assume that the proximity effect in the SIF system, with exchange field in the "z" direction, is weak. Thus, we linearise the Usadel equation (eq.\ref{eq:Usadel2}) with all anomalous components, which gives us,
\begin{equation}
\mathcal{D} \partial^2_x \check{f}= 2 E \check{f} - \left\lbrace  h \sigma_z , \check{f}  \right\rbrace = 2 E \check{f} + h (f_0 \sigma_z + f_3 \sigma_0).
\end{equation}
This is a matrix equation that contains three scalar equations. By solving the eigenvalue problem, we obtain,
\begin{subequations}
\begin{equation}
\partial^2_x f_\pm = k^2_\pm f_\pm, \quad k_\pm = \sqrt{\frac{2(E \mp h)}{i\mathcal{D}}} \; ,
\end{equation}
\begin{equation}
\partial^2_x f_2 = k^2_2 f_2, \quad k_2 =\sqrt{\frac{2E}{i\mathcal{D}}} \; .
\end{equation}
\end{subequations}
Here $f_\pm= f_0 \pm f_3$ and $\mathcal{D}$ is the diffusion constant. The $k_\pm$ values corresponding to singlet and short range triplet components depend on the exchange field. While $k_2$, corresponding to the long range triplet component is independent of $h$, as both electrons of the Cooper pairs are parallel to the exchange field. This means that the penetration depth of the $f_2$ component into the ferromagnet is independent of the exchange field, unlike for $f_0$ and $f_3$. Increasing the exchange field of the ferromagnet results in a decrease in the penetration length of the latter components whilst that of the long range component remains constant. This is one of the superconducting nanohybrid topics that has drawn most attention in recent years, mostly due to the wide range of potential applications~\cite{revbergeret}.   

The position dependent solutions of the anomalous components in the ferromagnet can take the exponential or trigonometric form,
\begin{equation}
f_\pm (E,x)=A e^{k_\pm x} + B e^{-k_\pm x}
\end{equation}
or
\begin{equation}
f_\pm (E,x)=A \sinh(k_\pm x)  + B \cosh(k_\pm x)
\end{equation}
and 
\begin{equation}
f_2 (E,x)=C e^{k_2 x} + D e^{-k_2 x}
\end{equation}
or
\begin{equation}
f_2 (E,x)=C \sinh(k_2 x)  + D \cosh(k_2 x) \; .
\end{equation}
Here $A$,$B$,$C$ and $D$ are constants that must determined by the boundary conditions at both side of the ferromagnet, two equations on the right side and another two on the left.   

Considering the case of a long ferromagnet ($L \gg \xi$), we can reduce the anomalous function expressions to just an exponentially decaying factor. We stipulate that very far away from the S there is no proximity effect i.e. $f_{\pm,2}(x \rightarrow \infty)=0$. Thus, we obtain,
\begin{subequations}
\begin{equation}
f_\pm (E,x)= B e^{-k_\pm x},
\end{equation}
\begin{equation}
f_2 (E,x)= D e^{-k_2 x}.
\end{equation}
\end{subequations}
This process is equivalent to imposing a boundary condition at the right side of the ferromagnet. Now we only have two constants (B,D) to be determined by the boundary condition of the left side. In this case that is the K-L relation (eq.\ref{eq:kl}). The geometry is described in fig.~\ref{fig:sfprox1}. 

\begin{figure}[h]
  \centering
  \includegraphics[width=0.5\columnwidth]{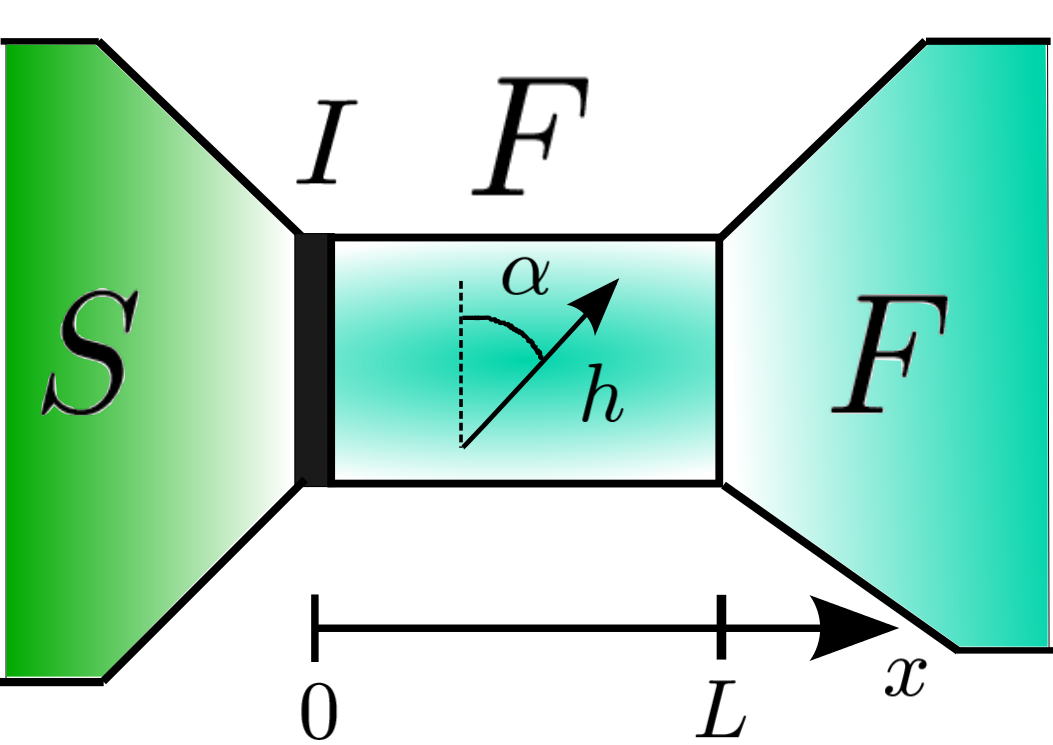}
  \caption{The SIFF tunnel junction. Here the thin ferromagnet is sandwiched between a superconductor and a F electrode.}
\label{fig:sfprox2}
\end{figure}

Another setup closely related with experimental work is for the ferromagnet to be in good contact with a ferromagnetic reservoir at the right end, as shown in fig.~\ref{fig:sfprox2}. Usually the thin ferromagnetic layer is etched from a big ferromagnetic electrode, being a smaller part of the same piece. Until now we have introduced two boundary conditions: Kupriyanov Lukichev, eq.\ref{eq:kl} and vacuum, eq.\ref{eq:vac}. The new boundary condition we introduce here corresponds to \textit{transparent interfaces}. In this case the GF and their derivatives are equal at both sides of the interface.
\begin{equation}
\check{G}_L \bigl|_{x=L-0}=\check{G}_R \bigl|_{x=L+0}, \quad \partial_x \check{G}_L \bigl|_{x=L-0}=\partial_x \check{G}_R \bigl|_{x=L+0} \; .
\label{eq:trans}
\end{equation}
In ferromagnetic reservoirs, as in normal metal ones, the anomalous components equal zero. This corresponds to the condition $f_{\pm,2}(x=L)=0$, which is set at a finite position, $x=L$. Using the trigonometric representation we obtain,
\begin{subequations}
\begin{equation}
f_\pm (E,x)=A (\sinh(k_\pm x)  - \tanh(k_\pm L)  \cosh(k_\pm x)) \; ,
\end{equation}
\begin{equation}
f_2 (E,x)=C (\sinh(k_2 x)  - \tanh(k_2 L)  \cosh(k_2 x) )\; .
\end{equation}
\end{subequations}
Again, only two constants are left to be determined by the equations of the boundary conditions at the left side. Applying the K-L relation, eq.\ref{eq:kl}, to model the tunnel interface at $x=0$ gives,
\begin{subequations}
\begin{equation}
\partial_x f_\pm \bigl|_{x=0} = 2 \kappa_t (g_S f_\pm - f_S) \bigl|_{x=0} \; ,
\end{equation}
\begin{equation}
\partial_x f_2 \bigl|_{x=0} = 2 \kappa_t g_S f_2 \bigl|_{x=0} \; .
\end{equation}
\end{subequations}
It is easy to see that in this case the $f_2$ component does not have a "source term", whilst for $f_\pm$, $f_S$ is given by eq.\ref{eq:gsfs}. This leads us to the following $f_{\pm,2}$ anomalous components in a $SIF$ structure for the linearised case,
\begin{subequations}
\begin{equation}
f_\pm =\frac{-2 \kappa_t f_s}{k_\pm - 2 \kappa_t g_S \tanh(k_\pm L) } (\sinh(k_\pm x)  - \tanh(k_\pm L)  \cosh(k_\pm x)) \; ,
\end{equation}
\begin{equation}
f_2 =0 \; .
\end{equation}
\end{subequations}
Remember that $f_0=(f_+ + f_-)/2$ and $f_3=(f_+ - f_-)/2$. Unlike the singlet and short range triplet components, the long range triplet component is not present. We have found the explicit expressions of the linearised GFs in a SIF system with a small proximity effect. This is the first example in which the anomalous short range triplet component is present.

\paragraph{Arbitrary polarization direction for ferromagnets}

We apply the Usadel equation to a ferromagnet with an arbitrary exchange field direction with small proximity effect. The ferromagnet is in contact with a superconductor but due to the tunnelling barrier between them, the anomalous component can be linearised. The exchange field of the ferromagnet is not parallel to the z axis and instead of the method mentioned in section\ref{sec:ferro}, where the GFs are rotated, we proceed with the original ones. We define $\textbf{ h}=h(0,\sin \alpha, \cos \alpha)$ and the linearised Usadel equation now reads,
\begin{equation}
\partial_x^2 \check{f}-\kappa_E^2 \check{f}- \frac{i \kappa_h^2}{2} \{ \textbf{ h} \boldsymbol{\sigma}, \check{f} \}=0 \; .
\end{equation}
Here $\boldsymbol{ \sigma}=(\sigma_1,\sigma_2,\sigma_3)$ and $\check{f}=f_0 + f_2 \sigma_2 + f_3 \sigma_3$. Also, 
\begin{equation}
\kappa_E^2=2E/\mathcal{D} \quad \kappa_h^2=2 h /\mathcal{D} \; .
\label{eq:kom}
\end{equation}
Again, we have a value of $\kappa$ that is independent of the exchange field and one that is dependent on it. So we obtain,
\begin{subequations}
\begin{equation}
\partial_x^2 f_0-\kappa_E^2 f_0-i \kappa_h^2 (\cos \alpha f_3 + \sin \alpha f_2)=0
\end{equation}
\begin{equation}
\partial_x^2 f_3-\kappa_E^2 f_3-i \kappa_h^2 \cos \alpha f_0 =0
\end{equation}
\begin{equation}
\partial_x^2 f_2-\kappa_E^2 f_2-i \kappa_h^2 \sin \alpha f_0 =0 \; .
\end{equation}
\end{subequations}
In order to solve this equation system, we must obtain the eigenvalues and eigenvectors, where the following determinant must be equal zero,
\begin{equation}
\left| \begin{array}{ccc}
\kappa^2-\kappa_E^2 & -i \kappa_h^2 \cos \alpha & -i \kappa_h^2 \sin \alpha \\
-i \kappa_h^2 \cos \alpha & \kappa^2-\kappa_E^2 & 0 \\
-i \kappa_h^2 \sin \alpha  & 0 & \kappa^2-\kappa_E^2 \end{array} \right|=0 \; .
\end{equation}
As we assumed that the general solutions reads,
\begin{subequations}
\begin{equation}
f_0=A_1 e^{-\kappa x} +A_2 e^{\kappa x}
\end{equation}
\begin{equation}
f_3=B_1 e^{-\kappa x} +B_2 e^{\kappa x}
\end{equation}
\begin{equation}
f_2=C_1 e^{-\kappa x} +C_2 e^{\kappa x} \; .
\end{equation}
\end{subequations}
This results in the following three eigenvalues $\kappa_0, \kappa_\pm$,
\begin{equation}
\kappa_0=\kappa_E \quad \kappa_\pm= \kappa_E^2 \pm i \kappa_h \; .
\end{equation}
The following relations are also extracted,
\begin{subequations}
\begin{equation}
A_0=0 \quad B_{1_0} \cos \alpha = - C_{1_0} \sin \alpha
\end{equation}
\begin{equation}
 B_{\pm}= \pm \cos \alpha A_\pm \quad C_\pm = \pm \sin \alpha A_\pm \; .
\end{equation}
\end{subequations}

The general solutions then reads,
\begin{equation}
\left( \begin{array}{ccc}
f_0  \\
f_3  \\
f_2  \end{array} \right)=
\left( \begin{array}{ccc}
0  \\
-\sin \alpha  \\
\cos \alpha  \end{array} \right) (A_0 \cosh \kappa_0 x + B_0 \sinh \kappa_0 x)
\nonumber
\end{equation}
\begin{equation}
\left( \begin{array}{ccc}
1  \\
\cos \alpha  \\
\sin \alpha  \end{array} \right) (A_+ \cosh \kappa_+ x + B_+ \sinh \kappa_+ x)+
\left( \begin{array}{ccc}
1  \\
-\cos \alpha  \\
-\sin \alpha  \end{array} \right) (A_- \cosh \kappa_- x + B_- \sinh \kappa_- x) \; .
\end{equation}
In order to obtain the anomalous components of the Green functions anomalous components in the arbitrary direction ferromagnet we now need to apply the b.c. corresponding to the two ends. Note that when $\alpha=0$, the exchange field aligns to the \textbf{ z} axis. The rotation of the GFs is an equivalent method to this one. As mentioned before, which might be considered intuitive, the long range triple component is not finite in this case. This is related to the fact that the source component of the superconductor is an anomalous singlet component, this means mixing between both components is required for the long range triplet to be finite.

\subsubsection{Ferromagnets with spatially rotating exchange field in superconducting hybrid structures}
\label{sec:rotate}

\begin{figure}[h]
  \centering
  \includegraphics[width=0.5\columnwidth]{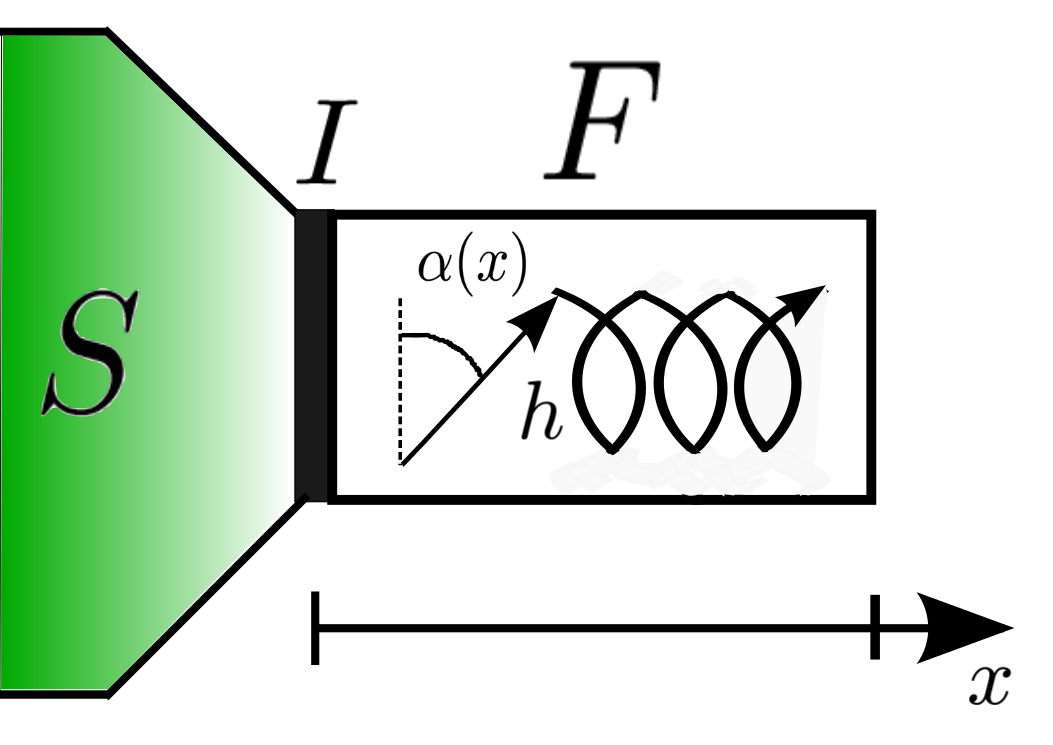}
  \caption{The SIF tunnel junction with a rotating exchange field. }
\label{fig:sfproxrot}
\end{figure}

We now consider a domain-wall type of structure, which consists of a F layer with a nonhomogeneous magnetization, i.e. position dependent magnetization, which is shown in fig.~\ref{fig:sfproxrot}. For simplicity we assume that the exchange field vector takes the form  $\textbf{ h}=h(0,\sin \alpha(x), \cos \alpha(x))$, as it rotates with position. The rotation angle has a simple position dependence $\alpha= Q x$. This means that at $x=0$ the exchange field vector is aligned parallel to the z axis and at $x=\pi/Q$ antiparallel. 

We want to calculate the condensate function in this layer, so we assume it to be in contact with a bulk superconductor. We also assume that due to a tunnelling barrier or a mismatch of the Fermi velocities between the superconductor and the ferromagnet, the condensate function in the F region is small. Finding the condensate functions now that the magnetization varies continuously in space is more difficult. This is when the rotation of GFs become an extremely useful tool. We introduce a rotation in Nambu and spin space that depends on position. We define a new matrix $\hat{\tilde{f}}$ related to $\hat{f}$ by a unitary transformation,
\begin{equation}
\hat{f}=\hat{R} \hat{\tilde{f}} \hat{R}^\dagger \; ,
\end{equation} 
where $\hat{R}=\exp[i \tau_3 \sigma_1 \alpha(x)/2]$. Performing this transformation the linearised Usadel equation for $\hat{\tilde{f}}$ now reads,
\begin{equation}
(\partial_x^2-Q^2/2)\hat{\tilde{f}} - \kappa_E^2 \hat{\tilde{f}} + \frac{i \kappa_h^2}{2} \{ \sigma_3, \hat{\tilde{f}} \} - \frac{Q^2}{2} \sigma_1 \hat{\tilde{f}} \sigma_1 + i Q \tau_3 \{ \sigma_1, \partial_x \hat{\tilde{f}} \}=0 \; .
\end{equation}

For the boundary condition it is important to show that the first position derivative of the anomalous component transforms in the following way,
\begin{equation}
\partial_x \hat{f} \Rightarrow \partial_x \hat{\tilde{f}} + i \tau_3 (Q/2) \{\sigma_1, \hat{\tilde{f}} \} \; .
\end{equation}

The eigenvalues $\kappa$ now obey the following equation,
\begin{equation}
[(\kappa^2-Q^2-\kappa_E^2)^2+4Q^2 \kappa_h^2](\kappa^2-\kappa_E)+\kappa_h^4 [\kappa^2 -Q^2-\kappa_E]=0 \; ,
\end{equation}
where $\kappa_{E,h}$ were defined in eq.~\ref{eq:kom}. Now we obtain two solutions for $\kappa$, $\kappa_Q^2= Q^2+\kappa_E^2$ and $\kappa_\pm$ that remains unchanged, provided $Q,\kappa_E \ll \kappa_h$. 
In the opposite limit where $Q\gg \kappa_h$, the eigenvalues $\kappa_\pm$ take the form
\begin{equation}
\kappa_\pm=\pm i Q [1 \mp i \kappa_h^2/\sqrt{2}Q^2] \; .
\end{equation}
This quantity is imaginary in the main approximation, which means that the function $\hat{\tilde{f}}$ oscillates fast in space with period $2\pi/Q$. In this case the second eigenvalue takes the form $\kappa^2=\kappa_E^2+\kappa_h^4/Q^2$. Therefore the limit of a very fast rotating magnetization ($\kappa_h/Q\rightarrow 0$) is analogous to the case of a normal metal. However, more interesting and realistic is the opposite limit, $Q,\kappa_E \ll \kappa_h$, where the long range penetration of the triplet component into the ferromagnet becomes possible. In the limit of large $\kappa_h$, the singlet component penetrates into the ferromagnet over a short distance of the order $\xi_F=1/\kappa_h$, while the long range triplet penetrates over a length $\sim 1/\kappa_Q$.

In conclusion the general solutions now reads,
\begin{subequations}
\begin{equation}
f_0=-A_+ e^{-\kappa_+ x}+ A_- e^{-\kappa_- x} \; ,
\end{equation}
\begin{equation}
f_3=A_+ e^{-\kappa_+ x}+ A_- e^{-\kappa_- x} \; ,
\end{equation}
\begin{equation}
f_2=B e^{\kappa_Q x} + \bar{B} e^{-\kappa_Q x} \; .
\end{equation}
\end{subequations}
This lead us to
\begin{equation}
\left( \begin{array}{ccc}
\tilde{f}_0  \\
\tilde{f}_3  \\
\tilde{f}_2  \end{array} \right)= 
\left( \begin{array}{ccc}
1  \\
\Gamma_+  \\
\Omega_+ \Gamma_+  \end{array} \right) A_+ e^{-\kappa_+ x}+\left( \begin{array}{ccc}
1  \\
\Gamma_+  \\
-\Omega_+ \Gamma_+  \end{array} \right) \bar{A}_+ e^{\kappa_+ x}
\nonumber
\end{equation}
\begin{equation}
+\left( \begin{array}{ccc}
1  \\
\Gamma_-  \\
\Omega_- \Gamma_-  \end{array} \right) A_- e^{-\kappa_- x}+\left( \begin{array}{ccc}
1  \\
\Gamma_-  \\
-\Omega_- \Gamma_-  \end{array} \right) \bar{A}_- e^{\kappa_- x}
\nonumber
\end{equation}
\begin{equation}
+\left( \begin{array}{ccc}
1  \\
\Gamma_Q  \\
\Omega_Q \Gamma_Q  \end{array} \right) A_Q e^{-\kappa_Q x}+\left( \begin{array}{ccc}
1  \\
\Gamma_Q  \\
-\Omega_Q \Gamma_Q  \end{array} \right) \bar{A}_Q e^{\kappa_Q x} \; .
\end{equation}
Here we define,
\begin{equation}
\Gamma_{\pm,Q}=\frac{\kappa_{\pm,Q}^2-\kappa_E^2}{i \kappa_h^2} \quad 
\Omega_{\pm,Q}=\frac{2\kappa_{\pm,Q} Q}{\kappa_{\pm,Q}^2-Q^2-\kappa_E^2} \Gamma_{\pm,Q} \; .
\end{equation}

After obtaining the general solutions for the system, we just have to apply the boundary conditions at the two ends in order to obtain the Green functions. This is the only system described up to now in which the long range triplet component is finite.

\subsubsection{Spin Filters in superconducting hybrid structures}
\label{sec:spinfilters}

So far, the boundary conditions we presented (eq.\ref{eq:kl}, eq.\ref{eq:trans} and eq.\ref{eq:vac}) are not spin-dependent. The first set of boundary conditions that describes spin dependent transport in normal systems has been proposed by Brataas, Nazarov, and Bauer~\cite{bnb}. These boundary conditions are able to describe interfaces with spin-dependent transmission, so-called \textit{spin filters}. They have been generalized for the superconducting structures by Cottet et. al.~\cite{cottet}. However, the later boundary conditions are only valid for the case of a small spin-polarization of the barrier. More recently Bergeret et. al.~\cite{BVV1,BVV2} derived from the tunnelling Hamiltonian effective boundary conditions for the description of spin-filtering barriers for arbitrary spin polarization. We shall present these conditions below and use them in several examples.    

%Thus far we have introduced the following boundary conditions: Kupriyanov-Lukichev, eq.\ref{eq:kl}, transparent interface, eq.\ref{eq:trans}, and vacuum, eq.\ref{eq:vac}. We also mentioned the Nazarov boundary condition that generalizes K-L for arbitrary transmission values and will be introduced in section\ref{sec:nazarov}. In the field of spin dependent transport, Brataas, Nazarov and Bauer~\cite{bnb} formulated a theory of an electronic circuit involving ferromagnetic elements with non-collinear magnetizations, which was based on the conservation of spin and charge current. This was later developed by Cottet et. al.~\cite{cottet} deriving spin-dependent boundary conditions for isotropic GFs for contact encompassing superconducting and ferromagnetic correlations. However, their use requires the knowledge of the full scattering matrix of the contact, which is not usually available for realistic interfaces. In the case of weakly polarized tunnel interfaces, the boundary conditions can be expressed in terms of a few parameters, i.e. the tunnel conductance of the interface and five conductance-like parameters accounting for the spin-dependence of the interface scattering amplitudes. However, this calculation requires really complex and sometimes cumbersome calculations. In order to avoid this, yet another model was proposed by Bergeret et. al.~\cite{BVV1,BVV2} to model tunnelling barriers of spin filtering materials, which generalizes the Kuprianov-Lukichev b.c. in a simple way. We shall describe and use this model in this text. 

%
\begin{figure}[h]
  \centering
  \includegraphics[width=0.5\columnwidth]{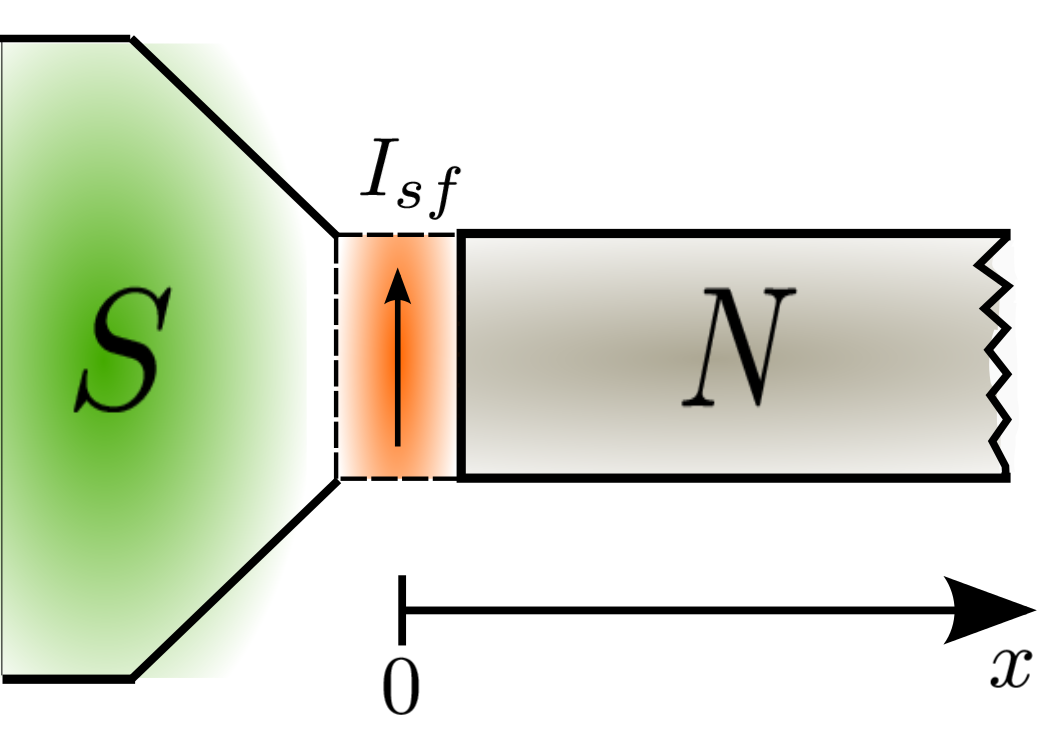}
  \caption{The S$I_{sf}$N  tunnel junction with a spin filter barrier. The normal metal is assumed to be infinitely long.}
\label{fig:snspin}
\end{figure}

The tunnelling amplitude for the spin up (down) case for the spin filter is given by $T_{\uparrow (\downarrow )}=\mathcal{T}\pm \mathcal{U}$. So $\mathcal{U}=(T_{\uparrow}-T_{\downarrow})/2$ and $\mathcal{T}=(T_{\uparrow}+T_{\downarrow})/2$ are the spin dependent and independent  tunnelling matrix elements, respectively. We also define the spin-filtering efficiency of the barrier as,
\begin{equation}
 P= \frac{T_{\uparrow}^2-T_{\downarrow}^2}{T_{\uparrow}^2+T_{\downarrow}^2}= \frac{2 \mathcal{T} \mathcal{U}}{\mathcal{T}^2+\mathcal{U}^2} \; . 
\end{equation}
In order to obtain the non spin filter barrier of section\ref{sec:kupri} both tunnelling amplitudes must be equal, $T_\uparrow=T_\downarrow$. This means that $ \mathcal{U}=0$ and $\mathcal{T}=T_\uparrow$, so the efficiency takes the value $P=0$, as expected. For the case of a fully polarized barrier we set the amplitude $T_\downarrow=0$, we specify that particles with spin down cannot tunnel. So $\mathcal{U}=\mathcal{T}=T_\uparrow /2$ with $P=1$, for a a fully polarized spin-filtering barrier. 

We introduce a new tunnelling matrix, $\check{\Gamma}=\mathcal{T}\hat{\tau}_{0}\hat{\sigma}_{0}+\mathcal{U} \hat{\tau}_{0}\mathbf{\hat{\sigma}}_{3}$. The new b.c. fro the GFs of the left side at the spin filter read,
\begin{equation}\label{eq:bcSF}
\breve{J}\bigl|_{x=SF}=\breve{G}_L \partial_x \breve{G}_L\bigl|_{x=SF}=-\frac{(R_I \mathcal{A} \sigma_L)^{-1}}{\mathcal{T}^2+\mathcal{U}^2}\left[\breve{G}_L,\check{\Gamma}\breve{G}_{R}\check{\Gamma}\right] ,
\end{equation}
with the equivalent for the GFs of the Right side written as,
\begin{equation}
\breve{J}\bigl|_{x=SF}=\breve{G}_R \partial_x \breve{G}_R\bigl|_{x=SF}=\frac{(R_I \mathcal{A} \sigma_R)^{-1}}{\mathcal{T}^2+\mathcal{U}^2}\left[\breve{G}_R, \check{\Gamma} \breve{G}_L \check{\Gamma} \right]_{x=SF}.
 \label{current}
\end{equation}
Here we define $\kappa_{t/R(L)}=(R_I \mathcal{A} \sigma_{R(L)})^{-1}$, as described in section\ref{sec:kupri}. Here the subscripts stand for Left and Right sides of the spin filter. The spectral matrix functions relate in the following way,
\begin{equation}
  \text{Tr} (\breve{G}_R \partial_x \breve{G}_R)\bigl|_{x=SF}= -\text{Tr} (\breve{G}_L \partial_x \breve{G}_L)\bigl|_{x=SF} \; .
 \label{current}
\end{equation}
The commutation property of the trace is required for this equality.
%here $\gamma= \sigma_R/ \sigma_L$, the conductivity mismatch and the trace is required for its commutative properties. From this expression we can observe that a difference in conductivities generates the same effect as an increase of barrier resistance as this prefactor to the commutator multiplies $\kappa_t$. In our calculation we usually neglect this difference and set $\gamma =1$. 

For clarity, we use the new b.c. to study the proximity effect in a simple $NI_{sf}S$ system with  a spin-filtering barrier ($I_{sf}$), corresponding to fig.~\ref{fig:snspin}. Here we assume a weak proximity effect in N. We obtain (omitting the R upperscript),
\begin{equation}
\partial_x f_N \bigl|_{x=0}=- r \kappa_t f_S \; .
\end{equation}
Here we have defined the spin filter parameter as $r=(2 T_{\uparrow} T_{\downarrow})/(T^2_{\uparrow}+T^2_{\downarrow})$. The latter is related to the spin-filter efficiency of the barrier $P$ by the expression $r=\sqrt{1-P^2}$. Therefore, for a fully polarized barrier $r=0$, while for a non-polarized one $r=1$. 

We obtain the value of the anomalous GF in the normal metal from the linearised Usadel equation,
\begin{equation}
\partial^2_x f_N= k^2_E f_N \; .
\end{equation} 
That, complemented with the boundary condition at $x=0$ gives us,
\begin{equation}
f(x)=\frac{r \gamma \kappa_t f_S}{k_E} e^{-k_E x} \; .
\end{equation}
Here $(k_E)^2= 2 i E / \mathcal{D}$. Thus, the amplitude of the induced condensate is proportional to the spin filter parameter $r$. In particular, the proximity effect is completely suppressed if the barrier is fully spin polarized ($r=0$), as in this case where we are only allowing tunnelling of particles with spin up. Although this result is quite obvious it has not been obtained previously without using this new b.c..

\begin{figure}[h]
  \centering
  \includegraphics[width=0.5\columnwidth]{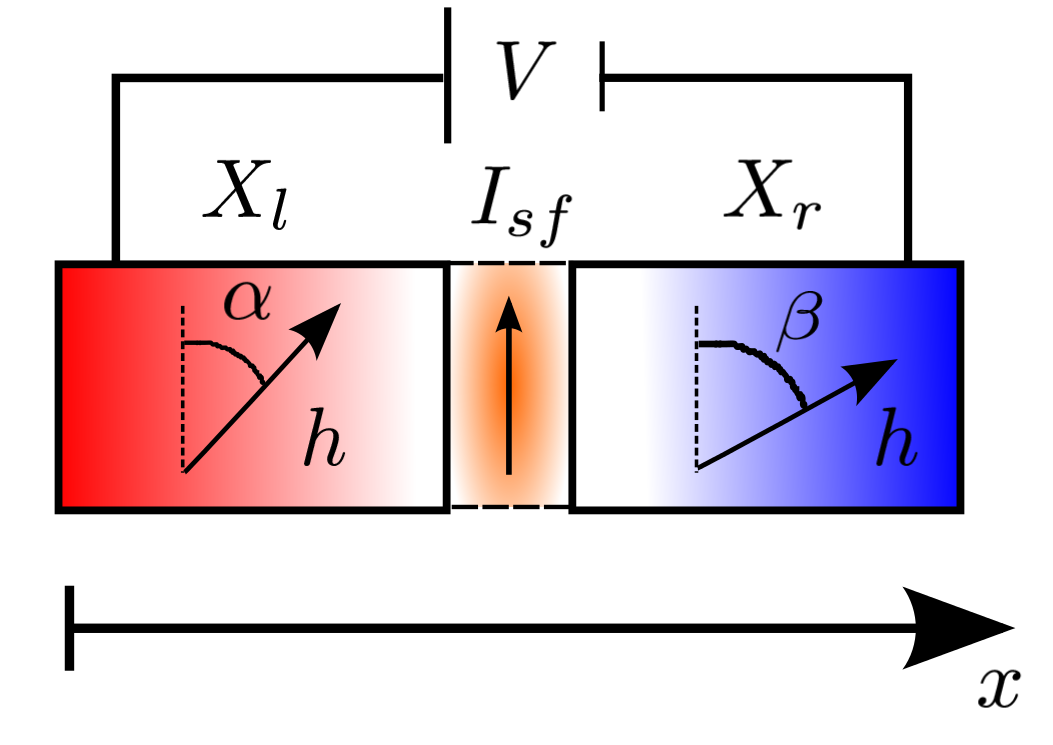}
  \caption{The $X_l/I_{sf}/X_r$ tunnel junction.}
\label{fig:xsfx}
\end{figure}

Other types of junctions can also be described using the general form of this new boundary condition. Consider a generic tunnel junction $X_l/I_{sf}/X_r$ as shown in fig.~\ref{fig:xsfx}, the left and right electrodes, $X_l$ and $X_r$, posses intrinsic ferromagnetic fields with their corresponding directions. The layer $I_{sf}$ ($x=0$) is a spin-filter barrier, i.e. a spin-dependent tunnelling barrier. A rotation in the (y,z) plane is described by the matrix
\begin{equation}
\check{R}_\alpha=\cos(\alpha/2)+i \tau_3 \sigma_1 \sin(\alpha/2),
\end{equation}
where $\alpha$ is the rotation angle with respect to the spin filtering axis. It is important to know that all directions, in this b.c., are defined with respect to the spin filtering axis. The generalized BCs now read,
\begin{equation}
\breve{J}\bigl|_{x=0}=\breve{G} \partial_x \breve{G}\bigl|_{x=0}=\frac{\kappa_t}{\mathcal{T}^2+\mathcal{U}^2}\left[ \check{R}^\dagger_\alpha \check{\Gamma} \check{R}_\beta \check{G}_r \check{R}^\dagger_\beta \check{\Gamma} \check{R}_\alpha,\check{G}_l \right].
\end{equation}
That can be written taken into account the commutation properties as,
\begin{equation}
\breve{J}\bigl|_{x=0}=\breve{G} \partial_x \breve{G}\bigl|_{x=0}=\frac{\kappa_t}{\mathcal{T}^2+\mathcal{U}^2}\left[ \check{\Gamma} \check{R}_\beta \check{G}_r \check{R}^\dagger_\beta \check{\Gamma}, \check{R}_\alpha \check{G}_l  \check{R}^\dagger_\alpha \right].
\end{equation}
The Greens functions $\check{G}_{l,r}$ correspond to the case of the magnetization vector oriented parallel to the "z" axis. In a similar way as we did with the ferromagnets, we are rotating the GF relative to the arbitrary (relative to \textbf{ z}) orientation of the exchange field.  Note that for the $\alpha=\beta=0$ case we obtain eq.\ref{eq:bcSF}. In order to have more compact notation we can define $\check{\Gamma}_{\alpha \beta}= \check{R}^\dagger_\alpha \check{\Gamma} \check{R}_\beta$.

One should have in mind that by going over to the quasiclassical Green functions we lose the spin dependence of the DoS in the normal state. In that case the retarded (advanced) Greens functions $\check{g}^{R(A)}_{l,r}$ in the ferromagnet have a trivial structure in spin-space, $\check{g}^{R(A)}_F=\pm \tau_3 \sigma_0$, so that the normalized density of states is the same for spin up and down. This approach is valid for electrodes with small spin-splitting at the Fermi level and was used for the calculation of the Josephson current through a $SI_{sf}S$ junction. However, if the spin-polarization of the electrodes at the Fermi Level is large enough one has to use this new boundary condition to compute the current. 

We emphasize that the derivation of the b.c. cannot be regarded as a microscopic derivation. However, it has been shown that this boundary condition give correct physical results, and hence they can be used, for example, for the calculations of the tunnel current in $S_MI_{sf}S_M$ junctions and for the study of the proximity effect in $I_{sf}S_M$ and other systems as shown in ref.~\cite{BVV1,BVV2}.

\subsubsection{Ferromagnetic superconductors}\label{sec:superferro}

\begin{figure}[h]
  \centering
  \includegraphics[width=0.5\columnwidth]{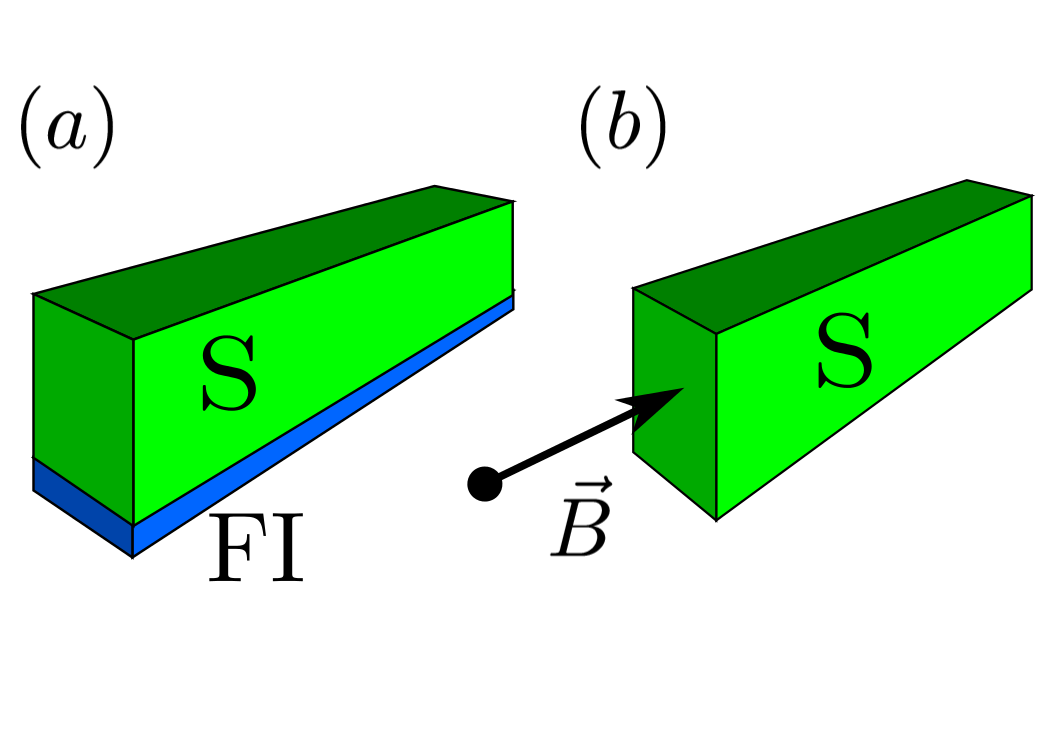}
  \caption{Two ways to create a "ferromagnetic superconductor": (a) by placing a ferromagnetic insulator in contact with a conventional superconductor, or (b) by applying an in-plane magnetic field. }
\label{fig:supfergeo}
\end{figure}

The \textit{ferromagnetic superconductors} play a major role in the calculations. Here the presence of a finite exchange field inside the superconductor generates a spin splitting in the DoS. This exchange field can be due to a Zeeman effect generated by an applied magnetic field~\cite{Tedrow1971,catelani08} (fig.~\ref{fig:supfergeo} (b)) parallel to the easy axis of the ferromagnet or from a magnetic proximity effect with either a ferromagnetic insulator~\cite{Moodera1990,Moodera2013,Tokuyasu88,xiong11} or with a thin ferromagnetic metallic layer~\cite{Bergeret2001, giazotto07} placed directly below the superconductor (fig.~\ref{fig:supfergeo} (a)). It shifts the energies of electrons with parallel and antiparallel spin orientations to opposite directions. This breaks the electron-hole symmetry for each spin separately, but conserves charge neutrality, as the total density of states remains electron-hole symmetric.

\begin{figure}[h]
  \centering
  \includegraphics[width=0.5\columnwidth]{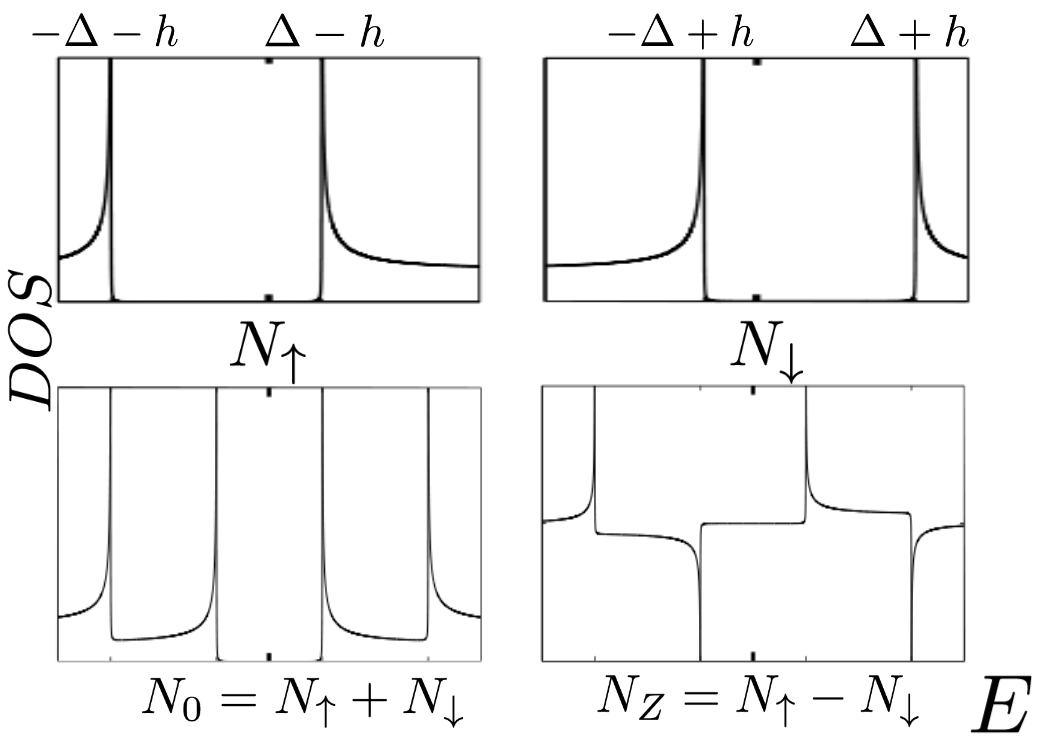}
  \caption{Spin up $N_\uparrow$, spin down $N_\downarrow$, symmetric $N_0= N_\uparrow+N_\downarrow$ and antisymmetric  $N_Z= N_\uparrow-N_\downarrow$ DoS of a ferromagnetic superconductor.}
\label{fig:supfer}
\end{figure}

In this formalism the main difference with a BCS superconductor is the non trivial spin space structure of the GFs. They correspond to those of a BCS superconductor shifted depending on the direction of the spin. Assuming that the exchange field is set in the \textbf{ z} direction they read,
\begin{subequations}
%\begin{align}
\begin{equation}
\breve{G}_S = \tau_3 \sigma_0 g +  \tau_3 \sigma_3 g_z + \tau_1 \sigma_0 f + \tau_1 \sigma_3 f_z \; , 
\label{eq:GFsf}
\end{equation}
\\
\begin{equation}
f_{(z)}=(f_\uparrow \pm f_\downarrow)/2, g_{(z)}=(g_\uparrow \pm g_\downarrow)/2 \; ,
\end{equation}
\\
\begin{equation}
(g_{\uparrow,\downarrow}, f_{\uparrow,\downarrow}) = \frac{1}{2}(E \pm h, \; i\Delta)/\epsilon, \quad \epsilon =
\sqrt{(E \pm h+i\eta)^2 - \Delta^2} \; . \label{uv}
%\end{align}
\end{equation}
\end{subequations}
Here we have omitted the Retarded upperscript and $\uparrow$,$\downarrow$ correspond to the orientation of the spin in the \textbf{ z} axis. In fig.~\ref{fig:supfer} we plot the density of states of all the components. First, we introduce $N_{\uparrow,\downarrow}=g^R_{\uparrow,\downarrow}-g^A_{\uparrow,\downarrow}$ that corresponds to the density of states of spin up and down particles. They are just BCS DoS, eq.\ref{eq:gsfs}, shifted by the exchange field in opposite directions (and divided by two). Next, we define $N_0=N_{\uparrow}+N_{\downarrow}$ as the regular DoS used up to now. It is just the sum of the two spins and is symmetric in energy.  The new DoS, $N_Z=N_{\uparrow}-N_{\downarrow}$, asymmetric in energy, is finite only in the case in which a finite exchange field is present. In other words, when the DoS of spin up and down are different, which is the reason it has not appeared earlier in the text.  

Superconductivity is suppressed in the presence of a exchange field and can never exceed the Chandrasekhar-Clogston limit, $h=0.707 \Delta_0$. A detailed study of the self consistent $\Delta$ dependence with  temperature and exchange field was presented in Sec.~\ref{sec:self}.

\newpage

\subsection{Superconducting Point Contacts}
\label{sec:sqpc}

\begin{figure}[h]
  \centering
  \includegraphics[width=0.5\columnwidth]{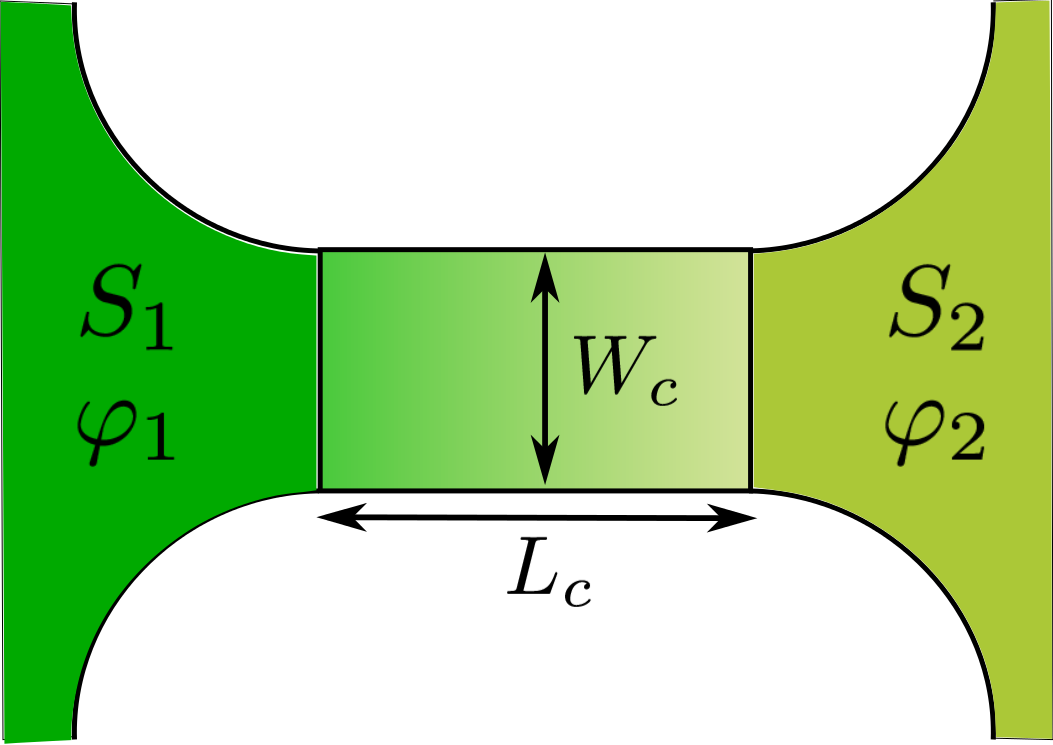}
  \caption{Schematic representation of a superconducting quantum point contact.}
\label{fig:sqpcsketch}
\end{figure}

\begin{figure}[h]
  \centering
  \includegraphics[width=0.5\columnwidth]{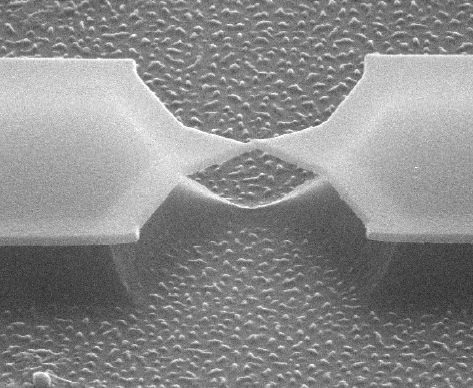}
  \caption{ Experimental picture of a quantum point contact.(From Ref.~\cite{Chauvin})}
\label{fig:sqpcexp}
\end{figure}

In all previous sections we have considered large planar diffusive junctions, \textit{i.e.} junctions with many conduction channels. It is however possible, to fabricate very small contacts between superconductors, with only few channels. Such junctions are called superconducting quantum point contacts (SQPC). As shown in fig.~\ref{fig:sqpcsketch}, the constriction in contact with the two electrodes has a length $L_c$ much smaller than the superconducting coherence length of the electrodes $\xi$. Also, the width of the constriction $W_C$, is of the order of the Fermi length $\lambda_F$ of the electrons in the system. This fact only allows a few conduction channels~\cite{fewconduc}. Inside the definition of a SQPC we can include many systems with very rich physics. With the recent advances in the fabrication of nanoscale devices, a closer comparison between the predictions of simple quasi-one-dimensional theories and experimental results is now at hand.  References~\cite{44} and \cite{55} provide two different examples of the progress made in the experimental field of superconducting quantum point contacts, fig.~\ref{fig:sqpcexp} shows an experimental setup.  It turns out that, in spite of its apparent simplicity, the case of a single-channel SQPC still contains nontrivial physical behavior under certain regimes. For example, the high transmission values allow nontrivial ABS located deep into the gap, which give rise to new interesting physics. 


The properties of quantum transport in a SQPC are similar to those of a waveguide with a potential barrier. Assume that we have a waveguide extended along the x direction with variable cross section, bounded by impenetrable potential walls and a constriction. Different channels in the waveguide have different energies that we compare with the maximum barrier height of the impenetrable barrier. If the energy is bigger than the height, the electrons traverse the constriction, otherwise they are reflected back. This means a certain value of open channels can pass the constriction. Each channel has it own transmission and reflection amplitudes, so we can obtain a channel dependent transmission coefficient. The adiabaticity implies an almost classical potential barrier, for open channels we have $T=1$ and for closed ones $T=0$ (this is strange as there is no potential barrier). The only exception is when the energy interval aligns with the top of the barrier.

The expression for the current in the SQPC takes the simple expression,
\begin{equation}
I=G_{Q}N_{OPEN}V\;.
\end{equation}
Here $V$ is the applied voltage difference, $N_{OPEN}$ stands for the number of
open channels and $G_{Q}=2_{S}e^{2}/(2\pi)$is
the so called conductance quantum.  Conductance appears
to be quantized in units of $G_{Q}$, this unit does not depend on
material properties, nanostructure size or geometry, as it is made
up from fundamental constants.

The conductance of the system, meaning the number of open channels,
is determined by the narrowest part of the waveguide. We can change
the shape of the waveguide and maintain the conductance if we do not
change the narrowest part of it. So if the distance between waveguides goes to infinity we obtain
a quantum point contact (SQPC) instead of a waveguide. What distinguishes waveguides is that they have a finite number of channels at each energy and the spectrum consist of discrete energy
branches. SQPCs have a infinite number of channels and the spectrum is continuous,
but due to the constriction only a finite number are transmitted.

\subsubsection{Nazarov Boundary Conditions}
\label{sec:nazarov}

Structures like SQPC usually show high values of transmission for which the Kupriyanov-Lukichev boundary conditions are not valid. In section\ref{sec:kupri} and \ref{sec:spinfilters} we dealt with tunnel and low transmission interfaces, now the structure shows arbitrary transmission channels and eigenvalues. In order to study this system we have to introduce the last boundary conditions of this manuscript, the \textit{Nazarov boundary conditions}~\cite{nazarovbc}, in which the matrix current reads,
\begin{equation}
\check{J}=\Sigma_n \frac{2\tau_{n}[\check{G_{L}},\check{G_{R}}]}{4-\tau_{n}(2-\{\check{G_{L}},\check{G_{R}\}})} \; .
\label{eq:Jnaz}
\end{equation}
Here the {}``{[}$\ldots${]}'' stands for anticonmutator and {}``\{$\ldots$\}'' for commutator. The scalars $\tau_n$ are transmission eigenvalues of transmission channels $n$. They describe the properties of the interface and take values from 0 to unity. Note that the summation prefactor is over all the channels. For simplicity we assume a single channel, once the result is obtained we can replace the transmission values of each channel and sum them. Bear in mind that the difficulty for using this b.c. comes from the fact that there is a denominator with a nontrivial structure in Keldysh space. 

In this case we are considering many transmission channels while for K-L we only have a single parameter. At the tunnelling limit, when the transmission in the junction is low $\tau_{n}<<1$, equation~\ref{eq:Jnaz} reduces to 
\begin{equation}
\check{J}=\Sigma_n \frac{\tau_{n}}{2}[\check{G_{L}},\check{G_{R}}] \; .
\end{equation}
By direct comparison we can state that $\kappa_t=\Sigma_n T\tau_n/2$. If the transmission values of the channels are low, there can be multiple channels in the K-L case as well, as long as the condition $\kappa_t \ll 1$ is fulfilled. More precisely, we can not talk about channels in this case as this concept is not present in tunnelling. Only with the inclusion of higher transmission values does the concept of transport channels arise. This fact does not change any of the theory related to the tunnelling case but allows us to connect both boundary conditions.

\newpage

\subsection{Critical temperature of a superconductor-ferromagnet multilayered structure}
\label{sec:crit}

\begin{figure}[h]\begin{center}
\includegraphics[width=0.6\columnwidth]{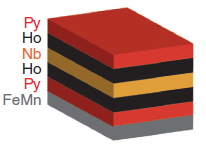}
\caption{An illustration of a Py(8 nm)/Ho/Nb/
Ho/Py(5 nm)/FeMn(5 nm) spin valve.(From Ref.\cite{aonat})}
\label{fig1N}
\end{center}
\end{figure}

In this section we use the quasiclassical equations to calculate the critical temperature $T_C$ of a ferromagnet-superconductor multilayer structure. Our main motivation is to describe the setup shown in Fig.\ref{fig1N} which corresponds to the  experiment of Ref.\cite{aonat}. It consist of a Py(8 nm)/Ho($d_{Ho}$)/Nb($d_{Nb}$)/Ho($d_{Ho}$)/Py(5 nm)/FeMn(5 nm) structure. The bottom Py layer is pinned by exchange bias to an antiferromagnetic layer, whereas the orientation of the free (top) Py layer can be switched by applying an in-plane magnetic field that is greater than its coercive field so that parallel or anti parallel states can be achieved. The Nb is interfaced by Ho, as this rare earth
helimagnet has previously been shown in \cite{18} and by Sosnin et al.~\cite{25} to be a spin-mixer.

\begin{figure}[h]\begin{center}
\includegraphics[width=0.6\columnwidth]{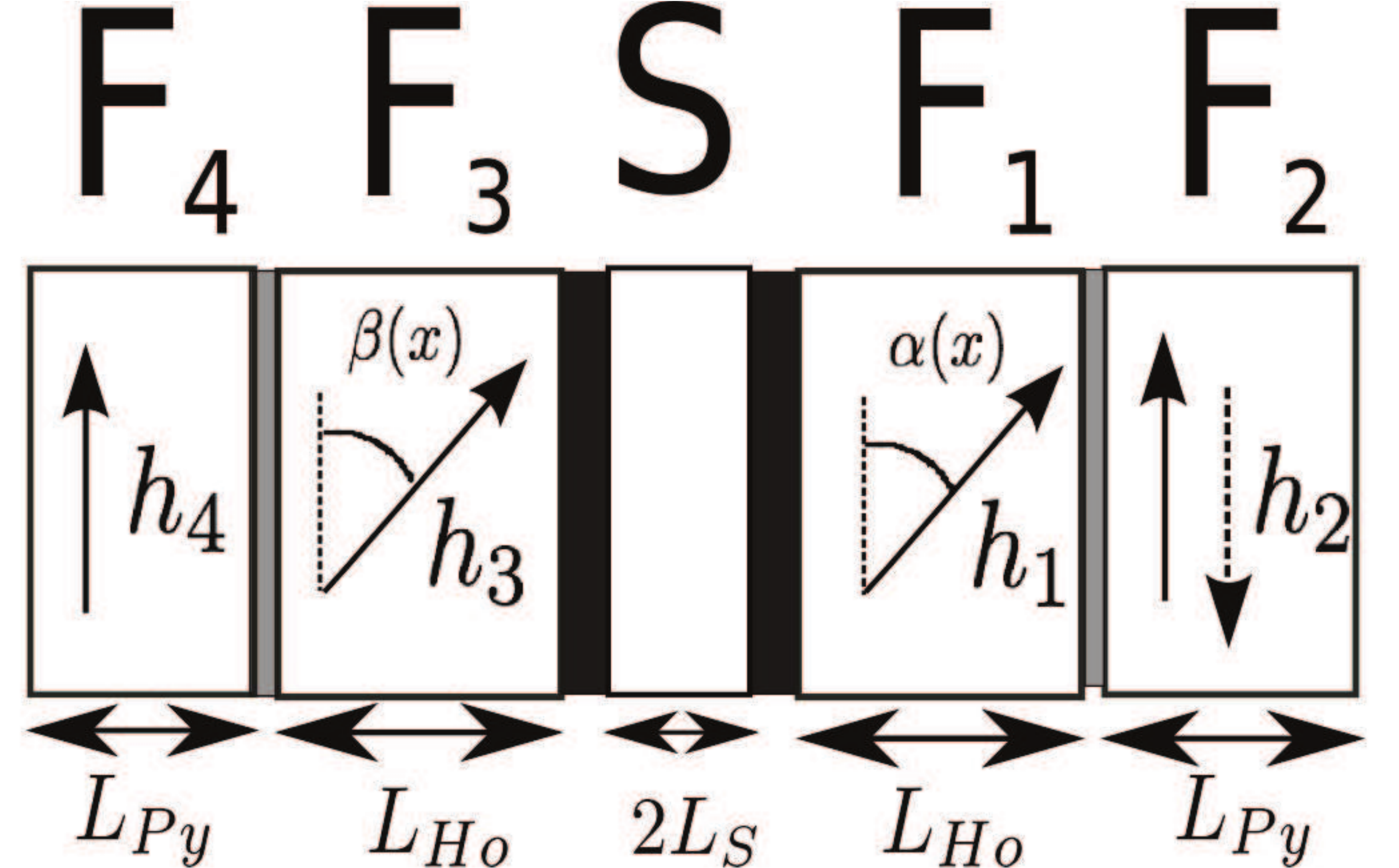}
\caption{The F$_4$F$_3$ISIF$_1$F$_2$ tunnel junction with non-collinear ferromagnets. We use it to model the Py/Ho/S/Ho/Py strcuture of Ref.\cite{aonat}(see Figure\ref{fig1N}).}
\label{geoo2} \vspace{-4mm}
\end{center}
\end{figure}

In order to model this experimental setup we consider the geometry show in Fig.~\ref{geoo2} and used the method described in section \ref{sec:self} for the calculation of  $T_C$.  The geometry under consideration consists of four ferromagnetic and one superconducting layers.  $L_s$ is the length of the superconducting layer,  while $L_{1/3}$ corresponds to the  Ho layers and $L_{2/4}$ Py layers (compare Fig. ~\ref{geoo2} and Fig. {fig1N}). Ho is a material with a spiral-like magnetization,  whereas  Py has a fixed exchange field direction. We model here the  rotating exchange field of Ho by an effective non collinear exchange field. We describe  the S$F_{1,3}$ interfaces by the parameter  $\kappa_t$ (that we introduce in eq.\ref{eq:kl}) and assume  a perfect transparent contact between $F_{4,2}$ and $F_{3,1}$. We focus here on the study of  the $T_C$ changes by switching the magnetization of the $F_2$ layer from parallel (P) to antiparallel (AP) with respect to $F_4$

 The exchange field in the $F_{1,3}$ layers is  in the yz-plane $h=(0,h\sin(\alpha(x)),h\cos(\alpha(x)))$. To obtain $\alpha(x)$ we impose the constraint that both magnetic orientations must be equal at the interface. Thus for P ($\alpha(x)=Q_1*(x+L_1+L_s)$) and AP ($\alpha(x)=Q_1*(x+L_1+L_s)+\pi$) cases $\alpha(x)$ is different. $\beta(x)=Q_3*(x-L_s-L_3)$ is obtained in the same way and for the P case the system is symmetric.  $Q_1=Q_3$ in this simulations. 

In accordance  to the experimental values, we assume that  the exchange field of $F_{1,3}$ is smaller than the one of $F_{2,4}$ and a symmetric structure,  hence $h_1=h_3$ and $h_2=h_4$, also $L_1=L_3$ and $L_2=L_4$, with $\kappa_{t1}=\kappa_{t2}$. 

The Usadel equation, eq.\ref{eq:FullUsadel}, for the superconductor in the Matsubara representation,
\begin{equation}
D \partial_x(\breve{g}_s \partial_x \breve{g}_s)-\omega_n[\hat{\tau}_3,\breve{g}_s]-i[\breve{\Delta},\breve{g}_s]=0
\end{equation}
In order to observe a significant change on $T_C$  the superconductor layer has to be thin enough. This justifies to work in the short limit approximation  ({\it i.e.} $L_S$ is shorter than the coherence length). In this limit the  Green functions  do not dependent on the spatial coordinate. Since we are interesting in temperatures close to $T_C$,  we also linearise the Greens functions. 

At the interfaces with the superconductor we use the Kuprianov-Lukichev boundary conditions, eq.\ref{eq:kl} 
\begin{equation}
\frac{\sigma_s}{\sigma_F} g_s \partial g_s=g_{F1} \partial g_{F1}=-\kappa_t [g_s,g_{F1}] \; ,
\end{equation}
which after linearisation have the form,
\begin{equation}
\partial \hat{f}_{F1}=-\kappa_{t1}(\hat{f}_{F1}-\hat{f}_s) \bigl |_{x=-L_s} \; ,
\end{equation}
\begin{equation}
\partial \hat{f}_{F3}=\kappa_{t3}(\hat{f}_{F3}-\hat{f}_s)\bigl |_{x=L_s} \; .
\end{equation}

By defining  $\gamma=\sigma_F/\sigma_s$  we obtain 
\begin{equation}
\hat{f}_s=\frac{\epsilon_{b3} \hat{f}_{F3}(L_s)+\epsilon_{b1} \hat{f}_{F1}(-L_s)+\Delta }{\epsilon_{b3}+\epsilon_{b1}+\Delta},
\end{equation}
where $\epsilon_{bj}=\gamma D \kappa_{tj}/4 L_s$, note that the length of the superconductor is $2L_s$ and not $L_s$. 

This last equations express $f_s$ in terms of the GFs of the ferromagnets at the boundary. This means, that all what remains to be done is to solve the linearised Usadel equation in  one of the sides of the junction, for example $F_1/F_2$, as we have done in section\ref{sec:deter}. Once the GFs are determine we solve the self-consistency equation \ref{eq:selfcon} and determine $T_C$. 
\begin{figure}[h]\begin{center}
\includegraphics[scale=0.23]{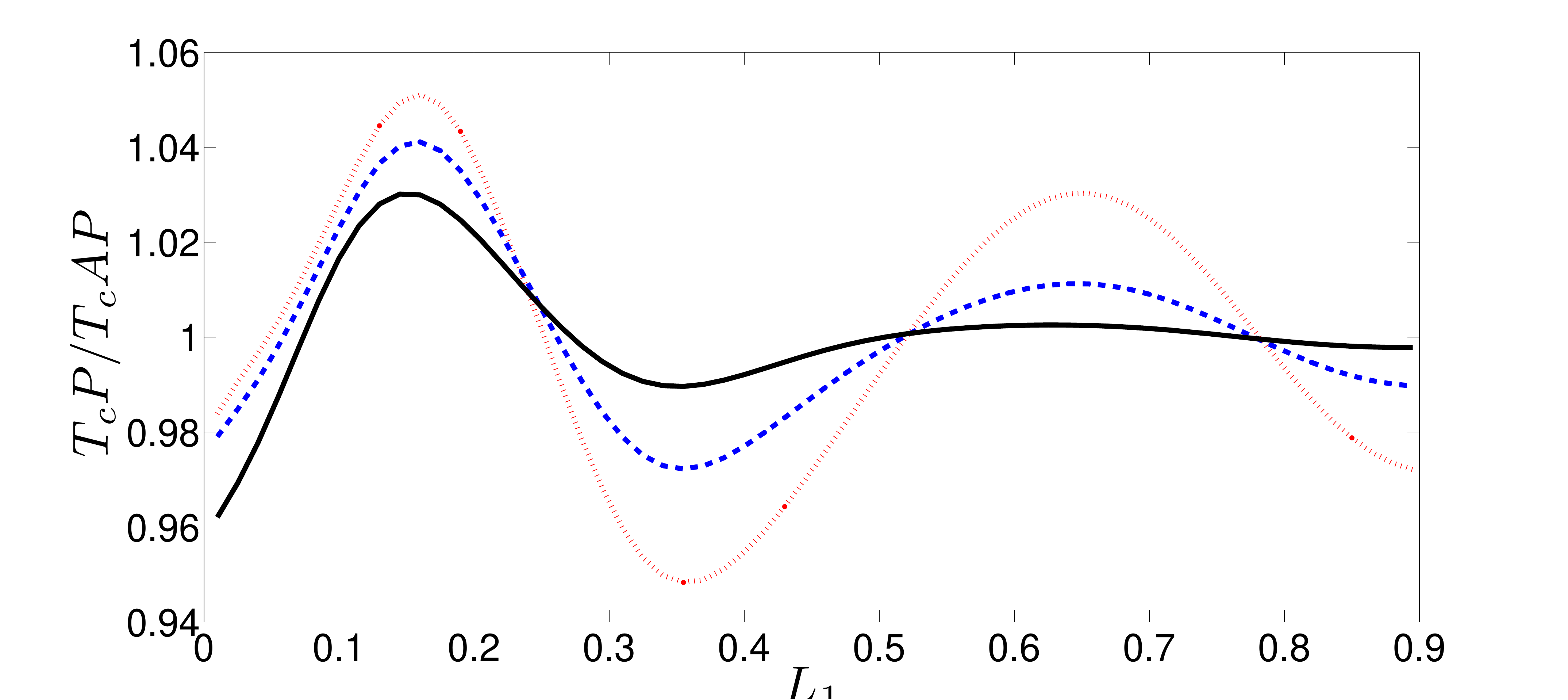}
\caption{Variation of the ratio of the critical temperatures with the length of the $F_1$ and $F_3$ layers in $\xi$ units as we vary $h_1$.  Correspondence with different plots: 12 (black solid),  20 (blue dashed) and 30 (red dotted). $h_2=2*h_1$ and $L_2= 30 nm$  }
\label{f2} \vspace{-4mm}
\end{center}
\end{figure}
As mentioned above, we focus here only on the parallel P and antiparallel AP, configurations and determine the ratio $T_c P/T_c AP$ versus the length of the Ho layer $L_1$. In fig.~\ref{f2} typical results of our theoretical calculations are depicted. The main feature of this figure is the fact that without the Ho layer this structure shows the usual spin valve effect. This means that the critical temperature of the parallel configuration is smaller than that of the antiparallel configuration $T_cP<T_cAP$. The inclusion of the Ho layer leads to the inverse spin valve effect, $T_cP>T_cAP$. Moreover, the ratio $T_cP/T_cAP$ oscillates from positive to negative values due to the oscillations of the superconducting condensate. 

Experimentally both the spin-valve (without Ho) and the inverse spin-valve behaviour (with Ho) predicted by our theory, have been proved. However, the oscillations of the ratio $T_c P/T_c AP$ with respect to the thickness of Ho have not been observed. Indeed, all samples with Ho show the inverse spin-valve effect.

The reason for the discrepancy between experiment and theory in this respect, is that the theory does not include the spin selectivity at the interfaces. In general, spin selectivity is not taken into account in the quasiclassical formalism.
\begin{figure}[h]\begin{center}
\includegraphics[width=0.6\columnwidth]{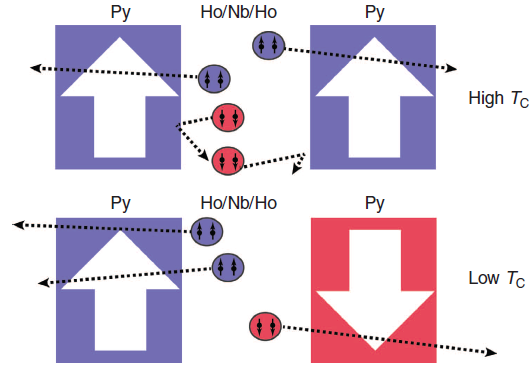}
\caption{A cartoon illustrating the
possible behaviour of spin-one triplet pairs in Py/Ho/Nb/Ho/Py/FeMn
F–S–F spin valves in the parallel (top) and antiparallel states (bottom).(From Ref.\cite{aonat})}
\label{fig5N}
\end{center}
\end{figure}

Fig.~\ref{fig5N} illustrates schematically how spin-selectivity at the interfaces works. If up$–$up and down$-$down triplet pairs, that are generated by the spin mixer, are equally involved in proximity coupling with the F layer, then the $T_C$ suppression would be independent of the magnetic orientation
of the spin valve. Instead, we assume that at the interface those
triplet pairs with spins parallel to the majority spin direction in a
Py are more likely to enter it. Spin conservation at the interface
therefore implies that opposite spin pairs are more likely to return
to the Nb layer. On the Nb side, these have a spatial range of at
least the singlet coherence length~\cite{24} and so, for the spin valve
systems considered here, can interact with the other S–F interface.
If this F layer is AP to the other F layer, then the pair has a higher
probability of entering it. In other words, the proximity
suppression of $T_C$ due to the presence of spin-one triplet pairs,
which is governed by the probability of pairs exiting S, is
enhanced in the AP state in comparison with the P state. This is
consistent with the observations. This explanation necessarily
requires both the spin-triplet pairs to be able to cross Nb and that
the F layers are spin selective to the spin of the triplet pairs. In
view of theory~\cite{32,33}, these assumptions are most likely to be
fulfilled in the samples: first, spin selectivity is an intrinsic
property of ferromagnets with large spin splitting at the Fermi
level~\cite{34} and, second, all triplet components of the condensate
induced at the SF boundary can propagate into a superconductor
a distance close to the superconducting coherence length~\cite{32}
$\xi=\sqrt{ D / T_C}$, which is at best $\sim 30–35$ nm in Nb where D
is the electron diffusivity in Nb. This effect leads to a finite magnetic moment in
the superconductor as predicted in refs.~\cite{32,33,35} and
observed in ref.~\cite{36}.

One of our near future goals is to add spin selectivity into the above calculation with the help of the boundary conditions presented in section\ref{sec:spinfilters}. This will allow for a better understanding of the $T_C$ behaviour in multilayered SF structures.

\newpage

\subsection{ The Hanle effect in a spin valve}
\label{sec:hanle}

\begin{figure}[h]
\begin{center}
\includegraphics[width=0.5\columnwidth]{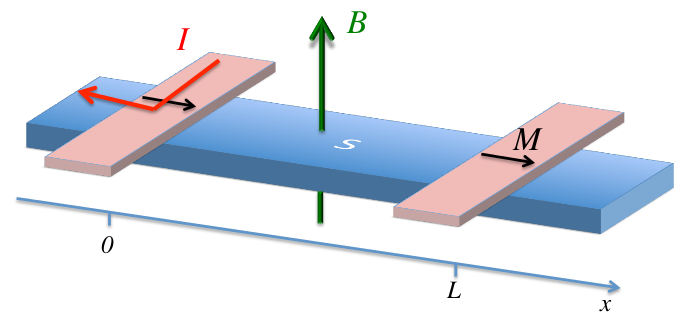}
\caption{ Schematic diagram of the system under study. A superconductor with an applied magnetic field $\bar{B}$ in the z direction, in contact with two ferromagnets polarized in x direction. One is located at $x=0$ and injects (polarized) current into the superconductor. The other, located at $x=L$, works as a detector of spin imbalance.
\label{fig-geometry}}
\end{center}
\end{figure}

\begin{figure}[h]
\begin{center}
\includegraphics[width=0.6\columnwidth]{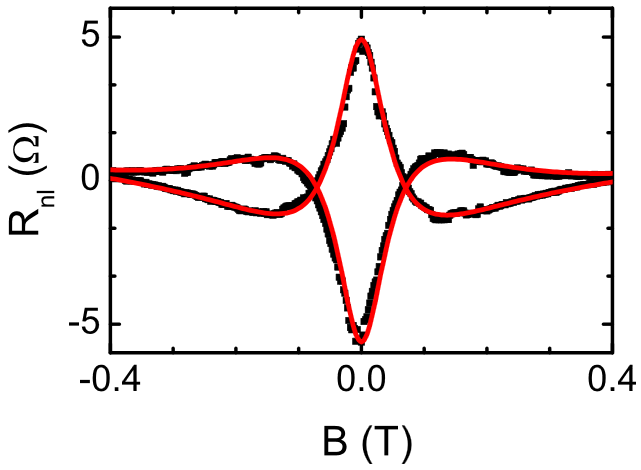}
\caption{ Measurement of nonlocal spin resistance versus perpendicular magnetic field. The dots are the experimental results and the solid line is the fitting. For parallel and anti-parallel magnetization of the injector/detector electrodes.(From Ref.\cite{vanwees})
\label{fig:hanlexp}}
\end{center}
\end{figure}

In this second example we study the spin diffusion in a normal metal. In particular, we analyse the geometry shown in figure \ref{fig-geometry}. It consists of a metal (S) that can be in the normal or superconducting state. At $x=0$ one injects a spin-polarized current from a ferromagnet whose magnetization points in $x$-direction.  A  second ferromagnet (detector), also polarized in $x$ direction, is situated at $x=L$.  Notice that the current flows towards negative $x$ and therefore no current is flowing into the second ferromagnet. A field $\textbf{ B}$ is applied perpendicular to the normal metal (in $z$ direction). In this situation, the injected spins undergo precession (Hanle effect) while diffusing from the injector towards the detector. The detector signal oscillates with the perpendicular field. The signal is maximum in the absence of field (no precession). It vanishes at fields when the average spin precession angle is of 90 degrees and it is minimum for 180 degrees. In fig.\ref{fig:hanlexp} we show how the spin signal precesses with the field.

 %%%%%%%%%%%%%%%%%%%%%%%%%%%%%%%%%%%%%%%%%%%%%%%%%%%%%%%%%%%%%%%%%%%
%%%%%%%%%%%%%%%%%%%%%%%%%
%%%%%%%%%%%%%%%%%%%%%%%%%%%%%%%%%%%%%%%%%%
%%%%%%%%%%%%%%%%%%%%%%%%%%%%%%%%%%%%%%%%%%%%%%%%%%%%%%%%%%%%%%%%%%%
  
 %%%%%%%%%%%%%%%%%%%%%%%%%%%%%%%%%%%%%%%%%%%%%%%%%%%%%%%%%%%%%%%%%%%
%%%%%%%%%%%%%%%%%%%%%%%%%
%%%%%%%%%%%%%%%%%%%%%%%%%%%%%%%%%%%%%%%%%%
%%%%%%%%%%%%%%%%%%%%%%%%%%%%%%%%%%%%%%%%%%%%%%%%%%%%%%%%%%%%%%%%%%%

Here we focus on the normal case where the spectrum is trivial and the retarded and advance GFs are very simple simple: $\check{g}^{R(A)}=\pm\tau_3$.  Thus, we only have to solve the kinetic equation for the distribution functions. By taking the trace of Eq.\ref{eq:kinetic}, multiplied by $\tau_0 \sigma_1$, $\tau_0 \sigma_2$, and $\tau_0 \sigma_3$, one obtains, 

 different traces we obtain:  
\begin{subequations}
\begin{align}\label{eq:spinim1}
 \mathcal{D} \partial_{xx}^2 f_{T1}= h f_{T2} + \frac{1}{\tau_f} f_{T1},
\\
  \mathcal{D} \partial_{xx}^2 f_{T2}= -h f_{T1} + \frac{1}{\tau_f} f_{T2},
 \label{eq:spinim2}
 \\
  \mathcal{D} \partial_{xx}^2 f_{T3}= \frac{1}{\tau_f} f_{T3}, 
 \label{eq:spinim3}
\end{align}
\end{subequations}
which describe the spin imbalance and
\begin{equation}
\label{chim}
  \mathcal{D} \partial_{xx}^2 f_{T}= 0,
\end{equation}
 the charge imbalance. According to eqs.\ref{eq:spinim1} and \ref{eq:spinim2}, the applied magnetic field (in z direction) leads to a precession (and decay) of the spin accumulation. The latter is given by:
\begin{equation}
 \mu_{Ti} =\int dE f_{Ti} \; , \quad \text{where} \quad i=0,1,2,3
\end{equation}
and can be obtained by integrating Eqs.\ref{eq:spinim1}-\ref{chim} over the energy.
The linear system of equations can solved analytically . In particular the general solution for the geometry shown in Fig. (\ref{fig-geometry}) is given by
\begin{subequations}
\begin{equation}
\mu_{T1}(x)=A e^{ \chi_+ x} + B e^{ \chi_- x},
\end{equation}
\begin{equation}
\mu_{T2}(x)=i(A e^{ \chi_+ x } - B e^{ \chi_- x}), \quad \text{for} \quad x<0
\end{equation}
\begin{equation}
\mu_{T1}(x)=C e^{-\chi_+ x} + D e^{-\chi_- x}+E e^{\chi_+ x} + F e^{\chi_- x},
\end{equation}
\begin{equation}
\mu_{T2}(x)=i(C e^{-\chi_+ x} - D e^{-\chi_- x}+E e^{\chi_+ x} - F e^{\chi_- x}), \quad \text{for} \quad 0<x<L
\end{equation}
\begin{equation}
\mu_{T1}(x)=G e^{-\chi_+ x} + H e^{-\chi_- x},
\end{equation}
\begin{equation}
\mu_{T2}(x)=i(G e^{-\chi_+ x} - H e^{-\chi_- x}), \quad \text{for} \quad x>L.
\end{equation}
\label{sol1}
\end{subequations}
Here $\chi_\pm^2=(\tau_f^{-1} \pm i h)/D$. Since the electrochemical potential should be finite at  $x\rightarrow \pm \infty$, the solutions for $x<0$ ($x>L$) contain only the exponentially decaying (increasing) terms. For $0<x<L$ the solution contains 4 terms so that the back flow of spins, generated by the detector, can be taken into account. These 8 coefficients (A,B,C,D,E,F,G,H) has to be obtained  from the boundary conditions at the interfaces with the ferromagnetic electrodes and the continuity conditions at $x=0$ and $x=L$. 
Before solving the problem we note that in the  well-known Bloch equations used in the literature\cite{vanwees} for describing the spin precession in a normal metal by the action of an external field
\begin{equation}
\label{Bloch}
 D \partial_{xx}^2 \vec{m} = \vec{h} \times \vec{m} + \frac{1}{\tau_f} \vec{m},
\end{equation}
are identical to the  Eqs. \ref{eq:spinim1}, \ref{eq:spinim2} and \ref{eq:spinim3} by making $ \vec{m}=(\mu_{T1}, \mu_{T2},\mu_{T3})$ and the exchange field $\vec{h}=(0,0,h)$. In other words, this equations that are derived form a microscopical model provide a proof for the phenomenological equation (\ref{Bloch}). 

Now we calculate the coefficients in Eqs.  (\ref{sol1}).
In order to describe the spin-injection at the SF interface and the voltage at the one at $x=L$ we used the effective  boundary conditions derived  in Ref.\cite{SFBC}.
After integration over energy we obtain 
at $x=0$ we obtain, 
\begin{subequations}
\begin{align}
\partial_{x} \mu_{T1}(0)= 2 \kappa_{t1} \left[ P_1 (\mu_T(0) - \mu_F) + \mu_{T1}(0) \right],
 \\
\partial_{x} \mu_{T2}(0)= 2 \kappa_{t1} \mu_{T2}(0),  
 \\
\partial_{x} \mu_{T}(0)= 2 \kappa_{t1} \left[  (\mu_T(0) - \mu_F) +P_1 \mu_{T1}(0) \right]\, ,
\end{align}
\end{subequations}
and  at the  detector ($x=L$),
\begin{subequations}
\begin{align}
\partial_{x} \mu_{T1}(L)= 2 \kappa_{t2}  \mu_{T1}(L), 
 \\
\partial_{x} \mu_{T2}(L)= 2 \kappa_{t2} \mu_{T2}(L),  
 \\
\partial_{x} \mu_{T}(L)= 2 \kappa_{t2} P_2 \mu_{T1}(L). 
\end{align}
\end{subequations}
Here, $\kappa_{t1(2)}= 1/(2 \sigma (R_{I1(2)}+R_F) A)$, where $A$ is the area of the contact. $R_{I1(2)}$, $R_F$ is the resistance per unit area of the detector (injector) interface and the ferromagnet, respectively. $P_{1(2)}$ is the polarization of the detector (injector) interface, that we assume to be equal to the polarization of the ferromagnets, $P_{1(2)}=P_{F1(2)}$. We also assume that the polarization is small, so $1-P_{1(2)}^2 \sim 1$. The term $\mu_F$ is the electrochemical potential of the injecting ferromagnet. Therefore we define $(\mu_T(0) - \mu_F)/(R_{I1}+R_F)= -e I$ as the injected current and assume that it is constant. Note that if $\mu_F=0$ and $\mu_T(0)$ is finite, there will be current flowing from the normal metal to the ferromagnet.  

As now the $\mu_{T1}$ and $\mu_T$ components are not coupled, the boundary conditions at the detector ($x=0$) for different polarizations read,
\begin{subequations}
\begin{equation}
\sigma A \partial_{x} \mu_{T1}(0)= \sigma A ( \partial_{x} \mu_{T1}(0^+)-\partial_{x} \mu_{T1}(0^-)) \nonumber
\end{equation}
\begin{equation}
=-e P I+ \frac{\mu_{T1}(0)}{R_{I1}+R_F},
\end{equation}
\begin{equation}
\sigma A \partial_{x} \mu_{T2}(0)=\sigma A ( \partial_{x} \mu_{T2}(0^+)-\partial_{x} \mu_{T2}(0^-))= \frac{\mu_{T2}(0)}{ R_{I1}+R_F}.
\end{equation}
\end{subequations}
For the injector ($x=L$),

\begin{subequations}
\begin{align}
\sigma A \partial_{x} \mu_{T1}(L)= \sigma A ( \partial_{x} \mu_{T1}(L^+)-\partial_{x} \mu_{T1}(L^-))=\frac{\mu_{T1}(L)}{R_{I2}+R_F},
 \\
\sigma A \partial_{x} \mu_{T2}(L)=\sigma A ( \partial_{x} \mu_{T2}(L^+)-\partial_{x} \mu_{T2}(L^-))= \frac{\mu_{T2}(L)}{ R_{I2}+R_F}. 
\end{align}
\end{subequations}
The physical interpretation of this boundary conditions is the following. At the $x=0$ position we inject polarized current in the x direction, $I_x(0)=P I$. This spin current diffuses to the left and to the right in the normal metal or flows back into the contact. The spin current into the contact is given by the spin accumulation in the contact position divided by the resistance of the interface and ferromagnet, $\mu_{T1}(0)/e(R_{I1}+R_F)$. The spin currents that diffuse in the normal metal are given by $\sigma A \partial_{x} \mu_{T1}(0^{\pm})/e$. 

As for the y axis polarized spin current, we are not injecting current in that direction so the equivalent $I_y(0)=0$. In the detector we are not injecting currents in any polarization direction, so $I_x(L)=I_y(L)=0$. 

We also impose continuity conditions of the different $\mu$ components,
\begin{subequations}
\begin{align}
 \mu_{T1}(0^+)= \mu_{T1}(0^-) \; \quad  \mu_{T2}(0^+)= \mu_{T2}(0^-)
 \\
 \mu_{T1}(L^+)= \mu_{T1}(L^-) \quad  \mu_{T2}(L^+)= \mu_{T2}(L^-).
\end{align}
\end{subequations}
We write the nonlocal resistance measured at the detector as
\begin{equation}
R_{S}(L)=\frac{P_2 \mu_{T1}(L)}{e I}.
\end{equation}
Using that $\sigma A/ \lambda_N= 1/R_N$, where $R_N$ represents the normal metal resistance and $\lambda_N=\sqrt{D \tau_f}$. $P_2$ being the polarization of the detector. We obtain,
\begin{equation}
R_{S}(L)=\frac{P_{1} P_2 (R_{I1}+R_F) (R_{I2}+R_F) \lambda_N}{R_{N}} 
\nonumber
\end{equation}
\begin{equation}
 \frac{ \text{Re}[\chi_+] \text{Re}[e^{-\chi_+ L}]}{(1+\frac{2 \text{Re}[\chi_+] (R_{I1}+R_F) \lambda_N}{R_{N}}) (1+\frac{2 \text{Re}[\chi_+] (R_{I2}+R_F) \lambda_N}{R_{N}})- \text{Re}[e^{-2 \chi_+ L}] }.
\end{equation}
This is the result obtained in the supplementary information of Ref.\cite{Otani} in the small exchange field limit. Hence, $  \text{Re}[\chi_+ e^{-\chi_+ L}] \approx\text{Re}[\chi_+] \text{Re}[e^{-\chi_+ L}]$. In the absence of external spin splitting field this formula reads,
\begin{equation}
R_{S}(L)=\frac{2 \chi_0 P_1 P_2 (R_{I1}+R_F) (R_{I2}+R_F) \lambda_N}{R_{N}((1+\frac{2 \chi_0 (R_{I1}+R_F) \lambda_N}{R_{N}})(1+\frac{2 \chi_0 (R_{I2}+R_F) \lambda_N}{R_{N}})-e^{-2 \chi_0 L})}.
\end{equation}
Here $\chi_0=\chi_\pm(h=0)$.

\newpage

\section{Summary}

In this chapter we have introduced the quasiclassical Green function technique which is the theoretical tool mainly used in this thesis. Several simple examples on how to use this technique and calculate physical quantities are presented in section\ref{sec:properties}. In sections\ref{sec:crit} and \ref{sec:hanle} we also present two interesting examples: In the first one we calculate the critical temperature of a FSF trilayer with spin-mixing interfaces and in the light of our results we discuss the experimental results of Ref.\cite{aonat}. In section\ref{sec:hanle} we derive from the quasiclassical formalism the spin-diffusion equations governing the spin-precession in normal metals. In particular we analyse the Hanle effect in a standard spin-valve configuration.

\newpage

\bibliographystyle{unsrtnat}
\renewcommand{\bibname}{Bibliography of Chapter 2} % changes default name Bibliography to References

%% file: 3112.tex
%%\documentclass[prb,aps,onecolumn,showpacs,superscriptaddress,floatfix]{revtex4-1}
%
%\documentclass{ucbthesis}
%
%\usepackage{amsmath}
%\usepackage{amssymb}
%\usepackage{amsfonts}
%\usepackage{mathptmx}
%\usepackage{graphicx}
%\usepackage{epstopdf}
%\usepackage{dcolumn}
%\usepackage{bm}
%%\usepackage{cite}
%
%
%
%\DeclareMathOperator{\tr}{\mathop{\mathrm{Tr}}}
%\DeclareMathOperator{\sgn}{\mathop{\mathrm{sgn}}}
%\DeclareMathOperator{\re}{\mathop{\mathrm{Re}}}
%\DeclareMathOperator{\im}{\mathop{\mathrm{Im}}}
%\DeclareMathOperator{\arctanh}{arctanh}
%\DeclareMathOperator{\const}{\mathop{\mathrm{const}}}
%\newcommand{\Eq}[1]{Eq.~(\ref{#1})}
%\newcommand{\Eqs}[1]{Eqs.~(\ref{#1})}
%\newcommand{\W}{\widetilde{W}}
%\hyphenpenalty=10 \hfuzz=1pt
%
%
%
%
%
%\begin{document}
%
%\chapter{Subgap charge transport in superconductor-ferromagnet junctions}

%\subsection{Introduction}

 Transport properties of hybrid structures consisting of
superconducting and non-superconducting materials have been studied extensively in the
last decades \cite{Nazarov_book}. 
 In particular, there is  a  renewed interest in the study of the subgap conductance of superconducting hybrid structures in the presence of  spin-dependent fields in view of the presumable detection of Majorana Fermions  \cite{Kouwenhoven2012}. As explained in section\ref{sec:proximityeffect}, at voltages smaller than the gap $\Delta$ the charge transport through a superconductor-normal metal (SN) junctions is dominated by the process of Andreev reflection. This effect generates a finite subgap conductance, as shown in experiments on SN structures.\cite{Kastalsky1991}  This behavior was discussed theoretically  in Refs.~ \cite{Hekking,VZK}.  It was demonstrated that the conductance of a  SNN$_\text{ e}$ structure, where N$_\text{ e}$ denotes a normal metal electrode,   shows a peak at  a voltage
smaller than the superconducting gap in the case of finite SN barrier
resistances  or if  N  is a diffusive metal \cite{Hekking,VZK}. A similar behavior is also predicted if one substitutes the normal by a ferromagnetic metal (F)\cite{Leadbetaer1999,Seviour1999,Yokoyama2005}.  

%This predictions were observed in experiments\cite{poirier,courtois}.

As discussed in section \ref{sec:ferro}, the Andreev reflection at a SF interface is suppressed since the incoming electron and reflected hole belong to different spin bands \cite{deJong}. Thus, this suppression is stronger the larger the exchange field h, \textit{i.e.} the spin-splitting at the Fermi level. In the
ferromagnet the coherence length  of the electron-hole pairs is given by the minimum
between the thermal  and the magnetic ($\sim\sqrt{\mathcal{D}/h}$) lengths. It has been shown in section\ref{sec:ferro} that in the diffusive case this is the characteristic length in a ferromagnet.  One expects
that by increasing the strength of the field $h$ the electron-hole coherence would be
suppressed and hence the subgap current reduced. As we show below,  this intuitive
picture does not hold always.  

In this chapter we investigate  the subgap transport properties of  a SIFN$_{\text{e}}$ structure. By solving the quasiclassical equations we first analyse the behaviour of the subgap current (the Andreev current)   as a function of the field strength for different values of the voltage, temperature and length of the junction. We show that there is a critical value of the bias voltage $V^*$ above which  the Andreev current
is enhanced by the spin-splitting  field. This unexpected behaviour can be explained as
the competition between two-particle tunnelling processes and decoherence mechanisms
originated from the temperature, voltage and exchange field respectively.   We also show
that at finite temperature  the Andreev current  has a peak for values of the exchange
field  close to the superconducting gap. Finally, we compute the differential conductance
and show that its measurement  can be used as an accurate way of determining the strength
of spin-splitting fields smaller than the superconducting gap. For a spin-splitting field, $h$, of the order of few $\Delta$, the DoS of the SF bilayer, (measured at the outer border of the ferromagnet) shows a peak at energy equal to $h$. 

We also analyse the subgap transport of a $SIFF^\prime$ structure and investigate the Andreev current. It is shown that all features studied in the $SIFN$ structure occur now
at the value of the ”effective field”, \textit{i.e.} the field acting
on the Cooper pairs in the multi-domain ferromagnetic region,
averaged over the decay length of the superconducting
condensate into a ferromagnet. In principle all these predictions can be verified by a usual tunnelling microscopy experiments.

%\newpage

\section{Andreev current and subgap conductance of SFN structures}
\label{sec:SFN}

\begin{figure}[h]
\begin{center}
\includegraphics[width=0.5\columnwidth]{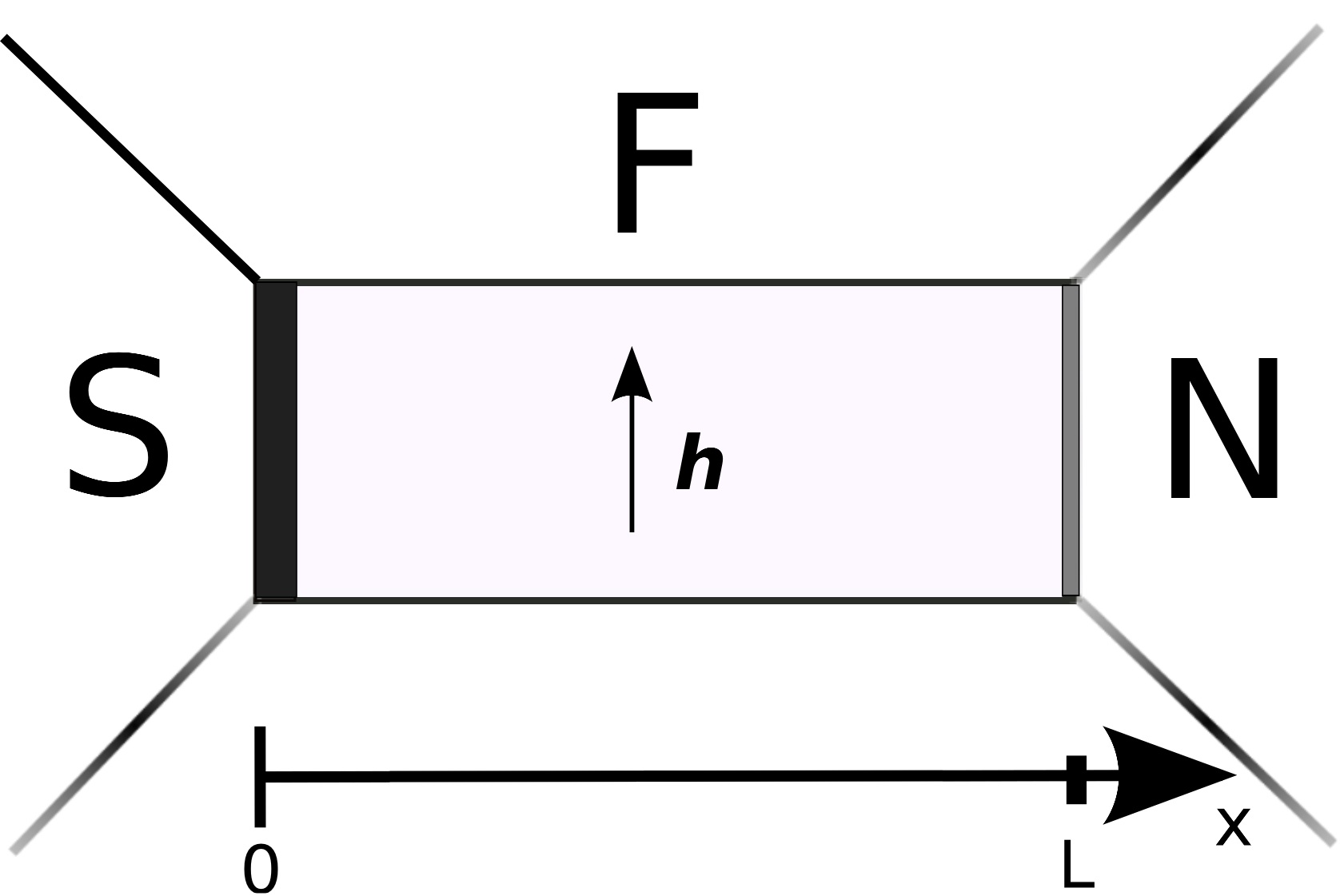}

\caption{Schematic representation of the SIFN structure under consideration. The interface at x = 0 it is an insulating barrier and the interface at x = L is fully transparent. h is the exchange field in the ferromagnet.(From ref.\cite{Ozaeta1})}\label{fig:geo}
\end{center}
\end{figure}
\begin{figure}[h]\begin{center}
\includegraphics[scale=0.4]{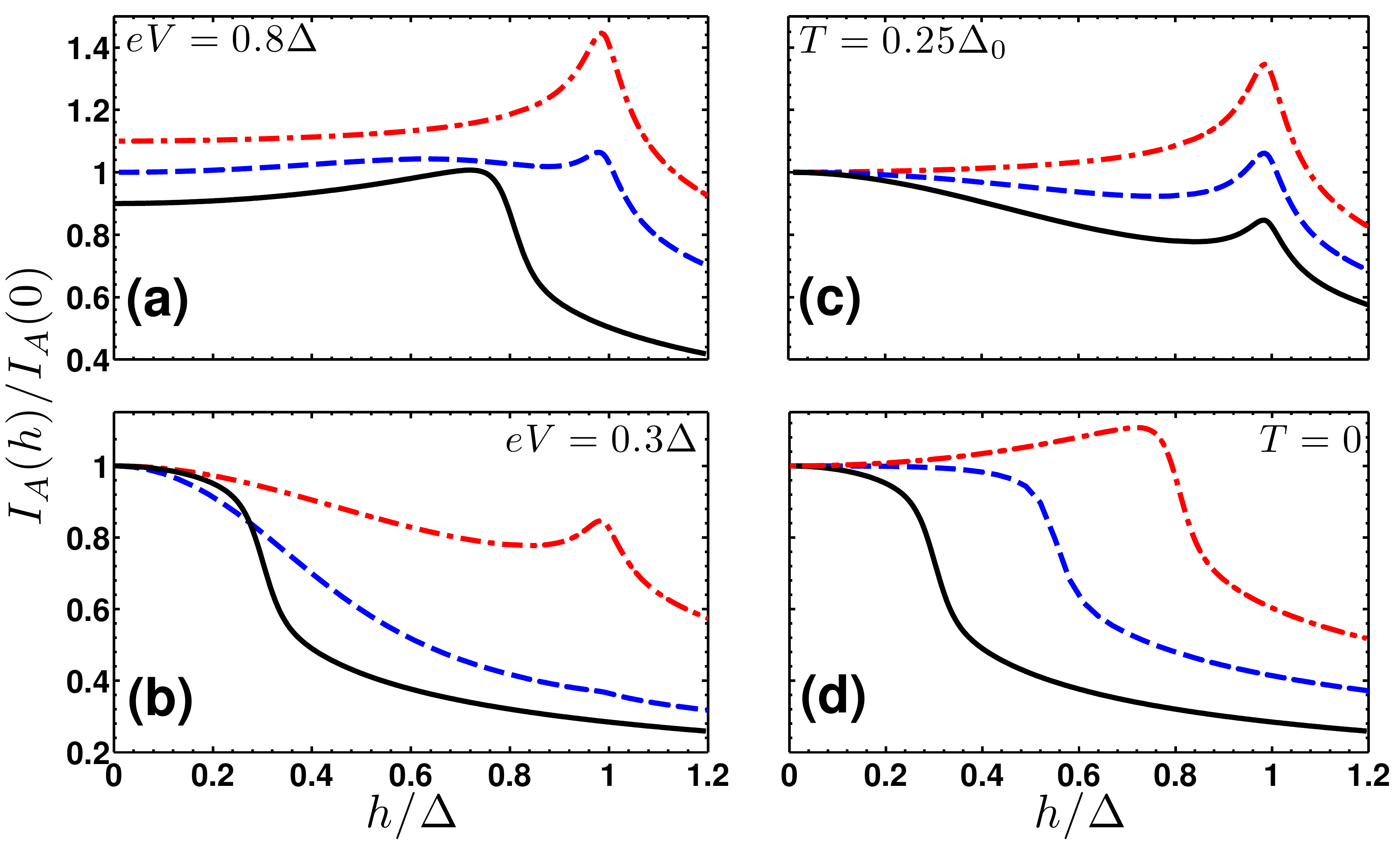}
\caption{ The $h$-dependence of the ratio $I_A(h)/I_A(0)$ for $L = 10 \xi$
and  $W = 0.007$. Left panels correspond to (a) $eV=0.8\Delta$ and (b) $eV=0.3\Delta$.
The different curves are for $T=0 $ (black solid line), $T=0.12 \Delta_0$ (blue dashed
line) and $T=0.25 \Delta_0$ (red point-dashed line).  The  right panels corresponds to (c)
$T=0.25\Delta_0$ and (d) $T=0$, while the different curves to  $eV=0.3 \Delta$ (black solid
line), $eV=0.55 \Delta$ (blue dashed line) and $eV=0.8 \Delta$ (red point-dashed line).
In  the (a) panel curves are vertically shifted with respect to each other for clarity.(From ref.\cite{Ozaeta1})}
\label{a1} \vspace{-4mm}
\end{center}
\end{figure}
%

%Model and basic equations
We consider the  ferromagnetic  wire F shown in fig.\ref{fig:geo}. Its  length,  $L$ is  smaller than the inelastic relaxation length. The wire is  attached at $x=0$ to a superconducting (S)  and at $x=L$ to a normal  (N$_\text{ e}$) electrode. The ferromagnet (F) can also describe a normal wire in a spin-splitting field $B$ (in which case $h = \mu_B B$, where $\mu_B$ is the Bohr magneton) or in proximity with an insulating ferromagnet \cite{Cottet2011}. We assume that the F wire is in the diffusive case, \textit{i.e.} the elastic mean free path is much smaller than the decay length of the superconducting condensate into the F region. For this reason we can use the Usadel equation, eq.\ref{eq:FullUsadel}, in the so called $\theta-$ parametrization (section\ref{sec:theta}), which reads,
\begin{equation}
\partial_{xx}^2\theta_\pm=2i\frac{E\pm h}{\mathcal{D}}\sinh\theta_\pm.\label{Usadel1}
\end{equation}
 Here the upper (lower) index denotes the spin-up (down) component.  The normal and anomalous Green functions are given by $g_\pm=\cosh\theta_\pm$ and $f_\pm=\sinh\theta_\pm$ respectively. Because of the high transparency of the  F/N$_\text{ e}$ interface  the functions $\theta_\pm$ vanish at $x=L$, \textit{ i.e.} superconducting correlations are negligible at the F/N$_\text{ e}$ interface.  We  consider a tunnelling barrier at  the SF interface and assume that its tunnelling resistance $R_T$ is  much larger than the normal resistance $R_F$ of the F layer.   Thus, by voltage-biasing  the N$_\text{ e}$  the voltage drop takes place at the  SF tunnel interface. To leading order  in  $R_F/R_T\ll 1$ the Green functions obey the Kupriyanov-Lukichev  boundary condition at $x=0$, eq.\ref{eq:kl},
\begin{equation}
\partial_x\theta_\pm |_{x=0}=\frac{R_F}{LR_T}\sinh[\theta_\pm |_{x=0} - \theta_S],\label{BC}
\end{equation}
where $\theta_S= \arctanh (\Delta/E)$ is the superconducting bulk value of the function $\theta$.  Once the functions $\theta_\pm$ are obtained one can compute the current through the junction. In particular, we are interested in the Andreev current, i.e. the current for voltages smaller than the superconducting gap due to Andreev processes at the SF interface. This current is given by eq.\ref{IA}, where we substitute the anomalous components corresponding to this system. The resulting expression reads,
\begin{equation}\label{I_A}
I_A=\sum_{j=\pm}\int_0^{\Delta}
\frac{n_-(E)\; dE / 2eR_T}{2W\alpha_j(E) - \sqrt{1-(E/\Delta)^2}\im^{-1}(\sinh\theta_j |_{x=0})},
\end{equation}
where $n_-(E)=\frac{1}{2}(\tanh[(E+eV)/2T]-\tanh[(E-eV)/2T])$ is the quasiparticle distribution function in the N$_\text{ e}$ electrode,  $\alpha_\pm(E)=(1/\xi)\int_0^L dx \;\cosh^{-2}[\re\theta_\pm(x)]$,  $W=\xi R_F / 2 L R_T$ is the diffusive tunnelling parameter \cite{KL, Chalmers}, and $\xi = \sqrt{\mathcal{D}/2\Delta}$ is the superconducting coherence length. Eq.~(\ref{I_A}) is the expression used throughout this article in order to determine the subgap charge transport \cite{footnote1}.

\begin{figure}[h]
\begin{center}
\includegraphics[width=0.7\columnwidth]{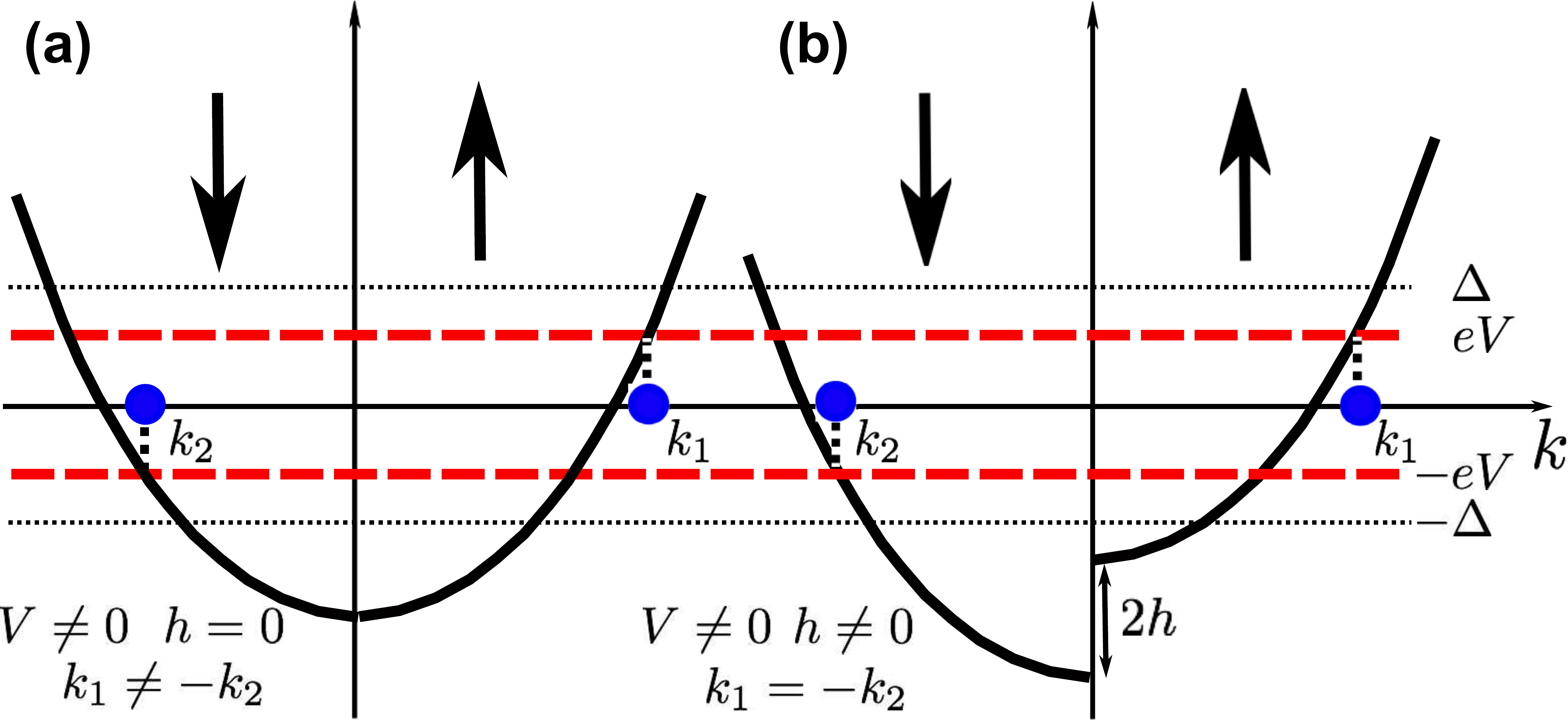}

\caption{ Schematic energy diagram  for a non magnetic (a) and magnetic (b) metal.  The thick solid parabolas are the dispersion of free electrons with spin up ($\uparrow$) and spin-down ($\downarrow$). 
The $k$ axis corresponds to the  Fermi level in the superconductor.   We consider quasi-electrons and -holes with energies $\pm eV<\Delta$ and momentum $k_{1,2}$ . Time-reversal pairing requires that $k_1=k_2$. In case of a normal metal [panel  (a)] this condition is satisfied only for $eV=0$ while  for a ferromagnet ($h\neq0$)  if $h=eV$.(From ref.\cite{Ozaeta1})}\label{par}
\end{center}
\end{figure}
%

%Results

We first compute the Andreev current  numerically by solving
Eqs.~(\ref{Usadel1}-\ref{I_A}). In Fig.~\ref{a1} we show the dependence of the Andreev
current on the exchange field $h$  for different values of the bias voltage and
temperature for a ferromagnetic F wire of the length $L=10\xi$.

We consider first the zero-temperature limit. For small enough voltages (\textit{ e.g} $eV=0.3\Delta$, black solid line in 
 Fig. \ref{a1}b) the Andreev current decays
monotonously  with increasing $h$. This behaviour  is the one expected, since  by
increasing $h$ the coherence length  of the Andreev pairs in the normal region is
suppressed, leading to a reduction of the subgap current. For  large enough voltages
(\textit{ e.g} $eV=0.8\Delta$ in Fig.~\ref{a1}) and keeping the temperature low, the Andreev current
first increases by increasing $h$, reaches a maximum at $h\approx eV$, and then drops by
further increase of the exchange field,  as it is shown for example  by the black solid
line in Fig.~\ref{a1}a.  A common feature of all the low- temperature curves in
Fig.~\ref{a1} is the  sharp suppression of the Andreev current at $h\approx eV$.

For large enough temperatures ($T=0.25\Delta_0$ in Fig.\ref{a1}c)  one observes a peak at
$h\approx\Delta$  [Fig.~\ref{a1}(c)].  The  relative height of this peak  increases with
temperature and voltage as one sees in Figs.~\ref{a1}a and \ref{a1}c respectively. In
the case of  large enough values of  $V$ and $T$, one is able to observe both  the
enhancement of the Andreev current by increasing $h$ and the peak at $h\approx\Delta$
(see for example blue dashed line in Fig~\ref{a1}a). For values of the exchange field larger than $\Delta$,  the Andreev current decreases by increasing $h$ in all cases . In principle all  the  behaviours of
the Andreev current  can be  observed by measuring the full electric current through the
junction as the single particle current is almost independent of $h$.

In order to give a physical interpretation of these results, we first recall the details
of the process of two-electron tunnelling that gives rise to subgap current~\cite{Hekking}
in diffusive systems in the absence of an exchange field. The value of this current is
governed by two competing effects. On the one hand, the origin of the subgap current is
the tunnelling from the normal metal to the superconductor of two electrons with energies
$\xi_{k_1}$ and $\xi_{k_2}$, respectively and momenta $k_1$ and $k_2$, that form a Cooper pair.
This process is of the second order in tunnelling and therefore involves a virtual state
with an excitation on both sides of the tunnel barrier. The relevant virtual state
energies are given by the difference $E_k - \xi_{k_1,k_2}$, where $E_k = \sqrt{\Delta^2 +
\xi_k^2}$ is the excitation energy of a quasiparticle with the momentum $k$ in the
superconductor. Typical values of $\xi_{k}$ are  $T$ or $eV$. Hence under subgap
conditions $T,eV \ll \Delta$, the virtual state energy is typically given by the
superconducting gap $\Delta$. However, when these characteristic energies become larger
and approach the value of the gap, the difference $E_k-\xi_{k_1,k_2}$ eventually vanishes. As a
result, the amplitude for two-electron tunnelling increases drastically, leading to a
strong increase of the Andreev current, accompanied by the onset of single-particle
tunnelling at energies above the gap $\Delta$. On the other hand, two-electron tunnelling
is a coherent process: the main contribution to two-electron tunnelling stems from two
nearly time-reversed electrons $k_1 \simeq -k_2$ located in an energy window of width
$\delta \epsilon \sim eV,T$ close to the Fermi energy, diffusing phase-coherently over a
typical distance $L_{coh} = \sqrt{\mathcal{D}/(\delta \epsilon)}$ in the normal metal
before tunnelling~\cite{Hekking}. This coherence length decreases upon increasing the
characteristic energies $\delta \epsilon \simeq eV,T$, thereby decreasing the Andreev
current.

We now turn to the effect of the exchange field $h$ on two-electron tunnelling. If $h$ is
nonzero, the majority and minority spin electrons at the Fermi level are characterized by different 
wave vectors $k_{F, \pm} = k_F \mp \delta k$, where $\delta k \sim h/v_F$ and $v_F$ is the Fermi velocity.  
In Fig. \ref{par} we show a schematic energy diagram. The wave vectors
$k_{F, \pm}$ are determined by intersection between the parabolas and the $k$-axis. For a given value of $eV\lesssim\Delta$ and in the absence of an exchange field  the relevant excitations with energies $\sim \pm eV$ and wave vectors $k_{1,2}$ are not time-reversed (see Fig. \ref{par}a) and  therefore their contribution to the current  is not coherent. However upon increasing $h$, $|k_1|\rightarrow |k_2|$, \textit{ i.e} the relevant excitations  become more and more coherent, leading to an additional increase of two-electron tunnelling. In particular when $h=eV$, $k_1=-k_2$ (\textit{ cf.} Fig. \ref{par}b).
If $T\rightarrow 0$  there are no occupied states for $\xi_k>eV$. 
Consequently  as soon as $h> eV$, the energy window around the Fermi level does not contain
time-reversed electrons. This leads  to the drop of the Andreev current shown for example in  Fig. \ref{a1}d.  In contrast,  for finite values of $T$,  there are thermally induced quasiparticles with energy $\sim \Delta$,  that become exactly time-reversed whenever $h=\Delta$. This leads to the maximum of the current at $h=\Delta$ when the temperature is finite (\textit{ cf.} Fig. \ref{a1}c). The effects are most clearly seen when plotting the
ratio $I_A(h)/I_A(0)$, as the Andreev pair decoherence effects due to temperature or
voltage are then divided out.

\begin{figure}[h]
\begin{center}
\includegraphics[width=\columnwidth]{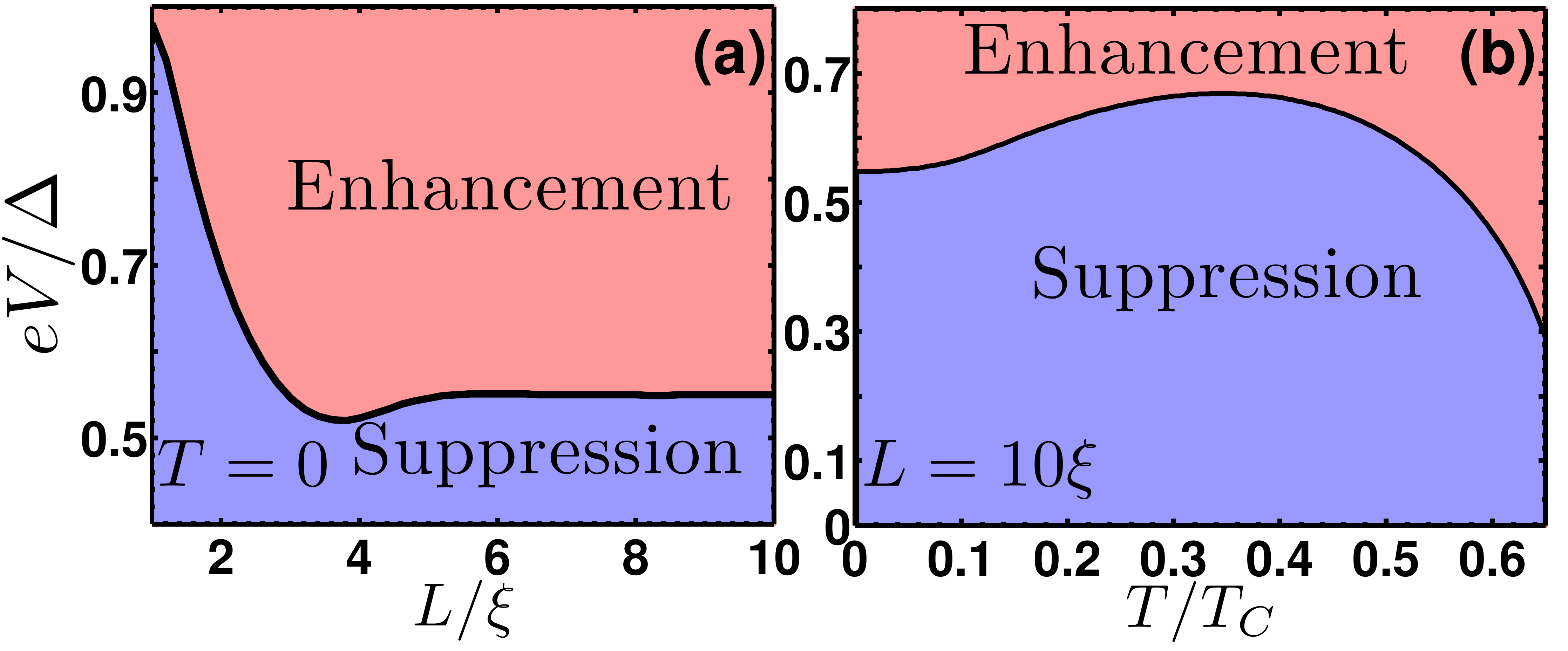}

\caption{ Voltage-junction length (a) and voltage-temperature (b)
diagrams.  The black solid line represents the values of $eV^*/\Delta$. For the range of
parameters situated  below this line the  Andreev current decreases in the presence of a
small  exchange field (suppression), while   in the region above the line  the current
increases (enhancement).  We set $W = 0.007$ in both panels, $T=0$ in panel  (a)  and
$L=10\xi$ in panel (b).(From ref.\cite{Ozaeta1})} \label{a4} \vspace{-4mm}
\end{center}
\end{figure}

A more quantitative understanding of the effects discussed above can be get by analysing
some limiting cases in which simple analytical expressions for the current can be
derived. We first focus the analysis on the zero-temperature limit. Due to the
tunnelling  barrier at the SF interface  the proximity effect is weak and hence  one
can linearise  Eqs.~(\ref{Usadel1}-\ref{BC}) with respect to $R_F/R_T \ll 1$. Due to the zero-temperature limit the integration of eq.\ref{I_A} is limited by the value of the voltage. For $R_F/R_T \ll 1$, the term proportional to $W$ in the denominator is neglected. This lead to the following expression of the Andreev current in this limit, 
\begin{equation}
I_A = \frac{W \Delta_0^2}{2 e R_T}\sum_{j=\pm}\int_0^{eV} \frac{dE}{\Delta_0^2 - E^2} \nonumber \re\left[ \sqrt{\frac{i \Delta_0}{E +j h}} \tanh \left( \sqrt{\frac{E + jh}{i\Delta_0}}\frac{L}{\xi} \right) \right].
\label{lin_IA11}
\end{equation}
For  a large exchange field, $h\gg\Delta_0>eV$  one can evaluate  this expression obtaining
\begin{equation}
I_A\approx \frac{R_F \Delta_0}{8eLR_T^2}\sqrt{\frac{\mathcal{D}}{h}}\log\left[\frac{\Delta_0+eV}{\Delta_0-eV}\right].
\end{equation}
Thus, the Andreev current  decays  as $h^{-1/2}$ for  large values of $h$ in accordance
with the numerical results (see Fig.~\ref{a1}).

In the case of small values of $h$, $h\lesssim eV<\Delta_0$,  one can evaluate
Eq.~(\ref{lin_IA11}) in   the  long-junction limit, i.e. when $L\gg\sqrt{D/h}$.   In this
case  the  Andreev current reads
\begin{equation}
I_A=\frac{\Delta_0 \xi R_F} {eLR_T^2}\sum_{j=\pm} \frac{\arctanh\left(\sqrt{\frac{eV+jh}{\Delta_0+jh}}\right)
+\arctan\left(\sqrt{\frac{eV-jh}{\Delta_0+jh}}\right)}{\sqrt{\Delta_0+jh}}.\label{ia_lj}
\end{equation}
This expression describes the two different behaviours obtained  in Fig.~\ref{a1} for
$h\leq eV$. For small voltages $I_A$ decreases by increasing the field $h$. However, for
large enough values of the voltage $I_A$ is enhanced by the presence of the field. From
Eq.~(\ref{ia_lj})  we can determine  the voltage $V^*$, at which the crossover between
these two behaviours takes place,  by  expanding the expression for the current up to
second order in  $h/eV\ll 1$, \textit{ i.e.} up to the first non-vanishing correction to
$I_A$ due to the  exchange field. This expansion leads to   the following transcendental
expression which determine the voltage $V^*$ at which the crossover takes place,
\begin{equation}
\left(\frac{\Delta_0}{eV^*}\right) ^{3/2}=\frac{3}{2}\left( \arctanh\sqrt{eV^*/\Delta_0}+\arctan\sqrt{eV^*/\Delta_0}\right).
\end{equation}
From here we get  $eV^*\approx0.56 \Delta_0$.  For  $V<V^*$ the Andreev current decays
monotonically with $h$ while for  $V>V^*$  it increases up to a maximum value at
$h\lesssim eV$. This  is in agreement with the numerical results in Fig.~\ref{a1}.

For an  arbitrary length  $L$ and finite temperature we have computed the value of $V^*$
numerically.    In Fig.~\ref{a4} we show the results. The solid black line gives the
values of $V^*$ as a function of $L$ and $T$ [the (a) and (b)  panels of Fig.~\ref{a4}
respectively]. The area below the black  curve corresponds to the range of parameters for
which the Andreev current is suppresses by the presence of a spin-splitting field, while
the area above the solid line corresponds to the range of parameters for which the
unexpected enhancement of the subgap current takes place. According to Fig.~\ref{a4}(a)
at $T=0$ the value of $V^*$ first decreases as $L$ increases, reach a minimum and then
grows again up to the asymptotic value $eV^*\approx 0.56\Delta_0$. Also the dependence of
$V^*$ on the temperature is non-monotonic having a maximum value at $T\sim 0.2 \Delta_0$.

\begin{figure}[t]
\begin{center}
\includegraphics[width=\columnwidth]{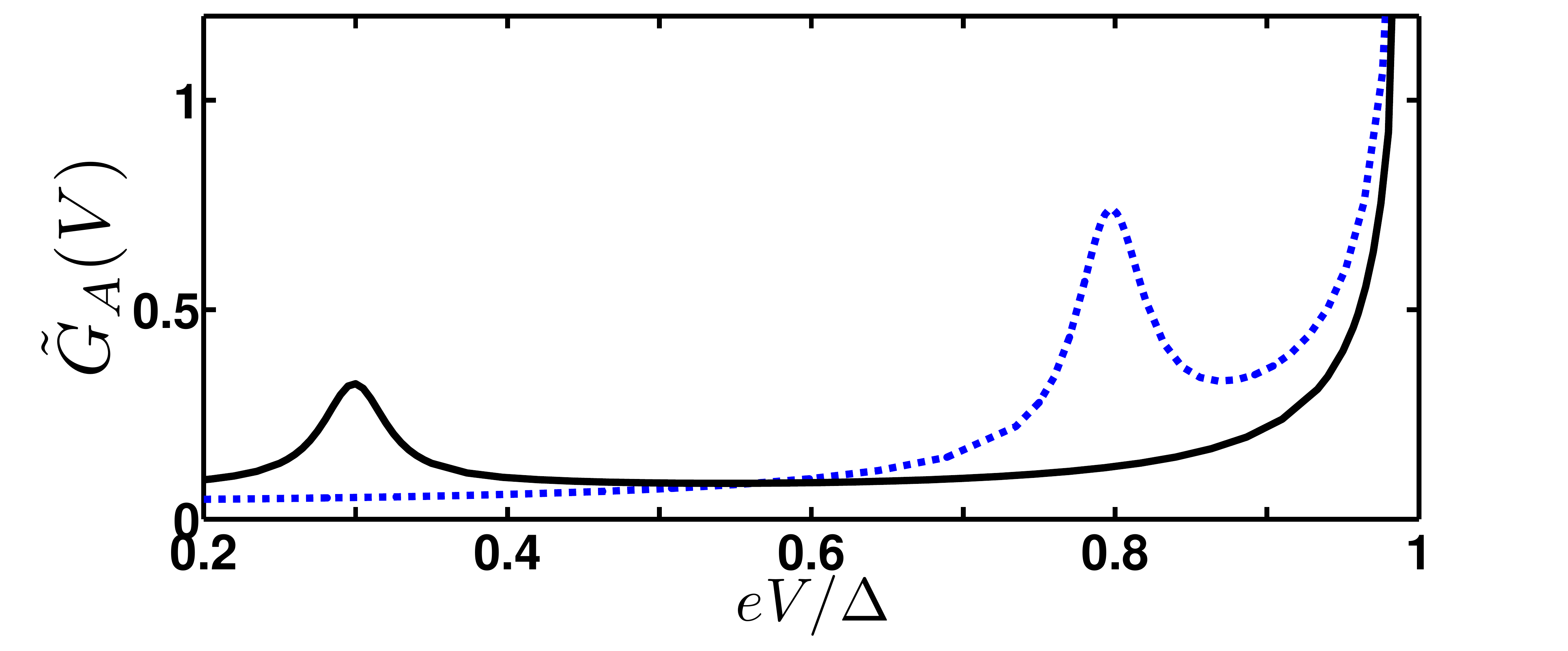}
\caption{ The bias voltage dependence of differential conductance  at $T=0$
for  fields: $h=0.3$ (black solid) and $h=0.8$ (blue dashed).   Here $\tilde{G}_A= 4 R_T
G_A $, $W=0.007$ and $L=10\xi$.(From ref.\cite{Ozaeta1})} \label{a2}\vspace{-5mm}
\end{center}
\end{figure}

 Small spin-splitting fields, as those studied in the present work,
can be created by applying an external magnetic field $B$,  in which case $h=\mu_B B$ or
by the proximity of a ferromagnetic insulator as discussed in
Ref.~\cite{Cottet2011}. It may be also an intrinsic exchange field  of weak
ferromagnetic alloys (see, for example, Ref.~\cite{small_h}). Such small exchange
fields are in principle difficult to detect.  However, as we show in Fig.~\ref{a2},  by
measuring   the  subgap differential conductance $G=dI/dV$ at low temperatures, one can
accurately determine the value of $h$.  At $T=0$  the conductance shows  two well defined
peaks, one at $eV=h$ and the other at $eV=\Delta$. These are related to  a sudden increase
of the coherence length  between the  electron-hole pairs in the ferromagnet and of the
two-particle tunnelling amplitude respectively. As we have seen above, at small voltages
$eV<h$ electrons with majority spins do not find time-reversed partners in the narrow
energy window around the Fermi energy, i.e. such pairs show weak coherence. By increasing
the voltage $eV > h$, the contribution of time-reversed electrons to the current
 gradually increases and consequently the differential conductance increases, reaching a
maximum at $eV=h$. Further increase of the voltage, $eV>h$,  leads to an  increasing
contribution to the current from non time-reversed electron-hole pairs and therefore to a
suppression of the coherent contribution to $G$.  At $h<eV\lesssim \Delta$ the
two-electron tunnelling amplitude increases as $(eV-\Delta)^{-1}$ due to virtual state
contributions with energies $eV$ close to the gap; as a result the conductance shows a
sharp increase. For $h \to 0$ (normal metal) the peak moves toward $eV \to 0$ (not shown here), 
which corresponds to the zero bias peak discussed, for example, in Ref. \cite{ZBA}.

\subsection{Detection of exchange fields larger than the superconducting gap}

In the present section we consider just a simple FS bilayer with a
transparent interface: wire F of a length $d_f$ (smaller than the
inelastic relaxation length \cite{Arutyunov, VH}) is attached at $x
= 0$ to a superconducting electrode by a transparent interface. We
show that the density of states (DOS) measured at the outer
border of the ferromagnet ($x=d_f$) shows a peak at the energy equal
to the exchange field for $d_f \gg \xi_f$ in case when
$h$ is of the order of few $\Delta$ \cite{SIFS2,
Buzdin_H}. Thus, $h > \Delta$ can be determined 
in experiments by measuring the DOS at the outer border of the
ferromagnetic metal in corresponding SF bilayer structure, which can
be done by scanning tunnelling microscopy (STM).

The DOS $N_f(E)$ normalized to the DOS in the normal state, can be
written as
\begin{equation}
N_f(E) = \left[ N_{f \uparrow}(E) + N_{f \downarrow}(E)\right]/2,
\label{DOS_full}
\end{equation}
where $N_{f \uparrow(\downarrow)}(E)$ are the spin resolved DOS
written in terms of spectral angle $\theta_f$,
\begin{equation}
N_{f \uparrow(\downarrow)}(E) = \re\left[\cosh\theta_{f
\uparrow(\downarrow)}\right].\label{DOS_spin}
\end{equation}
To obtain $N_f$, we use a self-consistent two-step iterative
procedure \cite{SIFS2, triplet}. In the first step we calculate the
pair potential coordinate dependence $\Delta(x)$ using the
self-consistency equation in the S layer in the Matsubara
representation. Then, using the $\Delta(x)$ dependence, we solve the
Usadel equations in the S layer,
\begin{equation}
\frac{\mathcal{D_s}}{2i} \partial_{xx}^2\theta_{s} = E
\sinh\theta_{s} + i\Delta(x) \cosh\theta_s\; ,\label{UsadelS}
\end{equation}
together with the Usadel equations in the F layer and
corresponding boundary conditions, repeating the iterations until
convergence is reached \cite{SIFS2}. At the outer border of the
ferromagnet ($x=d_f$) we have $\partial_x \theta_{f
\uparrow(\downarrow)} = 0$. At $x=0$ we use Kupriyanov-Lukichev
boundary conditions which in case of the transparent interface is
convenient to write as
\begin{subequations}\label{KL1}
\begin{align}
\gamma \partial_x \theta_f |_{x=0} &= \partial_x \theta_s |_{x=0} ,
\\
\xi_n \gamma_B \partial_x\theta_{f} |_{x=0} &=\sinh(\theta_{f} -
\theta_{s})|_{x=0}.
\end{align}
\end{subequations}
Here $\gamma = \sigma_f/ \sigma_s$, $\sigma_{f(s)}$ are the
conductivities of the F (S) layers correspondingly, $\xi_n =
\sqrt{\mathcal{D}/ 2 \pi T_c}$, $T_c$ is the critical temperature of
the superconductor, and $\gamma_B = d_f R_T/\xi_n R_F = \xi_s/ \xi_n
W$. The parameter $\gamma$ determines the strength of suppression of
superconductivity in the S layer near the interface (inverse
proximity effect). No suppression occurs for $\gamma = 0$, while
strong suppression takes place for $\gamma \gg 1$. In the numerical
calculations we assume small $\gamma \ll 1$. Since we consider the
transparent interface $R_F \gg R_T$ and contrary to
section\ref{sec:SFN}, $\kappa_t \gg 1$, therefore $\gamma_B \ll 1$. Notice that in the
\Eqs{UsadelS}-\eqref{KL1} we have omitted the subscripts
$\uparrow(\downarrow)$ because equations for both spin directions
are identical in the superconductor.

\begin{figure}[tb]
 \centering
  \includegraphics[width=0.6\columnwidth]{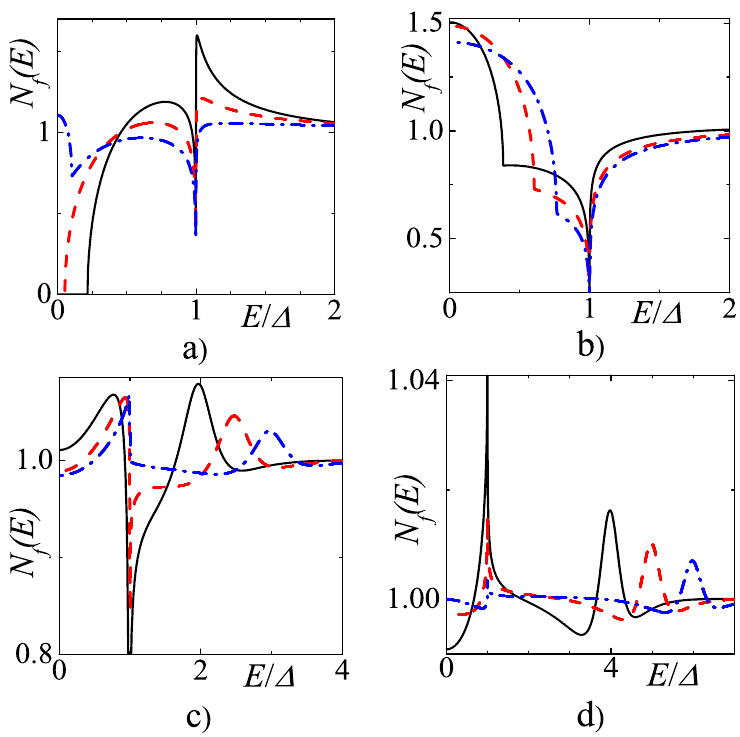}
 \vspace{-3mm} 
 \caption{ DOS $N_f(E)$ at the outer border of the F layer in the FS
bilayer calculated numerically for different values of the exchange
field $h$. Parameters of the FS interface are $\gamma
= \gamma_{B2} = 0.01$, $T = 0.1T_c$. Plots (a) and (b): $d_f/\xi_n =
1$; plots (c) and (d): $d_f/\xi_n = 3$. For plots (a) and (c) solid
black line corresponds to $h/\Delta = 2$, dashed red
line to $h/\Delta = 2.5$, dash-dotted blue line to
$h/\Delta = 3$. For plots (b) and (d) solid black line
corresponds to $h/\Delta = 4$, dashed red line to
$h/\Delta = 5$, dash-dotted blue line to
$h/\Delta = 6$.(From ref.\cite{triplet})} \label{DOS}
%\vspace{-5mm}
\end{figure}

In Fig.~\ref{DOS} we plot the DOS $N_f(E)$ at the outer border of
the F layer in the FS bilayer calculated numerically for different
values of the exchange field $h$ and for different F
layer thicknesses $d_f$. At large enough $d_f$ we see the peak at $E
= h$ [see Fig.~\ref{DOS} (c) and (d)]. For small $d_f$
the peak is not visible and DOS tends monotonously to unity for $E
> \Delta$ [see Fig.~\ref{DOS} (a) and (b)]. The amplitude of the
peak is decreasing with increasing $h$: peak is only
visible for $h$ of the order of few $\Delta$ (see
\cite{SIFS2} for details).

To better illustrate the conditions when the peak at $E =
h$ is visible in experiments we consider an analytic
limiting case. If the F layer is thick enough ($d_f \gg \xi_{f}$)
and $\gamma = 0$ in \Eq{KL1}, the DOS at the outer border of the
ferromagnet can be written as \cite{SIFS1, Cretinon}
\begin{equation}  \label{DOS_bound}
N_{f \uparrow(\downarrow)}(E) = \re[ \cos\theta_{b
\uparrow(\downarrow)} ] \approx 1 - \frac{1}{2}\re \theta_{b
\uparrow(\downarrow)}^2.
\end{equation}
Here $\theta_{b \uparrow(\downarrow)}$ is the boundary value of
$\theta_f$ at $x=d_f$, given by
\begin{equation}  \label{theta_bound}
\theta_{b \uparrow(\downarrow)}= \frac{8 F(E)}{\sqrt{F^2(E) + 1} +
1} \exp\left( -p\frac{d_f}{\xi_f}\right),
\end{equation}
where we use the following notations,
\begin{subequations}
\label{qef}
\begin{align}
p_{\uparrow(\downarrow)} &= \sqrt{2/h}\sqrt{-iE_R \pm ih}, \label{p}\\
F(E) &= \frac{\Delta}{-iE_R + \sqrt{\Delta^2 - E_R^2}}, \quad E_R =
E + i\eta.
\end{align}
\end{subequations}

From \Eqs{DOS_bound}-\eqref{theta_bound} we obtain for the full DOS
the following expression in the limit  $d_f \gg \xi_{f}$ and for $E
\geq \Delta$,
\begin{align}
N_f(E) = 1 + \sum_{j=\pm} \frac{16 \Delta^2 \cos(b_j) \exp(-b_j)}
{(E+\epsilon)(\sqrt{E+\epsilon} + \sqrt{2\epsilon})^2},\label{Exp}
\end{align}
\begin{equation}
b_j = \frac{2d_f}{\xi _{f}}\sqrt{\frac{|E +j
h|}{h}}, \quad \epsilon = \sqrt{E^2 -
\Delta^2}.\nonumber
\end{equation}
We can clearly see the exponential asymptotic of the peak at $E =
h$ from the \Eq{Exp}. We should keep in mind that
\Eq{Exp} is valid for large $d_f/\xi_f$, but nevertheless we may
qualitatively understand why we do not see the peak at $E =
h$ for small ratio of $d_f/\xi_f$: if this factor is
small the variation of the exponent $\exp\{-2(d_f/\xi_{f})\sqrt{|E -
h|/h}\}$ near the point $E =
h$ is also small. The peak is observable only for
$h$ of the order of a few $\Delta$. For larger
exchange fields the peak is difficult to observe, since the energy
dependent prefactor of the exponent in \Eq{Exp} decays as $E^{-2}$
for $E \gg \Delta$. Detecting this peak one can carefully measure the value of small
exchange filed $h > \Delta$ in the ferromagnetic
metal.

\newpage

\section{Andreev current and subgap conductance of spin-valve SFF structures}
\label{sec:SFFN}

%\section{Introduction}

%%%%%%%%%%%%%%%%%%%%%%%%%%%

%\section{Model and basic equations}

\begin{figure}[h]
\centering
\includegraphics[width=0.6\columnwidth]{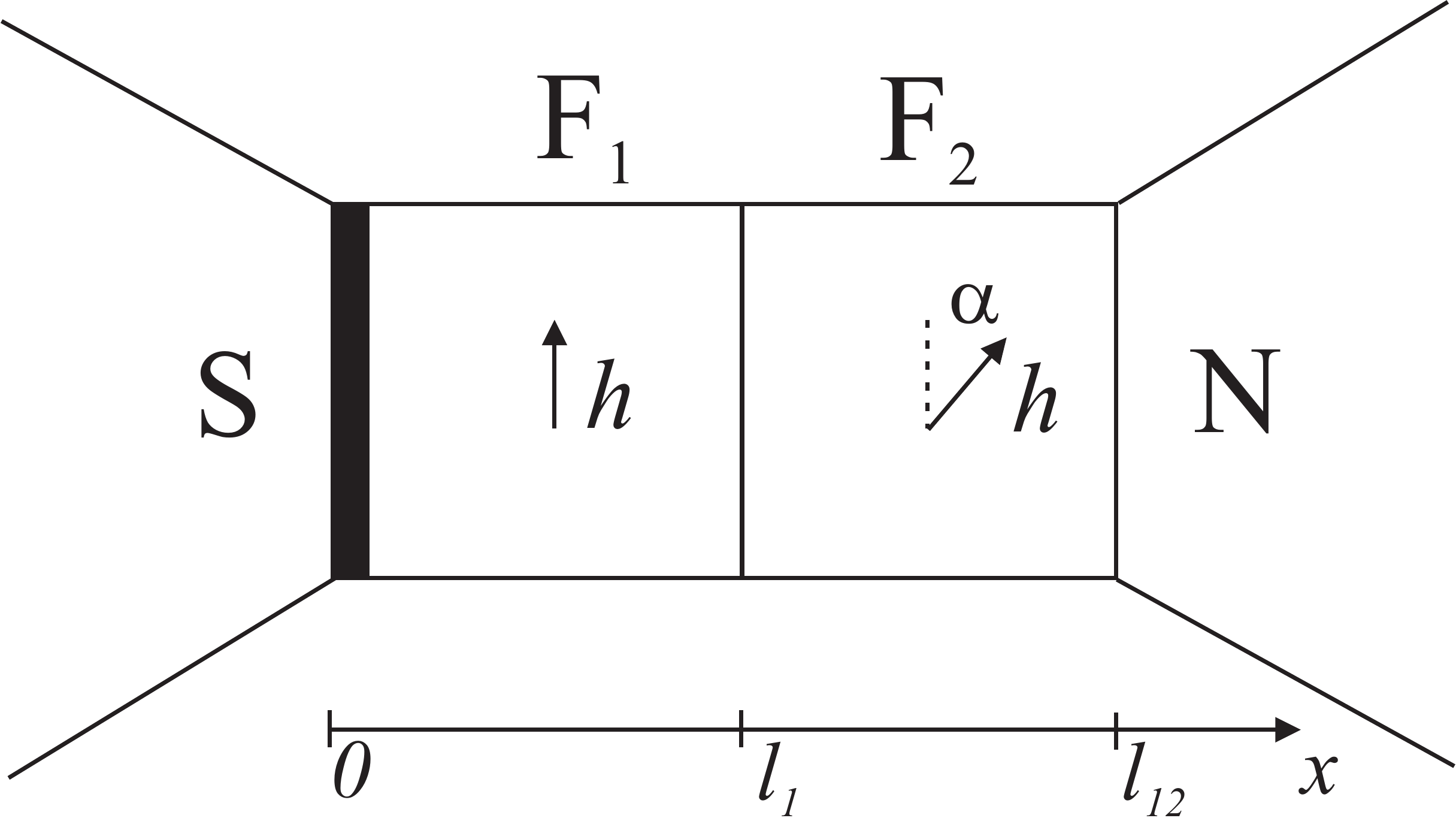}
\caption{The  SF$_1$F$_2$N junction.
The interface at $x=0$ corresponds to the insulating barrier (thick black line). Interfaces at $x=l_1$ and $x=l_{12}$ are fully transparent.
 $\alpha$ is the angle between the magnetization directions of  F$_1$ and F$_2$.(From ref.\cite{ao4})} \label{model3} \vspace{-4mm}
\end{figure}

The purpose of this section is
to consider a hybrid structure with a multi-domain ferromagnetic metal, instead of the mono-domain FIS hybrid system of the section\ref{sec:SFN}. The model of a SF$_1$F$_2$N junction we are going to study is depicted in Fig.~\ref{model3}
and consists of a ferromagnetic bilayer F$_1$F$_2$ of thickness $l_{12}=l_1+l_2$ connected
to a superconductor (S) and a normal (N) reservoirs along the $x$ direction.
As in the rest of the document we consider the diffusive limit, i.e the elastic scattering length $\ell$ is much smaller than
the decay  length of the superconducting condensate into a ferromagnet $\xi_h = \sqrt{\mathcal{D}/ 2 h}$
and the superconducting coherence length $\xi = \sqrt{\mathcal{D}/ 2 \Delta}$,
where $\mathcal{D}$ is the diffusion coefficient and $h$ is the value of the exchange field
(for simplicity we assume the same $\mathcal{D}$ in the whole structure).
We also assume that the F$_1$F$_2$ and F$_2$N interfaces are transparent, while the SF$_1$ is a tunnel barrier.
Thus, the two ferromagnetic layers are kept at the same potential as the voltage-biased normal reservoir.
The F$_1$F$_2$ bilayer can either model a two domain ferromagnet or an artificial hybrid magnetic structure. The applicability of this same structure for the purpose of cooling is studied in section\ref{sec:SFFN}.  

The magnetization of the F$_1$ layer is along the $z$ direction, while the magnetization of the
F$_2$ layer forms an angle $\alpha$ with the  one of the layer F$_1$. Both magnetization vectors lie in the $yz$ plane.
Correspondingly the exchange field vector in the F$_1$ is given by
$\textbf{ h} = (0, 0, h)$, and in the F$_2$ layer
by $\textbf{ h} = (0, h\sin\alpha, h\cos\alpha)$, where
the angle $\alpha$ takes values from  0 (parallel configuration)
to  $\pi$ (antiparallel configuration).

Under these conditions, the microscopic calculation of the electric current
through the structure requires solution of the quasiclassical equation
for the $8\times 8$ Keldysh-Green function $\breve{G}$ in the Keldysh $\times$ Nambu $\times$ spin space
in the F$_1$F$_2$ bilayer. That we obtain using eq.\ref{eq:FullUsadel},
\begin{equation}\label{Usadel3D32}
i\mathcal{D} \partial_x \breve{J} = \left[ \breve{H},
\breve{G}\right], \quad \breve{G}^2 = 1 \; .
\end{equation}
Here $\breve{H} = \tau_3\left(E  - \textbf{ h} \boldsymbol{ \sigma}\right)$ is the Hamiltonian, $\breve{J} = \breve{G}\partial_x \breve{G}$ is the
matrix spectral current,
$\boldsymbol{ \sigma} = (\sigma_1, \sigma_2, \sigma_3)$ are the Pauli matrices in spin space and $\tau_3$  in Nambu space. In \Eq{Usadel3D32} we neglect the inelastic collision term, assuming $l_{12}$ to be
smaller than the inelastic relaxation length \cite{Arutyunov}.

In the F$_1$ region  $\textbf{ h} \boldsymbol{\sigma} = h \sigma_3$ and the equation \Eq{Usadel3D32} has the form
\begin{equation}\label{Usadel32}
i\mathcal{D} \partial_x \breve{J} = \left[ \tau_3\left(E  - \sigma_3 h\right),
\breve{G}\right].
\end{equation}
In the F$_2$ region $\textbf{ h} \boldsymbol{\sigma} = h \sigma_3 \exp(-i \sigma_1 \alpha)$ and
it is convenient to introduce Green's functions rotated in spin space
\cite{Bergeret2002},
\begin{equation}\label{gauge32}
\widetilde{\breve{G}} = U^\dagger \breve{G} U, \quad U =
\exp\left( i \sigma_1 \alpha/2 \right).
\end{equation}
The  rotated function $\widetilde{\breve{G}}$ is then determined by \Eq{Usadel32}.

The \Eq{Usadel32} should be complemented by boundary conditions at the interfaces.
As mentioned above, we assume that
the F$_1$F$_2$ and F$_2$N interfaces are transparent and therefore the boundary conditions at $x=\l_1,l_{12}$ read, as in eq.\ref{eq:trans},
\begin{eqnarray}
\breve{G} \bigl |_{x=l_1 - 0}& =& \breve{G} \bigl |_{x=l_1 + 0}\label{bc132},\\
\partial_x \breve{G} \bigl |_{x=l_1 - 0}& =& \partial_x \breve{G}
\bigl |_{x=l_1 + 0}\label{bc1132},\\
\breve{G}\bigl|_{x=l_{12}-0} &=&\tau_3.
\end{eqnarray}
At $x=0$,  the SF$_1$ interface is a tunnel barrier, where the boundary
conditions are given by the relation, eq.\ref{eq:kl},
\begin{equation}\label{kupluk}
\breve{J} \bigl |_{x=0} = \kappa_t \left[
\breve{G}_S, \breve{G} \right]_{x=0}.
\end{equation}
Here $\breve G_S$ is the Green function of a bulk BCS superconductor and $\kappa_t$is the well-known prefactor of K-L boundary conditions. In the calculations we set a small inelastic scattering parameter, $\eta = 10^{-3} \Delta_0$, where
$\Delta_0$ is the superconducting gap at zero temperature. Below we omit
$\eta$ in analytical expressions for simplicity.

The electric current through the structure is given by the following expression\cite{LOnoneq,Belzig},
\begin{equation}\label{I}
I = \frac{g_N}{8 e} \int_0^{\infty} \tr \tau_3 \check{J}^K \; dE,
\end{equation}
where  $\breve{J}^K \equiv \left( \breve{G}\partial_x \breve{G} \right)^K =
\check{G}^R \partial_x \check{G}^K + \check{G}^K \partial_x \check{G}^A$.

In particular, we are interested in the Andreev current, i.e. the current for voltages smaller than the superconducting
gap due to Andreev processes at the SF$_1$ interface. It is given by the expression\cite{VZK,VBCH},
\begin{equation}\label{IA11}
I_A = \frac{1}{eR} \int_0^\Delta n_-(E) M_S(E) \re f_0 \; dE.
\end{equation}
where $M_S(E) = \Delta \Theta(\Delta - |E|)/\sqrt{\Delta^2 - E^2}$ is the condensate spectral function,
$\Theta(x)$ is the Heaviside step function and the function $f_0$ is the singlet component of $\hat{f}$ at $x = 0$.
This equation is used throughout the article to determine the Andreev transport. We neglect the contribution
to the Andreev current due to the partial Andreev reflection at the energies above the superconducting gap.
In the case of strong enough tunnel barrier at $x=0$ this contribution leads to negligible corrections\cite{VBCH}.

Because of the  low transparency of the tunnel SF$_1$
barrier, the proximity effect is weak and the  retarded Green function can be linearised (we omit the superscript $R$),
\begin{equation}\label{G_lin}
\check{G} \approx \tau_3 + \tau_1 \hat{f},
\end{equation}
where $\hat{f}$ is the $2 \times 2$ anomalous Green function in the spin space ($|\hat{f}|\ll 1$) that obeys  the linearised equation,
\begin{equation}\label{linearized32}
i\mathcal{D} \partial^2_{xx} \hat{f} = 2 E \hat{f} - \left\{ \textbf{ h}
\boldsymbol{\sigma}, \hat{f} \right \},
\end{equation}
where $\{\cdot, \cdot\}$ stands for the anticommutator.
The general solution of this equation has the form
\begin{equation}\label{f32}
\hat{f}(x) = f_0(x) + f_y(x) \sigma_2 + f_z(x) \sigma_3,
\end{equation}
where $f_0$ is the singlet component and $f_{3, 2}$ are the triplet components with respectively
zero and $\pm 1$ projections on the spin quantization axis\cite{Bergeret}.

Solving \Eq{linearized32} in the F$_1$ layer we obtain for the components of \Eq{f32},
\begin{subequations}\label{f_i}
\begin{align}
f_\pm(x) &= a_\pm \cosh (k_\pm x) + \frac{2 \kappa_t }{k_\pm} (u a_\pm - v)
\sinh (k_\pm x),
\\
f_2(x) &= a_y \cosh (k_2 x) + \frac{2 \kappa_t}{k_2} u a_2 \sinh (k_2 x),
\end{align}
\end{subequations}
where $f_\pm = f_0 \pm f_3$,  $a_i$ are the boundary values of $f_i$ at  $x=0$
($i$ stands for $+,-,2$) and the characteristic wave vectors are
\begin{equation}
k_\pm = \sqrt{\frac{2(E \mp h)}{i\mathcal{D}}}, \quad
k_2 =\sqrt{\frac{2E}{i\mathcal{D}}}.
\end{equation}
In the F$_2$ layer  the general solution has the form,
\begin{equation}
\widetilde{f}_i(x) = b_i \sinh \left[ k_i (x - l_{12}) \right],
\end{equation}
where $\widetilde{f}_i$ are the components of the rotated Green function, \Eq{gauge32}.
Using the boundary conditions at the F$_1$F$_2$ interface, Eqs. (\ref{bc132}-\ref{bc1132}) we obtain a set of six linear equations for the
six coefficients $a_i$ and $b_i$, that can be solved straightforwardly.
In particular we are interested in $f_0=(a_++a_-)/2$ which enters the equation for the Andreev current, \Eq{IA11}.
Since the analytical expression is cumbersome we do not present it here.

%%%%%%%%%%%%%%%%%%%%%%%%%%%%%%%%%%%%%%%%%%%%%%%%%%%%%%%%%%%%%%%%%%%%%%%%%%%

%\section{Results and discussion}

%
\begin{figure*}[t]
\begin{center}
\includegraphics[scale=0.40]{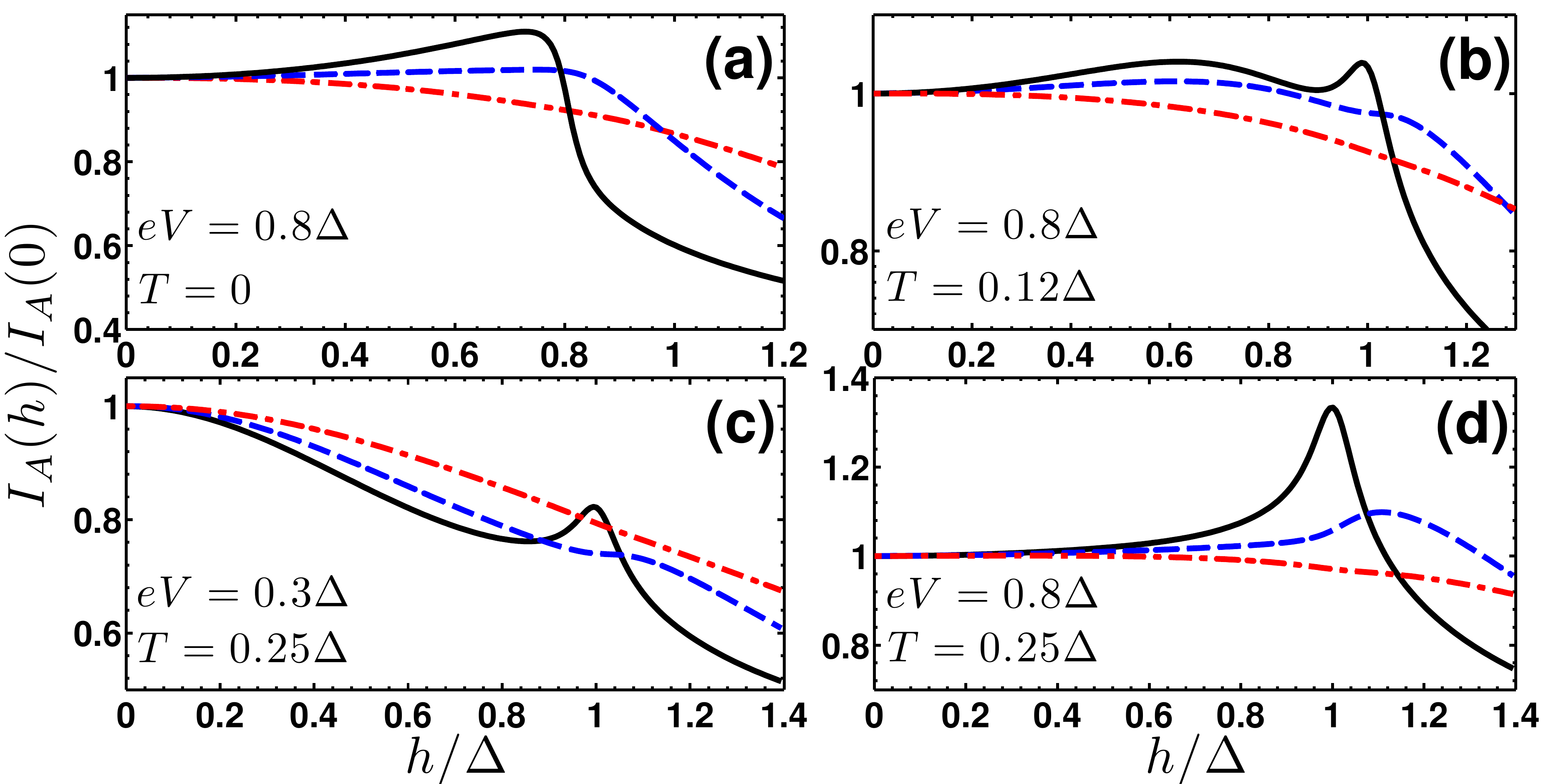}
\vspace{4mm}
\caption{The $h$-dependence of the ratio $I_A(h)/I_A(0)$ for $l_1 = \xi$ and $l_2 = 9\xi$, $\kappa_t = 0.007$, $\alpha = 0$ (solid
black line), $\alpha = \pi/2$ (dashed blue line) and $\alpha = \pi$ (dash-dotted red line). (a) $eV = 0.8 \Delta$, $T = 0$;
(b) $eV = 0.8 \Delta$, $T = 0.12 \Delta$; (c) $eV = 0.3 \Delta$, $T = 0.25 \Delta$; (d) $eV = 0.8 \Delta$, $T = 0.25 \Delta$.(From ref.\cite{ao4})}
\label{Fig1} \vspace{-4mm}
\end{center}
\end{figure*}

Now let us reconsider the features found in the SIFN system for the two-domain situation in case of $\alpha = \pi/2$
(dashed blue lines in Fig.~\ref{Fig1}) and $\alpha = \pi$
(dash-dotted red line in Fig.~\ref{Fig1}). As the $\alpha=0$ case is equivalent to the aforementioned system. The thickness of the F layers is chosen to be
$l_1 = \xi$ and $l_2 = 9\xi$, $l_1$ short enough for the
superconducting condensate penetrates both ferromagnetic layers and $l_2$ long enough for
the full development of the proximity effect in F$_1$F$_2$ bilayer
(at small values of $l_2$ the Andreev current is suppressed by the proximity of the normal
reservoir at $x = l_{12}$) \cite{Ozaeta2}.

Firs of all, we see that increasing $\alpha$ the features (peaks at $h \approx eV, \Delta$)
smear and their amplitude reduces. For $\alpha = \pi$ we do not see any more the enhancement of
the Andreev current. Secondly, we see shift of these peaks to the larger
values of $h$, which is explicitly seen for $\alpha = \pi/2$. The peak at $h \approx eV$ is
shifted to the right (Fig.~\ref{Fig1} (a), dashed blue line) as
well as the peak at $h \approx \Delta$ (Fig.~\ref{Fig1} (d), dashed blue line).
This can be explained as follows. The superconducting condensate penetrates
both ferromagnetic layers and feel the ``effective exchange field'' $\bar h$
acting on the Cooper pairs, averaged over the length $\xi_h$ \cite{Bergeret_h}.
The $\bar{h}(\alpha)$ is gradually reduced as $\alpha$ increases from $0$ to $\pi$.
As before the Andreev current peak is at $\bar{h}(\alpha) \approx \Delta$
which in the case of a finite $\alpha$ corresponds to larger values of the bare $h$, therefore
we observe shift of the Andreev current peak to the right.

\begin{figure*}[h]\begin{center}
\includegraphics[scale=0.35]{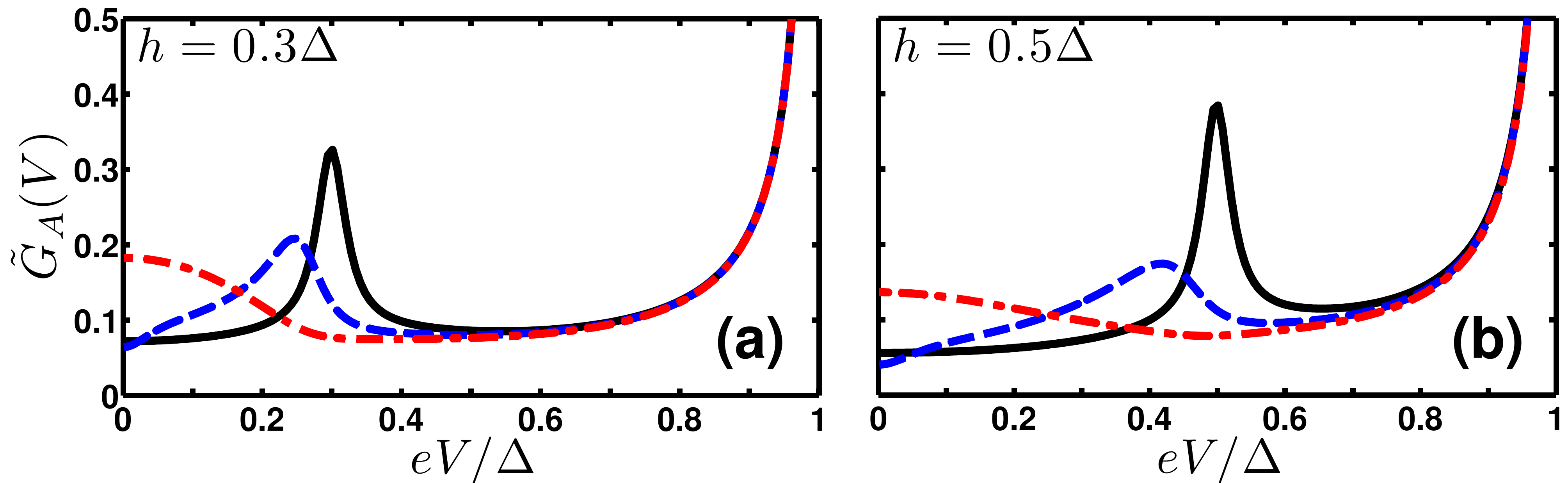}
\vspace{4mm}
\caption{The bias voltage dependence of the differential conductance at $T=0$ for exchange
fields (a) $h = 0.3 \Delta$ and (b) $h = 0.5 \Delta$ for $l_1 = \xi$ and $l_2 = 9\xi$, $\kappa_t = 0.007$, $\alpha = 0$ (solid
black line), $\alpha = \pi/2$ (dashed blue line) and $\alpha = \pi$ (dash-dotted red line).
Here $\tilde{G}_A= 4 R_T G_A$.(From ref.\cite{ao4})} \label{Fig2} \vspace{-4mm}
\end{center}
\end{figure*}

Let us now calculate the subgap differential conductance $G_A = dI_A/dV$ at zero temperature.
It is known that for a diffusive NIS junction the differential conductance at low temperatures has a
peak at $eV = \Delta$ and a zero bias
anomaly (ZBA) peak due to the impurity confinement and the electron-hole
interference at the Fermi level\cite{Kastalsky,Klapwijk,Hekking}.
It occurs at zero bias since for $V = 0$ the electron is perfectly retro-reflected as a hole
during the Andreev reflection process. Thus the electron and the reflected hole
interfere along the same trajectory and the interference effect strongly enhance the
subgap conductance at zero bias\cite{Klapwijk}.

For the FIS structures with $h < \Delta$ the ZBA peak is now shifted to the finite voltage
$eV = h$\cite{Ozaeta1,Leadbetaer1999}, see Fig.~\ref{Fig2}, solid black lines. This can be described as follows. Upon entering
of the Cooper pair into the ferromagnetic metal the
spin up electron in the pair lowers its potential energy by $h$, while the spin down
electron raises its potential energy by the same amount.
In order for each electron to conserve its total energy, the
spin up electron must increase its kinetic energy, while the
spin down electron must decrease its kinetic energy, to make
up for these additional potential energies in F\cite{Demler}. Therefore the electron-hole pair in F has now the
momentum mismatch, i.e. the electron is not perfectly retro-reflected. However, if $eV = h$ there is
a possibility for exact retro-reflection (and interference along the trajectory) of an electron to a hole with
a same kinetic energy equal to the Fermi energy.

\begin{figure*}[h]\begin{center}
\includegraphics[scale=0.35]{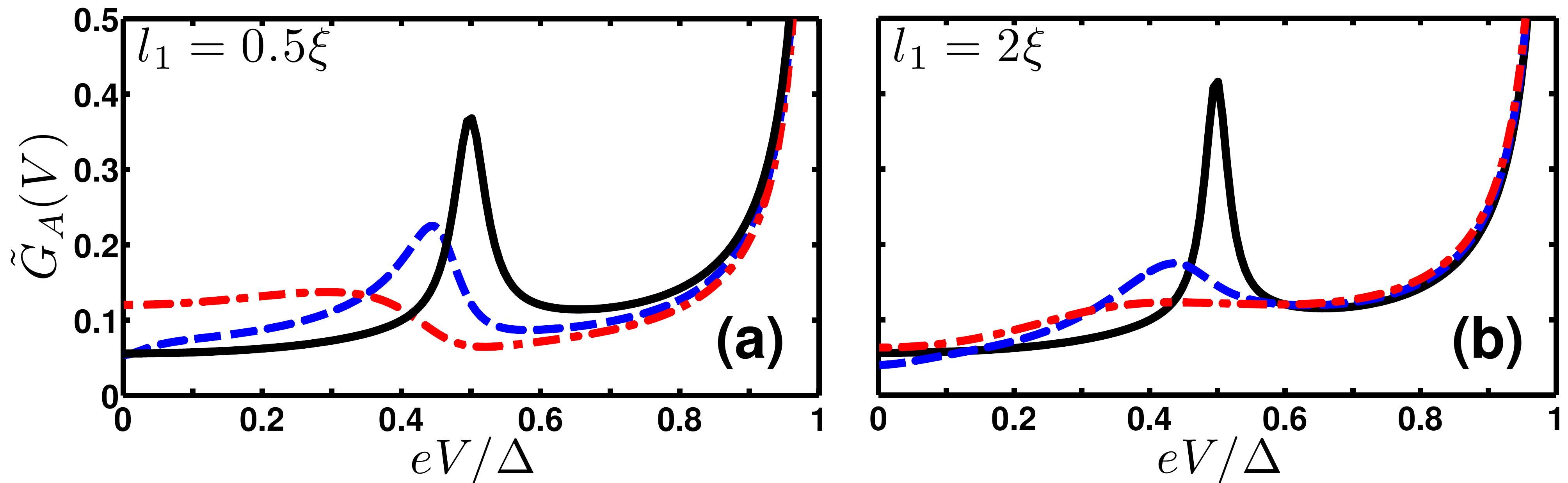}
\vspace{4mm}
\caption{The bias voltage dependence of the differential conductance at $T=0$ for exchange
field $h = 0.5 \Delta$ for (a) $l_1 = 0.5 \xi$ and (b) $l_1 = 2\xi$, $l_2 = 9\xi$, $\kappa_t = 0.007$, $\alpha = 0$ (solid
black line), $\alpha = \pi/2$ (dashed blue line) and $\alpha = \pi$ (dash-dotted red line).
Here $\tilde{G}_A= 4 R_T G_A$.(From ref.\cite{ao4})} \label{Fig3}
\vspace{-4mm}
\end{center}
\end{figure*}

In case of the two-domain ferromagnetic metal we now have the ZBA shift to the
``effective exchange field'' $eV = \bar{h}(\alpha)$. The ``effective field'' is smaller
than the bare $h$, $\bar{h}(\alpha) < h$, and therefore we observe the shift of the differential
conductance peak to the left.
We can explicitly see this for $\alpha = \pi/2$
(Fig.~\ref{Fig2}, dashed blue lines). For $\alpha = \pi$ the situation is more complicated as the effective
exchange field is rather small in the antiparallel configuration.
For $l_1 = \xi$ we observe a broad ZBA peak at $V = 0$ for $\alpha = \pi$ for both values of $h = 0.3 \Delta$
and $0.5 \Delta$ (Fig.~\ref{Fig2}, dash-dotted red lines).
For $l_1 \neq \xi$ the maximum is shifted from the zero bias, see Fig.~\ref{Fig3}, dash-dotted red lines.

%
%\subsection{Conclusions}

\newpage

\section{Summary}

In this chapter we have presented an exhaustive study of the Andreev current and conductance through SIFN$_\text{ e}$ and $SIFF^\prime$ hybrid structures in the presence of a spin-splitting field, $h$. This field is either the intrinsic exchange field of a ferromagnet or a spin-splitting field in a normal metal caused by either an external magnetic field or the proximity of an insulating ferromagnet\cite{Cottet2011}. We focus the study on weak fields, $h\leqslant\Delta$ and $h \gtrsim\Delta$. We find an interesting interplay between phase-coherent diffusive propagation of Andreev pairs due to the proximity effect and decoherence mechanisms originated from the temperature, voltage and exchange field respectively. This interplay leads to a  non-monotonic behaviour of the transport properties as a function of $h$. For very low temperatures and voltages  $eV\ll\Delta$  the Andreev current  decays monotonically  by increasing $h$ as expected. If one keeps the voltage low but now increases the temperature,  the Andreev
current shows a peak at $h\approx\Delta$.  An unexpected behaviour is obtained when the
voltage exceeds some critical value $V^*$. In this case, the Andreev current is enhanced
by the field  $h$  reaching  a maximum  at $h\approx eV$.  We show that the value of
$V^*$ depends on the length of the F wire and the temperature. In particular, for
zero-temperature  and  in the  long-junction limit, i.e. when the length of F is much
larger than the coherence length,  we show that  $eV^*\approx0.56\Delta_0$, where
$\Delta_0$ is the value of $\Delta$ at $T=0$.   We also compute the subgap conductance of
the system at low temperatures and small fields $h<\Delta$.  We show that it has a peak
at  $eV = h$. Thus, transport measurements of this type can be used to determine the
strength of a weak exchange or Zeeman-like field in the nanostructure.

Beyond the fundamental interest,  these results  can also be useful  for the implementation of  a
recent and  interesting proposal \cite{Cottet2011} which suggests a way to detect the  odd-triplet
component \cite{Bergeret2005} of the superconducting condensate induced in a normal metal
in contact with a superconductor and a ferromagnetic insulator. The latter  induces
an effective exchange field in the normal region. The amplitude of such induced exchange fields 
is smaller than the superconducting gap\cite{euo}. Therefore the proposed ferromagnet proximity system in Ref. \cite{Cottet2011} is a candidate to observe the phenomena described in the present work. In order for the calculation to be of wider application range, there is a method for detecting exchange fields larger than the superconducting gap.

%%%%%%%%%%%%%%%%%%%%%%%%%%%%%%%%%%%%%%%%%%%%%

We have also studied the Andreev current and the subgap conductance behaviour in SIFFN hybrid structures with arbitrary direction of magnetization of the F layers. This structure is known as a  ``superconducting spin-valve'' \cite{Karminskaya}. We show that the features we mention in section\ref{sec:SFN} of the Andreev current and subgap conductance in a SIFN system occur at the value of the ``effective exchange field'' $\bar{h}(\alpha) < h$. This is the field acting on the Cooper pairs in the multi-domain ferromagnetic region, averaged over the decay length of the superconducting condensate into a ferromagnet, $\xi_h$\cite{Bergeret_h}. Increasing $\alpha$ from $0$ to $\pi$ one gradually reduce the effective field $\bar{h}(\alpha)$.

%%%%%%%%%%%%%%%%%%%%%%%%%%%%%%%%%%%%
\bibliographystyle{unsrtnat}
\renewcommand{\bibname}{Bibliography of Chapter 3} % changes default name Bibliography to References

%\end{document}

%% file: 312.tex
Thermoelectric and thermomagnetic effects originate from the coupling between heat and charge currents. They have been known for almost two centuries, since the discovery of the Seebeck and Nersnst effects\cite{ashcroft} and have applications in the fabrication of detectors, power generators and coolers. 

One of the major challenges in the road to electronic nanodevices is to reduce and harvest waste heat. Further decrease in size and increase of the transistor speed go in parallel with very high levels of ohmic energy dissipation, and the breakdown of Moore´s law. The study of the coupling between charge and heat currents at the nanoscale is called "\textit{caloritronics}" (from calor, the Latin word for heat and electronics). If magnetic materials are involved in the nanodevices, the additional degree of freedom provided by the electron spins allows for further ways of heat management at the nanoscale. This shapes the field of \textit{spin caloritronics}.

In this chapter, we broaden the field of caloritronics by studying heat/charge transport in superconducting devices. We focus mainly in two effects: Electronic cooling in superconducting hybrid structures and the thermoelectric effect in superconducting ferromagnets.

%The field of spin caloritronics focuses on non-equilibrium phenomena related to the interaction of spins with heat currents in magnetic structures. The challenge for this field of condensed matter physics is to develop efficient devices to harvest waste heat. Another issue is the breakdown of Moore´s law,  further decrease in size and transistor speed go in parallel with very high levels of ohmic energy dissipation. The additional degree of freedom provided by the electron spin allows us to increase the thermoelectric figure of merit as it provides new functionalities. We broaden the theoretical frame of spin calitronics focusing in two areas: the electronic cooling in superconducting hybrid structures and the thermoelectric effect in superconducting ferromagnets.

\section{Electronic cooling in superconducting hybrid structures}

The main goal of this section is to study heat transport related to the electrons of the systems. Thermal energy is transferred  via finite heat current from a part of the circuit to another that heats up. The latter is usually a heat bath with constant temperature. This process is known as cooling. The first kind of thermoelectric cooling, was discovered more than 180 years ago (J.C.A. Peltier, 1834). A finite heat flux was generated by the application of an electric voltage. The thermoelectric effects have been used in thermoelectric generators, that work as heat engines, refrigerators, with no fluid or moving parts and thermocouples to measure temperature difference. During the last decade, solid-state refrigerators for low temperature applications and in particular, operating in the subkelvin temperature range have been intensely investigated. The motivation for this investigation is the successful development and implementation of ultrasensitive radiation sensors and quantum circuits. These require on-chip cooling\cite{pek2004} for proper operation at cryogenic temperatures. Solid-state refrigerators as the thermoelectric devices are based on circuits of solid materials. These typically have lower efficiency as compared to more traditional systems (e.g., Joule-Thomson or Stirling gas-based refrigerators), nevertheless, they are more reliable, cheaper and on top of all, easily scaled down to mesoscopic scale. This gives an unique opportunity for combining on-chip refrigeration with different microdevices and nanodevices. 

\begin{figure}[h]
\begin{center}
\includegraphics[width=0.5\columnwidth]{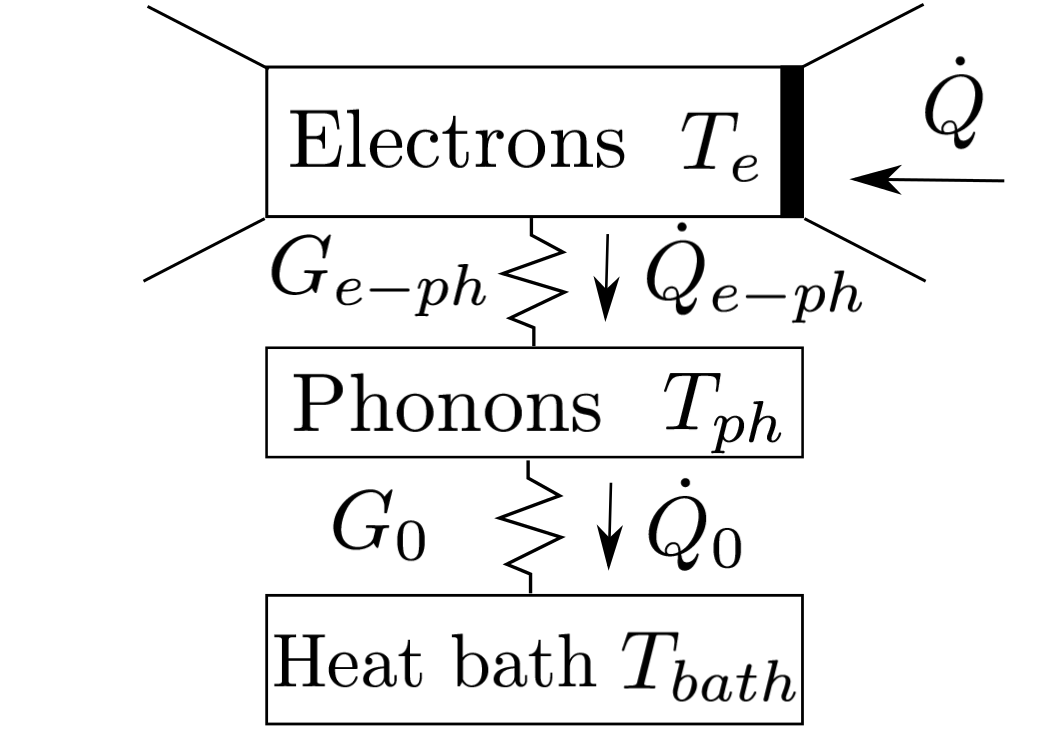}
\caption{Schematic picture of the system considered in the following calculations. The heat flow and thermal conductances between the studied electron system and the phonons in the lattice are indicated by arrows. } \label{fig:firstcool}
\end{center}
\end{figure}

The schematic idea of a setup studied in a typical cooling experiment is the one depicted in fig.\ref{fig:firstcool}. The main object is a diffusive metal wire connected to large electrodes acting as reservoirs where electrons thermalize quickly to the surroundings. The electrons in the wire interact among themselves and are coupled to the phonons in the film and to the radiation and the fluctuations in the electromagnetic environment. The temperature of the film phonons $T_{ph}$ can, in a nonequilibrium situation, differ from that of the substrate phonons and this can even differ from the phonon temperature in heat bath which is externally cooled. However, in this section we assume the phonon temperatures equal that of the bath, due to a good thermal conductance. The energy distribution function of electrons depends on each of this couplings and it is characterized by the electron temperature $T_e$. In this section we explain how $T_e$ can be driven much below the lattice temperature and how this can be exploited. 

%

%\subsubsection{Introduction to Cooling}

Superconductors can be also efficiently exploited for cooling purposes at cryogenic temperatures. They can be used both as passive and active elements. When using the superconductors as a passive element, properties like their low thermal conductivity and zero electric resistance are relevant. For example, a superconductor as a passive element can be used in one of the two arms in Peltier cooling\cite{rf:Leivon}.

\begin{figure}[h]
\begin{center}
\includegraphics[width=0.5\columnwidth]{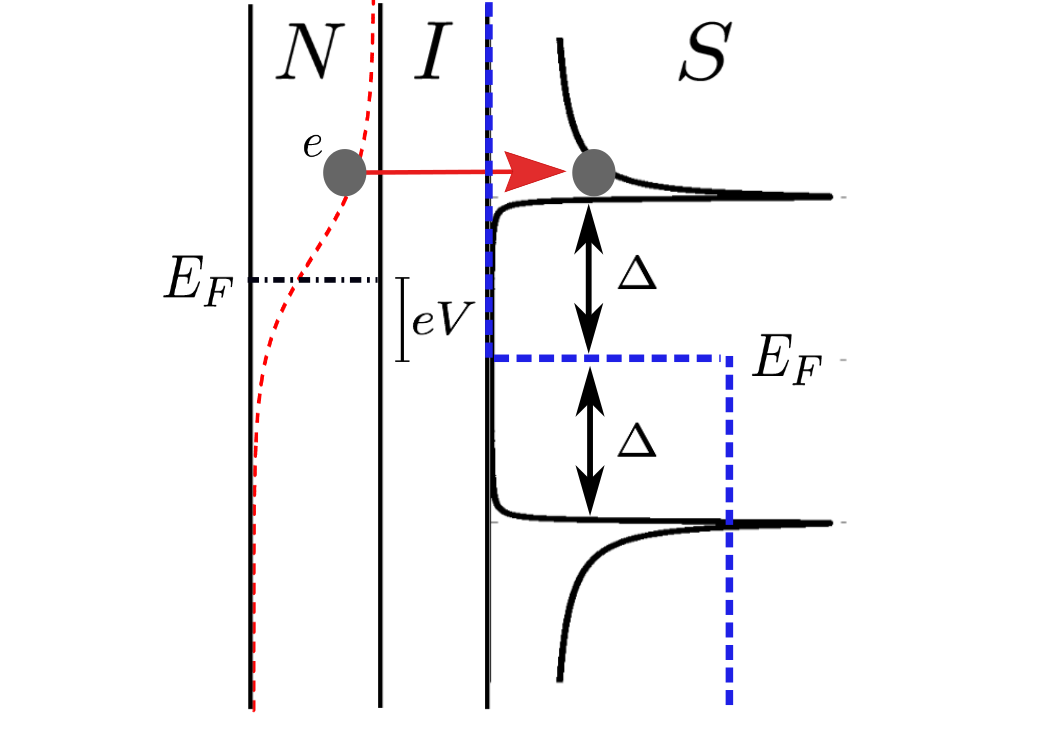}
\caption{Schematic picture illustarting the DoS at both sides of a SIN junction. At finite temperature only particles with high enough energies can tunnel to the empty states (red arrow). S is kept at zero temperature.} \label{fig:panelSIN}
\end{center}
\end{figure}

As an example for active element we consider a simple setup. It consists of a N/I/S (normal metal reservoir/insulator/ superconductor reservoir) tunnel junctions at a bias voltage $V$. The flow of charge current is accompanied by a heat transfer from N into S. Let us assume that we have a "hot" N that we want to cool by putting it in contact with a "cold" S. Particles are transmitted from the N to the empty states of the S. Particles with energies within the energy gap cannot be transmitted. Only those with energies $E>\Delta$, the higher energy or "hotter" ones, go to the S, as shown schematically in fig.\ref{fig:panelSIN}. Thus, the presence of the superconducting energy gap leads to a selective tunnelling of high-energy quasiparticles out of the normal metal.  \cite{Nahum, rf:Leivon} This phenomenon generates a heat current from the normal metal
to the superconductor (also referred to as ``cooling power''). 

%This cooling power is symmetric in voltage as depicted in fig.\ref{fig:NIScooling}.
%

The N/I/S junctions are indeed used for the realization of microcoolers.~\cite{Nahum,rf:Leivon,rf:Giazotto06,rf:Muhonen, Review_2} Present state-of-the-art experiments allow the reduction of the electron temperature in a normal metal lead from 300 to about 100 mK, offering perspectives for on-chip cooling of nano or micro systems, such as high-sensitive sensors, detectors and quantum devices.~\cite{rf:Clark,rf:ONeil,rf:Lowell, Ullom}

\begin{figure}[h]
\begin{center}
\includegraphics[width=0.5\columnwidth]{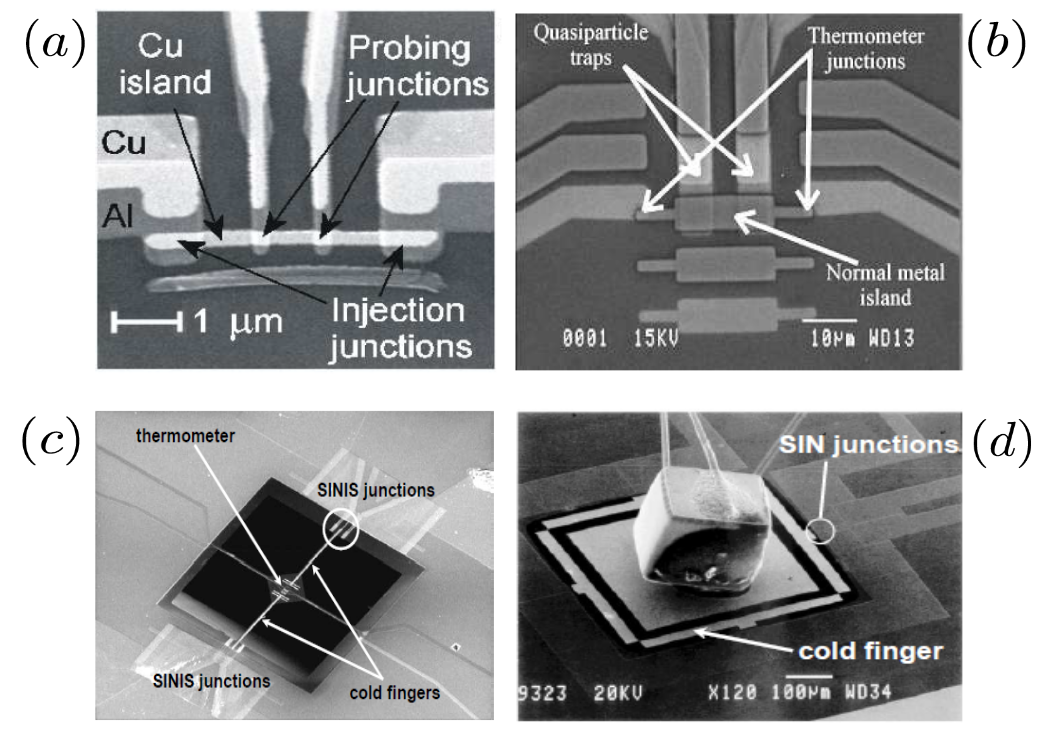}
\caption{ (a) Scanning electron micrograph
of a typical SINIS microrefrigerator. The structure
was fabricated by standard electron beam lithography combined
with Al and Cu UHV e-beam evaporation (from ref.\cite{pekola2004}). (b) SEM micrograph
of an Al/Al$_2$O$_3$/Cu SINIS microrefrigerator exploiting
large-area junctions ($\sim 10 \ \mu m^2$) with quasiparticle traps (from ref.\cite{pekola2000}). (c) SEM image of a Si$_3$N$_4$ membrane (in the center) with self-suspended
bridges. Two normal-metal cold fingers extending onto the
membrane are used to cool down the dielectric platform. The
Al/Al$_2$O$_3$/Cu SINIS coolers are on the bulk (far down and
top) and the thermometer stands in the middle of the membrane (from ref.\cite{manninen}). (d) SEM
micrograph of a NIS refrigerator device with a neutron transmutation
doped (NTD) Ge resistance thermometer attached
on top of it (from ref.\cite{clark}).} \label{fig:panelexp}
\end{center}
\end{figure}

The cooling power  of tunnel junctions depends on several parameters, some of them controllable.
For example the cooling power can be optimized by  controlling the  voltage across the junction. A  maximized cooling effect is reached
at a voltage bias just below the superconducting energy gap $\Delta$.
Larger values of voltage, $eV \gtrsim \Delta$, lead to a larger charge current $I$ through the junction and hence to larger values of  the
Joule heating power, i.e. to a negative  cooling power.
A  limitation of the performance of a NIS microcooler  arises also  from the fact that nonequilibrium quasiparticles injected into the superconducting electrode accumulate near the tunnel interface. \cite{Ullom, pekola2004, VH} As a consequence hot quasiparticles may tunnel  back  into the normal metal,
leading to a reduction of the cooling effect. \cite{VH,Jug}
In order to overcome this problem  a so called quasiparticle trap,\cite{traps} made of an additional
normal metal layer has been attached to  the superconducting electrode, removing hot quasiparticles
from the superconductor.  Recently, it was also shown that a small
magnetic field enhances relaxation processes in a superconductor and leads to significant improvement of the  cooling power in NIS junctions.\cite{Pekola_H} Improved cooling performance can be also achieved by proper tuning of the tunnelling resistances of the individual NIS tunnel junctions in a double junction SINIS cooling device. \cite{Chaudhuri} In fig. \ref{fig:panelexp} we show several experimental setups.

Another important limitation for NIS microcoolers arises from the intrinsic multi-particle nature of current transport in NIS junctions. As discussed in section\ref{sec:proximityeffect}, the transport
is governed not only by single-particle tunnelling but also by two-particle processes due to the
Andreev reflection.\cite{Andreev} The single-particle current and the associated heat
current are due to quasiparticles with energies $E > \Delta$. While at low temperatures or high junction transparencies, the charge transport in NIS junctions is dominated by the Andreev reflection, i.e. by subgap processes. The Andreev current $I_A$ does not transfer heat through the NS interface, as no quasiparticles are transferred. Rather, it generates the Joule heating $I_A V$. At low enough temperatures this heating  exceeds the single-particle cooling.\cite{Sukumar, BA, VB,rf:Rajauria}
The interplay between the single-par\-ticle tunnelling and Andreev reflection sets a
limiting temperature for the refrigeration $T_{min}$.\cite{VB}

One way to decrease $T_{min}$ is to decrease the NIS junction transparency.
However, large values of the contact resistance hinder carrier transfer and lead to
a severe limitation in the achievable cooling powers. In order to increase the junction transparency
and at the same time to reduce the Andreev current, it was suggested to use materials where the
proximity effect is suppressed, such as  ferromagnets, ferromagnetic insulators, and half-metals.  In particular  Giazotto \textit{et al.}
studied theoretically a  ballistic normal metal - ferromagnet - superconductor structure within a phenomenological model and predicted
an enhancement  of the cooling efficiency compared to  NIS junctions.\cite{Giazotto}
The  reason for that increase  lies in the suppression of the Andreev reflection due to the band structure of the
ferromagnetic metals. As explained in Section \ref{sec:andreev},  the electron involved in Andreev reflection and its time-reversed counterpart (hole) must belong to opposite spin bands; thus, suppression of the Andreev
current occurs in a FS junction and its intensity depends on the degree of the electron polarization at the Fermi level which is proportional
to the exchange field of the F layer.\cite{deJong1995,RevG, RevB, RevV}
Note  that   theoretical studies of electron cooling in SF proximity systems were performed only in the ballistic case, \cite{Giazotto, Burmistrova}
while real metallic systems are in the diffusive limit.
Moreover, ferromagnets show in general a  multi-domain structure that was not considered in previous  articles. Another way to suppress the Andreev reflection in a superconductor/two-dimensional electron gas nanostructure was also studied in Ref.~\cite{Giaprl}.

We now quantify these effects by introducing the quantity corresponding to cooling power $\dot{Q}$ in this formalism, i.e. 
the heat current flowing out of the wire to the reservoir at the same temperature as the heat sink.
As discussed above, the cooling power can be written as the sum of two different contributions,\cite{VB}
\begin{equation}\label{eq:P}
\dot{Q} = Q - I V = \dot{Q}_1 + \dot{Q}_A \; ,
\end{equation}
where
\begin{equation}
\dot{Q}_1 =  Q_1 - I_1 V \; , \quad \dot{Q}_A = - I_A V \; .
\end{equation}
The cooling power $\dot{Q}$ corresponds to the energy current flowing out of the wire to the reservoir minus the joule heating generated in the system by the electric current.  Here $Q_{1(A)}$ and $I_{1(A)}$ are
respectively the single-particle (Andreev) energy-current and the single-particle
(Andreev) charge-current. Notice, that the contribution of the Andreev processes
to the energy current vanishes, $Q_A = 0$. Therefore the Andreev process contributes
only to the Joule heating (\textit{i.e.} $\dot{Q}_A =- I_A V$), which is fully released in
the N electrode and leads to a severe reduction of the cooling power. It is clear that for equal temperature of the electrodes ($T_N = T_S$) and no bias voltage the cooling power vanishes.

\begin{figure}[h]
\begin{center}
\includegraphics[width=0.5\columnwidth]{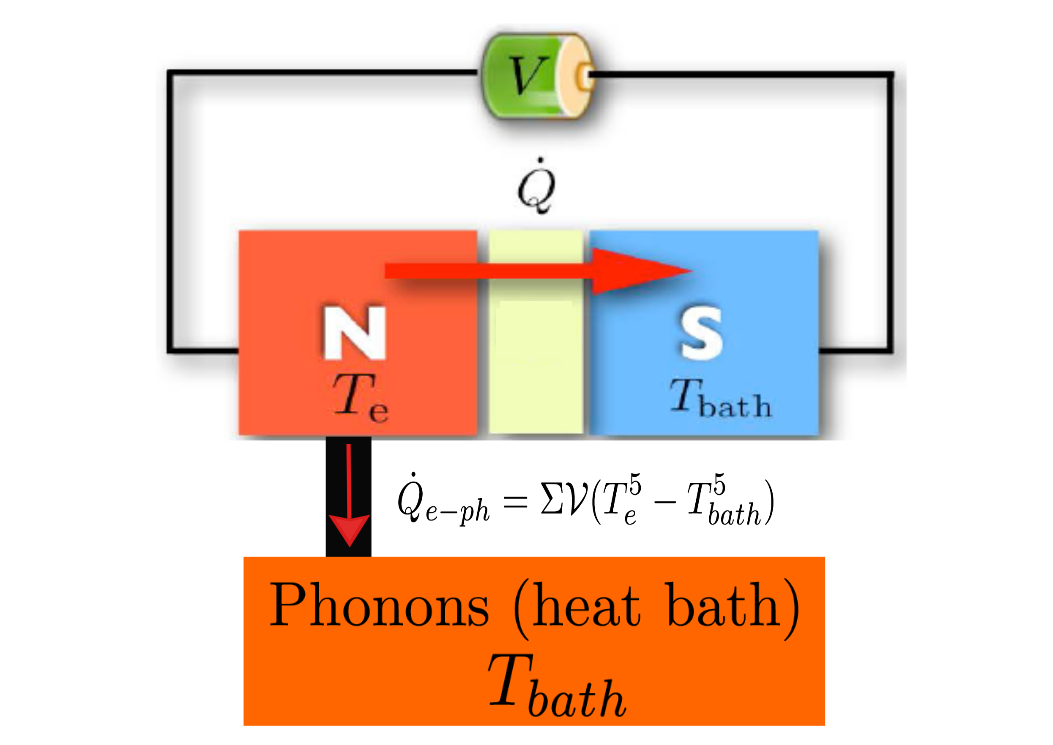}
\caption{Schematic description of the cooling process in SIN junctions. We are interesting in the Te of the normal metal, which due to the bias voltage is lowered by the superconductor (kept at the temperature of the bath, Tbath). The Te is also affected by the electron phonon coupling with the heat bath.} \label{fig:Te}
\end{center}
\end{figure}

Next, we introduce the concept of the final electron temperature of the normal
metal. So, we need to consider the mechanism of energy
transfer. Our purpose is to calculate which final electron temperature of the wire (smaller than the reservoir temperature) can be achieved  in the NIS system. We illustrated this process in fig.\ref{fig:Te}. In this case the electron temperature correspond to the temperature of the N. It depends on the efficiency with which heat generated in the electron population can be transferred to the low-temperature bath. Due to the voltage biasing some power is dissipated as heat, this power is
supplied initially to the electrons in the metal and is transmitted
to the bath by phonons. In thin films at low temperatures,
the wavelength of a thermal phonon is much less than the film
thickness so the temperature of the phonons is the same as the
temperature of the bath, the lattice temperature, $T_{bath}$ . Then the
electron temperature is determined by the rate at which electrons
can transfer energy to the phonons which is given by~\cite{wellstood}
\begin{equation}
\dot{Q}_{e-ph}=\Sigma \mathcal{V} (T_e^5-T_{bath}^5) \; .
\label{eq:phonon}
\end{equation}
Here $\mathcal{V}$ is the volume of N, $T_{e(T_{bath})}$ is the electron (bath) temperature and $\Sigma$ is a material dependent parameter.

%In most of the cases the results of the experiments have been fitted to clean-limit expressions, i.e. the heat current that flows into the phonon system is proportional to $ T_e^5$. The $T^5$ dependence comes from the fact that the Eliashberg functions scaling with $\omega^n$ translate into temperature dependences scaling as $T^{n+3}$. 

The final electron temperature $T_e$ is determined by the energy-balance equation 
\begin{equation}
\dot{Q}(T_e, T_{bath})+\dot{Q}_{e-ph}(T_e,T_{bath})=0 \; ,
\end{equation}
where we set the temperature of the superconductor reservoir ($T_S$) equal to the bath temperature $T_{bath}$. This is an idealized assumption since the presence of hot quasi-particles in S, that may strongly decrease the real cooling capability of the device, is expected. In real applications, however, one can overcome this problem through the exploitation of "quasi-particle traps"\cite{anghel}.  Note that in the absence of bias voltage there is no cooling and thus, $T_e=T_{bath}$. It is only when we apply a bias voltage that the N reaches lower temperatures. 

In order to quantify the efficiency of a refrigerator we introduce the coefficient performance ($\eta_{eff}$). It reads, 
\begin{equation}
\eta_{eff}=\frac{\dot{Q}_{max}}{I V_{opt}} \; .
\end{equation}
It is the ratio between the optimum cooling power and the total input power. Irreversible processes (e.g. thermal conductivity and Joule heating) degrade the efficiency of refrigerators and are essential elements that need to be evaluated for the optimization of any device.

%We now proceed to analyse different junctions for cooling. In particular we will study SIF$_1$F$_2$N and SI$_{sf}$N junctions. The simplest example is the SIN junction which has been extensively studied in the literature\cite{Nahum,rf:Leivon}. A way to suppress the Andreev reflection, and hence the Joule heating, is to place a F metal between the S and N electrodes, as it has been proposed in Ref.\cite{Giazotto}.

Besides the $SIN$ system other junctions can be used for cooling. In this chapter we present a study of different SF structures that can be used as coolers. In the first part we investigate we investigate heat and charge transport through a  diffusive SIF$_1$F$_2$N tunnel junction, where N (S) is a  normal (superconducting) electrode, I is an insulator layer and F$_{1,2}$ are   two ferromagnets with arbitrary direction of  magnetization. The flow of an electric current in such structures at subgap bias is accompanied by a heat
transfer from the normal metal into the superconductor, which enables
refrigeration of electrons in the normal metal.
We demonstrate that the
refrigeration efficiency depends on the strength of the ferromagnetic exchange field $h$ and the
angle $\alpha$ between the magnetizations of the two F layers.   As expected, for  values of $h$ much larger than the superconducting order
parameter  $\Delta$,   the proximity effect is  suppressed and the efficiency of refrigeration increases  with respect to a NIS junction.
However,  for  $h\sim \Delta$  the cooling power (i.e. the heat flow out of the
normal metal reservoir) has  a non-monotonic behaviour as a function of $h$ showing a minimum at $h\approx\Delta$.
We also determine the dependence of the cooling power on the lengths of  the ferromagnetic layers, the bias voltage, the temperature, the
transmission of the  tunnelling barrier and the magnetization misalignment angle $\alpha$.

In the second part of this chapter, we propose another more efficient way of Joule heating suppression by inserting a spin filter. A normal-metal/spin-filter/superconductor junction is proposed and demonstrated
theoretically. 
The spin-filtering effect leads to values of  the cooling power  
much higher than in conventional normal-metal/nonmagnetic-insulator/superconductor coolers and allows
 for an efficient extraction of heat from the normal metal.  
 We demonstrate that highly efficient cooling can be  realized in both ballistic and 
diffusive multi-channel junctions in which the reduction of the electron temperature from
300 mK to around 50 mK can be achieved. Our results indicate the practical usefulness of
 spin-filters for efficiently cooling detectors, sensors, and quantum
devices. 

\newpage

\subsection{Cooling in a SF$_1$F$_2$N junction}

%%%%%%%%%%%%%%%%%%%%%%%%%%%%%%%%%%%%%%%%%%%%%%%%%%%%%%%%%%%%%%%%%%%%%%%%%%%%
%%%%%%%%%%%%%%%%%%%%%%%%%%%%%%%%%%%%%%%%%%%%%%%%%%%%%%%%%%%%%%%%%%%%%%%%%%%%

%\section{Introduction}

In this section we  present a  quantitative  analysis of the thermal and electric transport using the expressions introduced in section\ref{sec:proximityeffect}. As it was shown in Ref.\cite{Giazotto}, a ferromagnetic interlayer in a SFN structure can, in principle, enhance the cooling power of structure due to the suppression of the Andreev Joule heating. In that work the ferromagnet was assumed to be monodomain, \textit{i.e.} its magnetization is spatially homogeneous. However, it is well known, that usual ferromagnets consists of many magnetic domains separated by domain walls. Therefore, it is important to understand how such a magnetic inhomogeneity may affect the cooling power.

\begin{figure}[h]
\begin{center}
\includegraphics[scale=0.25]{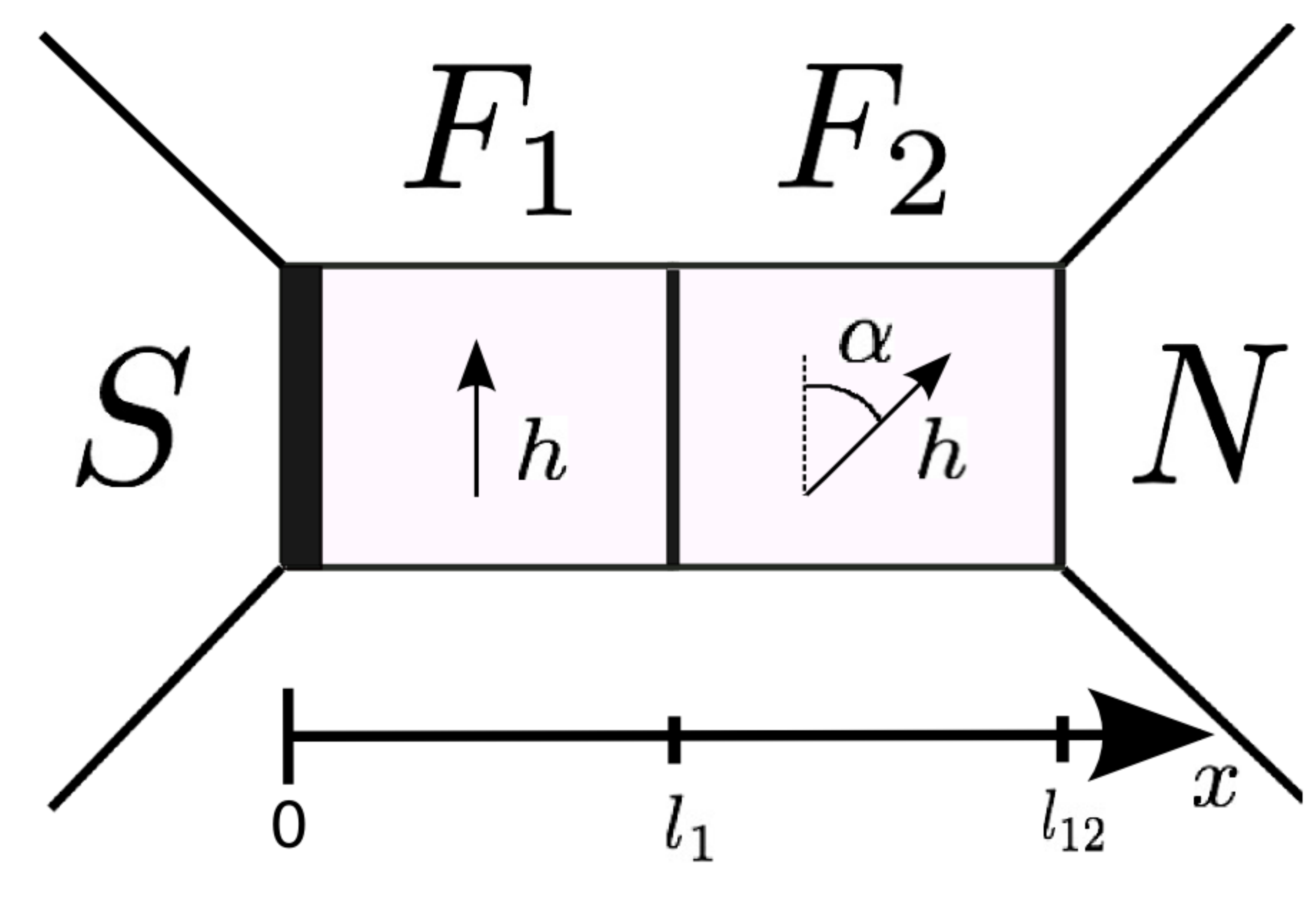}
\caption{The  SIF$_1$F$_2$N junction.
The interface at $x=0$ corresponds to the insulating barrier (thick black line). Interfaces at $x=l_1$ and $x=l_{12}$ are fully transparent.
 $\alpha$ is the relative angle between the magnetization directions of  F$_1$ and F$_2$. (From ref.\cite{rf:Ozaeta})} \label{model}
\end{center}
\end{figure}

In order to model such situation, we consider a simple structure consisting of two magnetic domains with arbitrary magnetization direction (so called ``superconducting triplet spin-valve''\cite{spin-valve}). As shown in Fig.\ref{model}. By using the Keldysh quasiclassical Green function formalism that is introduced in chapter\ref{ch:2}, we calculate the charge and heat currents and the cooling power of the structure.

%%%%%%%%%%%%%%%%%%%%%%%%%%%

%\section{Model and basic equations}

The SIF$_1$F$_2$N junction is depicted in Fig.~\ref{model}.
A ferromagnetic bilayer F$_1$F$_2$ of length $l_{12}=l_1+l_2$ smaller than the inelastic relaxation length\cite{Arutyunov} is connected
to a superconductor (S)  and a normal (N) reservoirs along the $x$ direction.
The  F$_1$F$_2$ bilayer can either model  a two domain ferromagnet or an artificial  hybrid magnetic structure.
We consider the diffusive limit, i.e the elastic scattering length $\ell \ll \min (\xi_h, \xi)$, where $\xi_h = \sqrt{\mathcal{D}/ 2 h}$
is the characteristic penetration length of the superconducting condensate into the ferromagnet, $h$ is the value of the exchange
field, $\xi = \sqrt{\mathcal{D}/ 2 \Delta}$ is the superconducting coherence length and
$\mathcal{D}$ is the diffusion coefficient  (for simplicity we assume the same $\mathcal{D}$ in the whole structure).

We also assume that the F$_1$F$_2$ and F$_2$N interfaces are transparent, while the SF$_1$ is a tunnel interface. Thus, the two ferromagnetic layers
are kept at the same potential as the voltage-biased normal reservoir.
The magnetization of the F$_1$ layer is along the $z$ direction, while the magnetization of the F$_2$ layer forms an angle $\alpha$ with the
one of the layer F$_1$. Both magnetization vectors lie in the $yz$ plane.
Correspondingly, the exchange field vector in the F$_1$ is given by
${\textbf{h}} = (0, 0, h)$, and in the F$_2$ layer
by ${\textbf{h}} = (0, h\sin\alpha, h\cos\alpha)$, where
the angle $\alpha$ takes values from  0 (parallel configuration)
to  $\pi$ (antiparallel configuration).

The Usadel equation, eq.\ref{eq:Usadel3D2}, for this system reads, 
\begin{equation}\label{Usadel3D}
i\mathcal{D} \partial_x \breve{J} = \left[ \tau_z\left(E  - {\textbf{h}} {\boldsymbol{\sigma}}\right),
\breve{G}\right], \quad \breve{J} = \breve{G}\partial_x \breve{G}, \quad \breve{G}^2 = 1,
\end{equation}
where ${\boldsymbol{\sigma}} = (\sigma_1, \sigma_2, \sigma_3)$ are the Pauli matrices in spin space.
In the F$_1$ region  ${\textbf{h} \boldsymbol{\sigma}} = h \sigma_3$ and the Usadel equation \Eq{Usadel3D} has the form
\begin{equation}\label{Usadel}
i\mathcal{D} \partial_x \breve{J} = \left[ \tau_z\left(E  - \sigma_3 h\right),
\breve{G}\right], \quad \breve{G}^2 = 1.
\end{equation}
In the F$_2$ region ${\textbf{h} \boldsymbol{\sigma}} = h \sigma_3 \exp(-i \sigma_1 \alpha)$.
It is convenient to introduce Green's functions rotated in spin-space,
\cite{Bergeret2002}
\begin{equation}\label{gauge}
\widetilde{\breve{G}} = U^\dagger \breve{G} U, \quad U =
\exp\left( i \sigma_1 \alpha/2 \right).
\end{equation}
The  rotated function $\widetilde{\breve{G}}$ is then determined by \Eq{Usadel}.

The Usadel equation \Eq{Usadel} should be complemented by boundary conditions at the interfaces.
As mentioned above, we assume that
the F$_1$F$_2$ and F$_2$N interfaces are transparent and therefore the boundary conditions at $x=l_1,l_{12}$ read, eq.\ref{eq:trans},
\begin{eqnarray}
\breve{G} \bigl |_{x=l_1 - 0}& =& \breve{G} \bigl |_{x=l_1 + 0}\label{bc1},\\
\partial_x \breve{G} \bigl |_{x=l_1 - 0}& =& \partial_x \breve{G}
\bigl |_{x=l_1 + 0}\label{bc11},\\
\breve{G}\bigl|_{x=l_{12}-0} &=&\tau_z.
\end{eqnarray}

At $x=0$,  the SF$_1$ interface is a tunnel barrier, and we
may use the Kupriyanov-Lukichev boundary conditions, eq.\ref{eq:kl},
\begin{equation}\label{kupluk}
\breve{J} \bigl |_{x=0} = \kappa_t \left[
\breve{G}_S, \breve{G} \right]_{x=0},
\end{equation}
where $\breve G_S$ is the Green function of a bulk BCS superconductor and $\kappa_t$ is defined in section\ref{sec:kupri}. We set $\eta \simeq 10^{-3} \Delta_0$ in this calculations, where $\Delta_0$ is the superconducting gap at $T=0$. In the following we omit the inelastic scattering rate $\eta$ in the analytical expressions for simplicity.

Because of the  low transparency of the SF$_1$
barrier, the proximity effect is weak and the  retarded Green function can be linearised (we omit the superscript $R$),
\begin{equation}\label{G_lin}
\check{G} \approx \tau_z + \tau_x \hat{f},
\end{equation}
where $\hat{f}$ is the $2 \times 2$ anomalous Green function in the spin space ($|\hat{f}|\ll 1$) that obeys  the linearised Usadel equation,
\begin{equation}\label{linearized}
i\mathcal{D} \partial^2_{xx} \hat{f} = 2 E \hat{f} - \left\{ {\textbf{h}
\boldsymbol{\sigma}}, \hat{f} \right \},
\end{equation}
where $\{\cdot, \cdot\}$ stands for the anticommutator.

The general solution of \Eq{linearized} has the form
\begin{equation}\label{f}
\hat{f}(x) = f_0(x) + f_2(x) \sigma_2 + f_3(x) \sigma_3,
\end{equation}
where $f_0$ is the singlet component and $f_3$, $f_2$ are the triplet components with respectively zero and $\pm 1$ projections on the spin
quantization axis (we choose the $z$-axis).

The charge  and energy  currents,  $I$ and $Q$ respectively, take the expressions from section\ref{sec:proximityeffect},
%can be obtained from\cite{LOnoneq, Belzig, Vinokur, Golubov}
%
\begin{subequations}
\begin{align}\label{I}
I &= \frac{g_N}{e} \int_0^{\infty} I_- \; dE, \quad Q =
\frac{g_N}{e^2} \int_0^{\infty} E I_+ \; dE,
\\
\label{Ipm} I_- &\equiv (1/8) \tr \tau_3 \check{J}^K, \quad I_+
\equiv (1/8) \tr \tau_0 \check{J}^K,
\\
\check{J}^K &\equiv \left( \breve{G}\partial_x \breve{G} \right)^K
= \check{G}^R \partial_x \check{G}^K + \check{G}^K \partial_x \check{G}^A.
\end{align}
\end{subequations}
where  $\breve{J}^K$ is the Keldysh component of the matrix current  defined in Eq. (\ref{Usadel3D})
and $\tau_0$ is the unitary matrix in Nambu space.  Next, we calculate the cooling power of the SIF$_1$F$_2$N junction as a function of  the different parameters by solving equations \ref{I} and \ref{Ipm}.

%%%%%%%%%%%%%%%%%%%%%%%%%%%%%%%%%%%%%%%%%%%%%%%%%%%%%%%%%%%%%%%%%%%%%%%%%%%

%\section{Results and discussion}

In the linear case both the electric and heat currents are determined by
the singlet component  $f_0$ of the anomalous Green function,
\Eq{f}, evaluated at the SF$_1$ interface ($x=0$).
Solving \Eq{linearized} in the F$_1$ layer we obtain for the components of \Eq{f},
\begin{subequations}\label{f_i}
\begin{align}
f_\pm(x) &= a_\pm \cosh (k_\pm x) + \frac{2\kappa_t}{k_\pm} (u a_\pm - v)
\sinh (k_\pm x),
\\
f_2(x) &= a_y \cosh (k_y x) + \frac{2\kappa_t}{k_2} u a_2 \sinh (k_2 x),
\end{align}
\end{subequations}
where $f_\pm = f \pm f_z$,  $a_i$ are the boundary values of $f_i$ at  $x=0$
($i$ stands for $+,-,2$) and
\begin{equation}
k_\pm = \sqrt{\frac{2(E \mp h)}{i\mathcal{D}}}, \quad
k_2 =\sqrt{\frac{2E}{i\mathcal{D}}}.
\end{equation}
In the F$_2$ layer  the general solution has the form,
\begin{equation}
\widetilde{f}_i(x) = b_i \sinh \left[ k_i (x - l_{12}) \right],
\end{equation}
where $\widetilde{f}_i$ are the components of the rotated Green function, \Eq{gauge}.

Using the boundary conditions at the   F$_1$F$_2$ interface, Eqs. (\ref{bc1}-\ref{bc11}) one obtain a set of six linear equations for the
six coefficients $a_i$ and $b_i$, that can be solved straightforwardly.

\begin{figure}[h]
\begin{center}
\includegraphics[scale=0.3]{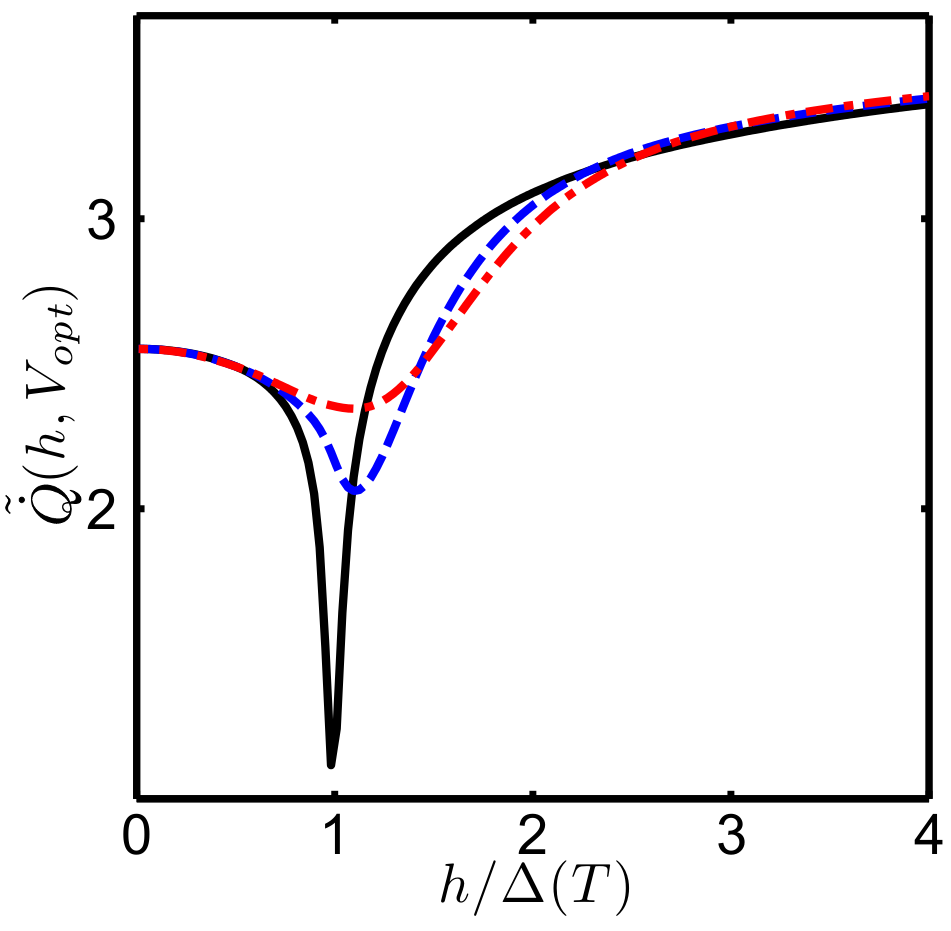}
\caption{ Cooling power versus exchange field for different
orientations of the exchange field vector in the second ferromagnetic layer F$_2$: $ \alpha= 0$ (black solid line),
$ \alpha= \pi /2$ (blue dashed line) and $ \alpha= \pi $ (red dash-dotted line), calculated at optimum bias;  $\kappa_t=7 \times 10^{-3}$,
$T=0.25 \Delta_0$, $l_1=\xi$ and $l_2=6 \xi$.  We have defined $\tilde{\dot{Q}}=10^{2}P(V_{opt})e^2R_0/\Delta^2_0$. (From ref.\cite{rf:Ozaeta})} \label{Pvsh10}
\end{center}
\end{figure}
\begin{figure}[h]
\begin{center}
\includegraphics[width=\columnwidth]{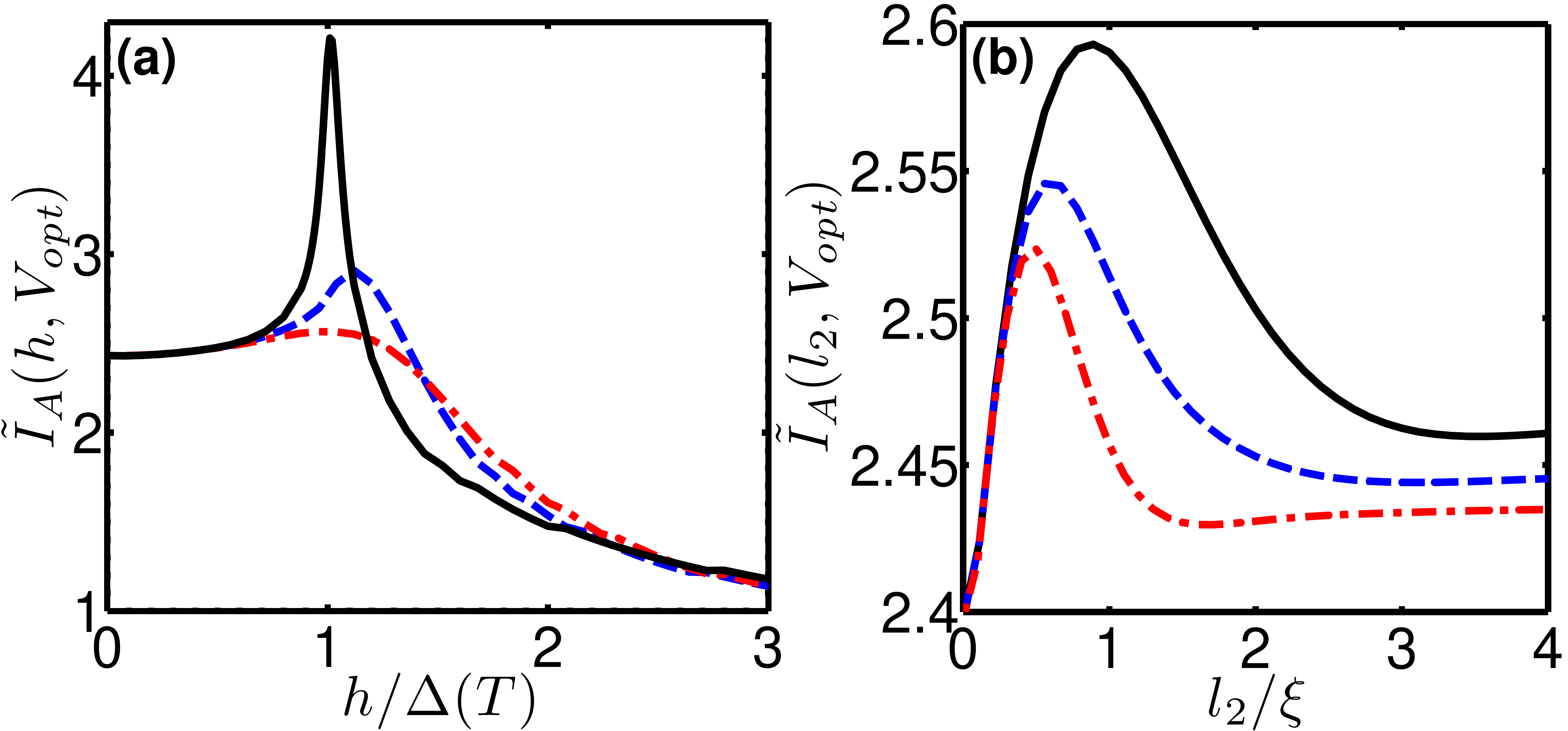}
\caption{ The Andreev current as a function of (a) the exchange field for  $l_2 = 6\xi$ and  as a function of (b)  the F$_2$  length  for
$h = 0.7 \Delta(T)$.  Different magnetic configurations are chosen:
$\alpha= 0$ (solid black line), $\alpha= \pi /2$ (dashed blue line),  $ \alpha= \pi $ (dash-dotted red line). The Andreev current is calculated at optimal bias;
$\kappa_t = 7 \times 10^{-3}$, $T=0.25 \Delta_0$, $l_1=\xi$.
We have defined $\tilde{I}_A=I_A(V_{opt})eR_0/\Delta_0$. (From ref.\cite{rf:Ozaeta})} \label{panelIA}  \vspace{-4mm}
\end{center}
\end{figure}

In particular we are interested in the value of the singlet component of the anomalous Green function
at $x=0$ which is given by  $f_0=(a_++a_-)/2$.
Once we obtain $f_0$ we compute the charge and energy currents from \eqref{I}. Finally, using eq.\ref{eq:P} we determine the cooling power.
In what follows we assume that the temperatures of the S and
N reservoirs to be equal, $T_S = T_N = T$, and neglect nonequilibrium effects in the ferromagnetic interlayer.\cite{VB}

The bias voltage between the S and N reservoirs is an easily adjustable
experimental parameter, so all the curves except those presented in Fig.~\ref{panelv} are calculated for optimal value
of the voltage bias $V_{opt}$, at which the cooling power reaches its maximum for given values of the
other parameters. In what follows, we assume the quantity $\kappa_t$ to be taken at $T=0$, allowing for its temperature
dependence in \Eqs{f_i} by means of corresponding temperature-dependent factors.
In the subsequent analysis the cooling power $P$ is given in units of  $\Delta^2_0/e^2 R_0$, where
$\Delta_0$ is the value of $\Delta$ at zero temperature and $R_0$ is the junction resistance at a fixed value $\kappa_t=10^{-2}$ of the
tunnelling parameter.

We first  study the dependence of the cooling  power on the strength of the exchange field $h$.
This dependence is shown in  Fig.~\ref{Pvsh10} for three different angles $\alpha=0,\pi/2,\pi$ between
the magnetizations of F$_1$ and F$_2$ layers at the optimum value of bias voltage. We have chosen the
values of the temperature and tunnelling parameter $\kappa_t$ such that the Andreev current role in the cooling
processes is essential (see Fig.~\ref{panelW}).\cite{VB} The thickness of the F layers is chosen to be $l_1=\xi $ and $l_2=6\xi$.

Depending on the value of $l_1/\xi_h$, where  $\xi_h=\sqrt{\mathcal{D}/2h}$ is the characteristic  penetration length of the
superconducting condensate into F$_1$, one identifies different behaviours.  If $l_1\gg\xi_h$, i.e. for large values of $h/\Delta(T)$
the amplitude of the superconducting condensate  in F$_2$ can be neglected, as well as  the dependence of the $f_0$ function on the angle $\alpha$.
Thus, in the limit $h/\Delta(T)\gg1$,   the  value of the cooling power  does not depend on $\alpha$. Moreover, this asymptotic value  is larger than
in the nonmagnetic case ($h=0$).  This is a consequence of the strong suppression  of the singlet correlations in F$_1$ due to the exchange field and
hence of the Joule heating associated to the  Andreev current [see eq.\ref{eq:P}].  Note that for the value of temperature used in
the figures $\Delta(T)\approx\Delta_0$ .

\begin{figure}[!htbp]
\begin{center}
\includegraphics[width=\columnwidth]{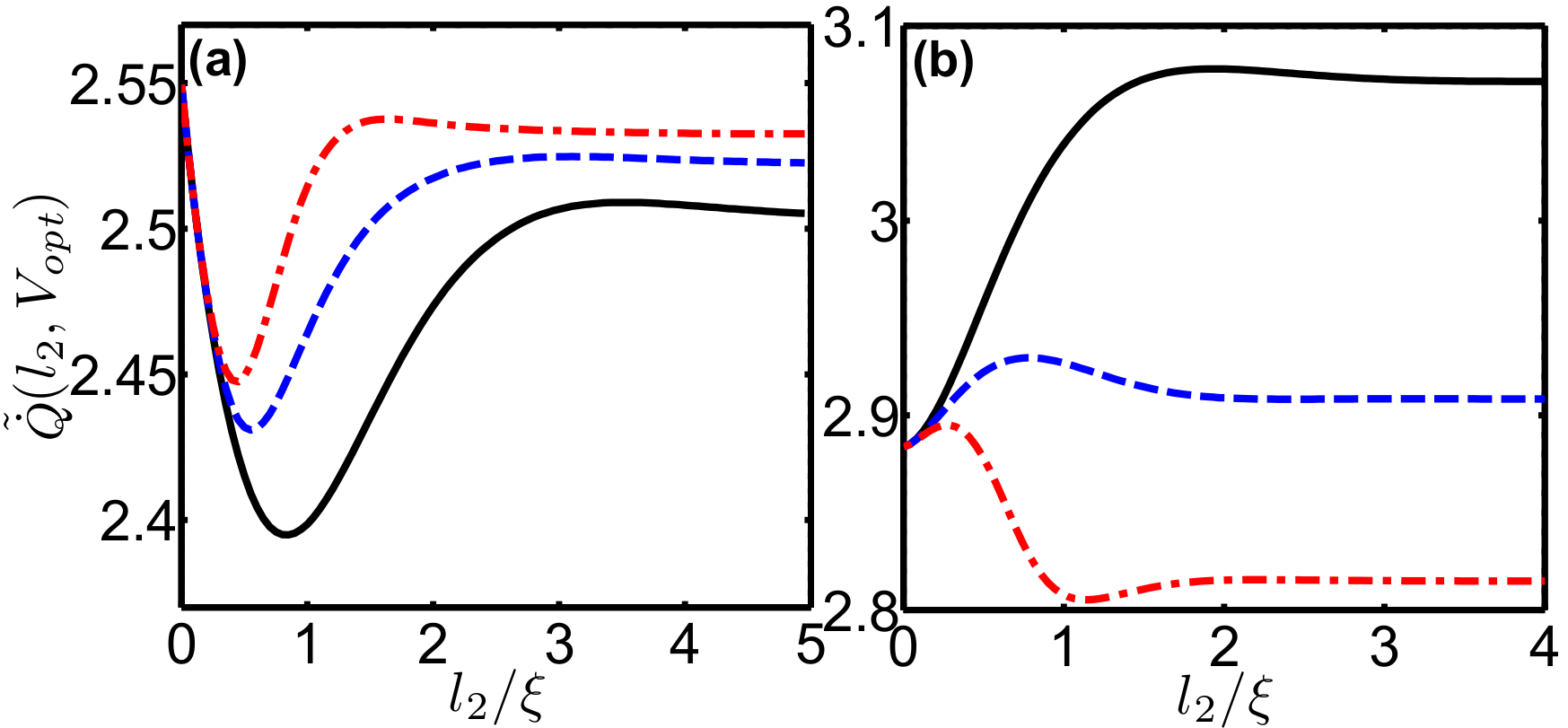}
\caption{ Cooling power versus  length $l_2$ of  the F$_2$ layer  for (a)  $h=0.7\Delta(T)$ and (b) $h=1.7\Delta(T)$.
We  consider different orientations of the exchange field vector in the second ferromagnetic layer F$_2$ with respect to the one in F$_1$:
$\alpha= 0$ (solid black line), $\alpha= \pi /2$ (dashed blue line),  $ \alpha= \pi $ (dash-dotted red line), and
calculate the cooling power  at optimal bias;
$\kappa_t = 7 \times 10^{-3}$, $T=0.25 \Delta_0$ and $l_1=\xi$. $\tilde{\dot{Q}}$ is defined in Fig.\ref{Pvsh10}. (From ref.\cite{rf:Ozaeta})} \label{panell2}  \vspace{-4mm}
\end{center}
\end{figure}

In the opposite limit, $l_1/\xi_h\ll 1$, the characteristic penetration length  depends weakly on $h$, and therefore the cooling power is also
$\alpha$-independent. However, by increasing $h$   the cooling power first decreases and reaches  a minimum.
This unexpected behaviour is qualitatively similar for all  magnetic configurations and is a consequence of the Andreev current peak at $h \approx \Delta(T)$
(for mono-domain case) in the  finite temperature and finite voltage regime, see Fig.~\ref{panelIA}(a), solid black line.
However,  there are quantitative differences between the mono-domain  ($\alpha=0$) and two domain  ($\alpha=\pi,\pi/2$) configurations.
For  $\alpha=0$,   $P(h)$ shows a minimum at $h \approx \Delta(T)$.   It is worth mentioning that   around
this minimum the cooling power of the SIF$_1$F$_2$N system is  lower than that of the NIS junction ($h=0$).
By increasing the angle $\alpha$ the minimum is less pronounced and shifts to larger values of $h\gtrsim\Delta(T)$. For these values of $h$
and for $l_1=\xi$ the superconducting condensate can penetrate both ferromagnetic layers.
Thus, the effective exchange field  $\bar h$ acting on the  Cooper pairs is a field, averaged over the length $\xi_h$.\cite{Bergeret_h}
The $\bar h(\alpha)$ is gradually reduced as $\alpha$ increase from $0$ to $\pi$. As before the
cooling power minimum  is at $\bar h (\alpha) \approx \Delta(T)$ which in the case of a finite $\alpha$ corresponds to larger values of the bare $h$.  The minimum
of the cooling power (Fig.~\ref{Pvsh10}), corresponds to   a  maximum of the Andreev current  [Fig.~\ref{panelIA}(a)].  The  unexpected nonmonotonic  behaviour of the Andreev current at small exchange fields $h \sim \Delta(T)$ is due to the competition between two-particle tunnelling processes and decoherence mechanisms as quantitatively explained in a recent work by the authors. \cite{elecozaeta}
\begin{figure}[!h]
\begin{center}
\includegraphics[scale=0.4]{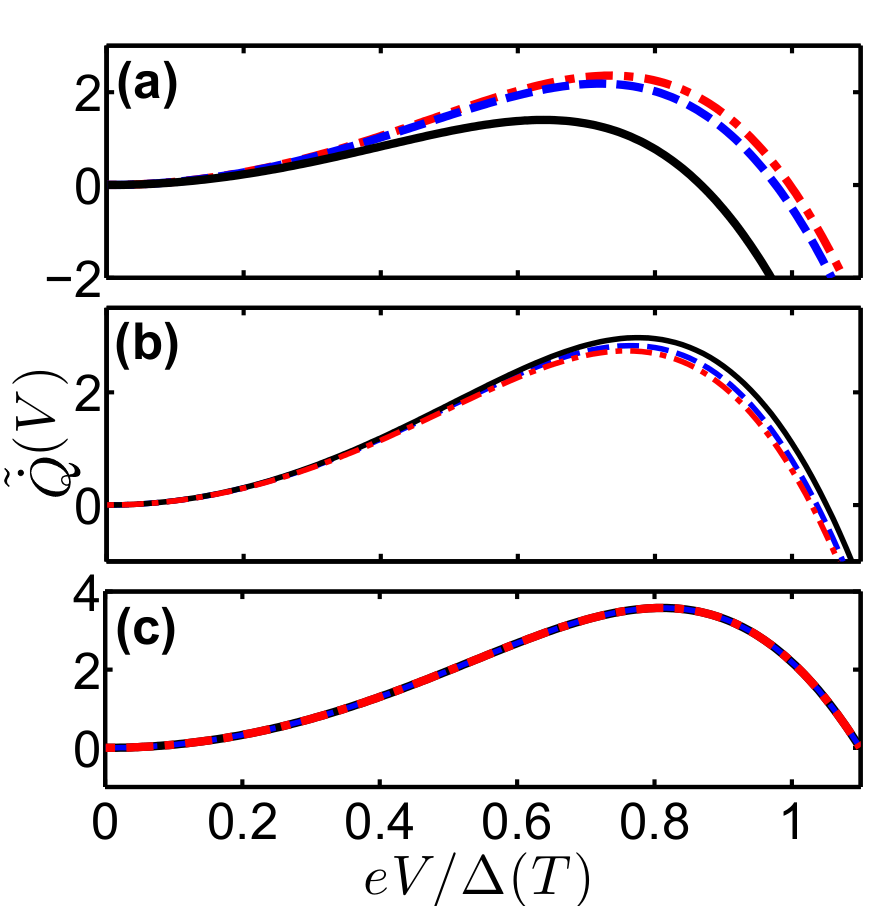}

\caption{Cooling power versus bias voltage  for $h=\Delta(T)$ (a),
$h=1.7\Delta(T)$ (b) and $h=8\Delta(T)$ (c)  for different
orientations of the exchange field vector in the second ferromagnetic layer F$_2$: $\alpha= 0$ (black solid line),
$\alpha= \pi /2$ (blue dashed line) and $\alpha= \pi$ (red dash-dotted line);
$\kappa_t=7 \times 10^{-3}$, $T=0.25 \Delta_0$, $l_1=\xi$ and $l_2=6 \xi$. $\tilde{\dot{Q}}$ is defined in Fig.\ref{Pvsh10}. (From ref.\cite{rf:Ozaeta})} \label{panelv}
\end{center}
\end{figure}

We analyse now the dependence of the cooling power on the length of the ferromagnetic bridge F$_1$F$_2$. To do this, we fix the thickness of F$_1$ at
$l_1=\xi$ and vary  $l_2$. Fig.~\ref{panell2} shows the $P(l_2)$ dependence for  two different values of the exchange field $h/\Delta(T)=0.7, 1.7$ and
different magnetic configurations $\alpha=0,\pi/2,\pi$. As expected   all curves tend to a finite asymptotic value  when $ l_2 \gg \xi$.
This value however depends on $\alpha$.

In the case of an exchange field smaller than the superconducting gap  [$h=0.7\Delta(T)$, see Fig.~\ref{panell2}(a)]
the cooling power first reduces monotonically to a minimum by increasing $l_2$, then enhances to a maximum and finally reduces to the asymptotic value.
Such behaviour is preserved for all magnetic configurations and it follows from the nonmonotonic  behaviour of the Andreev current, shown in Fig.~\ref{panelIA}(b).
Decrease of the Andreev current corresponds to the increase of the cooling power and vice versa. As shown in Fig.~\ref{panelIA}(b),  at large values of $l_2$ the Andreev current  increases
by decreasing $l_2$, reaches a maximum and finally decreases for $l_2 \lesssim \xi$. The strong suppression of the Andreev current for small values of $l_2$  is due to the proximity of the N reservoir at $x = l_{12}$.
On the other hand for larger values of $l_2$ the superconducting proximity effect in the ferromagnetic bridge is fully developed  and leads to an increase of the Andreev current. It is remarkable that the cooling power for $\alpha=\pi$ is larger than the one at $\alpha=0$ for all values of $l_2$.
In this case a lower effective exchange field $\bar h$ leads to larger values of the cooling power, due to the shift of the
minimum of  $P(h)$  observed in Fig.~\ref{Pvsh10}.

For an exchange field larger than $\Delta(T)$ [$h=1.7\Delta(T)$, see Fig.~\ref{panell2}(b)] the behaviour
of the cooling power as a function of $l_2$ strongly depends on $\alpha$. For a mono-domain magnet, $\alpha=0$,
the cooling power increases monotonically by increasing $l_2$ until it reaches the asymptotic value due to the suppression
of the Andreev current as in the ballistic case studied in Ref.~\cite{Giazotto}.  Similarly, in the antiparallel
configuration ($\alpha=\pi$), the cooling power first  increases by increasing $l_2$, however for a larger value of $l_2$
reaches a maximum and then decreases.  The presence of F$_2$ with a magnetization antiparallel to the one of F$_1$ leads
to a reduced effective exchange field of the F$_1$F$_2$ bridge.  Thus, the Andreev current contribution is enhanced with respect
to the one in the case $l_2=0$.  As intuitively expected the cooling power (Andreev current) reaches a minimum (maximum)
when $l_2\sim l_1=\xi$, i.e. when the average magnetization is minimized. Further increase of $l_2>\xi$ leads to a
suppression of the Andreev current and therefore to an increase of $P$ until the asymptotic values are reached.
Fig.~\ref{panell2}(b) also shows the intermediate case $\alpha=\pi/2$.

We now analyse the dependence of the cooling power on the bias voltage $eV$, tunnelling parameter $\kappa_t$ and temperature $T$.
In the subsequent analysis we consider  three different values of the exchange field
$h=\Delta(T),1.7\Delta(T), 8\Delta(T)$, and  three magnetic configurations $\alpha=0,\pi/2,\pi$.
We set $l_1=\xi$, short enough for the pair correlation
to be substantial in the F$_2$ layer [for $h=\Delta(T),1.7\Delta(T)$] and $l_2=6\xi$,  long enough to ensure
the asymptotic regime [see Fig.~\ref{panell2}].
Figs.~\ref{panelv}, \ref{panelW} and  \ref{panelT}  show  the cooling power as a function of
$eV$, $\kappa_t$ and $T$.    A common  feature of these figures is that the   range  of values of  $V$, $\kappa_t$ and $T$,  for which the cooling
power is positive increases by increasing $h$.
Also the magnitude of the cooling power increases  with $h$.  This is in agreement with the qualitative predictions of Ref.~\cite{Giazotto}.
Note that the  shape of all  curves in  Figs.~\ref{panelv}, \ref{panelW} and  \ref{panelT}
does not depend significantly on  the angle $\alpha$.

\begin{figure}[!h]
\begin{center}
\includegraphics[scale=0.4]{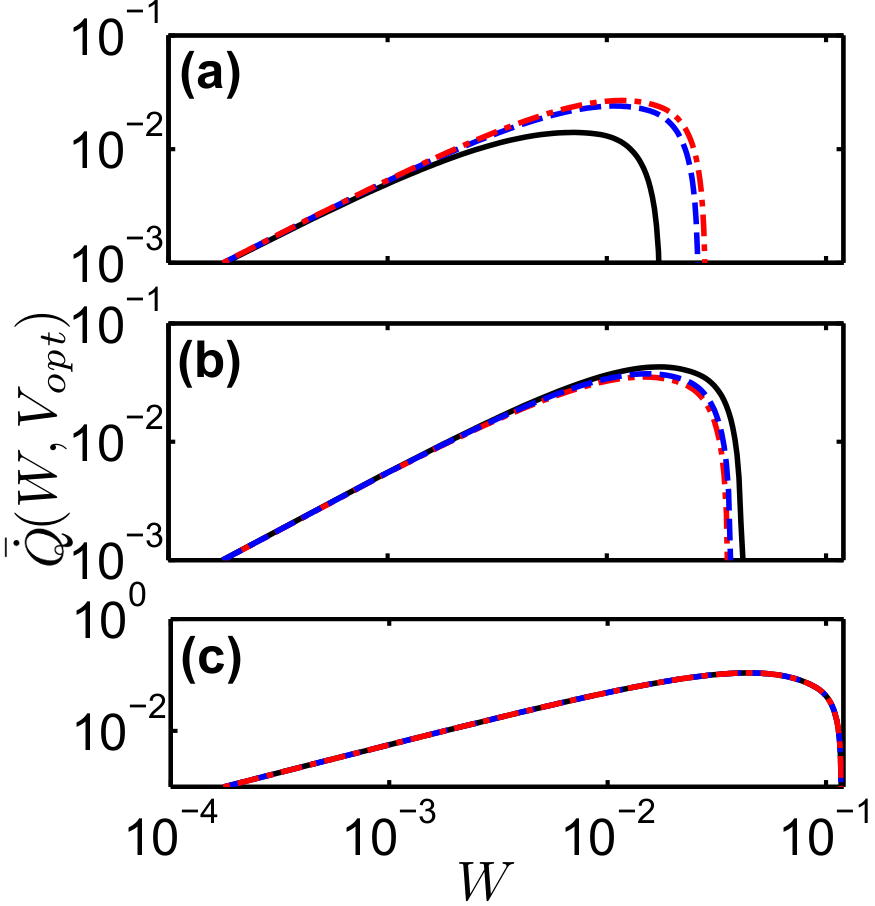}

\caption{ Dependence of the cooling power on the tunnelling parameter $\kappa_t$ for
$h=\Delta(T)$ (a), $h=1.7\Delta(T)$ (b) and $h=8\Delta(T)$ (c) and for different
orientations of the exchange field vector in the second ferromagnetic layer F$_2$: $\alpha= 0$ (black solid line),
$\alpha= \pi /2$ (blue dashed line) and $\alpha= \pi$ (red dash-dotted line). $P$ is calculated at optimum bias;
$T=0.25 \Delta_0$, $l_1=\xi$ and $l_2=6 \xi$. We have defined $ \dot{\bar{Q}}(\kappa_t,V_{opt})=\dot{Q}(\kappa_t,V_{opt})e^2R_0/\Delta^2_0$. Note the logarithmic scale. (From ref.\cite{rf:Ozaeta})} \label{panelW}
\end{center}
\end{figure}

Figs.~\ref{panelv} and \ref{panelW} show that for low values of $eV$ and $\kappa_t$, respectively, the cooling power depends only
weakly  on the relative magnetization  angle $\alpha$.  However, by increasing $eV$ and $\kappa_t$ the difference becomes appreciable,
in particular for $h\approx\Delta(T)$.

As shown in Fig.~\ref{panelv} at certain value of $eV_{opt}\lesssim 0.8\Delta(T)$,   the cooling power reaches its
maximum value $\dot{Q}_{max} = \dot{Q}(V_{opt})$. The $eV_{opt}$ value  is the one used as optimal bias value in the figures.
For voltages larger than this optimal value, the quasiparticle current $I$ and hence the Joule heating power
$IV$ increase drastically  leading to a rapid decrease of the cooling power. As can be seen from Figs.~\ref{panelv}, \ref{panelW} and  \ref{panelT}
the optimal voltage $V_{opt}$ depends on the temperature $T$, tunnelling parameter $\kappa_t$ and magnetic configuration angle $\alpha$.
For the exchange field equal to the superconducting gap the maximal cooling power $\dot{Q}_{max}$ is largest in the antiparallel
configuration, while for larger $h = 1.7 \Delta(T)$ the largest value $\dot{Q}_{max}$ is in the parallel configuration, in agreement with Fig.~\ref{Pvsh10}.

Fig.~\ref{panelW} shows that the cooling power has also a maximum as a function of $\kappa_t$. Increasing $\kappa_t$ the cooling power first linearly increases as single electron tunnelling dominates.
For larger values of the tunnelling parameter, the Andreev current heating dominates over the single-particle cooling and leads to a rapid decrease of the
cooling power, which tends to zero at a certain onset point. As the exchange field increases, the role of Andreev processes becomes less important, therefore
the onset shifts towards larger values of $\kappa_t$. This means that for higher exchange field in the ferromagnetic interlayer one may use weaker tunnel barriers
for the microcooler fabrication, which leads to higher amplitudes of the cooling power [see Fig.~\ref{panelW} (c)] and more effective electron refrigeration.

\begin{figure}[!b]
\begin{center}
\includegraphics[scale=0.4]{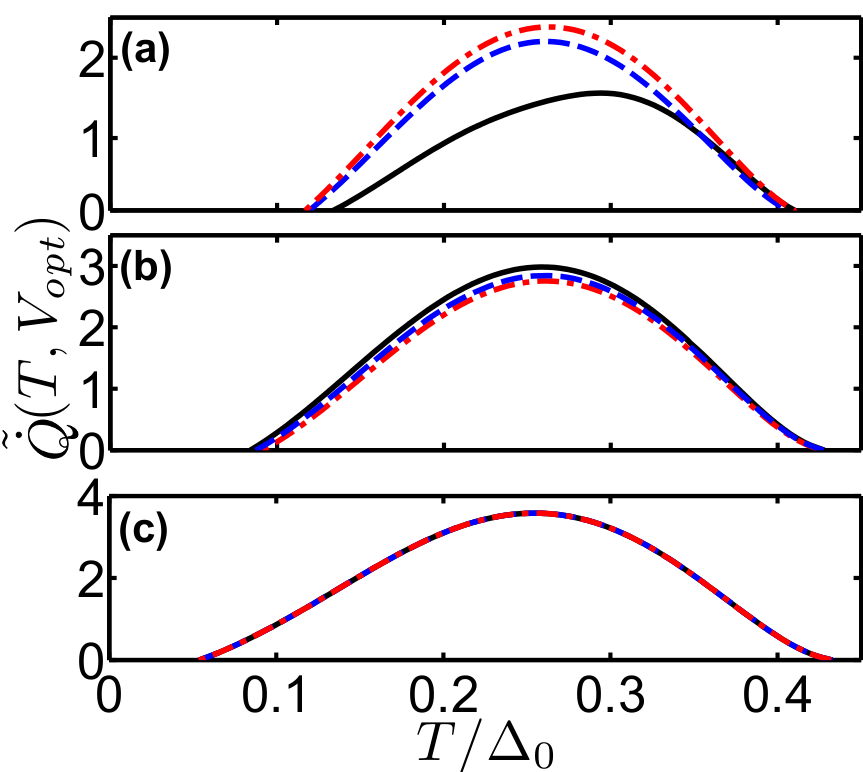}
\caption{ Temperature dependence of the cooling power for $h=\Delta_0$ (a), $h=1.7\Delta_0$ (b) and $h=8\Delta_0$ (c) and for different
orientations of the exchange field vector in the second ferromagnetic layer F$_2$:
$\alpha= 0$ (black solid line),
$\alpha= \pi /2$ (blue dashed line) and $\alpha= \pi$ (red dash-dotted line). $P$ is  calculated at optimum bias;
$\kappa_t=7 \times 10^{-3}$, $l_1=\xi$ and $l_2=6 \xi$. $\tilde{\dot{Q}}$ is defined in Fig.\ref{Pvsh10}. (From ref.\cite{rf:Ozaeta})} \label{panelT}
\end{center}
\end{figure}

In Fig.~\ref{panelT} we show the temperature dependence of the cooling power. At $T \gtrsim 0.42 \Delta_0 \approx 0.75  T_c$, where $T_c$ is the critical temperature of the superconductor,
the cooling power
becomes negative for all voltages. This value of the temperature holds for  a wide range of parameters.\cite{VB} The
existence of  such a maximal temperature is due to the increase of the number
of thermally excited quasiparticles which produce
enhanced  Joule heat. By lowering the  temperature  the cooling power at optimal bias
first increases and reaches a maximum.  At lower temperatures, the Joule heat due to
Andreev processes causes the cooling power to decrease. At a certain
temperature $T_{\textit{min}}$, the cooling power tends to zero, which defines
the lower limiting temperature for the cooling regime. As follows from
Fig.~\ref{panelT}, the temperature $T_{\textit{min}}$ decreases when increasing the
exchange field; this is because the Andreev current and the associated Joule heat
are suppressed by the exchange interaction in the ferromagnet.
Finally, one can see from Fig.~\ref{panelT}  that the minimum cooling temperature in the parallel $ T_{min}^P$ and antiparallel
$T_{min}^{AP}$  configuration satisfy:   $T_{min}^{AP} < T_{min}^P$ for $h=\Delta(T)$, while $T_{min}^{AP} > T_{min}^P$ for $h=1.7\Delta(T)$.
For $h=8\Delta(T)$ [Fig.~\ref{panelT}(c)]  $P(T)$ is almost independent on $\alpha$.

A common feature of Fig.~\ref{panelv}, \ref{panelW} and \ref{panelT} is that for rather small value of the exchange field,  $h=\Delta(T)$,  the antiparallel
configuration is more favourable for  cooling [see (a) panels]. For larger exchange field $h = 1.7\Delta(T)$, on the contrary, the
parallel configuration is favourable for cooling [see (b) panels].
As expected, in the case of strong enough ferromagnet [$h=8\Delta(T)$]  the thickness of F$_1$ layer $l_1 \gg \xi_h$ and the superconducting condensate practically does not penetrate into F$_2$ layer. Thus the cooling power is  $\alpha$-independent [see (c) panels of Figs.~\ref{panelv}, \ref{panelW} and \ref{panelT}].

%\section{Conclusions}
%
%In conclusion, we have developed a quantitative theory of charge and heat transport in normal metal - superconductor  tunnel junctions with an intermediate ferromagnetic bilayer. We assumed that the magnetizations
%of the ferromagnets form an angle $\alpha$ and  focused our study on  the cooling power of such a structure.

\newpage

\subsection{Spin-filter cooling}

In this section we theoretically  propose an alternative and novel N/I/S microcooler  
with  a ferromagnetic insulator as a tunnelling barrier which acts  as a spin-filter. This kind of spin-filter barrier was demonstrated 
  in experiments 
using europium chalcogenides tunnelling barriers.~\cite{rf:Moodera1}
The spin-filtering effect suppresses   
the Andreev reflection in a N/spin-filter(SF)/S junction as the one shown in Fig. 1(a).~\cite{rf:Kashiwaya,rf:Bergeret2}
 We show that this suppression leads  in both, ballistic and  diffusive, N/SF/S junctions  to   dramatic enhancement of the cooling power  
 which gives rise
to a dramatic reduction of the final achievable electron temperature.

%%%%%%%%%%%
%%%%%%%%%%%
%%%%%%%%%%%
%%%%%%%%%%%
\begin{figure}[h!]
\begin{center}
\includegraphics[width=0.5\columnwidth]{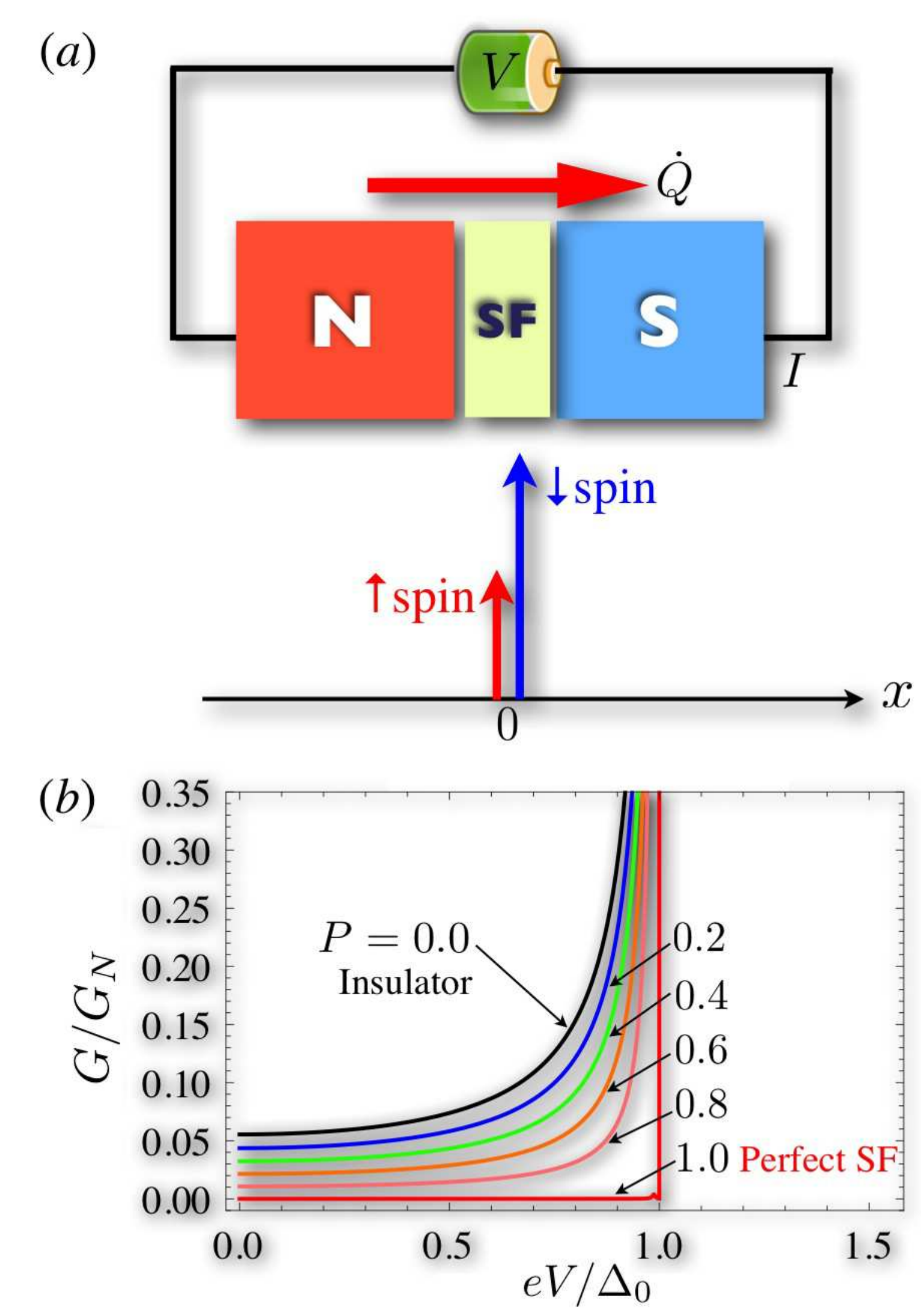}
\end{center}
\caption{(a) The schematic of a clean normal-metal/spin-filter/superconductor (N/SF/S)
cooler and the delta-function model of an SF barrier which allows the spin-selective
tunnelling and the suppression of the Andreev reflection.
(b) The differential conductance
$G$  for various values of $P$ vs the bias voltage $V$ at $T=0$K for an N/SF/S
junction with $t_\uparrow=0.1$. $G_N$ is the conductance of a N/SF/N junction, $\Delta_0$
is the superconducting gap at $T=0$K, and $P$ is the spin-filtering efficiency. (From ref.\cite{cooling2})
 }
\label{fig1S}
% \vspace{-6mm}
\end{figure}
%%%%%%%%%%%
%%%%%%%%%%%
%%%%%%%%%%%
%%%%%%%%%%%

%
%
%
%\section{Ballistic junction}
%
%
%
%

To begin with, we illustrate the basic cooling  mechanism using spin-filters in a one-dimensional clean N/SF/S junction [Fig.~\ref{fig1S}(a)]. As now we study clean and not diffusive contacts the Keldysh Usadel Greens Function formalism is not valid any more and we have to switch to Bogoliubov-de Gennes equation.

 The SF barrier can be modelled by a
spin-dependent delta-function potential [see  Fig.~\ref{fig1S}(a)], $i. e.$, $V_\sigma (x) = \left(
V + \rho_\sigma U \right) \delta(x)$, where $V$ is the spin-independent part of
the potential, $U$ is the exchange-splitting, and $\rho_\sigma=+(-)1$ for up (down)
spins.~\cite{rf:Kashiwaya,rf:Kawabata} The degree of the spin-filtering is characterized
by the spin-filtering efficiency $P=\left| t_\uparrow - t_\downarrow
\right|/(t_\uparrow+t_\downarrow)$, where $t_\sigma=1/\left[  1+ (Z+ \rho_\sigma S)^2
\right]$ is the transmission probability of the SF barrier for spin $\sigma$ with $m$,
$k_F$, $Z \equiv mV / k_F$, and $S \equiv m U / k_F$ being the mass of
electrons, the Fermi wave number, the normalized spin-independent and -dependent
potential barrier-height, respectively. For a perfect SF ($t_\uparrow > 0$ and
$t_\downarrow=0$), we get $P=1$.

The normal-reflection probability $B_\sigma$ and the Andreev-reflection probability $A_\sigma$ of the junction are obtained by solving the Bogoliubov-de Gennes equation
\begin{align}
\left[
\begin{array}{cc}
H_0-\rho_\sigma U \delta (x) & \Delta(x) \\
\Delta^* (x)  & -H_0+\rho_\sigma U \delta (x)
\end{array}
\right]
\Phi_\sigma (x)=
E
\Phi_\sigma (x)
,
\end{align}
together with the appropriate boundary conditions at the SF barrier ($x=0$),~\cite{rf:Kashiwaya}
where $H_0$ is the spin-independent part of the single-particle Hamiltonian, $i. e.$, $H_0= -\nabla^2/2 m+V \delta(x)-\mu_F$, $\Delta(x)=\Delta (T) e^{i \phi} \Theta(x)$ is a pair potential [$\phi$ is the phase of the pair potential and $\Theta(x)$ is the Heaviside step function], $\Phi_{\sigma} (x)$ is the eigenfunction, and the eigenenergy $E$ is measured from the chemical potential $\mu_F$.

 We first focus on  the spin-dependent charge-transport of the junction and address  the suppression of the Andreev reflection by the spin-filtering effect.
The voltage $V$ dependence of the differential conductance $G$ of the system can be calculated from the Blonder-Tinkham-Klapwijk formula,~\cite{rf:Blonder} 
\begin{equation}
G =(e^2/2 \pi  )
\sum_{\sigma=\uparrow,\downarrow}
\left.
\left(
  1- B_\sigma + A_\sigma
\right)
\right|_{E=eV}.
\end{equation}
In Fig.~\ref{fig1S}(b) we plot the spin-filtering efficiency $P$ dependence of $G/G_N$ vs $eV/\Delta_0$ for a junction with $t_\uparrow=0.1$ at $T=0$K, where
$G_N =( e^2/2 \pi ) (t_\uparrow+t_\downarrow)$ is the conductance of a N/SF/N junction and $\Delta_0\equiv\Delta(T=0$K).
When $P$ is increased, the sub-gap conductance for $|eV| \le \Delta_0$  is largely reduced.~\cite{rf:Kashiwaya}
Importantly if $P=1$, the Andreev reflection is completely inhibited, indicating that the SF would suppress the unwanted Andreev Joule heating.

%%%%%%%%%%%
%%%%%%%%%%%
%%%%%%%%%%%
%%%%%%%%%%%
\begin{figure}[!h]
\begin{center}
\includegraphics[width=8.7cm]{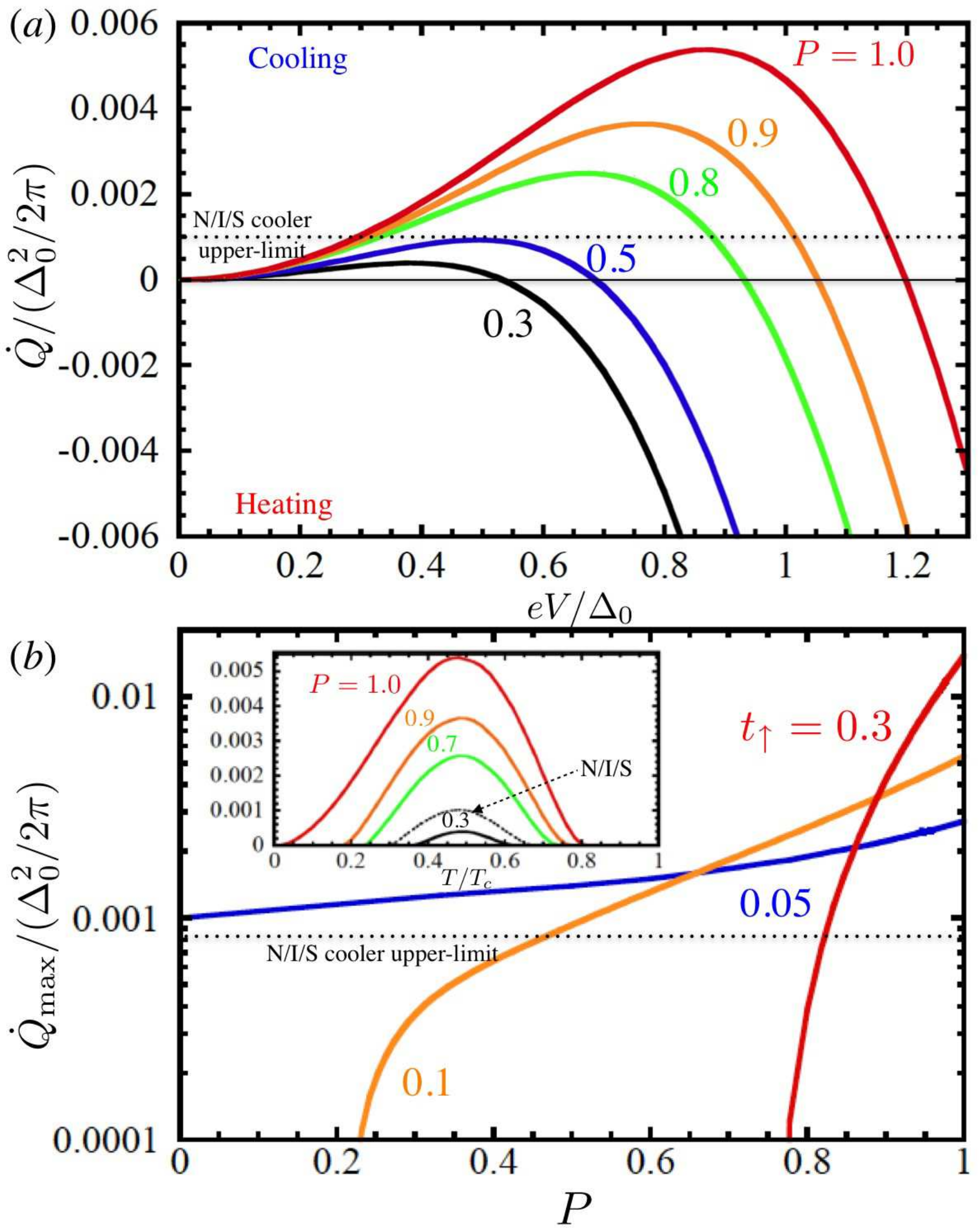}
\end{center}
\caption{(a) The cooling power ${\dot Q}$ vs the bias voltage $V$  of a clean N/SF/S refrigerator with $t_\uparrow=0.1$ at $T=0.5T_c$ for several spin filtering efficiencies $P$.
(b) The maximum cooling power $\dot{Q}_\mathrm{max}$ as a function of $P$ at $T=0.5T_c$ for several values of $t_\uparrow$.
The dotted line is the theoretical upper-limit of $\dot{Q}_\mathrm{max}  \approx 0.001 (\Delta_0^2/2 \pi )$ for N/I/S refrigerators, which is achieved in the case of $T/T_c \approx 0.5$ and $t_\uparrow=t_\downarrow \approx 0.05$.
Inset: The temperature $T$ dependence of $\dot{Q}_\mathrm{max}$ for several values of $P$.
The dashed curve shows $\dot{Q}_\mathrm{max}$ for an N/I/S refrigerator with $t_\uparrow=t_\downarrow = 0.05$. (From ref.\cite{cooling2})
 }
\label{fig2S}
\end{figure}
%%%%%%%%%%%
%%%%%%%%%%%
%%%%%%%%%%%
%%%%%%%%%%%

%%%%%%%%%%%
%%%%%%%%%%%

%%%%%%%%%%%

To see the benefit of the spin-filtering effect on the electron cooling,  we calculate the cooling power by using the Bardas and Averin formula.~\cite{rf:Bardas} This formula corresponds to the heat current in the N/I/S contact with arbitrary transparency of the insulator barrier for a clean contact. It is obtained from solving the Bogolyubov-de Gennes (BdG) equation. Here we modify it for a spin-filter barrier, it reads,
\begin{equation}
\dot{Q}
=
( e/ \pi )
\sum_{\sigma=\uparrow,\downarrow}
\int_{-\infty}^\infty d E
[ E ( 1- B_\sigma - A_\sigma )
-eV ( 1-B_\sigma +A_\sigma)
]
[ f(E-eV) -f(E)],
\end{equation}
where $f(E)$ is the Fermi-Dirac distribution-function.
The positive (negative) $\dot{Q}$ means the cooling (heating) of N.
In the calculation, we have determined $\Delta(T)$ by solving the BCS gap equation numerically.
The $P$ dependence of the cooling power $\dot{Q}$ vs the bias voltage $V$ for $t_\uparrow=0.1$ is shown in Fig.~\ref{fig2S}(a).
We have assumed that $T=0.5 T_c$, where $T_c$ is the superconducting transition temperature.
If we increase $P$, the cooling power $\dot{Q}$ is largely enhanced.
This  result is  attributed to the suppression of the Andreev reflection and the undesirable Andreev Joule heating.
Therefore we can conclude that the spin-filtering effect boosts dramatically  the cooling power with respect to  conventional N/I/S coolers.
In Fig.~\ref{fig2S}(b) we plot the spin-filtering efficiency $P$ dependence of the cooling-power
$\dot{Q}_\mathrm{max}$  at $T = 0.5 T_c$ and the optimal bias voltage $V=V_\mathrm{opt}$
in which $\dot{Q}$ is maximized as a function of $V$. The maximum
cooling-power $\dot{Q}_\mathrm{max}$ can be achieved in the case of the perfect SF
($P=1$) because of the complete suppression of the Andreev reflection. Notably for the
case of a large $t_\uparrow=0.3$ and $P=1$, the amount of heat extracted from N
can be about a factor of 15 larger than the theoretical upper-limit of
$\dot{Q}_\mathrm{max}$ for conventional N/I/S coolers [$\dot{Q}_\mathrm{max} \approx
0.001 (\Delta_0^2/2\pi )$], which can be achieved for $t_\uparrow =t_\downarrow
\approx 0.05$ and $T/T_c \approx 0.5$ [see the dotted line in Fig.~\ref{fig2S}(b)].~\cite{rf:Bardas}
It is crucial
to note that even for small $P$ values [$e.g.$, $P>0.0$ for $t_\uparrow =0.05$],
$\dot{Q}_\mathrm{max}$ overcomes the upper limit of N/I/S coolers. By calculating the
temperature dependence of $\dot{Q}_\mathrm{max}$, we also found that the
$\dot{Q}_\mathrm{max}$ is maximized at around $T \approx 0.5 T_c$ irrespective of the
value of $P$ [see the inset in Fig.~\ref{fig2S}(b)].

%
%
%
%\section{Diffusive junction}
%
%
%

%%%%%%%%%%%
%%%%%%%%%%%
\begin{figure}[!h]
\begin{center}
\includegraphics[width=6.5cm]{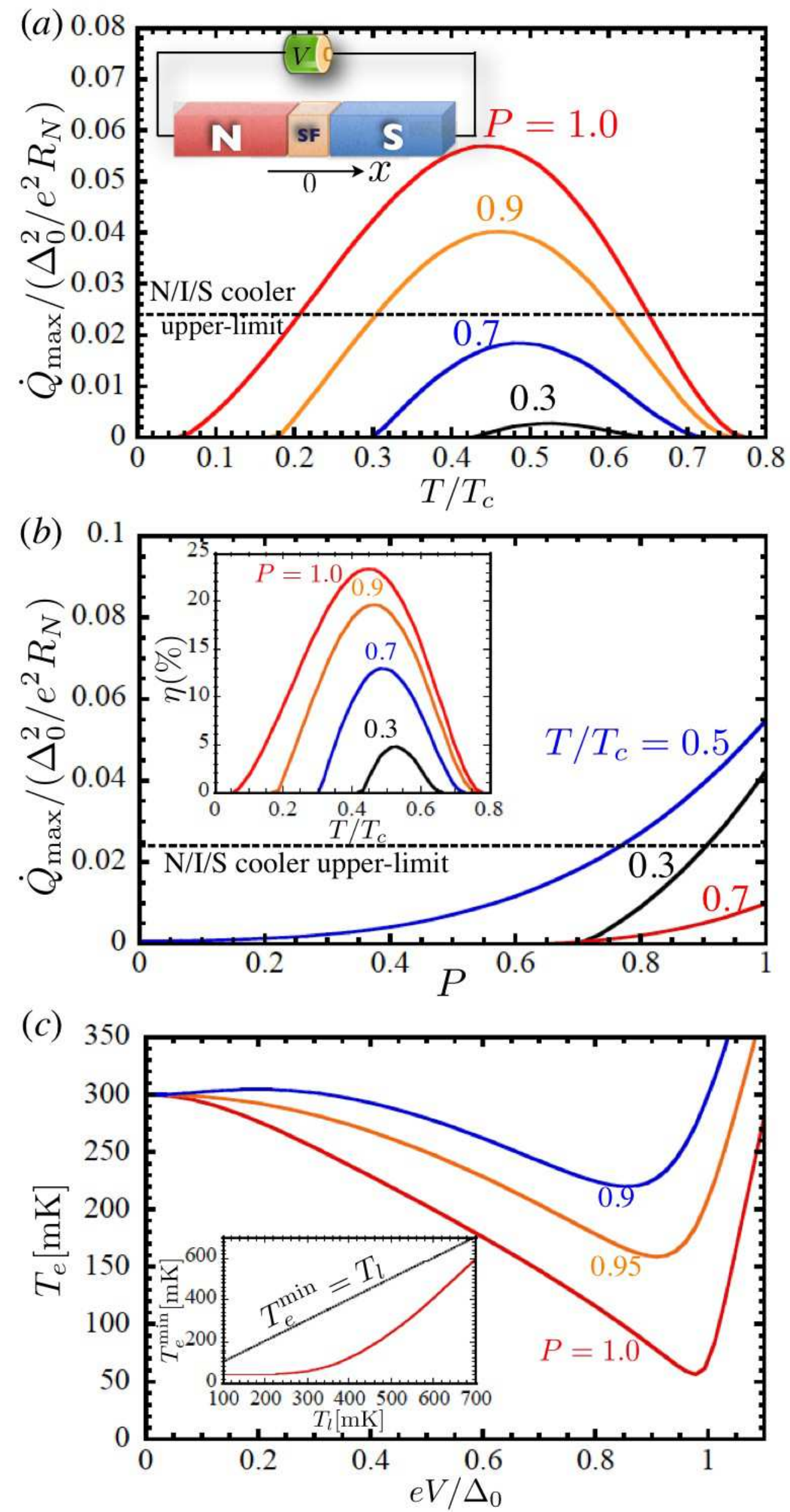}
\caption{
(a) The maximum cooling power $\dot{Q}_\mathrm{max}$  as a function of the temperature $T$ for a diffusive N/SF/S junction shown in the  inset and different values of $P$.
(b) The dependence of  $\dot{Q}_\mathrm{max}$ on  $P$ for $T/T_c$=0.7 (red), $T/T_c$=0.5 (blue), and $T/T_c$=0.3 (black).
Other parameters are  $\mathcal{A}=1 \ \mu \mathrm{m}^2$, $\sigma_N= 0.015 \ (\mu \Omega \mathrm{cm})^{-1}$ and $R_N=1.0 \ \mathrm{M} \Omega$.
Inset: The refrigeration efficiency  $\eta$  as a function of  $T$.
The horizontal dotted line in (a) and (b) is the theoretical upper-limit of $\dot{Q}_\mathrm{max}$ for diffusive N/I/S junctions, which is achieved in the case of $\mathcal{A}=1 \ \mu \mathrm{m}^2$, $\sigma_N= 0.015 \ (\mu \Omega \mathrm{cm})^{-1}$, $R_N=2.7 \ \mathrm{M} \Omega$, and $T/T_c=0.45$.
(c) The electron temperature $T_e$ as a function of $V$  for different values of $P$ and  $T_l=T_S=300$ mK.
Inset: the minimum electron temperature $T_e^\mathrm{min}$ as a function of the starting lattice temperature $T_l$ ($=T_e$ at $V=0$) for $P=1$. 
As a reference we show $T_e^\mathrm{min}=T_l$ line.
We have chosen  $\Sigma= 2 \times10^{-9} \ \mathrm{WK}^{-5} \mu \mathrm{m}^{-3}$, $\mathcal{V}=0.5 \ \mu \mathrm{m}^{3}$, and $\Delta=180 \ \mu \mathrm{eV}$. (From ref.\cite{cooling2})
}
\label{fig3S} \vspace{-4mm}
\end{center}
\end{figure}
%%%%%%%%%%%
%%%%%%%%%%%
%%%%%%%%%%%

We consider a more realistic N/SF/S diffusive junction in which the coherence length is much larger that the elastic mean free path, $l \ll \xi$. Here $\xi=\sqrt{D/2\Delta}$ is the superconducting coherence length and $D$ is the diffusion coefficient (in the following for simplicity we assume the same $D$ in the whole structure). In the previous section a perfect contact was considered, now elastic scattering by impurities is taken into account. We assume that the SF is a tunnel barrier and the N reservoir is infinite along the $x$ direction [see the inset in Fig.~\ref{fig3S}(a)].

The cooling power expression for diffusive contacts corresponds to eq.\ref{eq:P}. We only need to compute the charge and energy currents through the junction to obtain the cooling power. The general expression for the current is given by eq.\ref{eq:I}, complementing it with the spin-filter boundary condition eq.\ref{eq:bcSF}, we obtain,  
\begin{equation}\label{current}
I=\frac{1}{eR_{N}}\int d\epsilon
 n_- (\re g_S + r \re f_S \re f_0(x=0)  ) \; ,
\end{equation}
\begin{equation}\label{qcurrent}
Q=\frac{1}{e^2R_{N}}\int d\epsilon  \epsilon (n_+-n_-) ( \re g_S  - r \im f_S \im f_0(x=0) )) \; ,
\end{equation}
where $n = \tanh (E/ 2T_S)$ and $T_S$ are  the equilibrium quasiparticle distribution function and temperature in the superconducting
reservoir respectively.   The function $f_0(x=0)$ is the singlet component evaluated at the SF. According to the latter expressions the Andreev reflection is proportional to $r=\sqrt{1-P}$, the spin filter parameter.  Thus,  by increasing $P$ we expect   a suppression of  the unwanted Andreev Joule heating, {\textit{i.e.}} an enhancement of the cooling power, as it turns out from the following quantitative analysis.  In order to compute the currents $I$ and $Q$ we  assume a large SF barrier resistance $R_N$, such that 
 $\xi/(R_N \sigma_N \mathcal{A})\ll 1$. This assumption allows for a linearisation of the Usadel equation in the normal metal.~\cite{rf:Ozaeta} Solving for a semi infinite normal metal, the solution is built only from descending modes. We obtain the explicit expression of the components. 
\begin{equation}\label{anomalous}
f_{\pm}^R(x)=  \frac{r}{(R_N \mathcal{A} \sigma_N k_{\pm} + g_S )} e^{-k_{\pm} x}
\end{equation}%
where $k_\pm = \sqrt{2(E \mp h)/i\mathcal{D}}$, here $h$ is the exchange field of the material and $f_{\pm}=f_0 \pm f_z$.  Note that for a fully polarized system $P=1$ ($\mathcal{U}= \pm \mathcal{T}$), we no longer have proximity effect in the normal metal, as expected. On the other hand by making $P=0$, the expression of the anomalous Greens function corresponds to a Kupriyanov-Lukichev tunnelling barrier, as shown in Sec.\ref{sec:kupri}.

By following the procedure described above,  we compute  numerically  the cooling power $\dot{Q}$ of this system as a function of the different parameters.
We assume a junction area $\mathcal{A}=1 \ \mu \mathrm{m}^2$, a conductivity of N $\sigma_N= 0.015 \ (\mu \Omega \mathrm{cm})^{-1}$,~\cite{rf:ONeil} and $R_N=1.0 \ \mathrm{M} \Omega$.~\cite{rf:Moodera1}
In Fig.~\ref{fig3S}(a), we plot the maximum cooling power $\dot{Q}_\mathrm{max}$ as a function of temperature $T$.
As in the ballistic junction limit ${\dot Q}_{max}$ increases dramatically by increasing $P$ reaching much larger values than for a diffusive N/I/S cooler [the dotted line in Fig.~\ref{fig3S}(a) and (b)], which is achieved in the case of $\mathcal{A}=1 \ \mu \mathrm{m}^2$, $\sigma_N= 0.015 \ (\mu \Omega \mathrm{cm})^{-1}$, $R_N=2.7 \ \mathrm{M} \Omega$, and $T/T_c=0.45$.
Moreover, the window of positive values for  ${\dot Q}_{max}$  is larger for the larger $P$.  As for the N/I/S junctions, there is a maximum value of temperature, 
$T_{max} \sim 0.75 T_c$ for which cooling is achieved [{\textit{c.~f.}} Fig.~\ref{fig3S}(a)]. This maximum value  holds for a wide range of parameters.~\cite{rf:Vasenko,rf:Ozaeta} and is due 
to the increase of the number of thermally excited quasiparticles that contribute to Joule heat.   
By lowering the temperature from $T=T_{max}$,   the cooling power (at optimal bias $V_\mathrm{opt}$)  first increases, reaches a maximum and finally decreases due to the Joule heat produced by the Andreev processes.
At certain temperature the cooling power tends to zero, which defines the lowest  temperature for the cooling regime.

In Fig.~\ref{fig3S}(b) we show the dependence of the  cooling power on the spin-filter efficiency $P$.
For all temperatures ${\dot Q}_{max}$ increases  monotonically by increasing  $P$. 
The efficiency of a refrigerator is characterized by the ratio between the optimum cooling-power and total input power: $\eta=\dot{Q}_\mathrm{max}/I V_\mathrm{opt}$.
 The inset of Fig.~\ref{fig3S}(b) shows the temperature $T$ dependence of  $\eta$ for several $P$ values.
 For a fully polarized  SF barrier  ($P=1$), $\eta$ reaches up to  $23\%$, which is mach larger than that for N/I/S cooler ($\eta=15 \%$) and comparable to that for a half-metallic N/FM/S cooler with $P=1$.~\cite{rf:Giazotto06}
%Even for  lower spin-filter efficiency ($P=0.7$), as in the case of almost half-metallic N/FM/S junctions,  $\eta$ can exceed $13\%$ 
 We can then  conclude that the spin-filtering effect gives rise to highly efficient refrigeration.

We now determine the final electron temperature introduced in eq.\ref{eq:phonon}. Fig.~\ref{fig3S}(c) shows $T_e$ as a function of bias voltage for 4 different $P$ in the case of the starting lattice temperature $T_l=300$ mK.
The junction parameters were taken according to experimental values: $\Sigma= 2 \times10^{-9} \ \mathrm{WK}^{-5} \mu \mathrm{m}^{-3}$, $\mathcal{V}=0.5 \ \mu \mathrm{m}^{3}$ and assuming that Al is the superconductor ($\Delta=180 \ \mu \mathrm{eV}$).~\cite{rf:Giazotto02}
Fig.~\ref{fig3S}(c) shows a remarkable reduction of $T_e$, as we increase the value of voltage, the temperature tends to lower until it reaches a optimum voltage ($eV \sim \Delta_0$).
We observe that the increment of $P$ reduces drastically the minimum electron temperature, \textit{ i.~e.} $T_e^\mathrm{min} \sim 50$ mK for $P=1$.
In the inset of Fig.~\ref{fig3S}(c) we plot the minimum electron temperature $T_e^\mathrm{min}$ vs the starting lattice temperature $T_l$($=T_e$ at $V=0$).
The straight dotted line marked $T_e^\mathrm{min}=T_l$ as a reference.
The result indicates that in a wide $T$ range, we can effectively cool down the electron temperature of N.

\newpage

%% file: thermo2.tex
\section{Thermoelectric effects in ferromagnetic superconductor hybrid structures}

%Problem with notation eta inelastic parameter is already used!!!!

Thermoelectric effects, electric potentials generated by temperature
gradients and vice versa, are intensely studied because of their
possible use in converting the waste heat from various processes to
useful energy. The conversion efficiency is defined as $\eta=\dot{W}/\dot{Q}$, the
ratio of output power $\dot{W}$ to the rate of thermal energy consumed
$\dot{Q}$. In thermoelectric devices it typically falls short of
the theoretical Carnot limit and is low compared to other heat engines.
This fact has motivated an extensive search for better
materials. \cite{shakouri2011}

In electronic conductors a major contributor to thermoelectricity is given by the electron-hole asymmetry in the system\cite{ashcroftmermin}. Thus, semiconductors with their chemical potential tuned to the gap edge are usually used for this purpose. This thermoelectric effects are studied by the Mott formula, that describe materials where the charge carriers are in motion, so this motion is dependent of other carriers and other dynamics as phonons. In metals, where transport only occurs near the Fermi level, one can perform a Sommerfeld expansion that leads to the Mott relation\cite{cutler69}. This is much less general than the Mott formula expressed above. It predicts thermoelectric effects of the order $\sim{} T/E_0$, where $T$ is the temperature
and $E_0$ a microscopic energy scale describing the energy dependence
in the transport. This is usually a large atomic energy scale (in
metals, the Fermi energy), so that $E_0\gg{}T$ even at room
temperature and these effects are often weak. Larger electron-hole
asymmetries are however attainable in semiconductors, as the chemical
potential can be tuned close to the band edges, where the density of
states varies rapidly. \cite{shakouri2011,mahan1989-fmt}

The situation in superconductors is superficially similar to
semiconductors. The quasiparticle transport is naturally strongly
energy dependent due to the presence of the  energy gap $\Delta$,
which can be significantly smaller than atomic energy scales. However,
the chemical potential is not tunable in the same sense as in
semiconductors, as charge neutrality dictates that electron-hole
symmetry around the chemical potential is preserved. This implies that
the thermoelectric effects in superconductors are often even weaker
than in the corresponding normal state, in addition to being masked
by supercurrents \cite{ginzburg44,galperin02}.

This is not true for the case of a ferromagnetic superconductor, \textit{i.e.} a superconductor with an intrinsic exchange field. Here huge electron-hole asymmetry per spin can be achieved, but not charge. This idea is explained below in detail. 

\begin{figure}[h]
  \centering
  \includegraphics[width=0.7\columnwidth]{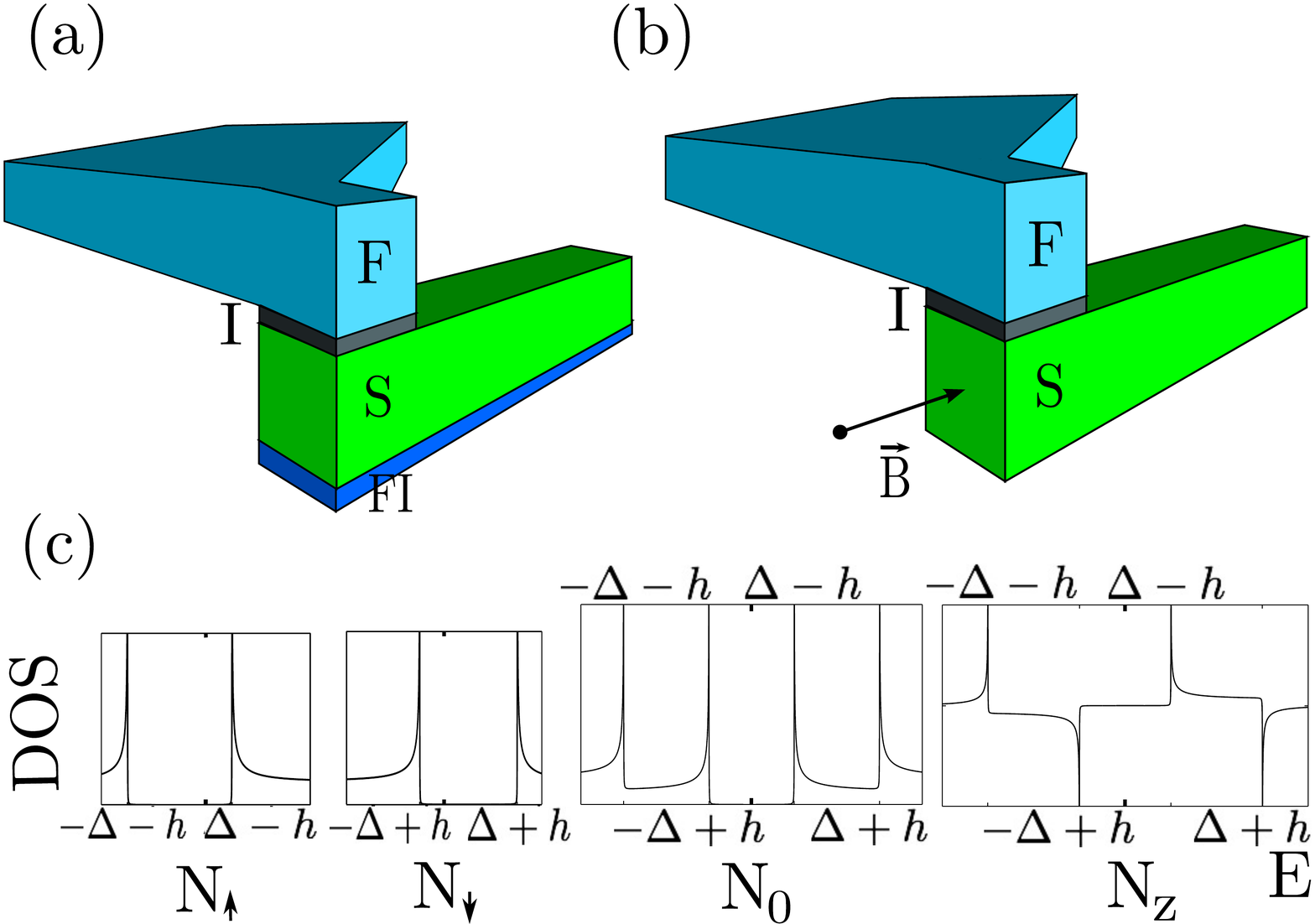}
  \caption{Top: Schematic systems studied in this work. In both of them
  a ferromagnet (F) is coupled via a tunnelling contact to a thin-film
  superconductor (S), whose tunnelling density of states is modified by
  an exchange field. In (a) the exchange field is induced by the
  proximity of a ferromagnetic insulator (FI), whereas in (b) it is
  induced by the Zeeman energy due to an applied magnetic field
  $\vec{B}$ parallel with the easy axis of the ferromagnet. Bottom:
  Tunnelling densities of states for spin $\uparrow/\downarrow$,
  averaged over spin ($N_0$), and the difference of them ($N_z$)
  obtained for an exchange field $h=\Delta/2$.(From ref.\cite{thermozaeta})}
\label{fig:system}
\end{figure}

\subsection{Huge thermoelectric effects in ferromagnet-superconductor junctions in the presence of a spin-splitting field}

In this section we show how the symmetry problem can be overcome in a
conventional superconductor by applying a spin-splitting field $h$. It
shifts the energies of electrons with parallel and antiparallel spin
orientations to opposite directions.  \cite{Tedrow1971} This breaks the electron-hole
symmetry for each spin separately, but conserves charge neutrality, as
the total density of states remains electron-hole symmetric. In this
situation, thermoelectric effects can be obtained by coupling the
superconductor to a spin-polarized system. A same type of a mechanism
was found to be present in proximity coupled multiterminal
superconductor-ferromagnet devices within the model of spin-active interfaces \cite{machon13}.

We propose that this effect can be realized in structures such as
shown schematically in Fig.~\ref{fig:system}: There, a ferromagnet
with a relatively large spin polarization is connected to a thin-film
superconductor via a tunnel contact. Moreover, we assume the presence
of a finite exchange field $h$ inside the superconductor. Such an
exchange field can result from a Zeeman effect due to an applied
magnetic field (Fig.~\ref{fig:system}b) \cite{Tedrow1971,catelani08}, or from a
magnetic proximity effect with either a ferromagnetic insulator
\cite{Moodera1990,Moodera2013,Tokuyasu88,xiong11} or with a thin ferromagnetic metallic
layer \cite{Bergeret2001, giazotto07} placed directly below the
superconductor (Fig.~\ref{fig:system}a). For simplicity, we assume
this exchange field to be collinear with the magnetization inside the
ferromagnet.

A standard tunnelling Hamiltonian calculation, as the one described in chapter\ref{ch:1} to obtain the current, is performed. This yields, for spin-$\sigma$ electrons from the ferromagnet the charge and heat
currents
\begin{subequations}
\begin{align}
  I_\sigma &= \frac{G_\sigma}{e} \int_{-\infty}^\infty dE N_\sigma(E)[f_F(E)-f_S(E)]
  \,,
  \\
  \dot Q_\sigma &= \frac{G_\sigma}{e^2} \int_{-\infty}^\infty dE (E-\mu_F) N_\sigma(E)[f_F(E)-f_S(E)]
  \,.
\end{align}
\end{subequations}
Here $N_{\uparrow/\downarrow}(E) =N_S(E \pm h) $ is the tunnelling
density of states (DOS) for spin $\uparrow/\downarrow$ particles
divided by the normal-state density of states at Fermi energy,
\cite{Tedrow1971} $N_S(E)=|E|/\sqrt{E^2-\Delta^2}\theta(|E|-\Delta)$
is the BCS DOS, $G_\sigma$ is the conductance through the junction for
spin $\sigma$ particles in the normal state, and $f_{F/S}(E)$ are the
(Fermi) distribution functions of electrons inside the ferromagnet and
the superconductor, respectively. We disregard the energy dependence
of the density of states inside the ferromagnet as well as the tiny
electron-hole asymmetry possibly existing in the
superconductor. Moreover, we fix the electrochemical potential of the
superconductor to zero and describe the applied voltage via the
potential $\mu_F=-eV$ in the ferromagnet.  Note that supercurrent
cannot flow into the ferromagnet, which prevents it from
short-circuiting this potential difference.

The spin-dependent densities of states $N_\sigma(E)$ are plotted in
Fig.~\ref{fig:system}c in the presence of a non-zero exchange
field. We can see that they break the symmetry with respect to
positive and negative energies for each spin. This symmetry breaking
allows for the creation of a large spin-resolved thermoelectric
effect, which can be converted to a spin-averaged effect via the spin
filtering provided by the polarization $P \equiv
(G_\uparrow-G_\downarrow)/(G_\uparrow+G_\downarrow)$. This can be seen
better by introducing the charge and spin currents
$I=I_\uparrow+I_\downarrow$ and $I_S=I_\uparrow-I_\downarrow$ as well
as the heat and spin heat currents $\dot Q=\dot Q_\uparrow+\dot
Q_\downarrow$ and $\dot Q_S=\dot Q_\uparrow-\dot Q_\downarrow$ along
with $N_0 \equiv (N_\uparrow+N_\downarrow)/2$, $N_z \equiv
N_\uparrow-N_\downarrow$,
\begin{subequations}
\label{eq:currents}
\begin{align}
I&=\frac{G_T}{e}\int_{-\infty}^\infty dE \left[N_0 +\frac{P N_z}{2}\right]\left[f_F-f_S\right]
\,,
\\
I_S&=\frac{G_T}{e}\int_{-\infty}^\infty dE \left[P N_0 +\frac{N_z}{2}\right] \left[f_F-f_S\right]
\,,
\\
\dot Q&=\frac{G_T}{e^2}\int_{-\infty}^\infty dE (E-\mu_F)\left[N_0 +\frac{P N_z}{2}\right] \left[f_F-f_S\right]
\,,
\\
\dot Q_S&=\frac{G_T}{e^2}\int_{-\infty}^\infty dE (E-\mu_F) \left[P N_0 +\frac{N_z}{2}\right] \left[f_F-f_S\right]
\,.
\end{align}
\end{subequations}
Here $G_T=G_\uparrow+G_\downarrow$ is the conductance of the tunnel
junction that would be measured in the absence of
superconductivity. The average density of states $N_0(E)$ is symmetric
and the difference $N_z(E)$ antisymmetric with respect to $E=0$ as
shown in Fig.~\ref{fig:system}c.  This means that they pick up a
different symmetry component of the distribution function difference
in Eqs.~\eqref{eq:currents} and eventually lead to a thermoelectric
effect. In this notation, exchanging the Fermi distribution functions by the "n" equilibrium distributions we could write, taking into account the cancellations due to symmetry,
\begin{subequations}
\label{eq:currents2}
\begin{align}
I&=\frac{G_T}{e}\int_{-\infty}^\infty dE \left[N_0 n_-(V,T_F) +\frac{P N_z}{2} (n_+(V,T_F)-n(0,T_S))\right]
\,,
\\
I_S&=\frac{G_T}{e}\int_{-\infty}^\infty dE \left[P N_0 n_-(V,T_F) +\frac{N_z}{2} (n_+(V,T_F)-n(0,T_S))\right]
\,,
\\
\dot Q&=\frac{G_T}{e^2}\int_{-\infty}^\infty dE (E-\mu_F)\left[N_0 (n_+(V,T_F)-n(0,T_S)) +\frac{P N_z}{2} n_-(V,T_F) \right]
\,,
\\
\dot Q_S&=\frac{G_T}{e^2}\int_{-\infty}^\infty dE (E-\mu_F) \left[P N_0 (n_+(V,T_F)-n(0,T_S)) +\frac{N_z}{2} n_-(V,T_F) \right]
\,.
\end{align}
\end{subequations}
Bear in mind that for no voltage difference between the reservoirs, $V=0 \rightarrow n_-=0$ and for no temperature difference (with $V=0$), $T_F-T_S=0 \rightarrow n_+-n=0$. From this notation it is easy to see that each current is divided in two parts, the regular one (proportional to $N_0$) and the one that generates the thermoelectric effect (proportional to $N_z$).

In order to grasp the size of the thermoelectric effects we assume
either a small voltage $V$ or a small temperature difference $\Delta
T/T=2(T_L-T_R)/(T_L+T_R)$ across the junctions and find the currents in
Eqs.~\eqref{eq:currents} up to linear order in $V$ and $\Delta
T/T$. Previous Eq. \ref{eq:currents2} is not limited and gives us the full solution of the system of the system up to first order. 

The charge and heat currents together with the spin and spin heat currents can be written in the most general  way as a function of the voltage, temperature, spin voltage $V_S$ and spin temperature $T_S$,
\begin{equation}
  \begin{pmatrix} I \\ \dot Q \\ I_S \\ \dot Q_S  \end{pmatrix} = \begin{pmatrix} \tilde{G} & \tilde{\alpha} &  \tilde{G}_{(S)} & \tilde{\alpha}_{(S)} \\ \tilde{\alpha} & \tilde{G}_{th} T & \tilde{\alpha}_{(S)} & \tilde{G}_{th(S)} T_S  \\   \tilde{G}^\prime & \tilde{\alpha}^\prime &  \tilde{G}_{(S)}^\prime & \tilde{\alpha}^\prime_{(S)} \\  \tilde{\alpha}^\prime & \tilde{G}_{th}^\prime T_S & \tilde{\alpha}_{(S)}^\prime & \tilde{G}_{th(S)}^\prime T_S \end{pmatrix} \begin{pmatrix} V \\ \Delta T/T \\ V_S \\ \Delta T_S/T_S \end{pmatrix} \; .\label{response1}
\end{equation}
Here we assume small spin voltage and temperature difference and $\Delta T_S/T_S$ is defined in a equivalent way to $\Delta T/T$. In this calculation we are only interested in the electrical potential difference and temperature dependence so the expression can be reduced to the following compact form,
\begin{equation}
  \begin{pmatrix} I \\ \dot Q \\ I_S \\ \dot Q_S  \end{pmatrix} = \begin{pmatrix} G & P \alpha  \\ P \alpha & G_{th} T \\ P G & \alpha \\ \alpha & P G_{th} T \end{pmatrix} \begin{pmatrix} V \\ \Delta T/T  \end{pmatrix} \; .\label{response2}
\end{equation}
\begin{figure}[t]
  \centering
  \includegraphics[scale=0.7]{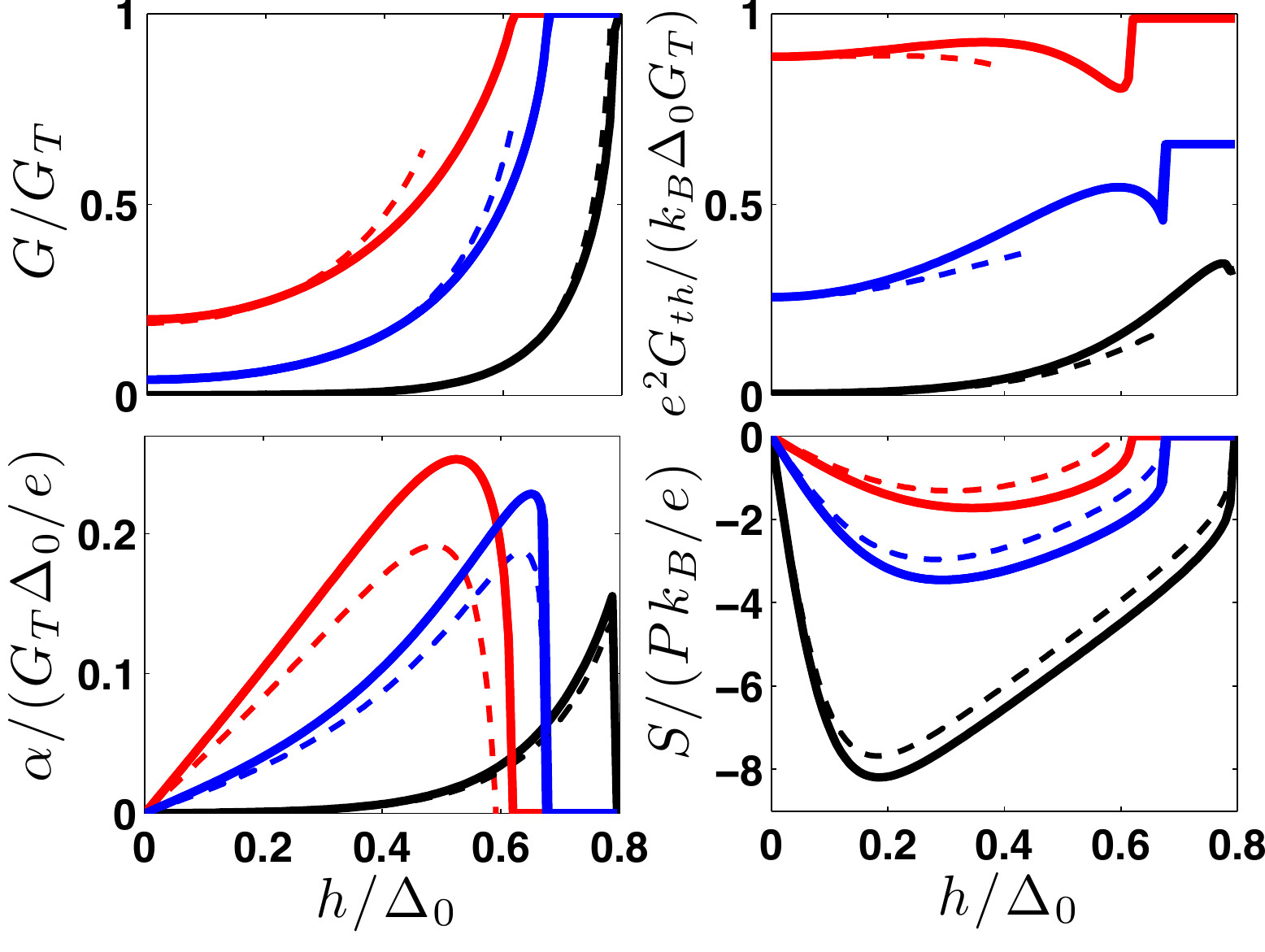}
  \caption{ Thermoelectric coefficients vs. exchange
    field $h$ at $ T/\Delta_0=0.1$ (black), 0.2 (blue) and 0.3
    (red). From top left to bottom right: conductance, heat
    conductance ($G_{th}^\prime = G_{th}/( G_T \Delta_0/e^2)$), thermoelectric coefficient, and thermopower. The solid lines are numerical integrals of Eqs.~\eqref{eq:numcoefs},
    the dashed lines are the approximations in
    Eqs.~(\ref{eq:coefs},8). The curves have been calculated for $\Gamma
    = 10^{-6}\Delta_0$ and $P=1$. $\Delta_0$ is the superconducting order parameter at $T=0$ and $h=0$.(From ref.\cite{thermozaeta}) }
  \label{fig:coefs}
\end{figure}
These response matrices  are expressed in terms of three coefficients,
\begin{subequations}
  \label{eq:numcoefs}
  \begin{align}
    G&=G_T \int_{-\infty}^\infty dE \frac{N_0(E)}{4 T \cosh^2\left(\frac{E}{2 T}\right)}
    \,,
    \\
    G_{th}&=\frac{G_T}{e^2} \int_{-\infty}^\infty dE \frac{E^2 N_0(E)}{4 T^2 \cosh^2\left(\frac{E}{2 T}\right)}
    \,,
    \\
    \alpha&=\frac{G_T}{2e} \int_{-\infty}^\infty dE \frac{E N_z(E)}{4 T \cosh^2\left(\frac{E}{2 T}\right)}
    \,.
\end{align}
\end{subequations}
Besides the main topic of this calculations, the thermoelectric effect, that we describe in detail below. We can obtain some conclusions from the already presented Eqs.~(\ref{response1}-\ref{eq:numcoefs}): First, the  matrices in Eq.~(\ref{response1}) obey the Onsager reciprocal relations \cite{onsager31,jacquod12,machon13}. For a generic thermoelectric response matrix $L$ describing response in a magnetic field $\vec{B}$ for magnetization $\vec{m}$ this relations read $L(\vec{B},\vec{m})=L^T(-\vec{B},-\vec{m})$. Furthermore, the coefficients satisfy a thermodynamic stability condition $\alpha^2/(TGG_{\text{th}})\le1$, due to Cauchy-Schwartz inequality. Second, when $N_Z=0$, i.e., when either no exchange field is applied ($h=0$) or when $\Delta=0$, there is no thermoelectric effect. The inversion of the exchange field changes the sign of the thermoelectric coefficients as $N_z(-h)=-N_z(h)$. Bear in mind that in the absence of spin polarization $P$ of the interface there is no  spin-averaged thermoelectric effect. Third, according to Eq.~(\ref{response1}), a finite spin-polarized current
can flow if there is a temperature difference across the junction. The spin-polarized current is finite even for a zero spin polarization $P=0$. This effect is the longitudinal analogue to the spin-Seebeck effect observed in metallic magnets \cite{uchida2008,adachi2013}.  

The exchange field $h$ dependence of the response coefficients from Eqs.~(\ref{eq:numcoefs}) is plotted in Fig.~\ref{fig:coefs}. The thermoelectric coefficient $\alpha$ increases linearly for small $h$ values. It reaches a maximum for $h < \Delta_0$ (here, $\Delta_0$ is the superconducting order parameter at $T=0$ and $h=0$). For high enough $h$ values the superconductivity is destroyed and it drops to zero. The thermal conductance $G_{th}$ has a similar non-monotonic behaviour. While the conductance G increases monotonically toward its normal-state value $G_T$. In the limit of low temperature $ T \ll \Delta-|h|$, the coefficients can be approximated in the following way
\begin{subequations}
\label{eq:coefs}
\begin{align}
G&\approx G_T\sqrt{2\pi\tilde\Delta} \cosh(\tilde h)e^{-\tilde\Delta}
\,,
\\
 G_{\text{th}}&\approx \frac{ G_T\Delta}{e^2}\sqrt{\frac{\pi}{2\tilde\Delta}}e^{-\tilde\Delta}\left[e^{\tilde h}(\tilde\Delta-\tilde h)^2+e^{-\tilde h}(\tilde\Delta+\tilde h)^2 \right]
\,,
\\
 \alpha &\approx \frac{G_T }{e}\sqrt{2\pi\tilde\Delta}e^{-\tilde\Delta}\left[\Delta\sinh(\tilde h)-h\cosh(\tilde h)\right]
\,,
 \end{align}
\end{subequations}
where $\tilde \Delta=\Delta/( T)$ and $\tilde h=h/( T)$. For
$h=0$, the expressions reduce to the standard results for the NIS
charge and heat conductance $G$ and $G_{\text{th}}$, \cite{Nahum94,rf:Leivon}
whereas $\alpha$ vanishes.

Rather than measuring the thermally induced current, the typical thermoelectric observable is the thermopower or the Seebeck coefficient $S=-P\alpha/(G T)$. This quantity is defined as the voltage $V$ observed due to a temperature difference $\Delta T$ in a open circuit so that $I=0$. It is calculated from Eqs.~\eqref{eq:numcoefs}. The Seebeck coefficient for this FIS junction is plotted in the lower right panel of Fig.~\ref{fig:coefs}. Although its qualitative behaviour is close to that of $\alpha$, it is quantitatively changed by the dependence of $G$ on $h$.

In the low temperature limit, $S$ can be obtained from
Eqs.~\eqref{eq:coefs}, $S\approx -\frac{P \Delta}{e
T}[\tanh(\tilde{h})-h/\Delta]$.  Thus, for low temperatures the
thermopower is maximized for $h= T {\text{arcosh}}(\sqrt{\tilde
  \Delta})$, where
\begin{equation}
\label{eq:thermopower1}
S_{\text{max}} \approx -\frac{1}{e} P\left[\frac{\Delta}{ T} -{\text{arcosh}}\left(\sqrt{\frac{\Delta}{ T}}\right)\right]
\,,
\end{equation}
It can hence greatly exceed $1/e$ and seems to diverge towards low
temperatures as $1/T$. In practice this divergence is cut off by
additional contributions beyond the standard BCS tunnel formula. These
are often described via the phenomenological ``broadening'' parameter
$\Gamma$ \cite{pekola2004}. Practical reasons for the occurrence of an
effectively non-zero $\Gamma$ are due to the fluctuations in the
electromagnetic environment \cite{pekola10}, the presence of Andreev
reflection \cite{rajauria08,laakso12}, or the inverse proximity effect
from the ferromagnet \cite{sillanpaa01,kauppila13}. The main effect
of the broadening parameter for the thermopower is to induce a finite
density of states inside the gap that in turn leads to a correction of
the charge conductance \eqref{eq:coefs} of the order $\delta G =
\frac{\Gamma}{\Delta} G_T$ (valid for $\Gamma \ll T \ll
\Delta$). The corrections for the other coefficients are less
relevant. Within this limit  we get for the thermopower
\begin{align}
  S&=P\frac{\Delta }{e T}\frac{h \cosh\left(\frac{h}{
        T}\right)-\Delta\sinh\left(\frac{h}{ T}\right)}{\Gamma
    e^{\Delta/( T)}\sqrt{\frac{
        T}{2\pi\Delta}}+\Delta\cosh\left(\frac{h}{ T}\right)}.
  \label{eq:thermopower2}
\end{align}
The result for $S$ is shown in the lower right panel of Fig.~\ref{fig:coefs}.

Typically in the literature, linear-response thermoelectric properties of metals are described by using the Sommerfeld expansion, which amounts essentially to expanding the electronic spectra around the Fermi energy. For the thermopower, this results into the celebrated Mott relation, 
\begin{equation}
S_{\text{Mott}}=\frac{\pi^2 T}{e} \partial_E \text{ln} G
\end{equation}
where $E$ denotes an energy dependence either of spectrum or of the scattering time in the sample. For the FIS thermopower, one would replace G by the (exchange field shifter) density of states $N_0(E)$ in this relation. However, the Mott relation does not work for the FIS thermopower described here in the limit $\Gamma \ll T,\Delta$ because major contributions to the energy integral do not come from the energy window within $\sim T$ around the chemical potential, as assumed in the Sommerfeld expansion, but rather from energies above the gap.

\begin{figure}[t]
  \centering
  \includegraphics[width=0.5\columnwidth]{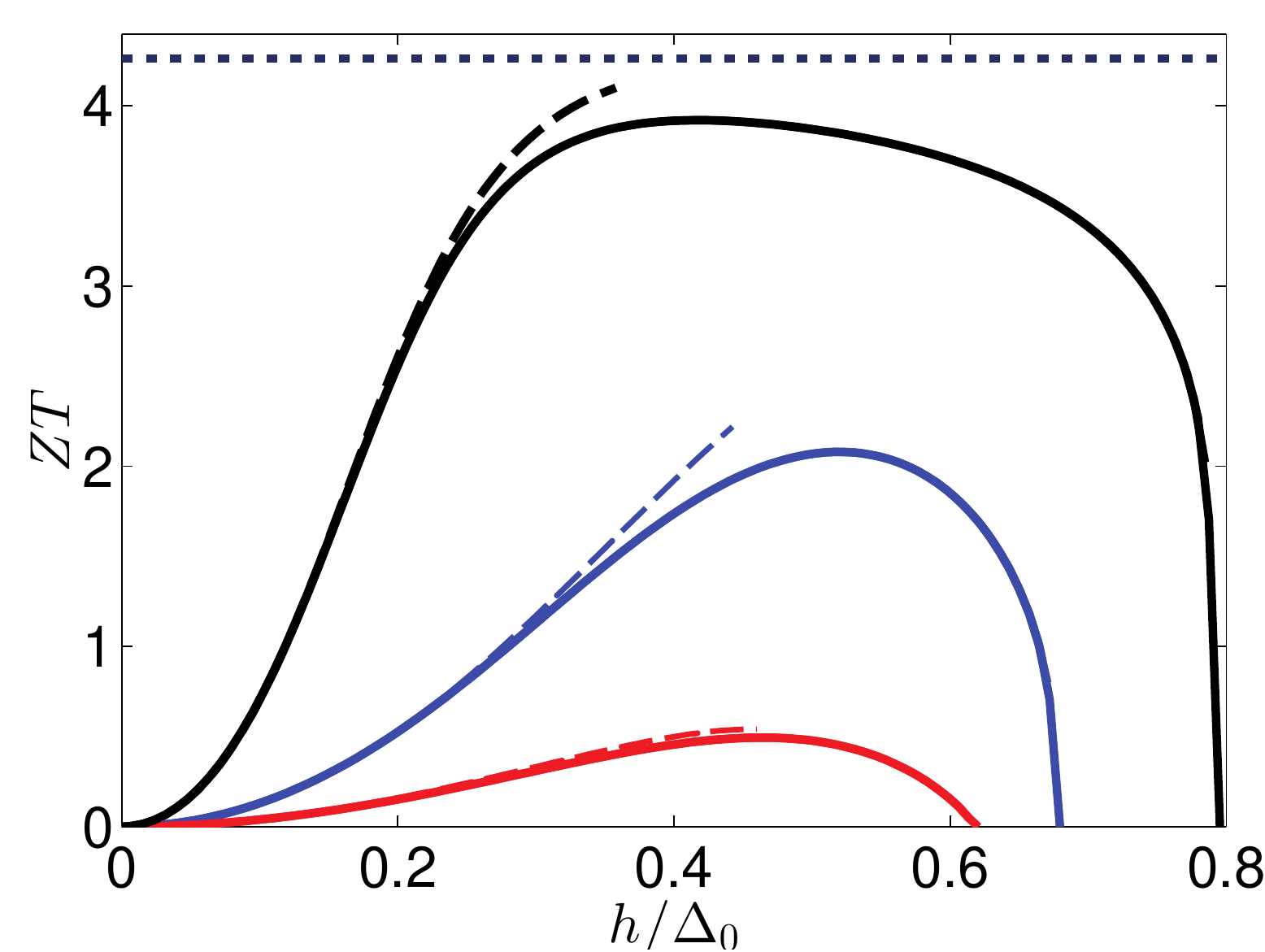}
 \caption{ Figure of merit ZT as a function of
    exchange field at $ T/\Delta_0=0.1$ (black), 0.2 (blue) and 0.3
    (red) and $P=0.9$. The solid lines are the exact results and the
    dashed line the results obtained from Eq.~\eqref{eq:zt}.
    The dotted line indicates the zero-temperature limit $ZT=P^2/(1-P^2)$.(From ref.\cite{thermozaeta})
  }
  \label{fig:ZT}
\end{figure}

The power conversion ability of thermoelectric devices is usually
characterized by a dimensionless figure of merit $ZT$, which can here
be related to the junction parameters by $ZT=S^2 G T/\tilde{G}_{\text{th}}$, where $\tilde G_{\text{th}}$ is the thermal conductance at zero
current \cite{foot}. At
linear response, $\Delta T\ll{}T$, this determines the efficiency at
maximum output power, $\eta=\eta_{CA}ZT/(ZT+2)$, where
$\eta_{CA}=1-\sqrt{T_{\text{cold}}/T_{\text{hot}}}$ is the Curzon-Ahlborn
efficiency \cite{curzon1975}. Best known thermoelectric bulk materials
have $ZT\lesssim{}2$, but better efficiencies are achievable in
nanostructures \cite{shakouri2011}.

Assuming that the thermal conductance is dominated by the electronic
contribution, we find at $ T \ll \Delta - |h|$
\begin{equation}
  ZT
  =
  \frac{
    P^2
  }{
    1 - P^2
    +
    \frac{
      \Delta^2
    }{
      \bigl[h\cosh\bigl(\frac{h}{T}\bigr) - \Delta\sinh\bigl(\frac{h}{T}\bigr)\bigr]^2
    }
  }
  \,,
\label{eq:zt}
\end{equation}
which is shown and compared to numerical results in Fig.~\ref{fig:ZT}.
For $ T \ll h$, we find $ZT=P^2/(1-P^2)$. For $P \rightarrow 1$
(half-metal injector), $ZT$ approaches infinity, and the efficiency
approaches theoretical upper bounds. From a practical point of view 
the main challenge in achieving large
values for $ZT$ is the fabrication of barriers with large spin-polarization $P$.

\begin{figure}[h]
 \centering
  \includegraphics[width= \columnwidth]{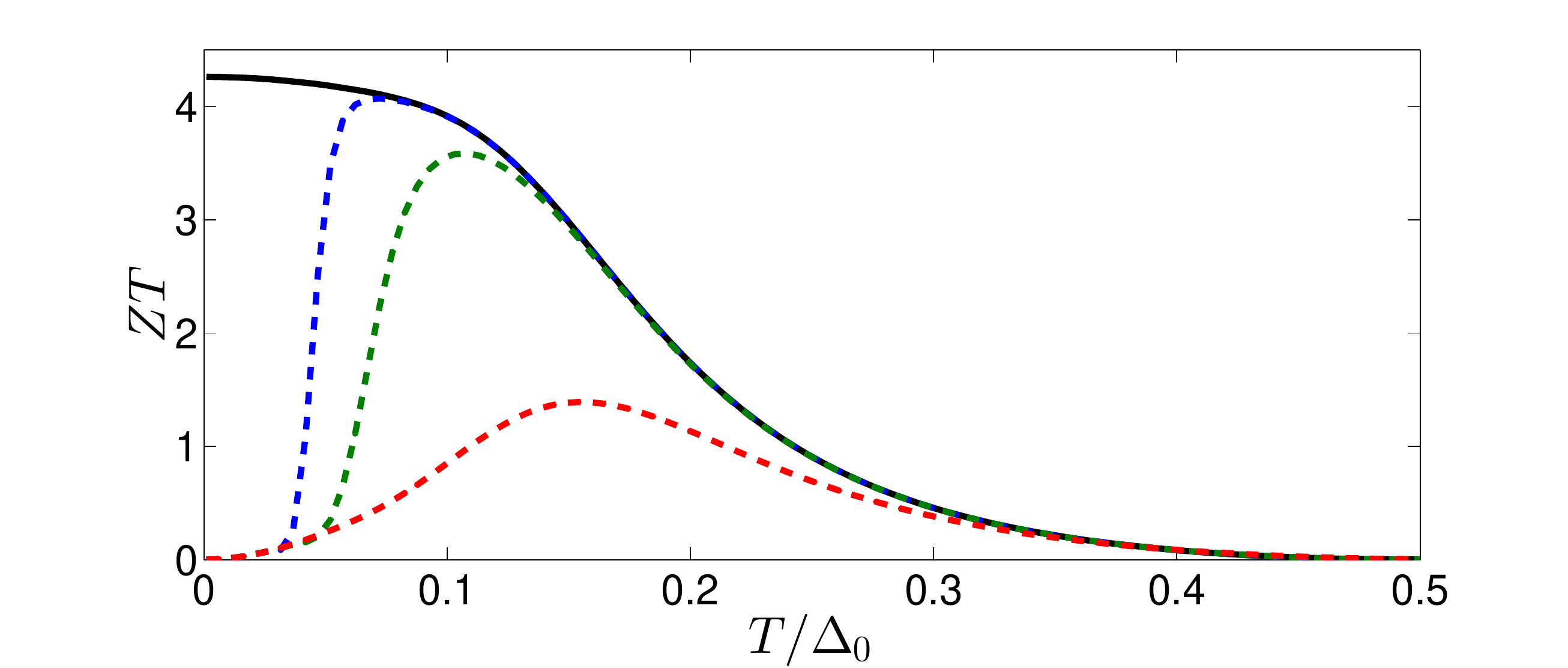}
  \caption{Figure of Merit (ZT) as a function of temperature for different values of the inelastic scattering parameter $\Gamma$. We plot for reference the ZT curve for $\Gamma=0$ as a solid line, the dashed lines represent finite values of $\Gamma$: 0.1 (red), $10^{-3} $ (green) and $10^{-5} $ (blue).(From ref.\cite{thermozaeta})
  }
  \label{fig:ztvst}
\end{figure}

The "broadening" parameter or inelastic scattering parameter $\Gamma$ plays a crucial role when determining ZT. We have considered the limiting case of vanishing $\Gamma$. Now we study the effect of a finite value in the power conversion ability of the system. In real systems, \textit{i.e.} experimental setups, the value of $\Gamma$ is always finite.\cite{tex} The following calculations are of special interest in order to study an experimental setup of the proposed junction. The expression now reads,
\begin{equation}
  ZT
  =
  \frac{
    P^2 (1+\Gamma G_T/(\Delta G))
  }{
    1 - P^2
    +
    \frac{
      \Delta^2 [1+\Gamma e^{\tilde{\Delta}} / \cosh (\tilde{h}) \sqrt{
        T/(2\pi\Delta)}]^2
    }{
      \bigl[h\cosh\bigl(\frac{h}{T}\bigr) - \Delta\sinh\bigl(\frac{h}{T}\bigr)\bigr]^2
    }
  }
  \,.
\end{equation}
Note that for a finite $\Gamma$ we no longer obtain the maximum value at $T=0$ but for finite temperatures as shown in Fig.\ref{fig:ztvst}. Increasing the value of $\Gamma$ the maximum value of $ZT$ decreases and it is obtained for higher temperature values. While in the absence of $\Gamma$ the figure of merit decays monotonically with increasing temperature, finite $\Gamma$ curves show a maxima at low temperatures, tending to zero for $T\rightarrow 0$ and high temperatures.

\begin{figure}[h]
\centering
  \includegraphics[scale=1.5]{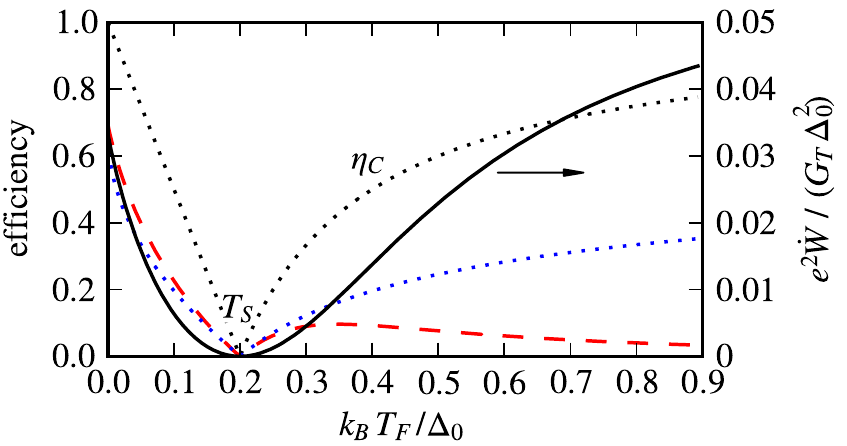}
  \caption{ Maximum power $\dot{W}=\max_V[-IV]$ generated by the FIS junction
   from a temperature difference $T_F-T_S$  (solid), 
   and the corresponding heat engine efficiency $\eta$ (dashed).
   We fix $P=1.0$, $T_S=0.2\Delta_0$, $h=0.6\Delta_0$, and
   $\Gamma=10^{-5}\Delta_0$. The linear-response result $\eta=\eta_{CA}
   ZT/(ZT+2)$ for $ZT=4.04$ and the Carnot efficiency $\eta_C=1-T_{\text{cold}}/T_{\text{hot}}$ are also shown (dotted).(From ref.\cite{thermozaeta})
  }
  \label{fig:nonlinear}
\end{figure}

Let us characterize the efficiency at temperature
differences for which the current dependence on temperature cannot be linearised. This means that we use the expressions for the currents Eqs.\ref{eq:currents2}. Figure~\ref{fig:nonlinear} shows the maximum extractable
power as a function of the temperature difference, together with the
conversion efficiency $\eta$. For a $1\,\mathrm{k\Omega}$ tunnel
junction to aluminium, the maximum power in this figure corresponds to
$\dot{W}\approx 1.5 \,\mathrm{pW}$. The efficiency can be rather high,
$\eta=0.7$, also when the extracted power is large.

Other known mechanisms in superconductors generating thermoelectric
signals in addition to the normal-state mechanisms include condensate
flow \cite{ClarkeEtal}. Here a superconductor carrying a supercurrent in a temperature gradient develops a charge-imbalance voltage. Another mechanism is electron-hole symmetry breaking by magnetic
impurities \cite{kalenkov2012}. Here the thermoelectric effect is caused by the violation of the symmetry between electronlike and holelike excitations due to the formation of subgap bound Andreev states in the vicinity of magnetic impurities. Thermopower significantly larger than the normal-state effect appears also in hybrid N/S systems.\cite{VolkovEtal} This is due to a deviation of the distribution functions from the equilibrium form generated by a temperature gradient. This changes the current flowing across the N/S interface. The efficiencies of this devices are much lower than the one presented in this calculation.

The cooling effect found in NIS junctions in the nonlinear regime
\cite{giazotto06,rf:Muhonen} is also similar to the effect described
here, if one substitutes the exchange field with a finite voltage $V
\approx \Delta/e$. Indeed, the extracted power found above is
comparable to the maximum cooling power of a NIS junction.  NIS
junctions, however, cannot be used for power conversion, as their
cooling power $\dot{Q}_{NIS}$ is a symmetric function of the bias
voltage. The effect of ferromagnetism on NIS cooling was also
discussed earlier, \cite{giazotto02,rf:Ozaeta,cooling2} but in
those works the exchange field was introduced in order to suppress the
Joule heating due to the Andreev current and did not affect the
density of the states of the superconductor. According to the results
the induced exchange field in the superconductor may lead to a larger
cooling efficiency than in NIS junctions.

%The results described above are obtained by assuming that the electron
%charge, spin and energy relax immediately after tunneling. This
%assumption can be lifted by considering the non-equilibrium state
%formed inside the ferromagnetic or the superconducting wire due to the
%biasing.  This can be described by generalizing the quasiclassical
%Green's function approach in Ref.~\cite{morten04} to the case of a
%superconductor in a spin-splitting field, and describing the effect of
%the finite spin polarization inside the ferromagnet via an effective
%boundary condition derived in Ref.~\cite{bergeret12}.  We have verified
%that the effects described above are qualitatively not affected by
%such corrections.  The details of this approach will be published
%elsewhere.

The assumption about equilibrium electron distributions in the above
model holds if the resistance $G_T^{-1}$ is large compared to the
quasiparticle boundary resistance $R_q\sim{}\rho_N\ell_{\text{in}}/A$,
\cite{HsiangEtal} which depends on the inelastic charge and spin
relaxation length $\ell_{\text{in}}$ ($A$ is the junction area and
$\rho_N$ the normal-state resistivity). In the opposite limit, the
kinetics of injection and relaxation of quasiparticles in the
superconductor (and the ferromagnet) need to be modelled
\cite{morten04,bergeret12} taking the split density of states into
account, which can alter the quantitative details.

We also note that in the geometry of Fig.~\ref{fig:system}(b), where the Zeeman field is
induced by a magnetic field, the orbital effect of the magnetic field
will also influence the form of the density of states and for large
fields it will eventually lead to a destruction of
superconductivity. We previously describe this phenomenon in detail in section\ref{sec:1.2}. Due to the application of an in-plane field, we have disregarded this effect in
the above calculation. In practice, to minimize this effect, the
magnetic field should be applied preferably in the longitudinal
direction of the wire \cite{Meservey1970}, as depicted in
Fig.~\ref{fig:system}(b).

\paragraph{Second order correction}

The results described above are obtained by assuming that the electron charge, spin and energy relax immediately around  the tunnel junction. This assumption can be lifted by considering the non-equilibrium state formed inside the ferromagnetic or the superconducting wire due to the biasing. This is the case when the conductance of the tunnel junction is high enough, so we do not have to limit this calculation to first order in tunnelling. Because charge and spin relaxation in ferromagnets typically occurs on very short lengths, we consider only the non-equilibrium state inside the superconductor. This system can be described by the quasiclassical Green function approach introduced in chapter\ref{ch:2}. For a superconductor in a spin splitting field, and describing the effect of the finite spin polarization inside the ferromagnet via a effective boundary conditions described in section\ref{sec:spinfilters}. 

The calculation of the charge and heat currents and their spin parts leads now to a full result with different orders in transmission. For example, we can write the charge current as a series of currents with different orders in transmission, 
\begin{equation}
I=I^{(1)}+I^{(2)}+ \ldots
\end{equation}
Here $I^1$ stands for the term in first order obtained in the eqs.\ref{eq:currents}, and $I^2$ is the correction that we introduce in this section. This becomes relevant for higher values of transmission. We limit ourselves to the study of this quantity although higher order terms for the current can be obtained. This kind of series expansion applies to all the currents in the system.

The terms in second order read, 
\begin{equation}
I^{(2)}= \frac{r G_T }{e} \int^{\infty}_{-\infty} dE [ n_- [({\mathcal{D}}_{L}^2-{\mathcal{D}}_{T3}^2) (N_0^2-N_z^2)        
+2 {\mathcal{D}}_{T3} (1-P^2) N_0 N_z+ {\mathcal{D}}_{L} [(P^2-2)N_0^2+ P^2 N_z^2]]
\nonumber
\end{equation}
\begin{equation}
+(n_+-n) P (N_z^2-N_0^2) {\mathcal{D}}_{T3}]/\bar{\lambda}_{FS} \; ,
\end{equation}
\begin{equation}
 I_S^{(2)}=\frac{r G_T }{e} \int^{\infty}_{-\infty} dE [n_- P (N_0^2-N_z^2) ({\mathcal{D}}_{T3}^2-{\mathcal{D}}_{L}^2- {\mathcal{D}}_{L})
+ (n_+-n) [2  {\mathcal{D}}_{L} (P^2-1) N_0 N_z- {\mathcal{D}}_{T3} ((P^2-2) N_z^2  
\nonumber
\end{equation}
\begin{equation}
+P^2 N_0^2)]] /\bar{\lambda}_{FS} \; ,
\nonumber
\end{equation}
\begin{equation}
Q^{(2)}=\frac{r G_T }{e^2} \int^{\infty}_{-\infty} dE E  [n_- P {\mathcal{D}}_{T3} (N_z^2-N_0^2)+ 
(n_+-n) [({\mathcal{D}}_{L}^2-{\mathcal{D}}_{T3}^2) (N_0^2-N_z^2)+2 {\mathcal{D}}_{T3} (1-P^2) N_0 N_z
\nonumber
\end{equation}
\begin{equation}
+ {\mathcal{D}}_{L} [(P^2-2)N_0+ P^2 N_z^2]]]/\bar{\lambda}_{FS} \; ,
\end{equation}
\begin{equation}
Q_S^{(2)}=\frac{r G_T }{e^2} \int^{\infty}_{-\infty} dE E [n_- [2  {\mathcal{D}}_{L} (P^2-1) N_0 N_z
- {\mathcal{D}}_{T3} ((P^2-2) N_z^2 +P^2 N_0^2)]]
\nonumber
\end{equation}
\begin{equation}
+(n-n_+)[P ({\mathcal{D}}_{T3}^2-{\mathcal{D}}_{L}^2-{\mathcal{D}}_{L}) (N_0^2-N_z^2)]]/\bar{\lambda}_{FS} \; .
\end{equation}
Here
\begin{equation}
\bar{\lambda}_{FS}= (\mathcal{D}^2_{L}-\mathcal{D}^2_{T3})
\end{equation}
and 
  \begin{eqnarray}
   \mathcal{D}_{L} &=& 1-({\text{Im}}\;f)^2+({\text{Im}}\;f_z)^2 -({\text{Im}}\;g)^2+({\text{Im}}\;g_z)^2-({\text{Re}}\;f)^2+({\text{Re}}\;g)^2-({\text{Re}}\;f_z)^2+({\text{Re}}\;g_{z})^2 \; , \\
   {\mathcal{D}}_{T3} &=& 2\left( {\text{Im}}\;g {\text{Im}}\;g_{z} - {\text{Im}}\;f {\text{Im}}\;f_z + {\text{Re}}\;g {\text{Re}}\;g_z-{\text{Re}}\;f{\text{Re}}\;f_z  \right) \; .  
\end{eqnarray}
In addition, the density of states that appear here correspond to; $N_0=g^R-g^A$ and $N_z=g_z^R-g_z^A$. The Green functions corresponding to those of a ferromagnetic superconductor. 

Assuming either a small voltage or temperature difference, our second order correction up to linear order in V and $\Delta T$ can be written in a compact way,
\begin{equation}
\begin{pmatrix} I^{(2)}  \\ \dot Q^{(2)} \\ I_S^{(2)}  \\ \dot Q_S^{(2)} \end{pmatrix} = \begin{pmatrix}\delta G & P \delta \alpha \\ P \delta \alpha & \delta G_{th} T \\ P \delta G_2 & \delta \alpha_2  \\ \delta \alpha_2 & P \delta G_{th2} T \end{pmatrix} \begin{pmatrix} V \\ \Delta T/T \end{pmatrix} \; .
\label{eq:response3}
\end{equation}
%
%and
%%
%\begin{equation}
%\begin{pmatrix}  I_S^{(2)}  \\ \dot Q_S^{(2)} \end{pmatrix} = \begin{pmatrix} P \delta G_2 & \delta \alpha_2  \\ \delta \alpha_2 & P \delta G_{th2} T \end{pmatrix} \begin{pmatrix} V \\ \Delta T/T \end{pmatrix} \;.
%\label{response4}
%\end{equation}
%%
The second order correction does not seem to satisfy the same symmetry as eq.\ref{response2}, first order calculation. As $\delta \alpha \not= \delta \alpha_2$. Note that in the same way as eq.\ref{response2}, eq.\ref{eq:response3} has in principle a $4\times 4$ matrix. We define 6 components for the linear response, 
\begin{equation}
\delta G=G_T \int_{-\infty}^\infty dE \frac{ (N_0^2-N_z^2)}{4 T \cosh^2\left(\frac{E}{2 T}\right)}
+\frac{2 {\mathcal{D}}_{T3} (1-P^2) N_0 N_z+D {\mathcal{D}}_{L} [(P^2-2)N_0^2+ P^2 N_z^2]}{4  T \cosh^2\left(\frac{E}{2 T}\right) \bar{\lambda}_{FS}}
\; ,
\end{equation}
\begin{equation}
\delta \alpha=\frac{G_T}{2e} \int_{-\infty}^\infty dE \frac{E  (N_z^2-N_0^2) {\mathcal{D}}_{T3}}{4  T \cosh^2\left(\frac{E}{2 T}\right) \bar{\lambda}_{FS}}
\; ,
\end{equation}
\begin{equation}
\delta G_2=G_T \int_{-\infty}^\infty dE \frac{  (N_0^2-N_z^2) ({\mathcal{D}}_{T3}^2-{\mathcal{D}}_{L}^2- {\mathcal{D}}_{L})}{4 T \cosh^2\left(\frac{E}{2 T}\right) \bar{\lambda}_{FS}}
\; ,
\end{equation}
\begin{equation}
\delta \alpha_2=\frac{G_T}{2e} \int_{-\infty}^\infty dE \frac{ E[2  {\mathcal{D}}_{L} (P^2-1) N_0 N_z- {\mathcal{D}}_{T3} ((P^2-2) N_z^2+ P^2 N_0^2) ]}{4 T \cosh^2\left(\frac{E}{2 T}\right) \bar{\lambda}_{FS}}
 \; ,
\end{equation}
\begin{equation}
\delta G_{th}=\frac{G_T}{e^2} \int_{-\infty}^\infty dE \frac{E^2 (N_0^2-N_z^2)}{4 T^2 \cosh^2\left(\frac{E}{2 T}\right)}
+\frac{E^2[2 {\mathcal{D}}_{T3} (1-P^2) N_0 N_z+ {\mathcal{D}}_{L} [(P^2-2)N_0^2+ P^2 N_z^2]]}{4 T^2 \cosh^2\left(\frac{E}{2 T}\right) \bar{\lambda}_{FS}}
\; ,
\end{equation}
\begin{equation}
\delta G_{th2}=\frac{G_T}{e^2} \int_{-\infty}^\infty dE \frac{E^2 (N_0^2-N_z^2) ({\mathcal{D}}_{T3}^2-{\mathcal{D}}_{L}^2- {\mathcal{D}}_{L}) }{4 T^2 \cosh^2\left(\frac{E}{2 T}\right) \bar{\lambda}_{FS}}
\; .
\end{equation}
These terms have similar form to the previously introduce linear response coefficients, eqs.\ref{eq:numcoefs}. From eq.\ref{eq:response3}, we observe that the general polarization $P$ prefactors in the correction are equivalent to those in eq.\ref{response1}. This means, for example, that the correction to the thermoelectric effect is only finite if the polarization $P$ is finite. We still require a finite exchange field $h$ in the superconductor for $\delta \alpha$ and  $\delta \alpha_2$ to be non zero. As in the $h=0$ case, $N_Z=0$, together with $\mathcal{D}_{T3}=0$. The second order corrections give a positive contribution to the thermoelectric effect. Due to an increase in the thermoelectric coefficients given by $\delta \alpha$ and $\delta \alpha_2$.

On the other hand, corrections to the charge and thermal conductances, $\delta G$, $\delta G_2$, $\delta G_{\text{th}}$ and $\delta G_{\text{th}}$, are finite in this $h=0$ limit. This means that, in the absence of polarization and exchange field, there is still a finite correction  $\delta G$ and $\delta G_{\text{th}}$ but no $\delta G_2$ and $\delta G_{\text{th}2}$. The $\bar{\lambda}_{FS}$ denominator does not generate divergences. In the limit $h=0$ it takes the value $\bar{\lambda}_{FS} = 1-({\text{Im}}\;f)^2 -({\text{Im}}\;g)^2-({\text{Re}}\;f)^2+({\text{Re}}\;g)^2$.

%%%%%%%%%%%%%%%%%%%%%%%%%%%%%%%%%%%%%%%%%%%%%%%%%

\subsection{Further works on the thermoelectric effect in SF structures}

Our ideas have motivated two further works in the field. In Ref.\cite{tjos} a Josephson junction with two superconductors $S_L$ and $S_R$ coupled through a ferromagnetic insulator (FI) and a non-magnetic (I) barrier is studied. The FI has different transmissivities for spin-up and spin down electrons and therefore acts as a spin-filter. An effective spin-splitting field $h$ in the left electrode $S_L$, that decays away from the interface over the superconducting coherence length, is generated by the FI. We assume the thickness of $S_L$ is smaller than the coherence length so that the induced h is spatially uniform.  While $S_R$ is prevented of such splitting by the thin I layer. The junction is temperature biased and a finite phase difference is set between superconductors. The study was focused on the static (i.e., time-independent) regime so that a dc Josephson current could flow in response to the applied thermal gradient but no thermovoltage developed across the junction. It has been predicted the occurrence of a giant thermophase in thermally-biased Josephson junctions based on FIs. This effect can be detected in a structure realizable with current state-of-the-art nanofabrication techniques and well-established materials. The very sharp thermophase response combined with the low heat capacity of superconductors could allow realizing ultrasensitive radiation detectors\cite{jos19} where radiation induced heating of one of the superconductors is detected via the thermophase. The presence of magnetic material also allows for adding a new control parameter to the experimental investigation of coherent manipulation of heat flow at the nanoscale\cite{jos27,jos28,jos29,jos30}.

In another recent paper\cite{tex}, our SFIN structure has been optimized in order to reach higher values of ZT. It has been demonstrated theoretically that a superconducting thermoelectric transistor which offers  unparalleled figures of merit of up to $\sim 45$ and Seebeck coefficients as large as a few $mV/K$ at sub-Kelvin temperatures can be built. The device is also phase-tunable, meaning its thermoelectric response for power generation can be precisely controlled with a small magnetic field.  Upon application of an external magnetic flux, the interferometer enables phase-coherent manipulation of thermoelectric properties whilst offering efficiencies which approach the Carnot limit.

\newpage

\section{Summary}
%\subsection{Conclusion}

In this chapter we have studied heat transport related to the electrons of the system. In particular cooling a particular section of our structure. Together with thermoelectric effects that originate from the coupling between heat and charge. The major contribution being given by the electron-hole asymmetry in the system. 

In the first section we have shown that the cooling power of a SIF$_1$F$_2$N junction shows a minimum value for $h\approx\Delta$. In the previous works as ref.\cite{Giazotto}, it has been suggested that  the larger the exchange field the more efficient the cooling.  In this case the enhancement of the cooling is due to the suppression of the Andreev processes and therefore suppression of the Joule heating, released in the normal metal electrode. However, our results have shown that this hypothesis is only valid  in the case of strong ferromagnets
[$h\gg\Delta$].  For weak ferromagnets with an exchange field  comparable to the superconducting
order parameter $\Delta$  the cooling power shows a non-monotonic dependence on $h$, with a minimum
at $h\approx\Delta$ (in mono-domain case) that corresponds to a maximum in the Andreev current $I_A$. Moreover, around this minimum the cooling power of the SIF$_1$F$_2$N structure is even lower than the one of the NIS junction. We have also shown that in the two-domain case, a finite value of  $\alpha$ shifts  the minimum  of cooling power to larger values of $h$ if the thickness of F$_1$ is comparable to the magnetic length $\xi_h$.  In this case,  the effective exchange field $\bar h$ acting on the Cooper pairs is gradually reduced as $\alpha$ increases from $0$ to $\pi$. The minimum then is at $\bar h \approx \Delta$ which corresponds to larger values of the bare $h$. Thus, for  exchange fields  $h \lesssim \Delta$ the  antiparallel magnetic configuration ($\alpha = \pi$) of magnetization leads to larger values of the cooling power.  Such  small exchange fields  can be realized in weak ferromagnetic alloys,\cite{small_h} or  in hybrid structures consisting of ferromagnetic insulators in contact with superconductors.\cite{Tokuyasu1988,Cottet2011} For values  of $h$ larger than $\Delta$,  the parallel configuration ($\alpha = 0$)  is the one that leads  to larger values of the cooling power.
For values of $h\gg\Delta$ the cooling is almost independent of $\alpha$. We also analyse the dependence of the cooling power on the bias voltage, the tunnelling parameter and the temperature.  The optimized values for more efficient cooling are shown in Figs.~\ref{panelv}, \ref{panelW} and \ref{panelT}.

Finally we have proposed, we have proposed a novel electron-refrigerator based on spin-filter  barriers. The  N/SF/S junction can suppress Andreev Joule heating effectively leading to cooling power values higher than those predicted for conventional N/I/S coolers. This junction can be build using using well known spin-filters barriers, as for example EuS/Al ($P \sim 0.86$)~\cite{rf:Moodera1}, EuSe/Al ($P \sim 1$)~\cite{rf:Moodera2} and GdN/NbN junction ($P \sim 0.75$).~\cite{rf:Senapati} This allow refrigeration efficiencies of $15\% \sim 23\%$. Another possibility for spin-filter materials are spinel ferrites ($e.g.,$ NiFe$_2$O$_4$)~\cite{rf:Luders,rf:Takahashi} or manganites ($e.g.,$ LaMnO$_{3+\delta}$ and Pr$_{0.8}$Ca$_{0.2}$Mn$_{1-y}$Co$_y$O$_3$).~\cite{rf:Satapathy,rf:Harada} The proposed junction is feasible so the derived results can be applied in efficient solid-based refrigerators. This devices have a wide range of applications, such as, superconducting X-ray detectors, single-photon detectors, magnetic sensors, NEMSs, and qubits.
%
%Summarizing the results of this section, we have develop a whole theory for out of equilibrium ferromagnetic superconductor. Namely, the kinetic equations and corresponding boundary conditions for a spin filter with arbitrary magnetization. Then, we applied the obtained results to a normal metal case in order to obtain the well known Maekawa formula for spin imbalance.

We have also shown that a junction between a conventional
superconductor in the presence of an exchange field and a ferromagnet
with polarization $P$ exhibits huge thermoelectric effects. The
thermopower diverges at low temperatures in the absence of limiting
effects, yielding a figure of merit $ZT \approx P^2/(1-P^2)$
and heat engine efficiencies close to theoretical upper bounds. Furthermore, we studied the dependence of ZT with the inelastic scattering parameter. Moreover, 
even in the case of $P=0$ our model predicts finite spin currents in the presence of a
temperature gradient, provided there is a
spin-splitting of the density of states. These mechanisms in principle can work also in semiconductors without requiring doping which typically  deteriorates the thermoelectric effects. We showed the second order correction to the currents of our all order calculation using the Keldysh Greens function formalism. This gives a positive correction to the thermoelectric coefficient and thus an increase in the thermoelectric effect.

\bibliographystyle{unsrtnat}
\renewcommand{\bibname}{Bibliography of Chapter 4} % changes default name Bibliography to References

%\end{document}

%% file: ch43.tex
%
%%\documentclass[prb,aps,onecolumn,showpacs,superscriptaddress,floatfix]{revtex4-1}
%
%\documentclass{ucbthesis}
%
%\usepackage{amsmath}
%\usepackage{amssymb}
%\usepackage{amsfonts}
%\usepackage{mathptmx}
%
%
%\usepackage{graphicx}
%\usepackage{epstopdf}
%\usepackage{dcolumn}
%\usepackage{bm}
%\usepackage{hyperref}
%
%\newcommand{\conv}{\stackrel{\circ}{,}}
%%\newcommand{\comment}[2]{\begingroup\em(#1 ---#2)\endgroup}
%\newcommand{\ket}[1]{\begingroup\lvert#1\rangle\endgroup}
%\newcommand{\bra}[1]{\begingroup\langle#1\rvert\endgroup}
%
%
%\begin{document}
%
%\chapter{Superconducting quantum point contact in a microwave field.}
%\label{ch:4}

In this section we analyse the supercurrent through a quantum point contact. In Section \ref{sec:sqpc} the concept of superconducting quantum point contact (SQPC) has been introduced. These are highly transmissive junctions generated by a constriction in a superconducting material. This has a length much smaller than the superconducting coherence length.

In order to study the Josephson effect in SQPC, two basic concepts of mesoscopic physics are used. Namely, conduction channels and Andreev bound states. In the normal state, the coherent transport through a mesoscopic system can be described in terms of the independent contributions of the eigenfunctions of the transmission matrix of the structure, known as conduction channels, and these contributions are determined by the corresponding transmission coefficients $\tau_i$. The index "i" denotes the channel. In the case  of a single-channel SQPC with transmission $\tau$, the energies of the ABSs are given  by\cite{Furusaki1991,Beenakker1991}
\begin{equation}
\label{eq-ABS}
E^{\pm}_\text{ A}(\varphi,\tau) = \pm E_\text{ A}(\varphi,\tau) = 
\pm \Delta \sqrt{1 - \tau \sin^2 (\varphi/2)} ,
\end{equation}
where $\Delta$ is the superconducting gap and $\varphi$ is the phase difference between the order parameters on both sides of the junction. Due to the high transmission values of the channels in the SQPC, the ABS are deep in the gap, far from the continuum.

The ground-state of the system is contributed to by all excitations energies: propagating quasiparticles above $\Delta$ and those of the ABS.  In order to obtain the expression for the supercurrent, that only depends on the phase of the superconductors $\varphi$, we focus on the last part. The energy shift per unit time is given by
\begin{equation}
\frac{dE}{dt}=\frac{\partial E_{A}(\varphi)}{\partial \varphi} \frac{d\varphi}{dt} \; .
\end{equation}
This theory is global gauge invariant, the time derivative of the superconducting phase is simply the potential $\dot{\varphi}=2eV$. The energy change per unit time is the power dissipated at the junction, which is the product of current and
voltage. In conclusion, the expression for the supercurrent reads,
\begin{equation}
I_{\text{A}} (\varphi) = (2e) \partial E_{\text{A}}/ \partial \varphi \; .
\label{eq:Ijosepandreev}
\end{equation}
In equilibrium, the two ABS carry opposite supercurrents $I^{\pm}_{\text{A}} (\varphi) = 
(2e) \partial E^{\pm}_{\text{A}}/ \partial \varphi$, which are weighted by 
the occupation of the ABSs (determined by the Fermi function). In the case of a multichannel SQPC, the supercurrent is simply given by the sum of the contributions from the individual channels.\cite{Beenakker1991}. 

%The expression of the current then reads,
%%
%\begin{equation}
%I^{\pm}_{\text{A}} (\varphi) = (2e/\hbar) \partial E^{\pm}_{\text{A}}/ \partial \varphi= \pm \frac{e\Delta}{2\hbar}\sum_{p}\frac{\tau_{p}\sin\varphi}{\sqrt{1-\tau_{p}\sin^{2}(\varphi/2)}}\label{eq:Ijosepandreev}
%\end{equation}
%%
%In the limit of low transmission, $\tau_p \ll 1$ the expression for the current reduces to the Josephson relation \textit{i.e.} $I(\varphi)=I_{c}\sin\varphi$. Here $I_{c}=\frac{e\Delta}{2\hbar}\sum_{p}\tau_{p}=\frac{\pi\Delta}{2e}G_{N}$. Note that in this limit the ABS a so close to the continuum that they are not considered independent states.  
%%
%\begin{figure}[h]
%  \centering
%  \includegraphics[width=0.5\columnwidth]{andrTbig}
%  \caption{Current versus phase difference for different transmission values: 1(red), 0.9 (green), 0.7 (blue), 0.5 (brown) and 0.3 (black).}
%\label{fig:abstra}
%\end{figure}
%%
%
%In fig.~\ref{fig:abstra} we plot the phase dependence this supercurrent corresponding to a single channel ($\tau$). For low  enough transmission values, the current shows the $\sin(\varphi)$ dependence. As we increase it, the phase symmetry is broken, the maxima is displaced to higher values of the phase and the reduction is more sudden. In the limiting case $\tau \rightarrow 1$ we observe an abrupt fall at $\varphi\sim \pi$, after a monotonic increase of the current. Also, the value of the current, as we could expect, increases with transmission.

This microscopic formulation of the dc Josephson effect has been
confirmed experimentally in the context of atomic contacts by Della Rocca
and coworkers.\cite{Rocca2007} In particular, these authors measured the current-phase relation (CPR) of 
an atomic contact placed with a tunnel junction in a small superconducting loop 
and found an excellent agreement with the theory using the independently determined
transmission coefficients. The supercurrent depends on the occupation of the ABS. Thus, one can imagine that by modifying this occupation by, for example, an external field, the supercurrent can be controlled. This is an issue that we explore in this section and for this purpose, we present here an extensive theoretical analysis
of the supercurrent and the dynamics of the ABSs of a SQPC under microwave irradiation.
The control of the supercurrent in a SQPC via microwave irradiation is relevant for the field of quantum computing. Since the ABSs of a SQPC have been proposed as a possible qubit.\cite{Desposito2001,Zazunov2003,Zazunov2005} In this proposal, a microwave field can be used to do the spectroscopy of the two-level system or to probe its quantum state by current measurements.

In the process of understanding the results of the exact theory, we wonder to what point the transport in a microwave-irradiated SQPC can be understood in terms of the dynamics of the ABSs. In principle we could extend the argument of the junction in equilibrium and derive the current in a similar way. For that purpose we study this system using the two-level Hamiltonians of a SQPC existing in the literature\cite{Zazunov2003}. Then, we compare this results with the exact theory. By this study we establish the range of validity of these two-level models. In order to carry out the computation within this model of any dc properties, a very complex and powerful tool was developed. This is valid for an arbitrary two-level system driven out of equilibrium by a periodic perturbation. This is described in section\ref{sec:TLM} and constitutes one of the main results of this chapter. In this section we show that with the Hamiltonian of Ref.~\cite{Zazunov2003} one can nicely reproduce the exact results at low temperatures and low radiation powers. Furthermore, we obtain analytical results for the supercurrent dips produced by microwave-induced transitions between the ABSs.

In combination with the two level model we use the Keldysh-Green function technique of chapter\ref{ch:2}. This method provides deep insight into the physics and quantitative predictions for arbitrary range of parameters. This formalism is specially useful for the case of strong fields, where the the current-phase relation is strongly distorted and the corresponding critical current does not follow a simple behaviour. It is also used to study the case of finite temperatures, when we allow transitions connecting the continuum of states outside the gap region and the Andreev states. This leads to enhancement of the critical current by irradiation of the microwave field.

The first microscopic analysis of a SQPC of arbitrary transparency under microwave irradiation was
reported by Shumeiko and coworkers.\cite{Shumeiko1993} These authors studied the
limit of weak fields and predicted the possibility to have a large suppression
of the current due to resonant transitions between the ABSs. Later, other
aspects of this problem, including the dynamics of the ABSs, have been addressed
focusing on the linear response regime.\cite{Gorelik1995,Gorelik1998,Lundin2000}
 Here, we present in detail this theory and, in particular, we show new analytical results that elucidate the origin of the microwave-enhanced supercurrents. This analytical expression also generalize the microwave-enhanced critical current theory presented by Eliashberg\cite{Eliashberg1970} in 1970.

%The quantum point contact, has been studied in depth in section\ref{sec:sqpc} of this thesis. There are few
%laboratories able to make it with great precision. This device gives
%us the chance to observe interesting physics of superconductors, when
%we apply a bias voltage or when we irradiate it with microwaves. For
%example, the discretization of transport channels and conductance
%or the Andreev Bound States. Two bound states for quasiparticles that
%can be found in the superconducting gap.  Due to the observed physics,
%applications for this type of device can be proposed, for example
%as qubit. As we have a easily tunable two level system, which has
%all the states below the Fermi level occupied. Varying the intensity
%of the field and the phase difference between superconductors we can
%change the current that flows through it.We must remark that the experimental activity
%in this field is in constant development, as we can see from Ref.\cite{key-21,key-22,key-23,key-24}.

\newpage

\section{The Josephson current through a point contact}

The charge transport in a SQPC it is studied now. This structure consists of two bulk superconductors with a gap $\Delta$ in contact by a single conduction channel of transmission $\tau$. In the absence of microwaves the supercurrent can be expressed as as sum of the contributions of the two ABSs,
\begin{equation}
I_\text{ eq}(\varphi)= I^-_\text{ A} n_\text{ F}(E^-_\text{ A}) + I^+_\text{ A} n_\text{ F}(E^+_\text{ A}) \; .
\end{equation}
Here $I^{\pm}_{\text{A}} (\varphi) = 2e \partial E^{\pm}_{\text{A}}/ \partial \varphi$ and $n_\text{ F}(E)$ are the Fermi distribution functions. This yields to,\cite{Haberkorn1978} 
\begin{equation}
\label{Ieq}
I_\text{ eq}(\varphi) = \frac{e\Delta^2}{2} \frac{\tau \sin\varphi}
{E_\text{ A}(\varphi)} \tanh \left( \frac{E_\text{ A}(\varphi)} {2k_\text{ B}T} \right) ,
\end{equation}
where $E_\text{ A}$ is defined in Eq.~(\ref{eq-ABS}) and $T$ is the temperature.
In the tunnel regime ($\tau \ll 1$), the ABS take the value of the gap edges and this expression reduces to the sinusoidal CPR given by the Ambegaokar-Baratoff formula:\cite{Ambegaokar1963} $I_\text{ eq}(\varphi)
= I_\text{ C} \sin\varphi$, with $I_\text{ C} = (e\Delta \tau/2) \tanh (\Delta/
2k_\text{ B}T )$. At perfect transparency ($\tau=1$), this expression reproduces
the Kulik-Omelyanchuk formula:\cite{Kulik1977} $I_\text{ eq}(\varphi) = I_0
\sin (\varphi /2) \tanh (\Delta \cos (\varphi /2)/ 2k_\text{ B}T )$. Here, 
$I_0 = e\Delta$ is the zero-temperature critical current for $\tau=1$
and we frequently use it to normalize the supercurrent 
in the different graphs. In the limit of zero temperature, only the lower ABS are occupied and contribute to the supercurrent, \textit{i.e.} $n_F(E)=0$ for $E>0$. Thus, Eq.~\ref{Ieq} reads  $I_\text{ eq}=I^+_\text{ A}$. For a finite temperature the upper ABS have a finite occupation, leading to a decrease of the total supercurrent. 

\begin{figure}[h]
  \centering
  \includegraphics[width=0.7\columnwidth]{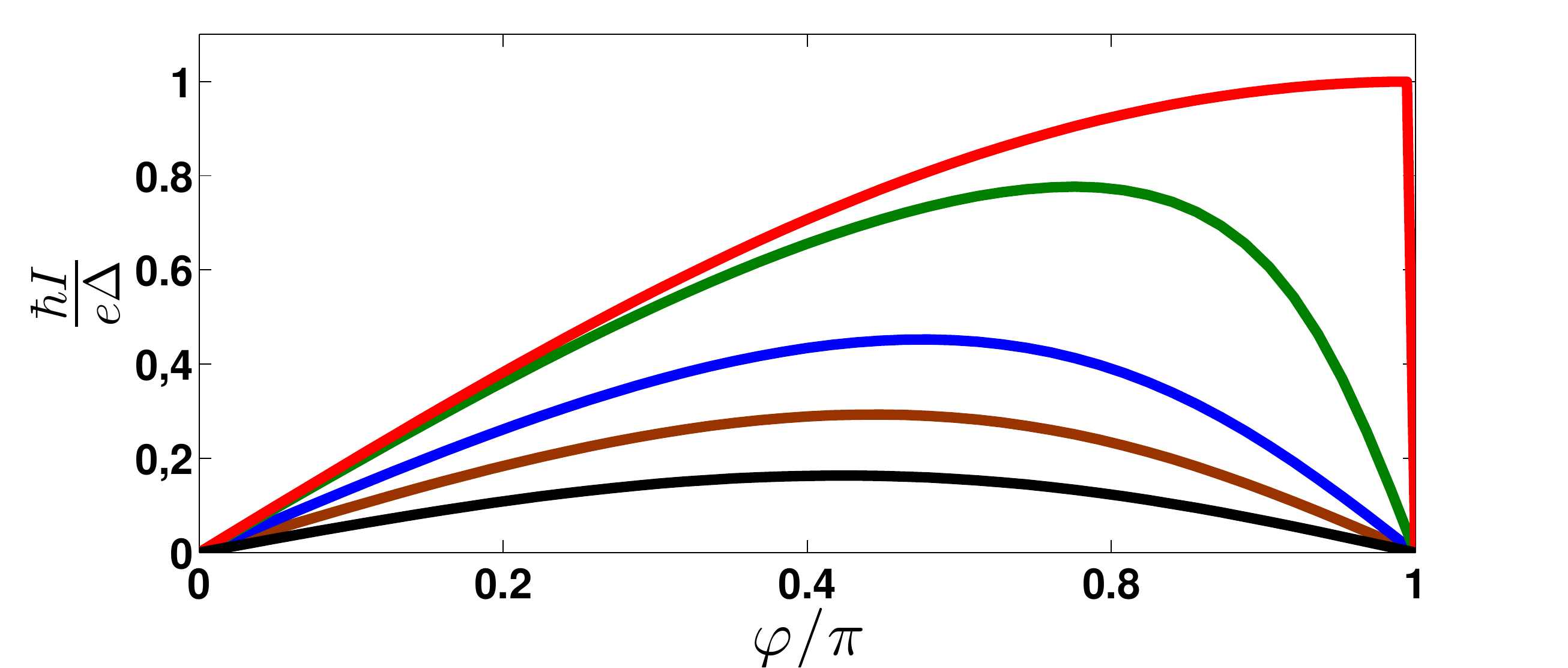}
  \caption{Current versus phase difference for different transmission values: 1(red), 0.9 (green), 0.7 (blue), 0.5 (brown) and 0.3 (black).}
\label{fig:abstra}
\end{figure}

In fig.~\ref{fig:abstra} we plot the phase dependence of the equilibrium supercurrent in a SQPC. For low  enough transmission values, the current shows the $\sin(\varphi)$ dependence. As we increase the transmission, the maxima is displaced to higher values of the phase and the decrease after is more sudden. In the limiting case of perfect transmission, $\tau \rightarrow 1$. We observe an abrupt fall at $\varphi\sim \pi$, after a monotonic increase of the current. Also, the average value for all $\varphi$ of the supercurrent, as we could expect, increases with transmission.

\section{The Josephson current in the presence of an "adiabatic" field}

We compute the supercurrent through a SQPC when it is subjected to a monochromatic microwave field of frequency $\omega$. We assume that the external radiation generates a time-dependent voltage  $V(t) = V_0 \sin \omega t$,\cite{Barone1982} where the amplitude $V_0$ depends on the power of the external radiation source , and eventually also on the polarization of the radiation. According to the Josephson relation, introduced in section\ref{sec:josep}, this voltage induces a time-dependent superconducting phase difference given by 
\begin{equation}
\phi(t) = \varphi + 2 \alpha \cos \omega t , \label{eq-phi}
\end{equation}
where $\varphi$ is the dc part of the phase and $\alpha = eV_0/ \omega$ is
a parameter that measures the strength of the coupling to the electromagnetic 
field.

The microwave-assisted supercurrent in SQPCs is often discussed in the framework
of the so-called adiabatic approximation.\cite{Barone1982} Within this approximation the ABSs
follow adiabatically the microwave field. This approximation does not take into
account the possible transitions between the ABSs. Therefore, the current at high frequencies or for highly transmissive
contacts, where the separation between the states can be rather small it is not described properly. 
Thus, the CPR in this approximation is obtained by replacing the stationary phase $\varphi$ 
in Eq.~(\ref{Ieq}) by the time-dependent phase $\phi(t)$ of Eq.~(\ref{eq-phi}).

The equilibrium CPR in the absence of microwaves, Eq.~(\ref{Ieq}), can be written as
\begin{equation}
I(\varphi)=\sum_{n=1}^{\infty}I_{n}\sin(n\varphi)\; .
\end{equation}
Here $I_n=(1/\pi) \int_0^{2\pi} d\varphi \; I_{\text{eq}}(\varphi) \sin(n\varphi)$ are the harmonics of the equilibrium CPR of Eq.~(\ref{Ieq}). In order to add the effect of the microwaves now we insert the time-dependent phase $\phi(t)$ to the expression,
\begin{equation}
I(\varphi)=\sum_{n=1}^{\infty}\frac{I_{n}}{2i}\{e^{in\phi(t)}-e^{-in\phi(t)}\} \; .
\end{equation}
We apply the Jacobi-Anger expansion that reads,
\begin{equation}
e^{i\alpha\cos(\omega t)}=\sum_{n=-\infty}^{\infty}(i)^{n}J_{n}(\alpha)e^{in\omega t} \; .
\label{eq:expcos}
\end{equation}
Here $J_{n}(\alpha)$ are the n-th order Bessel normal functions. Thus, the dc supercurrent in the adiabatic approximation can be expressed as,
\begin{equation}
I_\text{ ad}(\varphi,\alpha) = \sum_{n=1}^{\infty}I_n J_0(2n\alpha)
\sin(n\varphi), \label{Iad}
\end{equation}
where $J_0$ is the zero-order Bessel function of the first kind. Notice that the current in this
approximation does not depend explicitly on the radiation frequency. 

In the limit of low transmission $\tau \ll 1$ and zero temperature, the integral $I_n$ of the Fourier components can be calculated analytically. They read,
\begin{equation}
I_{1}=\frac{e\Delta\tau}{2};I_{n}=0 \; \text{if} \; n>1 \;.
\end{equation}
The supercurrent in this case then reads,
\begin{equation}
I(\varphi)=I_{1}J_{0}(2\alpha)\sin(\varphi_{0}) \; .
\end{equation}

\begin{figure}[t]
\begin{center}
\includegraphics[width=0.5\columnwidth]{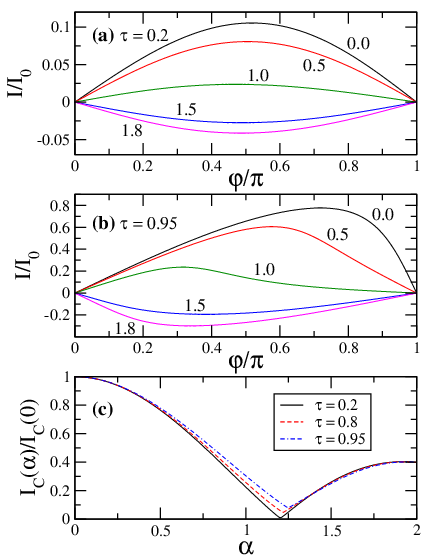}
\caption{ Panels (a) and (b): The current-phase relation in the 
adiabatic approximation for $\tau=0.2$ (a) and $\tau=0.95$ (b). The different
curves correspond to different values of $\alpha$ as indicated in the graphs. 
The current is given in units of $I_0=e\Delta$, where $\Delta$ is the value 
of the superconducting gap at $T=0$. (c) The zero-temperature critical current 
as a function of $\alpha$ for three different values of the transmission $\tau$.
Notice that critical current is normalized by its value in the absence of 
microwaves. (From Ref.\cite{paper1})\label{fig_Iad}}
\end{center}
\end{figure}

We illustrate the results of the adiabatic approximation in Fig.~\ref{fig_Iad} for the zero-temperature case. 
In particular, in the two upper panels we show the CPR (obtained from Eq.~\eqref{Iad}) 
for two different transmissions and several values of the $\alpha$ parameter 
(related to the microwave power). Panel (a) corresponds to the tunnel limit 
($\tau=0.2$) where the CPR is sinusoidal irrespective of the radiation power,
while in panel (b) we show the results for a high transmission of $\tau=0.95$.
In this latter case, the critical current is reached at different values of the
phase depending on the value of $\alpha$. Notice that no matter the value of the
phase $\varphi$, the magnitude of the supercurrent is always suppressed by the 
microwaves as compared with the zero-field result ($\alpha=0$), which is true at
any temperature. The critical current $I_C(\alpha)$, 
as one can see in Fig.~\ref{fig_Iad}(c), decays in a non-monotonic manner, which 
is governed by the Bessel function $J_0$.

\section{The QPC as an effective two-level system}\label{sec:TLM}

It is instructive to also analyse this system by restricting ourselves to the study of the contribution of ABSs present inside the gap, fig.\ref{figdos}. We ignore, for the moment, the contribution of the continuum part of the spectrum, at energies $|E|>\Delta$. This leaves us with a system with just the two ABS. What it is known as an effective two-level model. Previously, the dynamics of a SQPC under external ac fields were described with the help of a effective two-level model\cite{Zazunov2003}, that ensures charge neutrality. For this reason, we base the discussion here on the model put 
forward by Zazunov and coworkers.\cite{Zazunov2003} In this model, the 
SQPC is described with the following $2 \times 2$ Hamiltonian
%
% This can be done with
%the help of the two-level models that have been derived in 
%Refs.~\cite{Ivanov1999} and \cite{Zazunov2003} to describe the 
%dynamics of a SQPC under external ac fields. The models of these two references
%coincide in equilibrium, but the differ slightly when the phase depends
%on time. In particular, the model of Ref.~\cite{Zazunov2003} ensures
%charge neutrality, while the model of Ref.~\cite{Ivanov1999} does 
%not. For this reason, we shall based our discussion here on the model put 
%forward by Zazunov and coworkers.\cite{Zazunov2003} In this model, the 
%SQPC is described with the following $2 \times 2$ Hamiltonian
%
\begin{equation}
\hat H_\text{ B}(t) = \Delta e^{-i\hat\sigma_x r\phi/2}
\left(\cos\frac{\phi}{2} \hat\sigma_z +
r \sin\frac{\phi}{2}\hat\sigma_y\right)\;,\label{TLH-B}
\end{equation}
where $r=\sqrt{1-\tau}$ and $\phi(t)$ is the time-dependent phase given by 
Eq.~(\ref{eq-phi}). This Hamiltonian is written in the ballistic basis of 
right- and left moving electrons, which are eigenvectors of the current operator 
in the perfectly transmitting case ($\tau=1$). For the subsequent analysis it is 
more convenient to work in the instantaneous Andreev basis $\{\ket{+}_{\phi(t)}$, 
$\ket{-}_{\phi(t)}\}$, whose basis vectors are time-dependent. This is the basis 
where the Hamiltonian of Eq.~(\ref{TLH-B}) becomes diagonal in equilibrium. The 
Andreev basis is obtained from the ballistic basis by means of a transformation 
generated by the following unitary matrix
\begin{equation}
\hat R(t) = e^{-i \hat \sigma_x r\frac{\phi}{4}} e^{-i\frac{\pi}{4}\hat\sigma_z}
e^{-i\theta(\phi)\hat\sigma_y} ,
\end{equation}
where $\theta(\phi)=(1/2) \arctan [r \tan(\phi/2) ]$. With this transformation
the Schr\"{o}dinger equation for a state vector $\Psi(t)=(\alpha(t),\beta(t))^T$ 
becomes
\begin{equation}
i \partial_t \Psi(t)=\hat H_\text{ A}(t) \Psi(t)\; \label{schr},
\end{equation}
where
\begin{equation}
\hat H_\text{ A}(t) = E_\text{ A}(\phi(t))\hat{\sigma}_z -
\frac{r \tau \Delta^2 \sin^2(\phi(t)/2)} {4[E_\text{ A}(\phi(t))]^2}
\dot{\phi}(t) \hat{\sigma}_y , \label{tlm}
\end{equation}
and $\dot{\phi}(t) = \partial \phi(t) / \partial t$. 

The corresponding current operator can be written as
\begin{equation}
\hat I_\text{ A}(t) = 2E^{\prime}_\text{ A}(\phi(t))\hat{\sigma}_z +
\frac{r \tau \Delta^2 \sin^2(\phi(t)/2)} 
{E_\text{ A}(\phi(t))} \hat{\sigma}_x ,  \label{ctlm}
\end{equation}
where the prime in $E^{\prime}_\text{ A}$ means derivative with respect to the
argument (the time-dependent phase in this case).
To obtain the expectation value of the current at different 
times, Eq.~\eqref{schr} needs to be solved. Despite the apparent simplicity,
this task has nontrivial aspects: straightforward numerical approaches
run into problems, as both very fast ($t^{-1}\sim{}\omega$) and very slow 
($t^{-1}\sim{}E_\text{ A} - n\omega$) time scales can be simultaneously present. 
No closed-form analytical solution can be obtained either,\cite{Grifoni1998}
and the significantly nonlinear coupling to the drive makes it more difficult 
to derive approximations via standard routes.\cite{Bloch1940,Autler1955}

Focusing this analysis on time-averaged quantities, we can obtain accurate 
analytical and numerical results via a systematic Floquet-type approach. 
We are interested in two physical quantities: the dc current
\begin{equation}
\bar I=\lim_{t\to\infty}\frac{1}{t}\int_0^tdt'\Psi^\dagger(t') \hat I(t')\Psi(t')
\,, \label{avr_current}
\end{equation}
and the time-averaged populations of the Andreev levels
\begin{equation}
\bar p_\pm = \lim_{t\to\infty} \frac{1}{t} \int_0^t dt^{\prime}
\Psi^\dagger(t^{\prime}) \frac{\hat 1 \pm \hat \sigma_z}{2} \Psi(t^{\prime})
\label{avr_pop}\; .
\end{equation}
Below, we show how to obtain $\bar I$, although the method described
can as well be used to compute any other time-averaged quantity,
including $\bar p_\pm$.

We first introduce a modified Hamiltonian
\begin{equation}
  \hat H_\text{ A}(t,\chi)=\hat H_\text{ A}(t)+\chi\hat I_\text{ A}(t)\; ,\label{H_chi}
\end{equation}
where $\chi$ is a parameter conjugate to the observable, and it is set to
zero at the end of the calculation. The solution of the Schr\"{o}dinger 
equation $\Psi(t,\chi)$ can be formally written by introducing the time evolution 
operator $\hat U(t,0;\chi)$
\begin{equation}
\label{sol_schr}
\Psi(t,\chi)=\mathcal{ T} e^{-i\int_0^t dt'\hat H_A(t',\chi)}\Psi_0\equiv 
\hat U(t,0;\chi)\Psi_0 \,,
\end{equation}
where $\mathcal{ T}$ indicates time ordering and $\Psi_0$ is the state
vector at $t=0$. We define now the generating function:
\begin{equation}
  S(t,\chi)=\Psi_0^\dagger U(0,t;\chi=0)U(t,0;\chi)\Psi_0\,. \label{gen_func}
\end{equation}
One can easily check that the dc current defined in Eq.~\eqref{avr_current} 
can be written as
\begin{equation}
\bar I=\lim_{t\to\infty} \frac{i}{t}\partial_\chi S(t,\chi)|_{\chi=0}\;. 
\label{dc_curr}
\end{equation}
Thus, we only need to compute the function $S$, or, equivalently, the evolution 
operator $\hat U(t,0;\chi)\equiv \hat U(t;\chi)$.

Since the Hamiltonian is periodic in time with a period $T=2\pi/\omega$, i.e.,
$H_\text{ A}(t,\chi)=H_\text{ A}(t+T;\chi)$, we can define two periodic (Floquet) 
states $\text{ v}_{\pm}$ via the eigenvalue problem
\begin{equation}
  \hat U(T;\chi)\text{ v}_{\pm}(\chi)=e^{\pm iE(\chi)T}\text{ v}_{\pm}(\chi)
  \,.
\end{equation}
The symmetry of the two eigenvalues follows here from the fact that $\hat
H_\text{ A}(t,\chi)$ and $\text{ log}[U(T;\chi)]$ are traceless, and
$U(T;\chi)$ is unitary.  Moreover, from the periodicity of the
Hamiltonian it follows that
\begin{equation}
\hat U(nT;\chi)=\hat U(T;\chi)^n=\hat V(\chi)e^{iE(\chi)nT\hat\sigma_z}\hat V^{-1}(\chi)\;,
\end{equation}
where the eigenvectors $\text{ v}_{\pm}$ form the columns of the unitary
matrix $\hat V$. Replacing in Eqs.~(\ref{gen_func}) and (\ref{dc_curr}) $t$
by $nT$ and taking the limit $n\rightarrow\infty$ we now find the
derivative with respect to $\chi$:
\begin{equation}
\frac{1}{nT}\partial_\chi \hat U(nT;\chi)\stackrel{n\rightarrow\infty}
{\longrightarrow}i\hat V^{-1}(\chi)\hat\sigma_ze^{iE(\chi)nT\hat\sigma_z}
\hat V(\chi)\frac{\partial E(\chi)}{\partial \chi} \,.
\end{equation}
Thus, the dc current is given by
\begin{equation}
\bar I = -\Psi_0^\dagger \left( \text{ v}_+\text{ v}_+^\dagger-\text{ v}_-\text{ v}_-^\dagger
\right) \Psi_0 \left. \frac{\partial E(\chi)}{\partial \chi}\right|_{\chi=0}
\; .\label{Idc_num}
\end{equation}
This exact expression for the dc current is very useful for numerics. It is 
easy to compute and it handles the fast and slow time scales of the problem
separately. In order to obtain the dc current, one needs first to integrate
the Schr\"{o}dinger equation with the Hamiltonian of Eq.~(\ref{H_chi}) over one
period to find the $2\times2$ matrix $\hat U(T;\chi)$, then one computes its
eigenvalues $\pm{}E$ and eigenvectors $\text{ v}_\pm$, and finally the derivative
$\partial_\chi{}E(\chi)$ is computed via numerical differentiation.

\begin{figure*}[t]
\begin{center}
\includegraphics[width=\columnwidth]{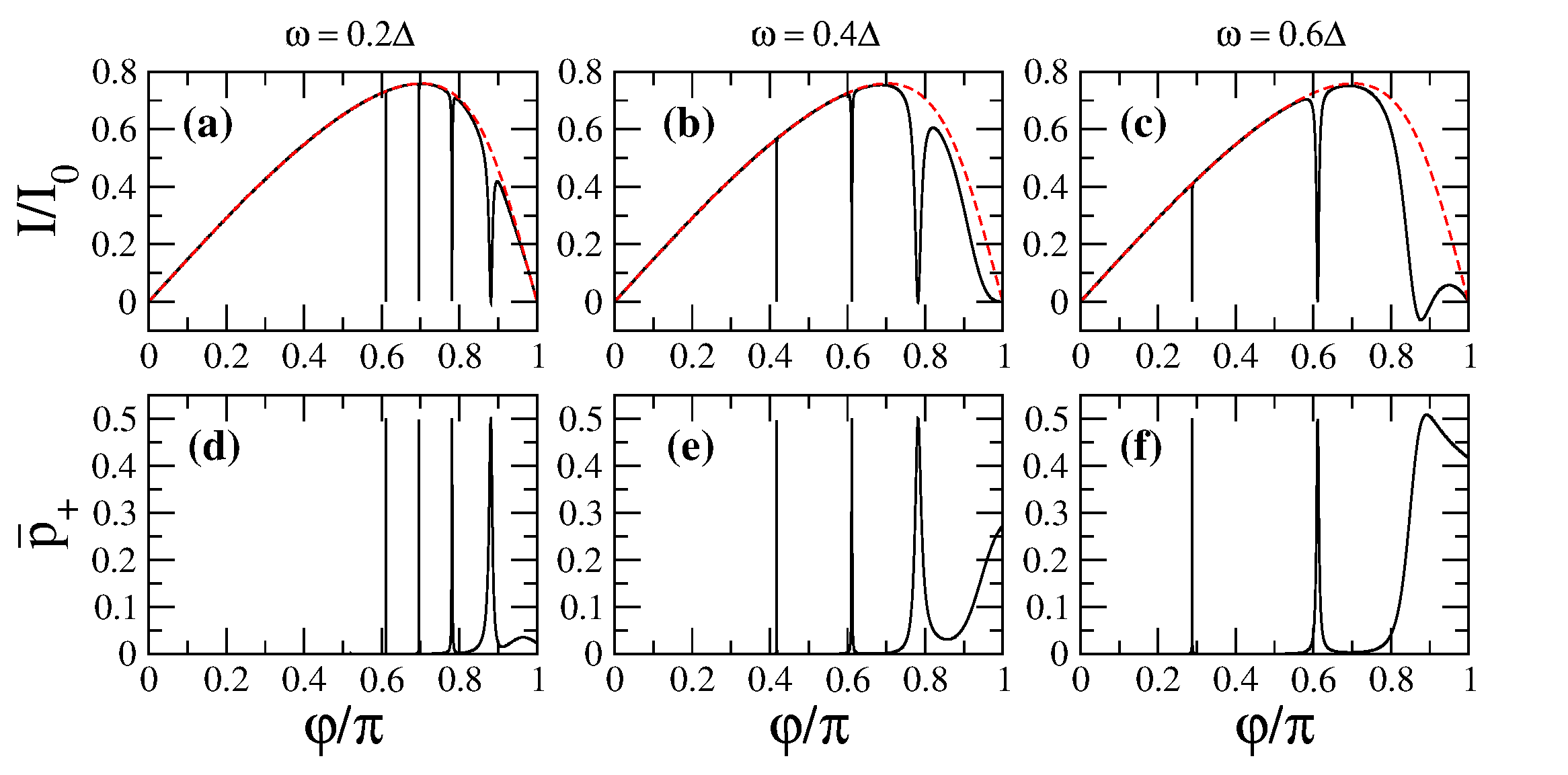}
\caption{ (a-c) Zero temperature supercurrent, in units of $I_0 =
e\Delta$, as a function of the phase for $\tau=0.95$, $\alpha=0.15$ and
three different values of the microwave frequency, as indicated in the upper
part of the graphs. The solid lines correspond the numerical results obtained
with the two-level model, while the dashed lines is the result obtained with 
the adiabatic approximation. (d-f) Time-average occupation of the upper ABS for 
the cases shown in the upper panels.} \label{2level-numerics}
\end{center}
\end{figure*}

In order to have a first impression of the results from this two-level model,
we show in Fig.~\ref{2level-numerics} a few examples of the CPR of a highly
transmissive channel ($\tau=0.95$) computed with the numerical recipe that
we have just described.\cite{note-IC} The upper panels of this figure show the CPR 
for a moderate power ($\alpha = 0.15$) and three different values of the microwave
frequency. For comparison, we also show the result obtained with the adiabatic
approximation of Eq.~(\ref{Iad}). As one can see, the main difference is the
appearance in the results of the two-level model of a series of dips at certain
values of the phases where the current is largely suppressed. It is easy to 
understand that such dips are due to transitions between the ABSs that are 
induced by the microwave field. These transitions enhance the population
of the upper ABS, which at zero temperature is empty otherwise, and at the 
same time they reduce the occupation of the lower ABS. This redistribution of
the quasiparticles results in turn in a suppression of the current. The 
microwave-induced transitions occur with the highest probability when the 
distance in energy between the ABSs (the Andreev gap) is equal to a multiple 
of the photon energy, i.e., when $2E_\text{ A}(\varphi) = n\omega$, where 
$n=1,2,\dots$ is the number of photons involved in the transition. If this
condition is expressed in terms of the phase $\varphi$, it adopts the following
from 
\begin{equation}
\varphi_n = 2\, \text{ arcsin} \sqrt{[1-(n \omega/2\Delta)^2]/\tau},
\; \; n= 1,2,\dots\; . \label{phin}
\end{equation}
A detailed analysis shows that this expression reproduces the positions of all
the dips appearing in the examples of Fig.~\ref{2level-numerics}.

This interpretation of the origin of the dips in the CPR can be corroborated
by a direct analysis of the occupations of the ABSs. Following the same 
numerical recipe, we have also computed the average occupation of the upper ABS,
$\bar{p}_+$, for the examples shown in the upper panels of Fig.~\ref{2level-numerics}.
The results can be seen in the lower panels of this figure and, as one can observe,
there is a nice one-to-one correspondence between the current dips and the 
enhancement of the population of the upper state. In particular, whenever the
upper states reaches a population equal to $1/2$, the current vanishes exactly.

The method described above is not only very convenient for numerical calculations,
but it also provides a route to obtain analytical results. In what follows, we
show how this method can be used, in particular, to gain a further insight into
the microwave-induced supercurrent dips. In order to obtain analytical results,
we must first rewrite Eq.~\eqref{Idc_num} in a more convenient form. In particular,
we would like to avoid the calculation of eigenvectors in this equation. This can 
be done by noting that the unperturbed Hamiltonian~\eqref{tlm} obeys
\begin{align}
\hat \sigma_x\hat H_\text{ A}\hat\sigma_x=-\hat H_\text{ A}\; .
\end{align}
Consequently, $\text{v}_-\propto \sigma_x\text{ v}_+$, and the dc current given 
by Eq.~\ref{Idc_num} can be written as
\begin{equation}
\bar I = -\text{ v}_+^\dagger \left(\hat\rho_0-\hat\sigma_x\hat\rho_0\hat\sigma_x \right)
\text{ v}_+\left.\frac{\partial E(\chi)}{\partial \chi}\right|_{\chi=0}\; ,\label{Idc_num2}
\end{equation}
where $\hat \rho_0 = \Psi_0\Psi_0^\dagger$. Using the expression for the
change of an eigenvalue due to a perturbation, we can finally write
\begin{equation}
\bar I=\left. \frac{\partial E(\chi,\mu)}{\partial \mu}\frac{\partial E(\chi,\mu)}
{\partial \chi}\right|_{\chi,\mu=0}\; ,\label{Idc_ana}
\end{equation}
where $ E(\chi,\mu)$ is an eigenvalue of the matrix
\begin{equation}
\hat M(\chi,\mu) = \frac{i}{T}\hat\omega(T,\chi) + 
\mu\left(\hat\rho_0-\hat\sigma_x\hat\rho_0\hat\sigma_x\right)\;, \label{matrixM}
\end{equation}
where $\hat\omega(T,\chi)\equiv\text{ log}[U(T,\chi)]$ and $\mu$ is an additional 
perturbation parameter. The problem is now reduced to finding the eigenvalues 
of a 2$\times$2 matrix.

As we have seen above, the dc current for weak fields only deviates from the
adiabatic result close to the resonant conditions $n\omega = 2E_\text{ A}$ (with
$n=1,2,3,\ldots$), where the transitions between the ABSs are more likely.
In order to study what happens close to these resonant situations, we can
consider the problem in a rotating frame, and rewrite the evolution operator 
$\hat U(t)$ defined in Eq.~\eqref{sol_schr} as
\begin{equation}
\hat U(t)=e^{-i\hat W_nt}\mathcal{ T}e^{-i\int_0^tdt^{\prime} \hat{\tilde H}_n(t^{\prime})}
\equiv e^{-i\hat W_nt}\hat{\tilde U}_n(t) \,, \label{int_pic}
\end{equation}
where $\hat W_n=n\omega \hat \sigma_z/2$ and the rotating-frame Hamiltonian
is
\begin{equation}
\hat{\tilde H}_n(t)=e^{i\hat W_nt}[\hat H_\text{ A}(t)-\hat W_n]
e^{-i\hat W_nt}\label{new_H}\;.
\end{equation}
The generating function can then be written as in Eq.~\eqref{gen_func}
simply by substituting $\hat{H}_\text{ A}$ by $\hat{\tilde H}_n$. The
additional exponential factors simply cancel out, and the Hamiltonian
$\hat{\tilde H}_n(t)$ remains periodic. Thus, we can proceed exactly as
we have done above.

The key idea that allows to obtain analytical results is the fact that
for weak fields ($\alpha\ll1$), the dynamics in the rotating frame are slow
($\hat{\tilde{H}}_n$ is small) around the corresponding resonance. For this
reason, we can use the Magnus expansion\cite{Magnus54} to determine the matrix 
$\hat \omega$ appearing in Eq.~\eqref{matrixM}:
\begin{align}
  \label{magnus}
  \hat \omega(T)=&-i\int_0^T dt_1\hat H_n(t_1)
  \\\notag
  &-\frac{1}{2} \int_0^T dt_1\int_0^{t_1}dt_2\left[\hat H_n(t_1),\hat H_n(t_2)\right]+\ldots
  \,.
\end{align}
This is essentially an expansion in the parameter $\lambda_n \sim 2n\pi
(E_\text{ A}-n\omega/2)/\omega$, which indeed is small close to a resonance.

We proceed now computing the dc current close to the first resonance 
$\omega=2E_\text{ A}$, assuming that initially the system is in its ground 
state $\Psi_0^\dagger=(0,1)$. We choose $\hat W=\omega\sigma_z/2$, and take 
only the first term of the expansion~\eqref{magnus}, expanding up to first 
order in $\alpha$ and $\chi$. The time integral is straightforward to evaluate, 
and we obtain
\begin{align}
  \frac{i}{T}
  \hat{\tilde \omega}_1(T,\chi)
  &\simeq
    \bigl[ E_\text{ A}-\omega/2+\chi E_\text{ A}'\bigr]\hat\sigma_z
  \\\notag
  &
  +
  \frac{r}{2E_\text{ A}^2}
  \alpha\bigl[
    (\Delta^2-E_\text{ A}^2)\omega/2-\chi (\Delta^2+E_\text{ A}^2) E_\text{ A}'\bigr]
    \hat\sigma_x\,.
\end{align}
Note that this expression is analogous with the well-known rotating
wave approximation, with the difference that by considering the
generating function, this formalism takes the time dependence of the
operator $\hat{I}(t)$ into account.  For the eigenvalues of the matrix
$\hat{\tilde M}_1=(i/T) \hat{\tilde \omega}_1+\mu\sigma_z$ we obtain
\begin{align}
  E^2 &= [E_\text{ A} - \omega/2 + \mu + \chi E_\text{ A}']^2
  \\\notag
  &+(r\alpha/2E_\text{ A}^2)^2[
  (\Delta^2-E_\text{ A}^2)\omega/2-
  \chi (\Delta^2+E_\text{ A}^2) E_\text{ A}']^2
  \,.
\end{align}
Finally, working in the limit $(\omega-2E_\text{ A})/\Delta\ll1$ for simplicity,
we find the dc current from Eq.~\eqref{Idc_ana}:
\begin{equation}
 \bar I _1(\varphi,\omega,\alpha)
  \approx
   -2eE_\text{ A}'\left(
    1
    -
    \frac{
      \omega_{r,1}^2
    }{
      (\omega-\omega_1)^2 + \omega_{r,1}^2
    }
  \right)
  \,,\label{I1}
\end{equation}
where $\omega_1 = 2E_\text{ A}$ is the resonant frequency equals the unperturbed
Andreev level spacing $2E_\text{ A}$ (up to first order in $\alpha$), and
$\omega_{r,1} = r \alpha \omega (\Delta^2-E_\text{ A}^2)/2E_\text{ A}^2$ is the
corresponding Rabi frequency. This expression tells us that the current 
vanishes exactly at the resonant condition $\omega = \omega_1$ and that
the width of the current dip is given by $\omega_{r,1}$, which is linear
in $\alpha$. Moreover, its form clearly suggests that the populations of 
the two states undergo Rabi oscillations with the frequency $\omega_{r,1}$, as
usual in two-state systems, and the time-averaged populations of the
ABS coincide at the resonance. As a consequence, the dc current drops
to zero at the resonance, a result that qualitatively coincides with the 
prediction in Ref.~\cite{Shumeiko1993}.

We can also determine the dc current at the higher resonances, for
example for $\omega\approx E_A$. In this case we work in the frame
corresponding to $\hat W_2=\omega\hat\sigma_z$. As the resonance is
due to two-photon processes, terms up to order $\alpha^2$ must be
taken into account, which requires including the first two terms in
Eq.~\eqref{magnus}. The computations are again straightforward, and up
to second order in $\alpha$ we obtain
\begin{align}
  \frac{i}{T}
  \hat{\tilde \omega}_2
  &
  \approx \bigl[E_\text{ A} - \omega + \alpha^2E_\text{ A}'' + \chi E_\text{ A}'
  +
  r^2 \alpha^2\omega\frac{(\Delta^2-E^2)^2}{12E_\text{ A}^4}\bigr]\hat\sigma_z
  \\\notag
  &
  - \frac{r\alpha^2E_\text{ A}'}{2E_\text{ A}^2}
\bigl[ \Delta^2\bigl(\frac{\omega}{E_\text{ A}}+1\bigr) -E_\text{ A}^2)\bigr]\hat\sigma_x\; .
\end{align}
For simplicity, we dropped terms of order $\alpha \chi$, which do not
essentially affect the form of the resonance. As above, the current is
obtained from the eigenvalues of $\hat{\tilde M}_2=(i/T) \hat{\tilde
\omega}_2+\mu\sigma_z$, and it adopts the form expression
\begin{equation}
\bar I_2(\varphi,\omega,\alpha) \simeq-2eE_\text{ A}'\left(1-\frac{\omega_{r,2}^2}
{(\omega-\omega_{2})^2+\omega_{r,2}^2}\right) \,,\label{I2}
\end{equation}
where $\omega_{2} = E_\text{ A} + \alpha^2E_\text{ A}'' +
r^2\alpha^2(\Delta^2-E_\text{ A}^2)^2/12E_\text{ A}^3$ and $\omega_{r,2} =
r \alpha^2(2\Delta^2- E_\text{ A}^2)E_\text{ A}'/2E_\text{ A}^2$.  Here, one can 
observe that the resonant frequency is shifted from the position 
$2\omega=2E_\text{ A}$ by two contributions: the first arises from nonlinearities, 
and the second is the Bloch-Siegert shift.\cite{Bloch1940}

One can also go further and compute the dc current around resonances
$n>2$, although this gets progressively more cumbersome as an
increasing number of terms are required in the Magnus expansion,
reflecting the increasing number of allowed multiphoton processes
generated by the nonlinearities. One can however see from
Eqs.(\ref{I1},\ref{I2}), and also check for higher resonances, that
the width of the resonances scales with $\alpha^n$. Moreover,
one can show that within this model, the time-averaged current
vanishes exactly at each resonance, i.e., $\bar{I}\rvert_{\omega=\omega_n}=0$.

\begin{figure}[t]
\begin{center}
\includegraphics[width=0.6\columnwidth,clip]{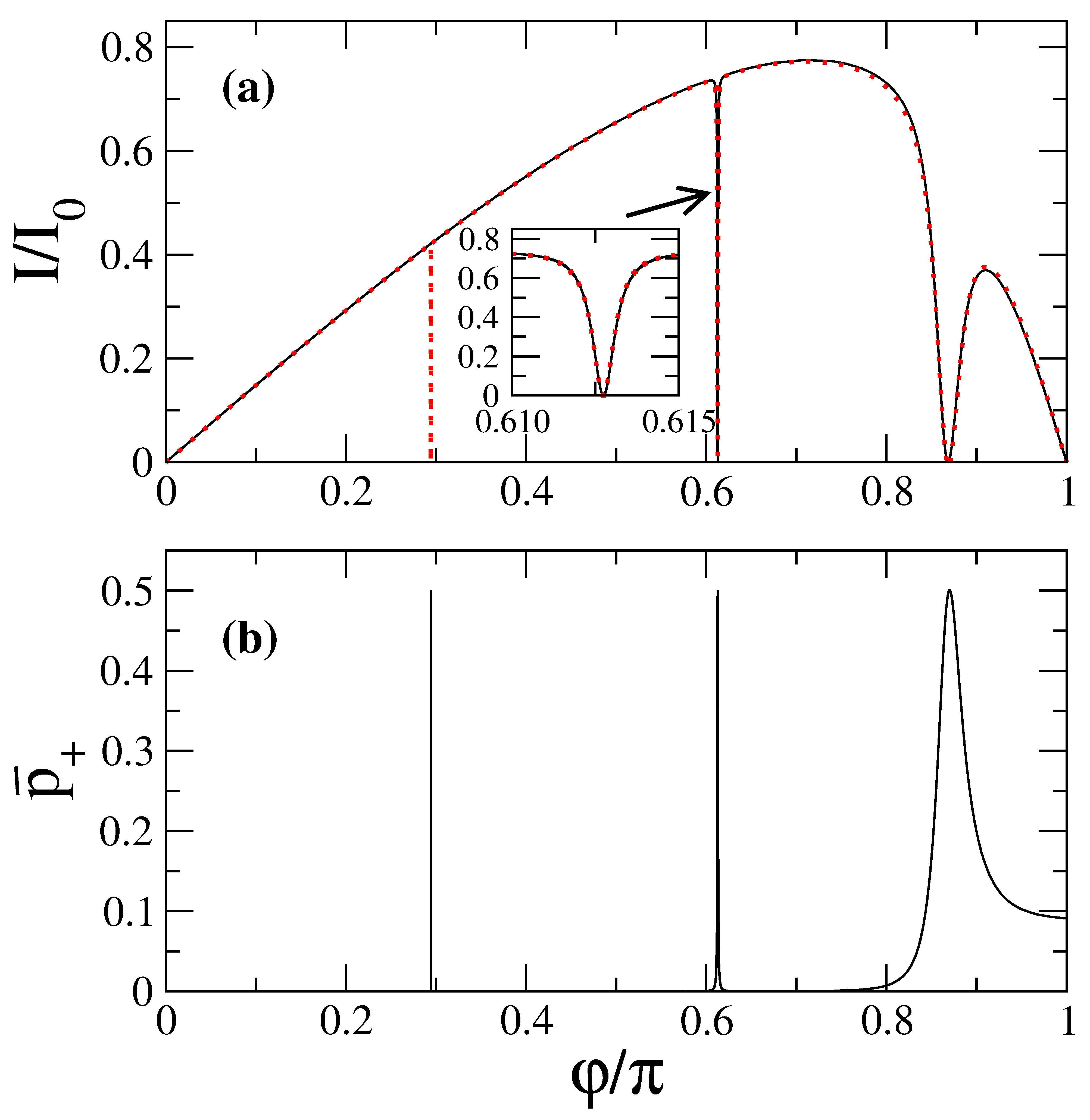}
\caption{\label{tlm-cpr}
 (a) Current-phase relation in the two-level model,
obtained from Eq.~\eqref{eq:tlm-compound-approx} (solid line) and 
numerics based on Eq.~\eqref{Idc_num} (dotted line). Parameters are 
$\omega=0.6\Delta$, $\tau=0.95$, and $\alpha=0.05$. The $n=3$ resonance 
is not included in the analytical approximation. Inset: close-up of the 
second resonance. (b) Time-averaged population $\bar{p}_+$ of the upper 
Andreev state for the same parameters. (From Ref.\cite{paper1})}
\end{center}
\end{figure}

The results Eqs.(\ref{I1},\ref{I2}) can be combined into a single
approximate expression
\begin{equation}
  \label{eq:tlm-compound-approx}
  \bar I
  \approx
  -2eE_A'
  \biggl(
    1 - \frac{\omega_{r,1}^2}{(\omega-\omega_{1})^2 + \omega_{r,1}^2}
  \biggr)
  \biggl(
    1 - \frac{\omega_{r,2}^2}{(\omega-\omega_{2})^2 + \omega_{r,2}^2}
  \biggr)
  \,.
\end{equation}
The quality of this approximation can be established by comparing it with
the exact numerical results. This is done in Fig.~\ref{tlm-cpr} where
we have considered the case of a weak microwave field ($\alpha=0.05$).
As one can see, there is an excellent agreement with the numerical 
results, apart from the fact that the numerics also include a dip 
produced by three-photon processes, which we have left out from the 
above approximation.

In spite of the simplicity 
of the two-level model considered here, such a model captures the essential 
physics of a microwave-irradiated SQPC and it provides accurate results
for not too high frequencies and up to moderate radiation power. As we
establish in the next section, the limitations of the two-level
model are related to the fact that it does not take into account the 
contribution of the the continuum of states outside the gap region.

\section{The contribution to the current from the continuum part of the spectrum}
\label{sec:conti}

Up to this point, we have analysed the supercurrent through a microwave irradiate SQPC assuming that the only contribution comes from the ABSs. This is true for a SQPC in equilibrium but it is not obvious that this should be the case in the presence of a microwave field. High frequencies or high radiation powers and specially finite temperatures is what allows this phenomena to happen. Therefore, in order to describe the complete phenomenology of irradiated SQPCs we must develop a fully microscopic theory. This is the goal of this section.

Our microscopic theory is based on the Keldysh-Green function approach introduced in section\ref{sec:quasi}.
In this approach the starting point is the expression for the quasiclassical
Green functions of the left (L) and right (R) electrodes, where we use the time dependent notation. The Green functions of the electrodes are position independent, as shown in section\ref{sec:spectral}. Due to the time dependent phase, eq.\ref{eq-phi}, we have to use the time dependent notation. For all this, the Green functions can be expressed in terms of the equilibrium Green functions $\check g(t-t^{\prime})$ as follows
\begin{equation}
\check G_{R (L)}(t,t^{\prime}) = e^{\pm i\phi(t) \hat \tau_3/2} 
\check g(t-t^{\prime}) e^{\mp i\phi(t^{\prime})\tau_3/2}\; ,
\label{eq-GF-leads}
\end{equation}
where $\phi(t)$ is the time-dependent phase given by Eq.~(\ref{eq-phi}). This time dependent Green functions are related to the until now used energy dependent ones in the following way,
\begin{equation}
\check g(t)= \int^{\infty}_{-\infty} \frac{dE}{2\pi} e^{-iE t} \check g(E) \; .
\end{equation}
Here the Retaded (R), Advanced (A) and Keldysh (K) components read,
\begin{eqnarray}
\hat g^{R(A)}(E)& =& g^{R(A)}(E)  \tau_3 + f^{R(A)}(E)  \tau_1\\
\hat g^{K}(E) &=& \left[\hat g^{R}(E) - \hat g^{A}(E)\right] \tanh(E/2k_\text{ B}T)
\end{eqnarray}
and
\begin{equation}
g^{R(A)}(E) = \frac{E}{\sqrt{(E \pm i\eta)^2-\Delta^2}} = 
\frac{E}{\Delta} f^{R(A)}(E) \; ,
\end{equation}
where $\eta$ describes the inelastic scattering energy rate within the relaxation 
time approximation and $T$ is the temperature.

Different authors have shown that the transport properties of a point contact
with an arbitrary time-dependent voltage can be described by making use of 
adequate boundary conditions for the full quasiclassical 
propagators.\cite{Zaitsev1998,Nazarov1999,Cuevas2001,Kopu2004} For the case of a single-channel SQPC of arbitrary transmission $\tau$ the Green functions of Eq.~(\ref{eq-GF-leads}) follow the Nazarov boundary conditions\cite{Nazarov1999}
\begin{equation}
\check J(t,t^{\prime}) = 2 \tau \left[\check G_L , \check G_R \right]_{\circ}
\circ \left[ 4 - \tau \left(2- \left\{ \check G_L , \check G_R \right\}_{\circ}
\right) \right]^{-1} (t,t^{\prime})  \label{matrixJ}.
\end{equation}
Here, the symbol $\circ$ denotes the convolution over intermediate time arguments. For low transmission values $\tau \ll 1$, this expression reduces to well known Kupriyanov Lukichev boundary condition,
\begin{equation}
\check J(t,t^{\prime}) = \frac{\tau}{2}\left[\check G_L , \check G_R \right]_{\circ} (t,t^{\prime})  \label{matrixJkl}.
\end{equation}
Finally, the electric current is obtained by taking the following trace
\begin{equation}
I(t) = \frac{e}{4} \text{ Tr} \tau_3 \hat J^K(t,t)\; ,
\end{equation}
where $\hat \tau_3$ is the third Pauli matrix in Nambu space. 

Due to the periodic time dependence of the phase, the 
Green functions $\check G_{L(R)}$, and any product of them, admit the following 
Fourier expansion 
\begin{equation}
\check G(t,t^{\prime}) = \sum^{\infty}_{m=-\infty} e^{im\omega t^{\prime}} 
\int \frac{dE}{2\pi} e^{-i E(t-t^{\prime})} \check G_{0m}(E) ,
\end{equation}
where $\check G_{nm}(E) \equiv \check G(E+n \omega, E+m \omega)$ are 3
the corresponding Fourier components in energy space, and $n,m$ are integer 
numbers. In particular, the Fourier components of $\check G_{L(R)}$ can be
easily deduced from Eq.~(\ref{eq-GF-leads}). Thus for instance, for the left
electrode $\check G_{nm}(E)$ is given by 
\begin{equation}
(\check G_L)_{nm} = \sum_l \check \Gamma_{nl} \check g_l \check \Gamma^{*}_{lm} ,
\end{equation}
where
\begin{equation}
\check \Gamma_{nm} = \left(
\begin{array}{cc}
\hat \Gamma_{nm} & 0\\
 0 & \hat \Gamma_{nm} \end{array} \right) ,
\hat \Gamma_{nm} = \left( \begin{array}{cc}
\mathcal{ P}_{nm} & 0\\
0 & \mathcal{ P}^{\ast}_{nm} \end{array} \right) .
\end{equation}
Here, $\mathcal{ P}_{nm}=(i)^{m-n}J_{m-n}(\alpha/2)e^{i\varphi/4}$, where $J_n$ is
the Bessel function of order $n$, and $\check g_n=\check g(E+n\omega)$ is 
the equilibrium Green function matrix with shifted argument. Here the Bessel functions are obtained from the previously introduced relations, eq.\ref{eq:expcos}. 

From this discussion, it is easy to understand that the current adopts the 
following general expression
\begin{equation}
I(t) = \sum^{\infty}_{m=-\infty} I_m e^{im \omega t} ,
\end{equation}
which means that the current oscillates in time with the microwave frequency and 
all its harmonics. These current components can be computed from the Fourier components
in energy space of $\check I$ in Eq.~(\ref{matrixJ}). From that equation, it is
straightforward to show that the Fourier components of $\hat I^K$ are given by
\begin{equation}
\hat J_{nm}^K = \sum_l[\hat A_{nl}^R \hat X_{lm}^K + \hat A_{nl}^K \hat X_{lm}^A] .
\label{eq-J}
\end{equation}
Here, we have defined the matrices $\check A_{nm} \equiv 2 \tau 
[\check G_L,\check G_R]_{nm}$ and  $\check X_{nm} \equiv [4\check 1-\tau(2-
\{\check G_L,\check G_R\})]^{-1}_{nm}$. Once the components of $\hat J_{nm}^K$ are obtained, one can compute the current. We are only interested here in the dc component, which reads
\begin{equation}
I(\varphi,\omega,\alpha) = \frac{e}{4} \int \frac{dE}{2\pi} \text{ Tr} 
\hat \tau_3 \hat J^K_{00}(E,\varphi,\omega,\alpha) \; .
\label{Idc}
\end{equation} 
The dc current can be calculated analytically in certain limiting cases like in 
the absence of microwaves, where it reduces to Eq.~(\ref{Ieq}), or in the tunnel 
regime or for very weak fields, as we show below.

%\paragraph{Tunneling limit }

Let us assume that transmission of the channel $\tau$ is very low. In this case the denominator of the boundary condition reduces to $\check X \equiv [4\check 1]^{-1}$. The expression for the current now reads,
\begin{equation}
I(\varphi,\omega,\alpha)=\frac{e\tau}{8}Tr\hat{\tau_{3}}\hat{A}_{00}^{K}
=\sum_{l=-\infty}^{\infty}\sum_{k=-\infty}^{\infty}\frac{-i}{4eR_{N}}\sin\varphi[J_{l}(\alpha/2)J_{k-l}(\alpha)J_{-k}(\alpha/2)]\int dE(f_{l}^{R}g_{{k}}^{K}+g_{{l}}^{K}f_{k}^{A}) \; . 
\end{equation}
The residue theorem cannot be applied to this expression due to the absence of simple poles. In order to deal with this inconvenience, we use the approximation $\frac{\omega}{\Delta} \ll 1$.
The value of the superconducting gap (in Al) is usually of the order
of 1 meV which is equivalent to 100 GHz. The frequency of the microwave
field is usually of the order of 1 GHz, two orders of magnitude or
more lower, which makes the approximation reasonable. Then we can assume that $E=\pm\Delta\pm l\omega\simeq\pm\Delta$, so we obtain,
\begin{equation}
I(\varphi,\omega,\alpha)=\sum_{l=-\infty}^{\infty}\sum_{k=-\infty}^{\infty}\frac{-i}{4eR_{N}}\sin\varphi[J_{l}(\alpha/2)J_{k-l}(\alpha)J_{-k}(\alpha/2)]
\nonumber
\end{equation}
\begin{equation}
\int dE[(f^{R}(E))^{2}-f^{R}(E)f^{A}(E)]tanh(\frac{E+k\omega}{2T})+[f^{R}(E)f^{A}(E)-(f^{A}(E))^{2}]\tanh(\frac{E+l\omega}{2T}) \; .
\end{equation}
Neglecting the anomalous cross terms, we obtain simple poles in the expression and the residue theorem can be applied. We end up with the following expression for the dc current of a SQPC under microwave irradiation in the tunnelling case,
\begin{equation}
I(\varphi,\omega,\alpha)=\sum_{l=-\infty}^{\infty}\sum_{k=-\infty}^{\infty}\frac{\pi\Delta}{4eR_{N}}\sin\varphi J_{l}(\alpha/2)J_{k-l}(\alpha)J_{-k}(\alpha/2)\left[\tanh(\frac{\Delta+k\omega}{2T})+\tanh(\frac{\Delta+l\omega}{2T})\right] \; .
\end{equation}
However, for arbitrary transmission and radiation power one needs 
to evaluate Eq.~(\ref{Idc}) numerically. In the next subsections we present the 
results for the dc current of this microscopic theory and we compare them with 
those obtained from the two-level model of section \ref{sec:TLM}.

\subsection{The zero-temperature limit}

% in particular, to make a comparison with the two-level model that we present in section \ref{sec:TLM} and

We now focus on the results of the exact theory at zero temperature. This allows to establish the range of validity of this formalism and to make a comparison with the two-level model that we present in section \ref{sec:TLM}.   

\begin{figure}[t]
\begin{center}
\includegraphics[width=0.8\columnwidth,clip]{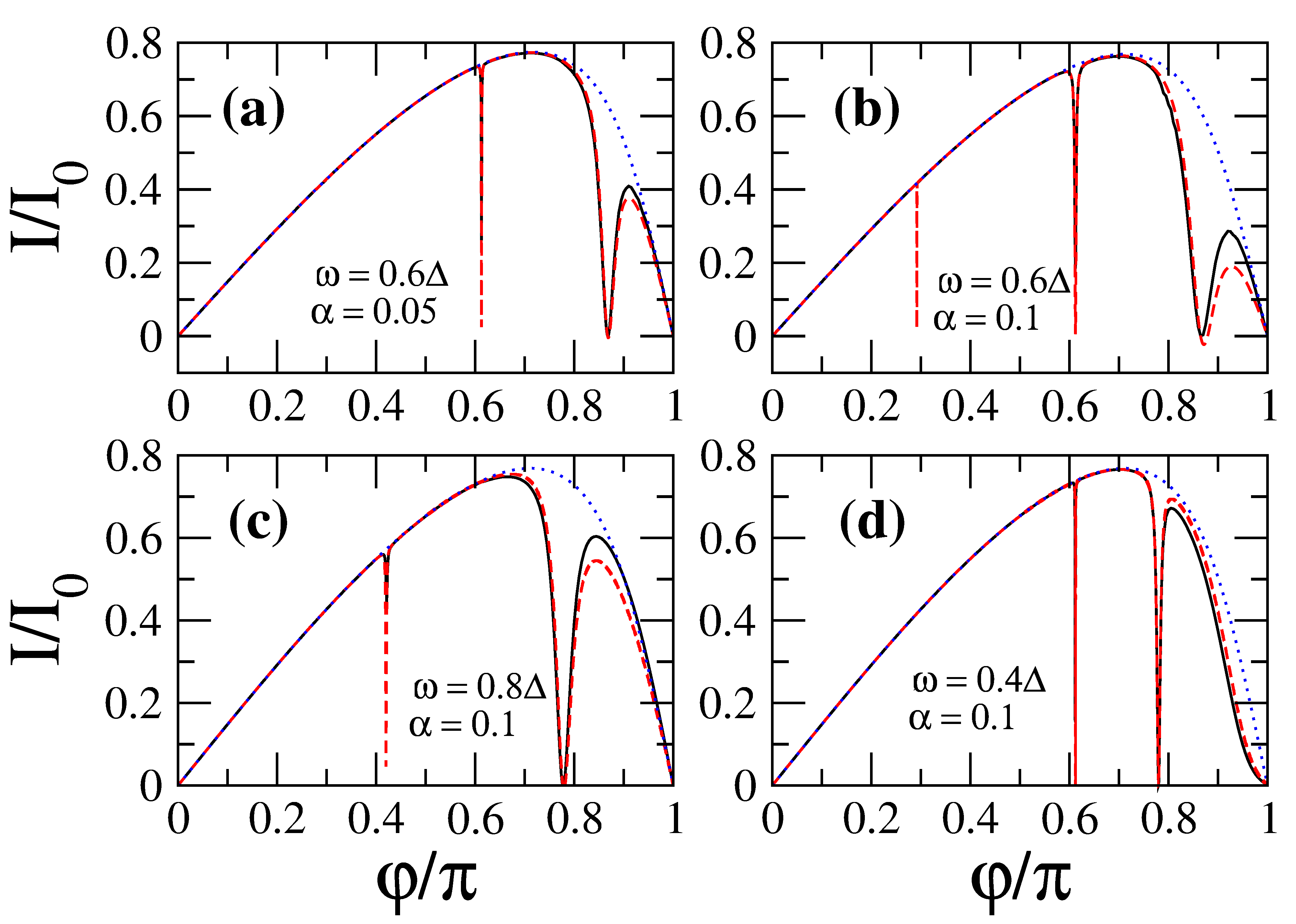}
\caption{\label{comparison1}
 Four examples of the zero-temperature current-phase relation for $\tau=0.95$ 
obtained from the microscopic model (solid lines), the two level model (dashed lines) 
and the adiabatic approximation (dotted lines). The parameters characterizing the 
microwave field are indicated in the different panels. (From Ref.\cite{paper1})}
\end{center}
\end{figure}

In Fig.~\ref{comparison1} we show several examples of the CPR calculated with
the microscopic approach (solid lines) for a highly conductive channel ($\tau=0.95$) 
for several frequencies and low values of the radiation power ($\alpha \ll 1$). For 
comparison, we also show the results of both the two-level model (dashed lines)
and the adiabatic approximation (dotted lines). As one can see, the main deviation
from the adiabatic results is the appearance of a series of dips, as discussed in
section III. These features, which originate from the microwave-induced transitions
between the ABSs, are accurately reproduced by the two-level model (both the position
and the width of the dips). There is a small discrepancy between the exact result
and those of the two-level model for phases close to $\pi$, i.e., when the level 
spacing between the ABS is very small. This is understandable since the model
assumes that $ \dot \phi(t) \ll 2E_\text{ A}$, which is not fulfilled when
$\varphi \sim \pi$ and $\tau$ is close to 1. Notice also that for the high-order
dips (due to high-order photonic processes), the current suppression for the two-level
model is larger than in the case of the exact theory. The reason is the additional
broadening introduced by the finite inelastic scattering rate used in the
calculations with the microscopic theory, which in this case is $\eta= 10^{-4}\Delta$.

\begin{figure}[t]
\begin{center}
\includegraphics[width=0.6\columnwidth,clip]{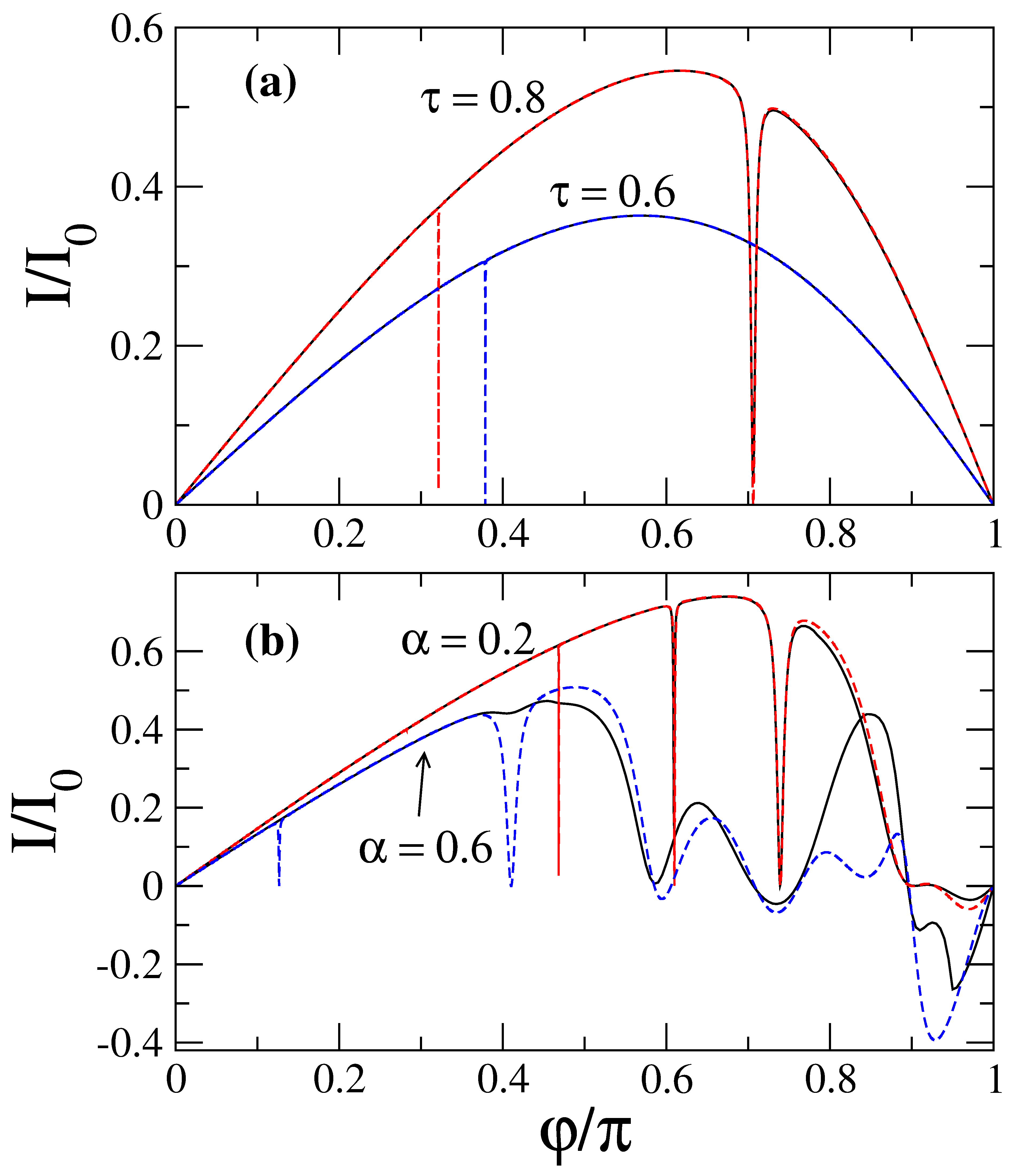}
\caption{\label{comparison2}
 (a) The current-phase relation for $\alpha=0.1$, $\omega=0.6\Delta$
and two values of the transmission coefficient, $\tau=0.8$ and $\tau=0.6$. 
(b) The current-phase relation for $\omega=0.3\Delta$, $\tau=0.95$ and two values
of the $\alpha$, 0.2 and 0.6. In both panels the solid lines correspond to the microscopic
theory and the dashed lines to the two-level model. (From Ref.\cite{paper1})}
\end{center}
\end{figure}

The good agreement between the microscopic theory and the two-level model in these
examples can be easily understood. At zero temperature, the lower ABS is fully occupied,
while the upper one is empty. Therefore, for small values of $\alpha$ and $\omega<\Delta$ 
transfer of quasiparticles between the continuum and the ABSs is not possible. The agreement 
between these models is further confirmed in Fig.~\ref{comparison2}(a), where the CPR 
is shown for $\omega=0.6\Delta$, $\alpha=0.1$ and two lower values of the transmission
($\tau=0.6$ and $0.8$). In this case, the agreement is almost perfect for all phases.
The reason is that now the smallest energy gap between the ABSs, which occur at 
$\varphi=\pi$, is large enough to avoid the overlap of the levels in the presence of the 
microwave field. If the transmission is further reduced, no transitions can occur
between the Andreev states and the adiabatic approximation becomes exact.
 
From the discussion above, we can conclude that the two-level model provides an
excellent description of the supercurrent at zero temperature and for weak fields
($\alpha \ll 1$). However, as the radiation power increases, the situation changes.
This is illustrated in Fig.~\ref{comparison2}(b), where we show the CPR for a highly 
conductive channel ($\tau=0.95$), a frequency $ \omega = 0.3\Delta$
and two values of $\alpha$. As one can see, the deviations between the 
results of the two-level model and the microscopic theory become more apparent
as the power increases. The main reason for this discrepancy is the occurrence
of multiphotonic processes, which become more probable as the power increases.
These processes induce quasiparticle transitions between the ABSs and the continuum 
part of the energy spectrum, which are not included in the two-level model.
 
\begin{figure}[t]
\begin{center}
\includegraphics[width=0.5\columnwidth,clip]{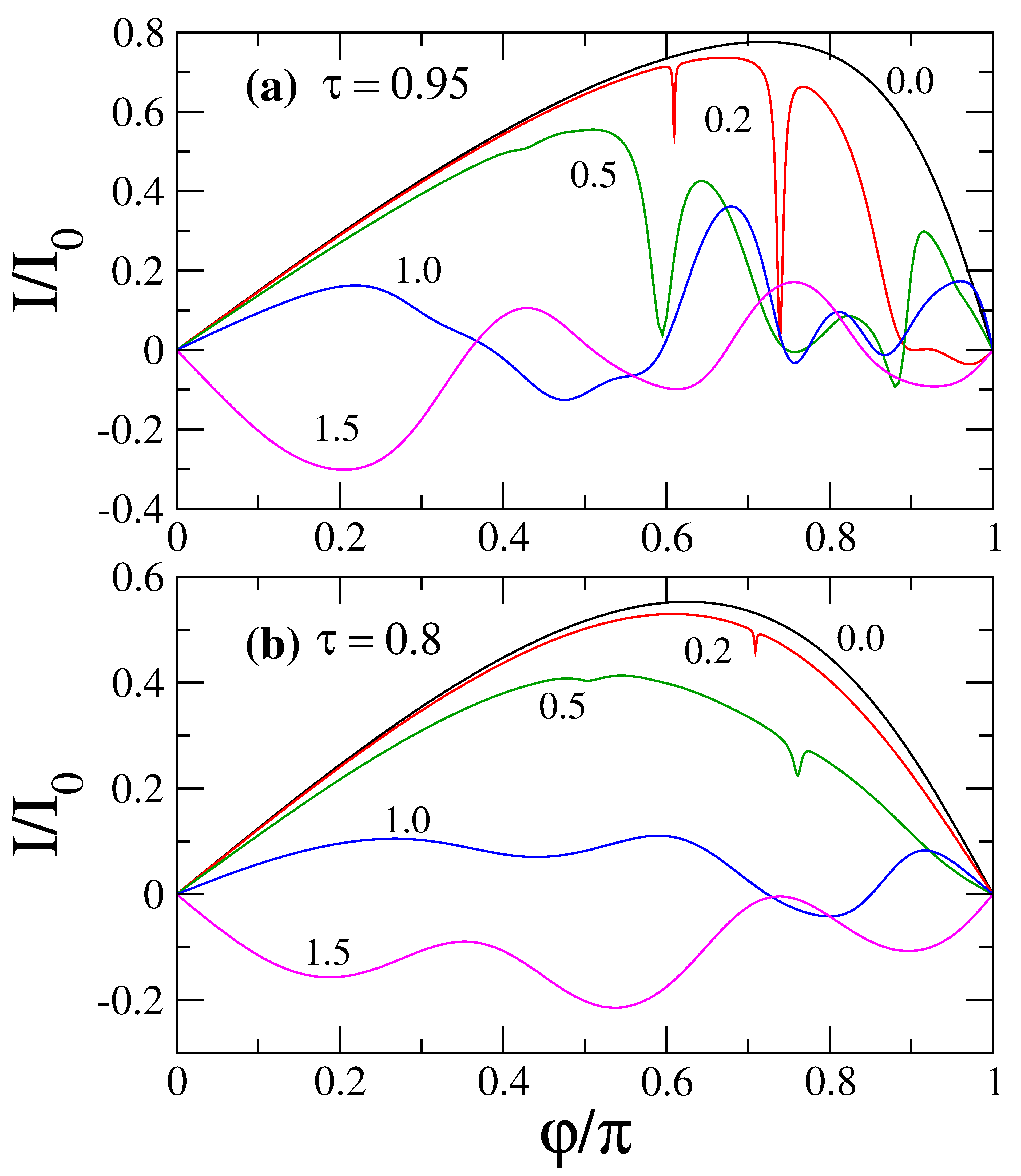}
\caption{\label{CPR-power} 
The zero-temperature current-phase relation for $\omega=0.3\Delta$ 
and two values of the transmission: (a) $\tau = 0.95$ and (b) $\tau = 0.8$.
The different curves correspond to different values of $\alpha$, as indicated
in the graphs. The inelastic broadening used in these calculations is $\eta =
10^{-3} \Delta$. (From Ref.\cite{paper1})}
\end{center}
\end{figure}

As one could already see in Fig.~\ref{comparison2}(b), as the radiation power
increases the supercurrent dips broaden and the CPR acquires a very rich 
structure. We illustrate this fact in more detailed in Fig.~\ref{CPR-power} 
where we show the evolution of the CPR with $\alpha$ for two values of the 
transmission (0.95 and 0.8) and a frequency $ \omega = 0.3\Delta$. Notice
that as the power increases, the dips disappear, the CPRs become highly non-sinudoidal,
and in some regions of the phase the current can reserve its sign. These results
are clearly at variance with those found within the adiabatic approximation
(see section II). They are a consequence of a complex interplay between the
dynamics of the ABSs, which are broadened by the coupling to the microwaves,
and the multiple transitions induced between the ABSs and the continuum of
states. This very reach behaviour has also important implications for the critical
current, which for high transmission strongly deviates from the standard behaviour
described by the adiabatic approximation. This is discussed below in detail.
Finally, it worth stressing that the values of $\alpha$ used in
Fig.~\ref{CPR-power} are easily achievable, as demonstrated in the context
of atomic contacts,\cite{Chauvin2006} semiconductor nanowires\cite{Doh2005}
or graphene junctions.\cite{Heersche2007,Jeong2011} Therefore, these results
indicate that the microscopic theory presented here is always necessary to 
describe the experimental results of highly transmissive junctions at sufficiently 
high power, no matter high low the microwave frequency is.

In a typical experimental situation the QPC consists  of few  conducting channels with different transmissions. According to Eqs. (\ref{matrixJ}, \ref{Idc}) the total dc current is given by the sum of the contributions of each of the channels. In Fig. \ref{fig:3chan} we show the CPR for a three-channel point contact  similar to the one of the QPC studied in the experiments of Ref. \cite{Chauvin2006}. The set of transmission coefficients is $\tau={0.17,0.6,0.97}$.  Due to the contribution from the two low transiting channels the current at the first resonance does not vanished completely, but nevertheless it still shows a dip.

% \begin{figure}
%  \includegraphics[width=\columnwidth,clip]{3chan}
%  \caption{\label{3chan}
%    (Color online) .}
%\end{figure}

\subsection{Finite temperature: Enhacement of the supercurrent}\label{MMb}

We now turn to the analysis of the supercurrent at finite temperature, which
is carried out within the microscopic  model. The new ingredient at 
finite temperature is the fact that the ABSs are neither fully occupied 
nor fully empty, which means that quasiparticle transitions between the 
continuum of states and the bound states are possible, even for 
frequencies $ \omega < \Delta$. As we see below, this has important
consequences.

\begin{figure}[t]
\begin{center}
\includegraphics[width=0.6\columnwidth,clip]{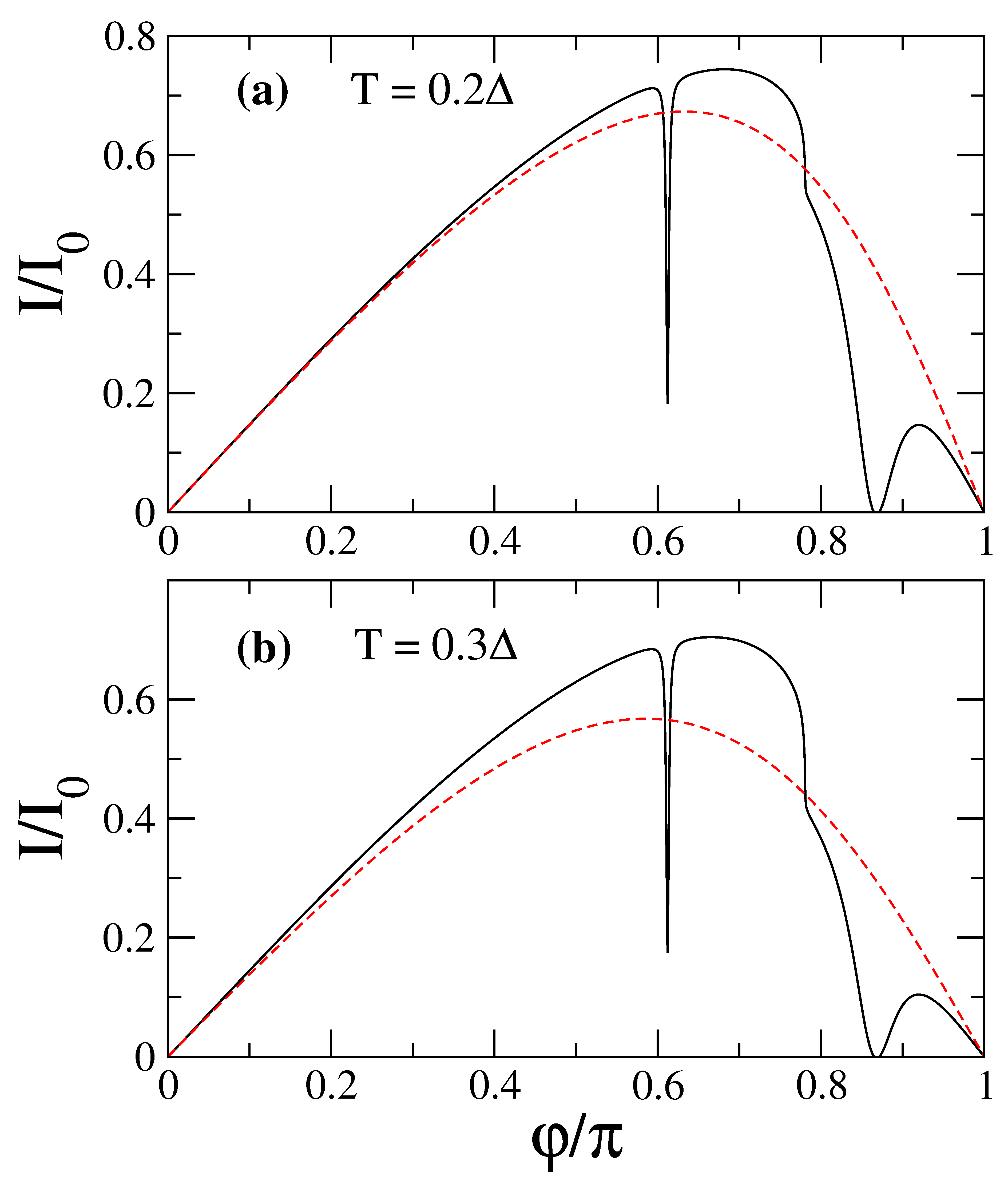}
\caption{\label{CPR-temp1}
 The current-phase relation for $ \omega=0.6\Delta$, $\alpha=0.1$,
$\tau=0.95$ and two different temperatures: (a) $ T=0.2\Delta$ and (b) $ T=0.3\Delta$.
The solid lines in both panels correspond to the results of the microscopic theory and
the dashed lines to the supercurrent in the absence of microwaves ($\alpha=0$). (From Ref.\cite{paper1})}
\end{center}
\end{figure}

In Fig.~\ref{CPR-temp1} we show the CPR for $\omega=0.6\Delta$, $\alpha=0.1$,
$\tau=0.95$ and two different temperatures. For comparison, we also show the 
results in the absence of microwaves (dashed lines). Apart from the dips, whose
origin has been already discussed in detail, one can observe that at certain value 
of the phase ($\varphi \approx 0.78\pi$) the current is suppressed. Notice 
that the suppression is larger as the temperature increases. Moreover, for phases
smaller than $\varphi$ the supercurrent exceeds its value in the absence of 
microwaves. In other words, for $\varphi<\varphi$ there is an enhancement of 
the supercurrent induced by the microwave field. The origin of this enhancement 
is the promotion of quasiparticles from the continuum below $-\Delta$ to the lower 
ABS by the microwave field. Of course, there is another contribution, which is
identical, coming from transitions connecting the upper state and the continuum
above $+\Delta$. These processes can only occur if the field frequency is larger 
than the distance in energy between the gap edges and the nearest ABS, i.e., 
if $ \omega > \Delta-E_\text{ A}(\varphi)$, and they become possible at finite 
temperature because the lower state is not fully occupied and the upper state
is not fully empty. For the parameters of Fig.~\ref{CPR-temp1} the previous 
condition is satisfied if $E_\text{ A}(\varphi)>0.4\Delta$, which corresponds to a phase 
$\varphi<0.78\pi$. Obviously, the phenomenon of microwave-enhanced supercurrent 
cannot be described by the two-level model since this models ignores the contribution 
of the continuum part of the spectrum.

%\subsection{Enhancement of the supercurrent}

In order to confirm this interpretation of the origin of the microwave-enhanced supercurrent, we have derived analytical results describing this phenomenon in the limit of weak microwave fields. We are interested in the correction to the current due to the microwave field which
is responsible for this variation. The perturbative analysis to lowest order in the parameter 
$\alpha$ shows that the supercurrent can be written as
\begin{equation}
I(\varphi) = I_\text{ eq}(\varphi) + \delta I(\varphi) , \label{anacurr}
\end{equation}
where $I_\text{ eq}$ is the equilibrium supercurrent given by Eq.~\eqref{Ieq}, and the correction $\delta I$ contains several contributions of order $\alpha^2$. The approach described in section\ref{sec:conti} happens to be very cumbersome in order to obtain analytical results. This is the reason why we develop an alternative approach to the problem. 

%Method of dressed GFs

The following results have been obtained with the help
of the method known as Hamiltonian approach, which for SQPCs has been 
shown to be equivalent to the microscopic theory described at the beginning of this 
section.\cite{Cuevas1996,Cuevas2001,Bergeret2005} In this approach, a point 
contact is described in terms of a tight-binding-like Hamiltonian and the transport
properties are calculated following a perturbative approach, where the coupling 
between the electrodes is treated as the perturbation. One of the advantages of this method is that it allows to obtain the density of states (DOS) at the contact. Moreover, a perturbative analysis (in the field) is
much simpler than using Eq.~\eqref{Idc}.

In this approach a single-channel SQPC can be described in terms of the following 
tight-binding-like Hamiltonian
\begin{equation}
\hat H = \hat H_L + \hat H_R + \sum_{\sigma} \left\{ t \hat c^\dagger_{L \sigma}
\hat c_{R \sigma} + t^{\ast} c^\dagger_{R \sigma} \hat c_{L \sigma} \right\},\label{tbh}
\end{equation}
where $\hat H_{L,R}$ are the BCS Hamiltonians describing the left (L) and 
right (R) electrodes and the last term describes the coupling between the
electrodes. In this last term, $t$ is a hopping element that determines
the transmission of the contact. 
 
In this model the current evaluated at the interface between the two electrodes
adopts the form
\begin{equation}
I(t) = ie \sum_{\sigma} \left\{ t \langle c^\dagger_{L \sigma}
\hat c_{R \sigma} \rangle - t^{\ast} \langle c^\dagger_{R \sigma} \hat c_{L \sigma}
\rangle \right\} .
\end{equation}
This expression can be rewritten in terms of the Keldysh Green functions as
follows
\begin{equation}
I(t) = e\text{Tr}\bigl [ \tau_3\bigl( 
\hat t \hat{G}_{RL}^K -\hat t^\dagger \hat{G}_{L R}^K\bigr)\bigr](t,t)
\;, \label{tbc}
\end{equation}
where $ \tau_3$ is the corresponding Pauli matrix, Tr denotes the trace
in Nambu space, and $\hat t$ is the hopping matrix in Nambu space given by
\begin{equation}
\label{hopping-matrix}
\hat t = \left( \begin{array}{cc}
t e^{i \phi(t)/2} & 0 \\
0 & -t^{\ast} e^{-i \phi(t)/2} \end{array} \right) .
\end{equation}
Here, $\phi(t)$ is the time-dependent superconducting phase given by 
Eq.~(\ref{eq-phi}).

In order to determine the Green functions appearing in the current expression,
we follow a perturbative scheme and treat the coupling term in Hamiltonian
(\ref{tbh}) as a perturbation. The unperturbed Green functions describe the 
uncoupled electrodes in equilibrium. Thus for instance, the retarded and
advanced functions are given by 
\begin{equation}
\hat g_{jj}^{R(A)}(E) = \frac{-i}{W} \frac{1}{\zeta^{R(A)}(E)} 
\begin{pmatrix} E & \Delta\\ \Delta & E
\end{pmatrix} \; ,
\end{equation}  
where $j=L,R$, $\zeta^{R(A)}=\sqrt{(E+i\eta)^2-\Delta^2}$, and $W$ is an energy
scale related to the normal density of state at the Fermi energy. The full
Green functions can then be determined by solving a Dyson equation, where 
the retarded and advanced self-energies are simply given by the hopping
matrix of Eq.~(\ref{hopping-matrix}).

\begin{figure}[t]
\begin{center}
\includegraphics[width=0.7\columnwidth,clip]{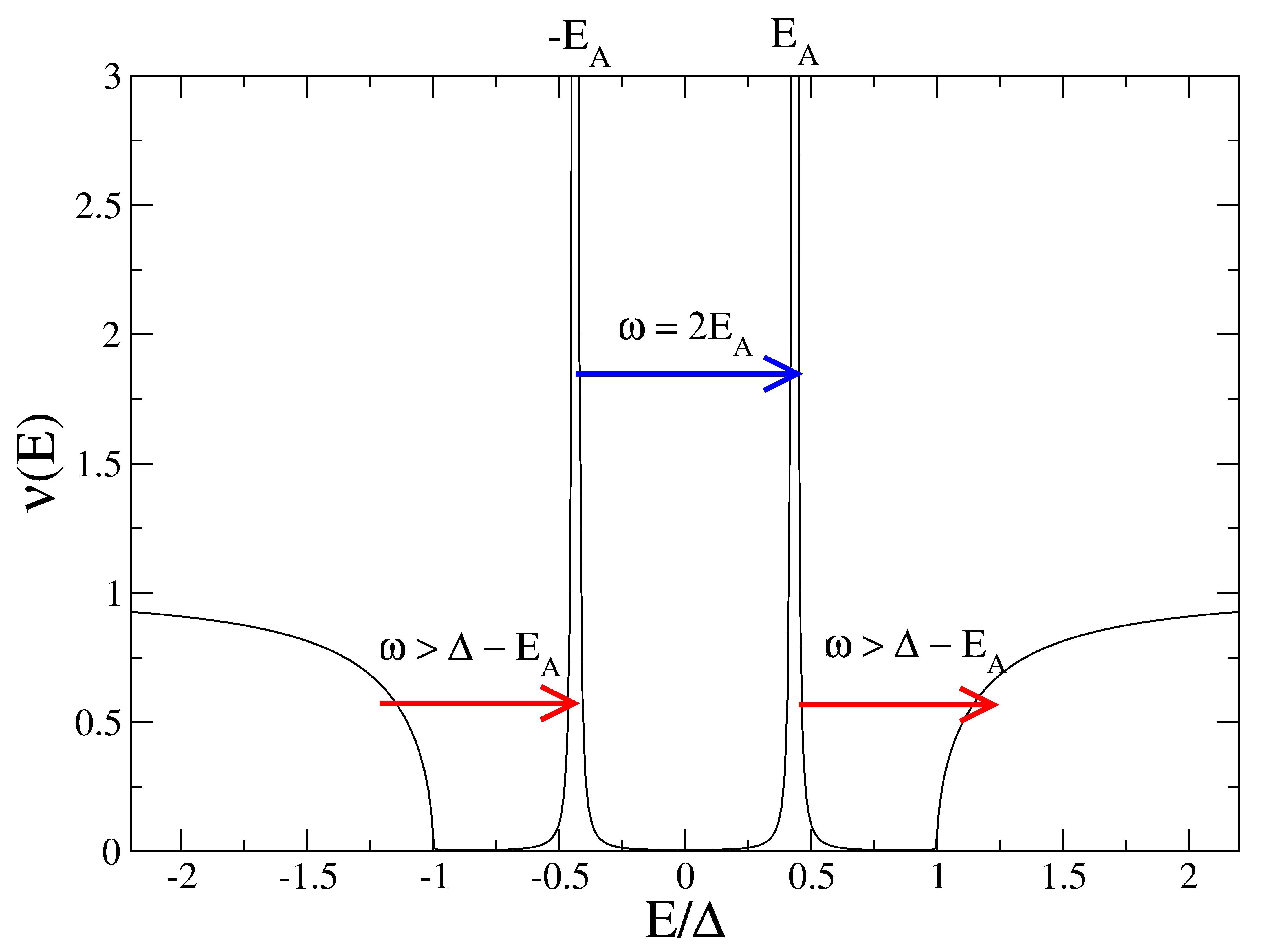}
\caption{\label{figdos}
 The local density of states at the contact in the absence of 
microwaves, as defined in Eq.~\eqref{tbdos}, as a function of energy for 
$\tau=0.95$, $\varphi=3\pi /4$ and $\eta=10^{-3}\Delta$. The lower 
arrows represent the microwave-induced transitions between the continuum
part of the spectrum and the Andreev bound states which are responsible for the 
supercurrent enhancement at finite temperatures. The upper arrow indicates the
resonant transition between the ABSs, which suppresses the supercurrent. (From Ref.\cite{paper1})}
\end{center}
\end{figure}

Since we are interested in the limit of weak fields ($\alpha \ll 1$), we
can expand the phase factors in Eq.~(\ref{hopping-matrix}) as follows
\begin{equation}
e^{i \phi(t)/2} \approx e^{i \varphi/2} \left( 1 + \alpha \cos \omega t +
\frac{1}{2} \alpha^2 (\cos \omega t)^2 + \cdots \right) .
\end{equation}
Moreover, for the perturbative treatment in $\alpha$ it is convenient to use
the full Green functions of the contact in the absence of microwaves ($\alpha=0$),
$\hat G_{ij}$. It is straightforward to show that these functions can be
expressed as 
\begin{equation}
\hat G_{LL}^{R(A)} = \frac{-i\zeta^{R(A)}}{W(1+\beta)\xi^{R(A)}}
\begin{pmatrix}E\pm i\eta & E_g^*\\E_g  &E \pm i\eta \end{pmatrix}\;\label{GLL}
\end{equation}
\begin{equation}
\hat G_{RL}^{R(A)} = \frac{-t}{W^2(1+\beta)\xi^{R(A)}}
\begin{pmatrix}a^{R(A)} & b^{R(A)}\\-b^{R(A)*} & -a^{R(A)*} \end{pmatrix}\;,
\end{equation}
where $E_g=\Delta(1+\beta e^{i\varphi})(1+\beta)$, $\beta=(t/W)^2$, $\xi=E^2-E_A^2$, 
$a=E^2e^{-i\varphi/2}-\Delta E_g^*e^{i\varphi/2}$ and $b=E(E_ge^{-i\varphi/2}-\Delta 
e^{i\varphi/2})$. Similar expressions hold for $G_{RR}$ and $G_{LR}$. These Green 
functions are now the zero-order propagators of the perturbation theory. It is
important to emphasize that by substituting these functions in the current 
expression of Eq.~(\ref{tbc}) and identifying the transmission coefficient as 
$\tau=4\beta/(1+\beta)^2$\cite{Cuevas1996}, one obtains the expression 
for the equilibrium current of Eq.~(\ref{Ieq}). On the other hand, from the
previous expressions one can determine the local density of states at the
contact and in the absence of microwaves, which is defined as $\nu_j(E) = 
W(1+\beta)(i/2)(\hat G_{jj}^R-\hat G_{jj}^A)_{11}$ ($\nu_L = \nu_R$ in the
symmetric contacts). This density of states is given by Eq.~(\ref{tbdos})
and it is shown in Fig.~\ref{figdos}.

Going into the energy representation, the 
first correction to the current of Eq.~(\ref{tbc}), which is of order $\alpha^2$, 
contains the following three terms 
\begin{equation}
\delta I = e \int \frac{dE}{2\pi}\text{ Tr}\bigl[\hat t^{(2)}\hat{ G}_{RL}^{(0)} +
\hat t^{(0)}\hat{ G}_{RL}^{(2)}+\hat t^{(1)}\hat{ G}_{RL}^{(1)}\bigr]- 
L\leftrightarrow R  \label{pert_c}\,,
\end{equation}
where the superindices denote the order of perturbation in $\alpha$. To obtain a 
complete analytical expression for an arbitrary value of the field frequency is 
quite cumbersome. Instead, we concentrate on the parameter range where the current
enhancement takes place. For that purpose, we focus on frequency values far from
the resonant condition $ \omega=2E_\text{ A}$ and close to $\Delta-E_\text{ A}$.
In this region, it turns out that the second term in Eq.~(\ref{pert_c}) is 
proportional to the parameter $\Delta/\eta$. All the other terms give a contribution 
which depends only weakly on the frequency. Assuming a small inelastic scattering 
rate, one can approximate the correction to the current by Eqs.~(\ref{eq-deltaI}) 
and (\ref{eq-deltan}), where $\rho_j=(i/2)W(1+\beta) [\hat G_j^R
-\hat G_j^A]_{1,2}$, $\tilde\rho_j=(i/2)W(1+\beta)[\hat G_j^R-\hat G_j^A]_{2,1}$ , 
and $F_n=F(E+n\omega)$. 

This correction to the current can be written in 
the spirit of Eq.~\eqref{Ieq} as follows
\begin{equation}
\label{eq-deltaI}
\delta I_\text{ enh}(\varphi) = I^-_\text{ A}(\varphi) \delta n^-(\varphi) 
+ I^+_\text{ A}(\varphi) \delta n^+(\varphi) ,
\end{equation}
where let us remind that $I^{\pm}_\text{ A} (\varphi) = 2e \partial 
E^{\pm}_\text{ A}/ \partial \varphi$ give the contribution of the states to the
equilibrium supercurrent, and $\delta n^{\pm}(\varphi)$ are the corrections to 
the occupations of the ABSs due to application of the microwave field. These 
corrections can be written as
\begin{eqnarray}
\label{eq-deltan}
\delta n^{\pm}(\varphi) & = & \frac{\alpha^2 \tau}{8} \left[ \text{ Re} \left\{ e^{i\varphi} 
\rho_L(E^{\pm}_\text{ A}) \tilde{\rho}_R(E^{\pm}_\text{ A} \pm \omega) \right\} + \right.
\nonumber \\
& & \left. \nu(E^{\pm}_\text{ A}) \nu(E^{\pm}_\text{ A} \pm \omega) \right] [F_0- F_{\pm 1}] .
\end{eqnarray}
Here, $F_n$ is the distribution function with shifted arguments $F_n = 
\tanh[(E+n\omega)/2k_\text{ B}T]$, $\nu (E)$ is the density of states at the 
contact in the absence of microwaves, and $\rho_j$ and $\tilde\rho_j$ are the 
real part of the anomalous Green functions on the left (L) and (R) side o the 
interface ($j=L,R$) without field. 
Eq.~(\ref{eq-deltan}) has a very appealing form and it tells us that 
the occupations of the ABSs can be changed by microwave-induced transitions 
connecting these states between the continua below and above the gap. These
transitions are illustrated in Fig.~\ref{figdos}, where we also present an
example of the density of states of the contact in the absence of microwaves,
$\nu (E)$. This density of states is given by 
\begin{equation}
\nu(E) = \text{ Re} \left\{ \frac{E \sqrt{(E+i\eta)^2-\Delta^2}}
{(E+i\eta)^2-E_\text{ A}^2} \right\}\; ,\label{tbdos}
\end{equation}
where the poles correspond to the ABSs and, as one can see in Fig.~\ref{figdos},
there is no longer singularities at the gap edges $E=\pm \Delta$.

\begin{figure}[t]
\begin{center}
\includegraphics[width=0.75\columnwidth,clip]{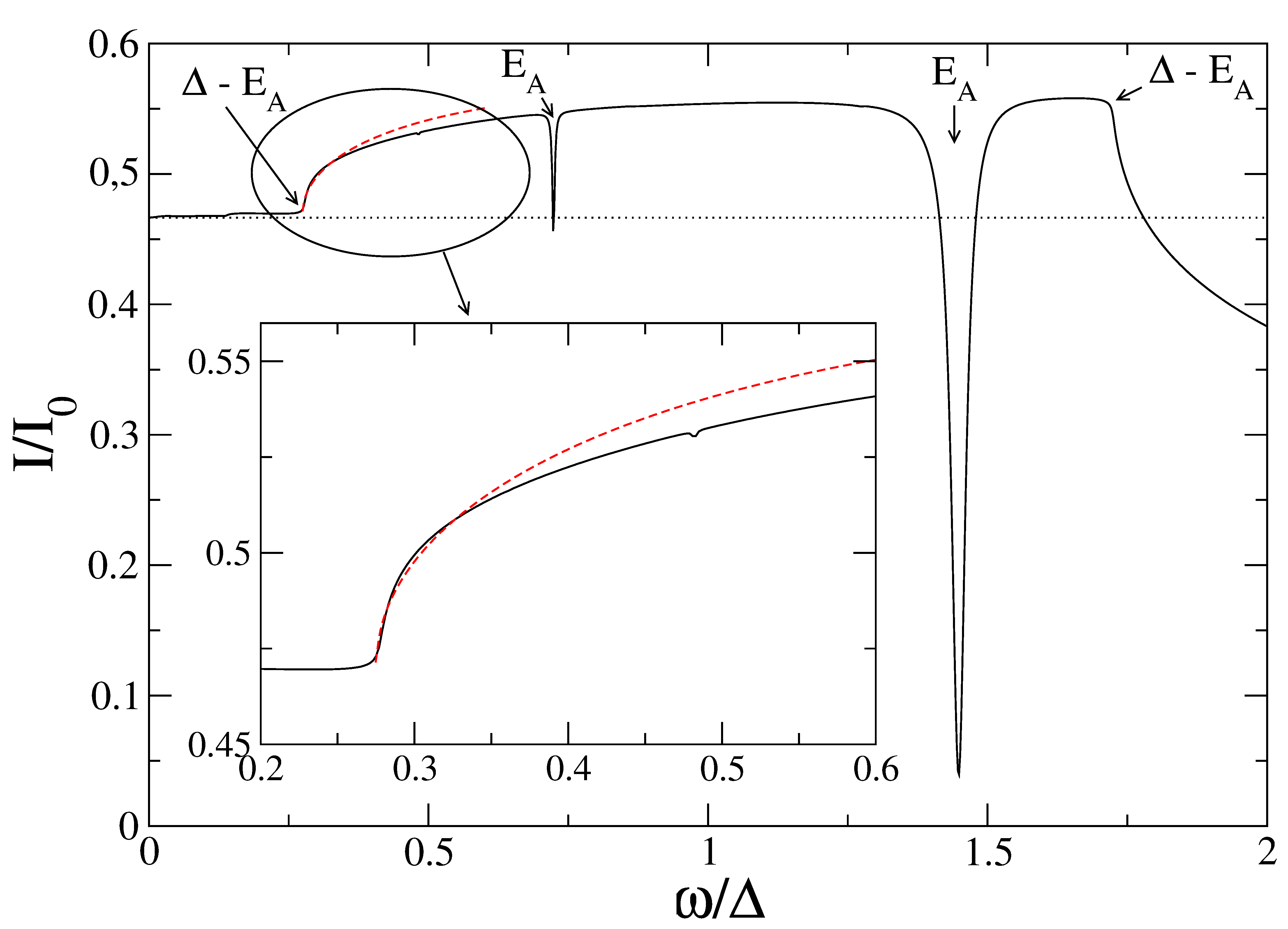}
\caption{\label{figwdp}
 The dc Josephson current as a function of the frequency $\omega$ 
of the microwave field for a fixed value of the phase $\varphi=\pi/2$, and 
$\alpha=0.1$, $\tau=0.95$ and $T=0.4\Delta$. (From Ref.\cite{paper1})}
\end{center}
\end{figure}

From Eq.~(\ref{eq-deltan}) one can show that the transitions between the 
continuum of states below $-\Delta$ and the lower ABS increases the population
of this state ($\delta n^{-} > 0$), while the photon processes connecting
the continuum above $+\Delta$ and the upper ABS decrease the
occupation of this state ($\delta n^{+} < 0$). As one can see from 
Eq.~(\ref{eq-deltaI}), both type of processes give a positive contribution to
the current at finite temperatures and thus, they are responsible for the 
supercurrent enhancement. Indeed, due to the electron-hole symmetry of this
problem, terms in Eq.~(\ref{eq-deltaI}) give the same contribution to the 
current. Finally, the correction to the current due to these microwave-induced 
transitions involving the continuum can be written as
\begin{eqnarray}
\delta I_\text{ enh}(\varphi) &  = & \alpha^2 \left( 2e E^{\prime}_\text{ A}
\right) \frac{\tau}{16} \times \label{corrcurr} \\
& & \frac{\sqrt{(E_\text{ A}+\omega)^2 - \Delta^2} \sqrt{\Delta^2-E_\text{ A}^2}} 
{\eta  \omega E_\text{ A}(2E_\text{ A}+\omega) } \times \nonumber \\
& & \hspace*{-1.6cm} \left[ E_\text{ A} \omega + \Delta^2 ( 1+ \cos \varphi) \right]  
\left[ F_1- F_0 \right] \Theta\left(|E_\text{ A}+\omega |-\Delta\right).
\nonumber
\end{eqnarray}
This expression gives a positive contribution to the supercurrent and it explicitly 
shows that the enhancement can only take place when $ \omega > \Delta-E_\text{ A}$. This
analytical result reproduces the exact results obtained with the microscopic approach
in the limit of weak fields and in the range of frequencies where the transitions
between the ABSs cannot take place. This is illustrated in Fig.~\ref{figwdp} where we 
show the supercurrent for a fixed value of the phase ($\varphi = \pi /2$) as 
a function of the frequency for $\tau = 0.95$, $\alpha = 0.1$ and $k_\text{ B}T = 
0.4\Delta$. As one can see, the exact result (solid line) remains constant for small
frequencies. Then, at $ \omega = \Delta - E_\text{ A}$ there is rise of the 
supercurrent due to the onset of the transitions connecting the ABSs with the 
continuum of states. This increase of supercurrent is well described by the analytical
result of Eq.~(\ref{corrcurr}) (dashed line). At higher frequencies, one can 
observe the dips due to the transitions between the ABSs. The dip at $ \omega = 
E_\text{ A}$ corresponds to a two-photon process, while the one at $ \omega =
2E_\text{ A}$ is produced by a single-photon process. Finally, at $ \omega =
\Delta + E_\text{ A}$ the supercurrent starts to decrease due to microwave-induced
transitions between the continuum below $-\Delta$ and the upper ABS and similar
ones between the continuum above $+\Delta$ and the lower ABS. These transitions,
which can also occur at zero temperature, tend to increase the occupation of the
upper state and to reduce the population of the lower one, which results in a
reduction of the net supercurrent.

As one can see in Fig.~\ref{CPR-temp1} and ~\ref{figwdp}, the maximum supercurrent
sustained by the junction, i.e.\ the critical current, can also be enhanced by the 
microwave field at finite temperatures. A microwave-enhanced critical current was
first reported in experiments on superconducting microbridges\cite{Wyatt1966,Dayem1967}
and explained by Eliashberg\cite{Eliashberg1970} in 1970 in terms of the stimulation 
of the superconductivity in the electrodes, which were made of thin films. Such
a stimulation, and the corresponding microwave-enhanced critical current, only 
occur at temperatures very close to the critical temperature. Enhancements at
much lower temperatures were reported in the 1970's in the context of SNS 
structures\cite{Notarys1973,Warlaumont1979}, and they have been recently explained
in terms of the redistribution of the quasiparticles induced by the field.\cite{Virtanen2010} 
In this case, for the enhancement to occur, the temperature must be of the order
of the minigap in the normal wire, which can be much lower than the critical 
temperature of the superconducting leads. 

\begin{figure}[!h]
\begin{center}
\includegraphics[width=0.6\columnwidth,clip]{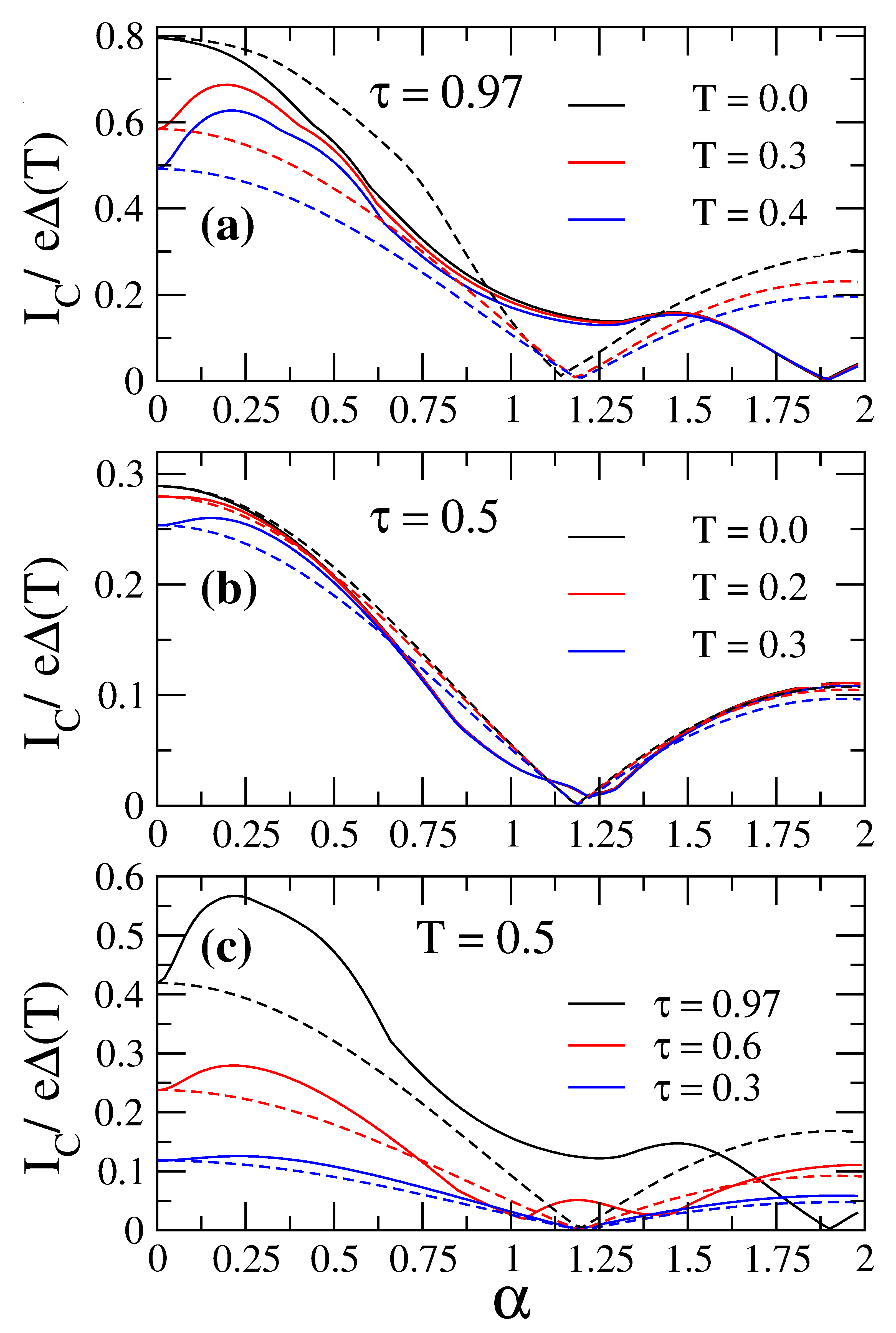}
\caption{\label{Icritical}
 The critical current as a function of $\alpha$ for $ \omega =
0.6\Delta$. The different curves correspond to different values of the temperature
and the transmission as indicated in the graphs. The solid lines correspond to
the exact results, while the dashed lines show the results of the adiabatic 
approximation. In the three panels the critical current has been normalized by
$e\Delta(T)$, where $\Delta(T)$ is the gap at the corresponding temperature. (From Ref.\cite{paper1})}
\end{center}
\end{figure}

As discussed above, in the case of a point contact the mechanism is similar to
that of diffusive SNS structures,\cite{Virtanen2010} but it involves discrete 
ABSs, rather than a continuous band of ABSs, as in the case of 
diffusive proximity structures. For this reason, the enhancement of the critical
current in SQPCs is expected to occur at intermediate temperatures, when $k_\text{ B}T$
is of the order of the energy distance between the ABSs and the gap edges ($\Delta
- E_\text{ A}(\varphi_\text{ max})$), where $\varphi_\text{ max}$ is the phase value 
at which the supercurrent reaches its maximum. This is illustrated in 
Fig.~\ref{Icritical}, where we show the critical current as a function of $\alpha$
for different temperatures and several values of the transmission. Thus for instance,
panel (a) shows the critical current for a highly transmissive channel ($\tau = 
0.97$) and three values of the temperature. Notice first that at finite temperatures,
the critical current at finite $\alpha$ ($\alpha \lesssim 0.5$) exceeds the value
in the absence of microwaves ($\alpha=0$). Notice also that as $\alpha$ increases,
the critical currents clearly deviate from the behaviour described by the adiabatic
approximation, which is shown here as dashed lines.
It is also important to emphasize that the microwave-enhancement of the critical current 
is not exclusive of high conductive channels and it persists up to relatively low
transmissions, as we show in Fig.~\ref{Icritical}(b-c). The relative enhancement
of the critical current is larger the larger is the temperature.
It is also worth remarking that at sufficiently
high power, the critical current depends only weakly on
the temperature.

\newpage

\section{Summary}

In this chapter we have performed a theoretical analysis of the supercurrent in a phase-biased SQPC in a microwave field. It was shown that the transport properties of the structure depend strongly on the value of the microwave frequency $\omega$. If this is not high enough to induce transitions between the ABSs or between the ABSs and the continuum of states outside the gap. The standard adiabatic approximation gives a proper description of the transport phenomena. Nevertheless, for the case when $ \omega$ is comparable to the Andreev gap (energy distance between the ABSs), quasiparticle transitions between the ABSs can occur. This can lead to a large suppression of the supercurrent at the corresponding values of the phase difference. It was shown in these calculations that this phenomenon can be nicely explained within a two-level model that describes the dynamics of the ABSs.\cite{Zazunov2003} This model indicates that the enhancement of the occupation of the uppper ABS by resonant transitions form the lower state is what causes the suppression of the supercurrent. Furthermore, at low temperatures and weak fields, this model is quantitatively correct if the following conditions are fulfilled: (i) the microwave frequency is not high enough to induce transitions connecting the ABSs and th continuum of states, and (ii) the Andreev gap is large compared to the broadening acquired by the ABSs by means of the coupling to the electromagnetic field. We conclude by showing that when the microwave transitions between the ABSs and the continuum of states are activated (due to finite temperatures, high frequencies or high radiation powers), the two-level Hamiltonian is not enough and a full microscopic theory is required to describe the supercurrent. This theory was developed in the calculations and we predicted the following effects. In the finite temperature case it is possible to enhance both the supercurrent and the critical current by the application of a microwave field. This effect is caused by the quasiparticle transitions between the ABSs and the continuum of states, which increase the occupation of the lower Andreev state, reducing the population of the upper one. In the case of high powers, the current-phase relation is strongly distorted. It can become highly non-sinusoidal exhibiting sign changes in the region between $0$ and $\pi$. Finally, the critical current can exhibit large deviations from the standard adiabatic approximation described by a Bessel-function behaviour, when plotted as a function of the radiation power. We have to mention a related work\cite{Bergeret2010} where the theory of the supercurrent through a microwave-irradiated SQPC in the framework of the Keldysh-Green function technique is developed.

%A complete solution of this problem, valid for arbitrary range of parameters, has only been reported very recently.\cite{Bergeret2010} In this latter work, a theory of the supercurrent through a microwave-irradiated SQPC in the framework of the Keldysh-Green function technique is developed. This theory allowed to put forward new predictions like the evolution of the CPR with the radiation power or the the possibility to enhance the critical current at finite temperatures by irradiating the junction.

It is interesting to understand the connection between the results and the experiments. In order to explain most of the experimental results related to the effect of microwaves on the supercurrent of a point contact, the adiabatic approximation is enough. The reason for this is the typical frequency used in experiments. This is relatively low ($ \omega \ll \Delta$) and does not allow any transitions between the ABSs. Nevertheless, there are no fundamental limitations to study the parameter regime where we predict the occurrence of novel effects presented in this chapter. Such as, supecurrent dips in the current-phase or the microwave-enhances critical current. Note that this effects are easier to observe the smaller the Andreev gap becomes(much smaller than $\Delta$). That is the reason to use highly transmissive point contacts.  The ideal experimental system  where to test these predictions is a superconducting atomic contact for many reasons. First, these contacts are designed to sustain a reduced number of channels, which facilitates the comparison with the theory. Second, it is possible to determine independently the set of transmission eigenvalues $\{\tau_i \}$,\cite{Scheer1997} which has allowed to establish a more direct comparison between  theory and experiment for many different transport  properties.\cite{Cron2001,Chauvin2006,Rocca2007}. Third, it is possible to tune,  at least to a certain extent, the transmission coefficients and, in particular, to achieve very transmission coefficients, as demonstrated in the context of Al atomic contacts.\cite{Scheer1997,Chauvin2006,Rocca2007} Lastly, the current-phase relation in this systems is amenable to measurements,\cite{Rocca2007}. Measurements of the transport properties of superconducting atomic contacts under microwave irradiation have already been performed.\cite{Chauvin2006,note-exp}

\begin{figure}[h]
\begin{center}
\includegraphics[width=0.5\columnwidth,clip]{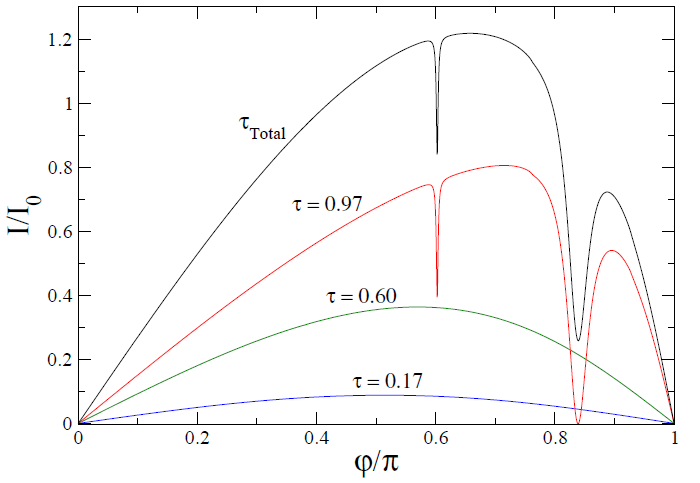}
\caption{\label{fig:3chan}
 The zero-temperature current-phase
relation for a point-contact consisting of three channels with
transmissions $\tau=0.17$,$0.6$,$0.97$, for $\alpha=0.1$ and $ \omega=0.6 \Delta$.
The dashed lines show the contribution of each channel, while
the solid line corresponds to the total current.. (From Ref.\cite{paper1})}
\end{center}
\end{figure}

In a experiment in superconducting atomic junctions, even at the level of a 
single-atom contact, one often has the contribution of several conduction
channels. It may happen that the presence of low transmission channels can hide the results presented in this chapter. In Fig. \ref{fig:3chan} we show the CPR for a junction with three conducting channels with transmissions $\tau = 0.17, 0.6, 0.97$. The total current that flows through the junction is obtained by adding the contribution of each channel, according to Eq. \ref{matrixJ}. The result is shown in  Fig. \ref{fig:3chan}. Here we still observe the dips at the resonances corresponding to the channel with the highest transmission ($\tau=0.97$). But the current does not vanish completely at the dips due to the contribution of the other low transmissive channels.

Note that the major problem when comparing the theory presented in this chapter and the experiments is that we have assumed a phase-biased junction. The fact is that in reality the phase across the junction may undergo fluctuations (both classical and quantum), which can affect the value of the critical current or the shape of the current-phase relation. These depend on the details of the electromagnetic environment seen by the point contact. Hence, in order to compare quantitatively the theory with the experiments. We may require in some cases to combine the results with a description of the phase fluctuations. For classical fluctuations, this can be done as in Ref.~\cite{Chauvin2007}. Here an extension of the resistively shunted 
junction using the microscopic current-phase relation as a starting point is used.

Let us conclude by saying that in this chapter we have shown that the application of 
microwaves to one of the simplest superconducting systems, namely a SQPC, leads to 
a very rich phenomenology, which has remained largely unexplored. Specifically, we have shown the role of a microwave field as an ideal tool to make direct spectroscopy of the Andreev bound states of a superconducting junction. The ideas presented in this chapter, open the way to understand the effect of a microwave field on the supercurrent of highly transmissive superconducting weak links.

%%%%%%%%%%%%%%%%%%%%%%%%%%%%%%%%%%%%
\bibliographystyle{unsrtnat}
\renewcommand{\bibname}{Bibliography of Chapter 5} % changes default name Bibliography to References

%\end{document}

%% file: conc_SB.tex
This thesis deals with the study of charge and heat transport in superconducting nanoscopic hybrid structures, which  consist of conventional (s-wave) superconductors in contact with normal metal, ferromagnets or insulators. We have presented  a number of novel transport  phenomena and proposed new devices as building blocks for nanocoolers and thermoelectric elements.  

In the first chapter,  we have presented a  brief introduction of conventional superconductivity and the proximity effect between superconductors and non-superconducting materials.  Being in electric contact,  the two materials influence each other and,  if the structure is in equilibrium,  their spectrum is modified  on the scale of the coherence length away from the interface. The proximity effect, resulting from the fundamental process of the Andreev reflection, is also introduced in the first chapter. 

In chapter 2, we present the  main theoretical  tool used in this thesis:  the quasiclassical Keldysh Green function formalism. All central quantities, such as the charge and heat currents as well as the critical temperature, are introduced in terms of the quasiclassical Greens functions. As an example of how this formalism can be used, we  calculate the critical temperature of a  $FSF$  trilayer  with spin-mixing  interfaces and study the influence of the triplet pairing on the critical temperature. This is a relevant issue in the context of the emerging technology of spintronics with superconductors, which aims to overcome large heat losses. This structure provides evidence of a spin selectivity of the ferromagnet to the spin of the triplet Cooper pair. In a second example we use  the quasiclassical formalism for  studying the Hanle effect in a typical spin-valve structure.  From the microscopic model we are able to derive the results of phenomenological formulations of the precession of the spin-accumulation  in normal metals. 

In chapter 3,  we analyse  the subgap transport in an $SIF$ structure. We demonstrate that, in contrast with the common knowledge,  the Andreev current for some critical value of the voltage is enhanced by the exchange field. This unexpected behaviour can be explained as the competition between two-particle tunnelling processes and decoherence mechanisms originating from the temperature, voltage, and exchange field.  We also demonstrate that if the ferromagnet consists of two domains  the same features occur  at the value of the exchange field corresponding to the average one. This results are relevant for the purpose of building accurate exchange field detectors.

In chapter 4, we study the thermal transport and its coupling to the  charge transport in different nanostructures. We begin analysing two hybrid cooling devices at the nanoscale : an $SIF_1F_2N$ structure with an arbitrary magnetization difference between $F_1$ and $F_2$  and an $S/I_{SF}/N$ structure with a spin filter interlayer.  For the first system, in the weak ferromagnet case, the cooling power shows a non-monotonic dependence on $h$. This is related to the maximum in the Andreev current discussed in chapter 3. We have also shown that different orientations of magnetizations can shift the minimum of cooling power towards larger values of the exchange field. Furthermore, the cooling power has a nontrivial dependence on the magnetization direction. For the $S/I_{SF}/N$  system, we propose a novel way of  refrigeration based on spin-filter barriers. This device can suppress Joule heating due to Andreev reflections very effectively, leading to cooling power values higher than those of $NIS$ coolers. This high efficiency cooling is demonstrated for both ballistic and diffusive multi-channel junctions. The proposed junctions are feasible to make, so the derived results have a wide range of applications, ranging from detectors to qubits.

Also in chapter 4, we address  thermoelectric effects in hybrid superconducting structures. We demonstrate an unexpectedly large thermoelectric effect in the superconductor, provided that its density of states is  spin-splitted.  This can be achieved either by applying an external magnetic field or placing a ferromagnetic insulator in contact with the superconductor. Additionally, a spin polarized current needs to be injected. If the spin polarization is close to 100\%,  the thermopower the heat engine efficiency are close to the Carnot limit. We also demonstrate that  the spin-splitting of the density of states  can generate a finite spin current if a temperature gradient is present in the system. Our work suggests to build a very precise heat detector. 

%Such as the recently proposed interferometer that enables phase-coherent manipulation of thermoelectric properties with efficiencies that approach the Carnot limit.

In chapter 5 we present an exhaustive  theory for the microwave irradiated superconducting quantum point contact(SQPC). Transport properties in this system depend strongly on the frequency of the microwaves,$\omega$, and the dynamics of the Andreev bound states(ABSs). We demonstrate  that the supercurrent through the SQPC can be greatly suppressed if $\omega$ is comparable to the Andreev gap (energy distance between Andreev bound states). In a first step,  we describe this phenomenon using an effective two-level model. When the microwave transitions between ABSs and the continuum of states are activated,  the full microscopic Keldysh formalism is required to describe the supercurrent. With its help we demonstrate that at finite temperature it is possible to enhance the supercurrent by  applying a microwave field.  We derive an analytical expression for the current that generalizes the result presented several years ago by Eliashberg for the description of microwave-enhanced critical current. 

This thesis does not only present a number of new predictions and phenomena, but also opens new directions of research.  We plan to investigate nonequilibirum properties of superconducting hybrid junctions with non-collinear magnetizations and long-range triplet correlations. 

Future directions of our work relate to the study of superconductors with a spin-splitted density of states in contact with spin filters barriers with arbitrary direction of magnetization. The existing spin transport theory for diffusive superconductors is limited to the study of conventional superconductors and collinear configurations of  spin injectors. We plan to go beyond that theory and study the completely unexplored case of non-collinear ferromagnets. Effects, such as the Hanle effect and the coupling between spin and heat transport, have to be re-addressed in the presence of superconudcting correlations. These studies will further help to set up the basis for the development of spintronics, caloritronics, and spin caloritronics devices using superconductors.  

%One of the  ideas is to study  superconductors with a spin-splitted density of states in contact with spin filters barriers with arbitrary direction of magnetization. The existing spin transport theory in diffusive superconductors has been developed for conventional superconductors and collinear configurations of the spin-injectors. The non-collinear case is completely unexplored and effects like the Hanle effect or the coupling between spin and heat transport in such structures has to be reconsidered. From such study on expect to provide important predictions for the development of spintronics, caloritronics and spin-caloritronics devices. 

%% file: app2.tex
\section{The quasiclassical approximation}
\label{sec:quasiap}

In this Appendix we present a brief derivation of equations for the quasiclassical Green functions.  We start with the general Hamiltonian describing hybrid systems: 
\begin{equation}
\hat{H}=\hat{H}_{BCS}+\hat{H}_{imp}+\hat{H}_{so}+\hat{H}_{sf}+\hat{H}_{ex} \; .
\label{eq:hamilap}
\end{equation}
The $\hat{H}_{BCS}$ term describes superconducting materials. However, for the $\Delta=0$ case it also describes normal metals. It reads,
\begin{equation}
\hat{H}_{BCS}= \Sigma_{\{ p,s \}} a_{sp}^\dagger \left( \xi_p \delta_{pp^\prime} + eV  \right) \delta_{ss^\prime} a_{s^\prime p^\prime}- \Delta \left( a^\dagger_{\bar{sp}} a^\dagger_{s^\prime p^\prime} + c.c. \right) \; .
\end{equation} 
The summation is carried out over all momenta $(p,p^\prime)$ and spins $(s,s^\prime)$ (the notation $\bar{s}$, $\bar{p}$ means inversion of both spin and momentum), $\xi_p=p^2/2m-E_F$ is the kinetic energy counted from the Fermi energy $\epsilon_F$, $V$ is a smoothly varying electric potential. The superconducting order parameter $\Delta$ must be determined self-consistently. 

The term $\hat{H}_{imp}$ describes the interaction of the electrons with nonmagnetic impurities. It reads,
\begin{equation}
\hat{H}_{imp}= \Sigma_{\{ p,s \}} a_{sp}^\dagger U_{\text{imp}} \delta_{ss^\prime} a_{s^\prime p^\prime} \; .
\end{equation}
The potential $U_{\text{imp}}=U(p-p^\prime)$ describes the interaction of the electrons with nonmagnetic impurities. 

The term $\hat{H}_{so}$ that describes the possible spin orbit interaction reads,
\begin{equation}
\hat{H}_{s.o.}= \Sigma_{\{ p,s \}} a_{sp}^\dagger U_{\text{s.o.}} a_{s^\prime p^\prime} \; .
\end{equation}
The potential $U_{\text{s.o.}}$ is described by\cite{abrigorkov},
\begin{equation}
U_{\text{s.o.}}=\Sigma_i \frac{u_{s.o.}^{(i)}}{p_F^2} (\textbf{p} \times \textbf{p}^\prime) \textbf{S}
\end{equation}
Here the summation is performed over all impurities. $\textbf{p}$ is the momentum vector and  $\textbf{S}=(\sigma_1,\sigma_2,\sigma_3)$ the matrix vector. The term $\hat{H}_{s.f.}$ is similar and can be derived from this expression. 

 In order to describe the ferromagnetic region we use a simplified model. This effect is caused by the electron-electron interaction between electrons belonging to different bands that can correspond to localized and conducting states. Only the latter participate in the proximity effect. If the contribution of free electrons strongly dominates, the exchange energy is caused mainly by free electrons. Using the value of the exchange field $h$, the term describing magnetic fields reads,
\begin{equation}
\hat{H}_{ex}= -\Sigma_{\{ p,s \}} a_{sp}^\dagger \textbf{h} \textbf{S} a_{s^\prime p^\prime} \; .
\end{equation}
Here we define $\textbf{h}=(h_x,h_y,h_z)$, where $h_i$ is the value of the exchange field in the "i" direction.

Starting from the Hamiltonian eq.\ref{eq:hamilap}, one can derive the Eilenberger and Usadel equations. Initially these equations have been derived for $2 \times 2$ matrix Green functions $g_{n,n^\prime}$, where indices $n,n^\prime$ relate to normal $(g_{11},g_{22})$ and anomalous or condensate $(f_{12},f_{21})$ Green functions. These functions describe the singlet component. In the case of a non-homogeneous magnetization considered in this thesis one has to introduce additional Green functions depending on spins describe not only the singlet but also the triplet component. These matrices depend  not only on $n.n^\prime$ indices but also on the spin indices $s,s^\prime$, and are $4\times 4$ matrices in the spin and Gorkov space (usually the $n, n^\prime$) space is called the Nambu or Nambu-Gorkov space). 

In order to define the Green functions in a customary way it is convenient to write the Hamiltonian eq.\ref{eq:hamilap} in terms of new operators $c^\dagger_{nsp}$ and $c_{nsp}$ that are related to the creation and annihilation operators $a^\dagger_s$ and $a_s$ by the relation (we drop the index $p$ related to momentum)
\begin{equation}
(n=1) \quad c_{1s}=a_s \quad (n=2) \quad c_{2s}=a^\dagger_{\bar{s}} \; .
\end{equation}
These operators (for $s=1$) were introduced by Nambu\cite{nambuop}. The new operators allow one to express the anomalous averages $\langle a_\uparrow a_\uparrow \rangle$ introduced by Gorkov as the conventional averages $\langle c_1 c_2^\dagger \rangle$ and therefore the theory of superconductivity can be constructed by analogy with a theory of normal systems. Thus, the index $n$ operates in the particle-hole (Nambu-Gorkov) space, while the index $s$ operates in the spin space. In terms of the $c_{ns}$ operators the Hamiltonian can be written in the form
\begin{equation}
H=\Sigma_{\{p,n,s \}} c_{ns}^\dagger \mathcal{H}_{(n n^\prime)(s s^\prime)}c_{n^\prime s^\prime} \; ,
\end{equation}
where the summation is performed over all momenta, particle-hole and spin indices. The matrix $\check{\mathcal{H}}$ is given by
\begin{equation}
\check{\mathcal{H}}=\frac{1}{2} \left( \left[ (\xi_p \delta_{pp^\prime}+eV)+U_{imp}\right]\tau_3 \otimes \sigma_0+\tilde{\hat{\Delta}} \otimes \sigma_3- \textbf{h} \tau_3 \textbf{S} + \Sigma_i \frac{u_{s.o.}^{(i)}}{p_F^2} (\textbf{p} \times \textbf{p}^\prime) \textbf{S} \right) \; . 
\end{equation}
The matrices $\tau_i$ and $\sigma_i$ are the Pauli matrices in the particle-hole and spin space respectively; $i=0,1,2,3$. The matrix order parameter equals $\tilde{\hat{\Delta}}=\tau_1 \text{Re}\Delta-\tau_2 \text{Im}\Delta$. 

 We introduce the time ordered matrix Green functions (in the particle-hole$\times$spin space) in the Keldysh representation,
\begin{equation}
\check{\textbf{G}}(t_i,t_k^\prime)=\frac{1}{i} \langle T_C \left( c_{ns}(t_i) c_{n^\prime s^\prime}^\dagger(t^\prime_k)\right)\rangle \; .
\label{eq:green}
\end{equation}
In the Keldysh representation the time coordinates have subindices ($k,j$) that take the values $+$ and $-$. These correspond to the upper and lower branches of the contour C, running from $-\infty$ to $\infty$ and back to $-\infty$.

In the Keldysh space $G$ is a $2\times 2$ matrix. The four elements of this matrix are not independent and can be reduced to three. Thus, in the Keldysh space the matrix  Green functions read,
\begin{equation}\label{kelmatap}
\breve{\textbf{G}} = \begin{pmatrix} \check{\textbf{G}}^R & \check{\textbf{G}}^K \\
0 & \check{\textbf{G}}^A
\end{pmatrix} \; .
\end{equation}
The upper scripts stand for Retarded, Advanced and Keldysh components. The Retarded and Advanced components contain information about the spectral properties of the system, such as the density of states (DoS). Whereas the Keldysh component describes how the states are occupied, \textit{i.e.} it contains the distribution function.  

The Dyson equation\cite{5ap} it is a perturbative expansion up to first order but can be expanded to higher orders. In the Keldysh space after the unitary transformation reads, for the Retarded and Advanced Green functions,
\begin{equation}
\textbf{ G}^{R,A}=\textbf{G}_0^{R,A}+ \textbf{G}_0^{R,A}  \Sigma^{R,A} \textbf{ G}^{R,A} \; .
\label{eq:dysonap}
\end{equation}
Here $\Sigma$ is the self-energy, $\textbf{G}_0$ the unperturbed GF and $\textbf{G}$ the perturbed one. It will be shown that this equation is only valid for the homogeneous case and that in order to study hybrid systems a Boltzman type of equation is required. In this case we define the self energy as (omitting the R subscript),
\begin{equation}
\Sigma=-\tilde{\hat{\Delta}}\otimes \sigma_3+\textbf{h}\tau_3\hat{\textbf{S}}-\check{\Sigma}_{\textit{imp}}-\check{\Sigma}_{\textit{so}}-\check{\Sigma}_{\textit{sf}} \; .
\end{equation}
We can solve eq.\ref{eq:dysonap} for the perturbed GFs $\textbf{G}$ and we obtain the so-called Gorkov equation\cite{gorkoveq}, which in coordinate space reads:
\begin{equation}
(\check{\textbf{G}}_0^{-1}-\tilde{\hat{\Delta}}\otimes \sigma_3+\textbf{h}\tau_3\hat{\textbf{S}}-\check{\Sigma}_{\textit{imp}}-\check{\Sigma}_{\textit{so}}-\check{\Sigma}_{\textit{sf}}) (r_1,t_1,r_2,t_2) \otimes \check{\textbf{G}}(r_2,t_2,r_1^\prime,t_1^\prime)=\delta(r_1,t_1,r_1^\prime,t_1^\prime)
 \label{eq:gorkovap}
\end{equation}
Here $\otimes$ represents convolution over coordinates, while $\check{\Sigma}_{\textit{imp}}$, $\check{\Sigma}_{\textit{s.o.}}$ and $\check{\Sigma}_{\textit{s.f.}}$ are the self-energies. In principle, this equation is valid for any self-energy but in this case these are given in the Born approximation by
\begin{eqnarray}
\check{\Sigma}_{\textit{imp}}=N_{\textit{imp}} u_{\textit{imp}}^2 \tau_3 \langle \check{\textbf{G}}  \rangle \tau_3 , \quad 
\langle \check {\textbf{G}}  \rangle= \mu \int d\xi_p \int \frac{d\Omega}{4\pi} \check{\textbf{G}} \\
\check{\Sigma}_{\textit{s.o.}}=N_{\textit{imp}} u_{\textit{s.o.}}^2 \langle \check{\textbf{G}}  \rangle_{\textit{s.o.}} \\
\check{\Sigma}_{\textit{s.f.}}=N_{\textit{imp}} u_{\textit{s.f.}}^2 \tau_3 \langle \check{\textbf{G}}  \rangle_{\textit{s.o.}} \tau_3, \quad
\langle \check {\textbf{G}}  \rangle_{\textit{s.o.}}= \mu \int d\xi_p \int \frac{d\Omega}{4\pi} (\textbf{n} \times \textbf{n}^\prime) \check{\textbf{S}}  \check{\textbf{G}} \check{\textbf{S}} (\textbf{n} \times \textbf{n}^\prime) \; .
\end{eqnarray}
Here $N_{\textit{imp}}$ is the impurity concentration, $\mu$ is the density of states at the Fermi level and $\textbf{n}$ is a unit vector parallel to the momentum. The free Green function of a non-superconducting bulk material in the Nambu space reads,
\begin{equation}
\check{\textbf{G}}_0^{-1}(r_1,t_1,r_1^\prime,t_1^\prime)=\delta(r_1-r_1^\prime) \delta(t_1-t_1^\prime)  \left[ i  \partial t_1 + \frac{1}{2m} (\nabla_{r1}-i \textbf{A})^2 \tau_3 + E_F \right] \; .
\end{equation}
For example, the Retarded and Advanced GFs are given from Eq.\ref{eq:gorkov} of a homogeneous superconductor after Fourier transform reads:
\begin{equation}
\check{\textbf{G}}^{R(A)}(\omega,\textbf{p})= \left[ (\omega\pm i \eta) \tau_3-\tilde{\hat{\Delta}}\otimes \sigma_3- E_F  \right]^{-1} \; .
\label{eq:gorkovhomoap}
\end{equation}
$\eta$  here  is the Dynes parameter\cite{eta1,eta2,eta} that  describes the inelastic scattering energy rate within the relaxation time approximation. In principle, the Gorkov equation can be used to describe hybrids structures consisting of different materials and interfaces. However, dealing with full double coordinate Green functions  becomes very cumbersome\cite{w43,w64} and in several cases the solutions are impossible to be found.

For simplicity in the formulation of the equations of motion within the quasiclassical approximation, we restrict ourselves to equilibrium and stationary nonequilibrium situations. Although the quasiclassical approximation can be applied to time dependent problems. The Green function oscillates as a function of the relative coordinate $|\textbf{r} - \textbf{r}^\prime|$ on a scale of the Fermi wavelength $\lambda_F$. This is much shorter than the characteristic length scales in the typical problems in superconductivity, $\xi_0=v_F/\Delta$ and $\xi_N=v_F/2\pi T$. Moreover, in those problems, it is important to study the phase of the two-electron wave function, which depends on the center of mass coordinate. For these reasons it is possible, and sufficient for most applications, to integrate out the dependence on the relative coordinate. This was first understood by Eilenberger\cite{w81} and by Larkin and Ovchinikov\cite{w82}. We should take into account that the reduction is not allowed for another class of mesoscopic effects, e.g. weak localization and persistent currents, which are controlled by the phase coherence of the single electron wave function, contained in the relative coordinates. However, these effects are usually much weaker than those related to superconductivity. On the other hand, the reduction is possible also for problems involving Andreev reflection, since the essential information is again contained in the difference of electron and hole wave-vectors close to the Fermi surface.

\subsection*{Gradient expansion}
\addcontentsline{toc}{subsection}{Gradient expansion}

When integrating over the difference of variables, the convolution $\otimes$ in the Gorkov equation, eq.\ref{eq:gorkovap}, requires some care. It can be expressed, after the Fourier transformation, as a Taylor series
\begin{equation}
(A\otimes B)(\textbf{ p},R, E)= \exp \frac{i}{2}(\partial^A_R \partial^B_\textbf{ p}-\partial^A_\textbf{ p} \partial^B_R) A(\textbf{ p},R, E) B(\textbf{ p},R, E) \; ,
\end{equation}
where $R$ refers to the center of mass coordinate. In the problems to be discussed we can neglect short-range oscillations, hence we expand this expression up to linear order. To proceed, first we subtract the Gorkov equation, eq.\ref{eq:gorkovap} from its conjugated form. We observe that the Green functions and the self energies are linear combinations of Pauli matrices not including the unit matrix. This simplifies the equation of motion to
\begin{equation}
[i \textbf{ p} \hat{\boldsymbol{\partial}}_R+E\tau_3-i \hat{\Delta}-\hat{\Sigma}, \hat{\textbf{G}}]-[ \hat{\boldsymbol{\partial}}(\hat{\Delta}+ \hat{\Sigma}+e \phi + \mu), \nabla_p \hat{\textbf{G}}] + [\nabla_p \hat{\Sigma},  \hat{\boldsymbol{\partial}}_R \hat{\textbf{G}}]=0 \; .
\label{eq:3.7w}
\end{equation}
This form is much simple than the original. We note that it still accounts for particle-hole asymmetry, which is necessary e.g. for the description of the thermoelectric effect.

\section{Quasiclassical Greens functions}

We want to ignore the information contained in the fast oscillations of the full Greens functions $\hat{\textbf{G}}$ as a function of $|R_1 - R_2|$, which produces, after Fourier transformation, a pronounced peak at $|\textbf{ p}|=p_F$. On the other hand, we have to pay attention to the dependence on the transport direction, i.e. on the direction of the velocity, $\textbf{ v}_F$, at the Fermi surface. To make this explicit we write $\hat{\textbf{G}}(\epsilon,\textbf{ v}_F,R,E)$ where $\epsilon=p^2/(2m)-\mu$ depends on the magnitude of the momentum. The quasiclassical Greens functions is then defined by
\begin{equation}
\hat{g}(R, \textbf{ v}_F,E) \equiv \frac{i}{\pi} \int d\epsilon \hat{\textbf{G}}(\epsilon,R, \textbf{ v}_F,E) \; ,
\end{equation}
where the integration contour has two parts covering both half planes\cite{w71}. Form eq.\ref{eq:3.7w} we get, now setting $p=p_F$, the Eilenberger equation of motion for quasiclassical Green functions,
\begin{equation}
-[\textbf{ v}_F \hat{\partial},\hat{g}(R, \textbf{ v}_F,E)]=[-iE\tau_3+\hat{\Delta} + \frac{1}{2\tau} \langle \hat{g}(R, \textbf{ v}_F,E) \rangle_{}v_F, \hat{g}(R, \textbf{ v}_F,E)] \; .
\label{eq:3.9w}
\end{equation}
Here $\hat{g}$ reads,
\begin{equation}
\hat{g}=\begin{pmatrix} g & f \\ f^\dagger & -g \end{pmatrix} \; ,
\end{equation}
which is still linear combination of three Pauli matrices $\tau_{1,2,3}$ with $f^\dagger$ being the time-reversed counterpart of $f$. This symmetry can be used for convenient parametrization of the Eilenberger equation.

As the right hand side of eq.\ref{eq:3.9w} vanishes (in contrast to the Gorkov equation), it only defines the Greens functions up to a multiplicative constant. The constant can be fixed by the following argument, $\hat{g}$ is a linear combination of the Pauli matrices so the square of the Green functions is proportional to the unit matrix $\hat{g} \hat{g}=c \hat{1}$. From eq.\ref{eq:3.9w} we see that the proportionality constant c is space independent. If we now consider a system containing a sufficiently large superconductor, we can identify a region "deep inside the superconductor". Here $\hat{g}$ equals its bulk values, which can be calculated by performing the steps used in the quasiclassical approximation form the know solution of the homogeneous Gorkov equation. This procedure yields $c=1$, i.e. the Green functions are normalized,
\begin{equation}
\hat{g} \hat{g}=\hat{1} \ \text{i.e} \ g^2+f f^\dagger=1 \; .
\label{eq:w3.11}
\end{equation}
A more general mathematical derivation of the normalization condition can be found in ref.\cite{w83}.

From the knowledge of the retarded and advanced Greens functions alone, we can calculate energy-dependent quantities like the density of states
\begin{equation}
N(R,E)=N_0 \text{Re}[\langle g^R(R, \textbf{ v}_F,E) \rangle] \; .
\end{equation}
This and similar expression including the off-diagonal Greens functions, will be denoted in the following as spectral quantities. In thermal equilibrium, the distribution functions are known and these quantities also determine all properties of the system. Equivalently, the thermal Green function may be calculated in Matsubara imaginary time technique. Then, the spectral quantities are found by analytic continuation.

We need in addition, information on how the quasiparicle states are occupied i.e. about distribution functions. To evaluate those under nonequilibrium conditions we will use the Keldysh technique. Both techniques are described in numerous references, so we will not rederive them but rather explain the basic idea.

\section{Matsubara representation}
\label{sec:matsubaraap}

The Matsubara Green function technique\cite{w86} has been developed to describe many-body systems in equilibrium at finite temperature\cite{w75,w76}. In thermal equilibrium, the eigenvalues of physical observables do not depend on (real) time. In order to calculate thermal expectation values, we trace over all states using the Boltzmann weight $\exp(-H/t)$, which can be viewed as an analytic continuation of the time-evolution operator $\exp(iHt)$ to imaginary direction $\tau=it$. It is sufficient to know this operator (and with it the Green function) only in the interval $0<\tau<1/T$. By Fourier transformation and exploiting the fermionic symmetry one sees that all the necessary information is contained in the Greens functions defined for a discrete set of energies $E=i \omega_n$, proportional to the Matsubara frequencies $\omega_n= (2n+1) \pi T$ with integer values of n.

For those frequencies, in a bulk superconductor, the Green functions reads
\begin{equation}
g_\omega=\omega/\Omega, f_\omega=\Delta/\Omega \  \text{with} \ \Omega=\sqrt{|\Delta^2|+\omega^2} \; .
\end{equation}
This form is frequently used as a boundary condition. The Green functions in imaginary times show usually no singular or oscillatory structure and behave rather smooth and monotonic. In addition, $g_\omega$ is real-valued in the absence of field or phase gradients, this two facts simplify the numerical calculations in thermal equilibrium.

Expectation values of physical quantities can be expressed via Green functions. To calculate thermal averages we rotate onto the imaginary time axis and perform the quasiclassical approximation. Similarly, the self-consistency equation for the pair potential can be obtained from the corresponding expression for Gorkov Green functions. The integrations leading to the quasiclassical Green function can be performed and the self-consistency relation reads,
\begin{equation}
\Delta=\lambda N_0 2 \pi T \Sigma_\omega \langle f_\omega \rangle \; .
\end{equation}

\section{The dirty limit and the Usadel equation}

In most cases the superconducting material has strong elastic impurity scattering and it is described by the so-called "dirty limit". Here conductors with high amount of impurities, scatter electrons and randomize their trajectories, which implies that the elastic scattering length fulfils $l \ll \xi$. The requirement is that the elastic scattering self-energy dominates all other terms in the Eilenberger equation. In this limit the electron motion is diffusive and the Green functions are nearly isotropic. This allows to expand the Green functions in spherical harmonics
\begin{equation}
\hat{g}(E,R,\textbf{ v}_F)=\hat{G} (E,R)+ \textbf{ v}_F  \hat{\bar{g}}(E,R) \; .
\end{equation}
We denote the angular average of the quasiclassical Green function by a capital letter. Along this calculation we use the energy representation. In the expansion we assumed $\textbf{ v}_f  \hat{\bar{g}} \gg \hat{G}$. From the normalization condition, eq.\ref{eq:w3.11}, it follows that $\hat{G}^2(E,R)=1$ and the anticommutator $\{ \hat{G}(E,R), { \hat{\bar{g}}} (E,R) \}=0$. Angular averaging of eq.\ref{eq:3.9w} leads to,
\begin{equation}
-\frac{1}{3} v_F^2 [ \hat{\boldsymbol{\partial}}_R,{ \hat{\bar{g}}}(E,R) ]=[-iE \tau_3 + \hat{\Delta}, \hat{G} (E,R)] \; ,
\label{eq:w3.21}
\end{equation}
while averaging eq.\ref{eq:3.9w} after multiplication by $\textbf{ v}_F$ yields
\begin{equation}
{ \hat{\bar{g}}}(E,R)=-\frac{l}{v_F} \hat{G} (E,R) [ \hat{\boldsymbol{\partial}}_R,\hat{G} (E, R)] \; .
\label{eq:w3.22}
\end{equation}
Here the condition $\hat{g}(R)v_F/l \ll -i E  \tau_3 + \hat{\Delta}$ has been used. Inserting eq.\ref{eq:w3.22} into eq.\ref{eq:w3.21} leads to the Usadel equation\cite{w91},
\begin{equation}
\mathcal{D}[ \hat{\boldsymbol{\partial}}_R,\hat{G} (E, R)[ \hat{\boldsymbol{\partial}}_R,\hat{G} (E,R)]]=[-i E \tau_3 +\hat{\Delta}, \hat{G} (E,R)] \; ,
\label{eq:usadelap}
\end{equation}
where $\mathcal{D}=v_F l^2 /3$ is the diffusion constant. This equation is much simpler than the original Eilenberger equations and has been widely applied to describe properties of mesoscopic proximity system. The matrix current becomes in the dirty limit
\begin{equation}
\hat{J}=\frac{ \sigma_N}{4 e} T \int dE \hat{G}(E,R) [ \hat{\boldsymbol{\partial}}_R,\hat{G} (E,R)] \; ,
\end{equation}
where $\sigma_N=2 e^2 N_0 \mathcal{D}$ is the normal state conductivity.

For the Keldysh Greens functions the reduction to the dirty limit can also be performed, similar to the procedure outlined above. It reads,
\begin{equation}
\mathcal{D}[ \hat{\boldsymbol{\partial}}_R,\check{G}(E,R)[ \hat{\boldsymbol{\partial}}_R,\check{G} (E,R)]]=[-i E \check{\tau_3}+\check{\Delta},\check{G} (E,R)] \; ,
\label{eq:w3.26}
\end{equation}
where
\begin{equation}
\check{\tau_3}=\begin{pmatrix} \tau_3 & 0 \\ 0 & \tau_3 \end{pmatrix} \quad \check{\Delta}=\begin{pmatrix} \hat{\Delta} & 0 \\ 0 & \hat{\Delta} \end{pmatrix} \; .
\end{equation}
The Keldysh technique also works for time-dependent situations. The normalization condition holds, $\check{G}\check{G}=\check{1}$, which in terms of the components implies
\begin{equation}
\hat{G}^R\hat{G}^R=\hat{G}^A\hat{G}^A=\hat{1}, \ \text{and} \ \hat{G}^R\hat{G}^K+\hat{G}^K\hat{G}^A=0 \; .
\end{equation}
As a consequence of the second relation $G^K$ can be parametrized as 
\begin{equation}
\hat{G}^K=\hat{G}^R \hat{f}- \hat{f} \hat{G}^A \; .
\end{equation}
From the Keldysh Usadel equation, eq.\ref{eq:w3.26} we obtain a kinetic equation of motion for the distribution function $\hat{f}$,
\begin{equation}
\mathcal{D}[\nabla^2 \hat{f}+(\hat{G}^R \nabla \hat{G}^R ) \nabla \hat{f}-\nabla \hat{f} (\hat{G}^A \nabla \hat{G}^A - \nabla(\hat{G}^R (\nabla \hat{f}) \hat{G}^A) ]- (\hat{G}^R [\hat{\Delta}, \hat{f}]-[\hat{\Delta}, \hat{f}]\hat{G}^A)
\nonumber
\end{equation}
\begin{equation}
+i E (\hat{G}^R [\hat{f}, \tau_3]-[\hat{f}, \tau_3]\hat{G}^A)=0 \; .
\end{equation}
The values $G^{R(A)}$ satisfy the respective components of the Usadel equation. Since the kinetic equation has only two independent entries and it is a linear equation, we can assume $\hat{f}$ to be diagonal. Returning to the definitions, we can relate it to the distribution functions for electrons and holes,
\begin{equation}
\hat{f} = \begin{pmatrix} 1-2f_e & 0 \\ 0 & 2 f_h-1 \end{pmatrix} \; .
\end{equation}
Here the energy is measured from the chemical potential of the superconductor. In thermal equilibrium, for instance in the reservoirs at voltage $V_R$, it can be expressed via Fermi function,
\begin{equation}
\hat{f} = \begin{pmatrix} 1-2f(E) & 0 \\ 0 & 2 f(-E)-1 \end{pmatrix}= \begin{pmatrix} \tanh \left( \frac{E+eV_R}{2 T} \right) & 0 \\ 0 & \tanh \left( \frac{E-eV_R}{2 T} \right) \end{pmatrix} \; .
\end{equation}
For practical purposes, it is convenient to split the distribution function into odd and even components with respect to the Fermi surface $\hat{f}=n_+ + n_- \tau_3$. Here, $n_+$ is related to difference in energy parametrized by a different effective temperature, whereas $n_-$ is related to a difference in particle number parametrized by a shift of the effective chemical potential. The form of this two parameters is
\begin{equation}
n_\pm= \frac{1}{2} \left[ \tanh \left( \frac{E+eV_R}{2 T} \right) \pm  \tanh \left( \frac{E-eV_R}{2 T} \right)       \right] \; .
\end{equation}
Here this two values obey the relation $2f(E)=1-n_+-n_-$.

\section{The $\theta$ parametrization}
\label{sec:theta}

For better handling of the Usadel equation, both numerically and analytically, we will parametrize the Greens functions in a convenient way by making use of the normalization condition.
\begin{subequations}
\begin{align}
\check{G}^R(E)=\begin{pmatrix} \cos(\theta) & \sin(\theta) \exp(i \chi) \\ -\sin(\theta) \exp(-i \chi) & -\cos(\theta) \end{pmatrix} 
\nonumber
 \\
\check{G}^A(E)=\begin{pmatrix} -\cos(\theta^*) & \sin(\theta) \exp(i \chi) \\ -\sin(\theta^*) \exp(-i \chi^*) & \cos(\theta^*) \end{pmatrix} 
\end{align}
\end{subequations}
Here $\theta(E)$ and $\chi(E)$ are complex functions and $"*"$ upperscript denotes complex conjugation. It is also possible to parametrize using $\sinh$  and $\cosh$ functions. In the present convention, we are required to choose the gauge of $\Delta$ such that for a imaginary pair potential $\Delta=|\Delta|$ and $\check{\Delta} \propto \tau_1$. The Usadel equation, eq.\ref{eq:usadelap}, in a normal metal can be written using this notation as
\begin{equation}
\mathcal{D} \partial^2_x \theta= -2 i E \sin \theta+ \frac{\mathcal{D}}{2} (\partial_x \chi)^2 \sin 2 \theta \; , \quad \tau_3 \check{J}=2 \sin^2 \theta \partial_x \chi \; .
\end{equation}
In imaginary time, it is more convenient to parametrize\cite{w102}
\begin{equation}
\check{g}_\omega=\begin{pmatrix} \cosh(\theta) & \sinh(\theta) \exp(i \chi) \\ \sinh(\theta) \exp(-i \chi) & -\cosh(\theta) \end{pmatrix} \; .
\end{equation}
Now $\theta$ and $\chi$ are real value functions. The Usadel equation in a normal metal now reads,
\begin{equation}
\mathcal{D} \partial^2_x \theta= 2 \omega  \sinh \theta+ \frac{\mathcal{1}}{2} (\partial_x \chi)^2 \sinh 2 \theta \; , \quad \tau_3 \check{J}_\omega= \sinh^2 \theta \partial_x \chi \; .
\end{equation}
Consider a good interface to a superconducting reservoir with $\Delta e^{i \phi_0}$, the boundary conditions, eq.\ref{eq:trans}, will now read
\begin{subequations}
\begin{align}
\theta=\theta_S=\text{arctanh}(\Delta/E) \; ,
\\
\chi=\phi_0 \; .
\end{align}
\end{subequations}
For a normal reservoir we set $\theta=0$. While the value corresponding to $\chi$ is arbitrary as it is related to $\sin(\theta)$ terms. In the presence of tunnelling barriers, we apply the Kupriyanov Lukichev boundary conditions, eq.\ref{eq:kl}. For a normal metal layer connected to a superconductor it reads,
\begin{equation}
\partial_x \theta= \kappa_t \sinh(\theta-\theta_S) \; .
\end{equation}

We must take into account that this parametrization has also several downsides: $\theta$ is unbounded, $\xi$ can undergo rapid spatial changes where $\theta$ is small and hyperbolic functions can lead to spurious solutions. For all this reasons and even if it is more difficult to handle, instead of the $\theta$ parametrization we can use the Ricatti parametrization. The latter can be in practice more useful for numerical work\cite{pt52,pt53,pt54}.

\section{Two-times Green functions}

Let us assume that we have a Josephson junction, Left and Right superconductors have a well defined phase and we apply a bias voltage. We set the phase $\varphi_{L}=-\varphi_{R}=\varphi/2$ and the voltage $V_{L}=-V_{R}=V$, where the subindex denotes the superconducting reservoir. According to the Josephson relation, for any given voltage the value of the total phase of the superconducting reservoir is the intrinsic one plus the one induced by the voltage. For right and left superconductors respectively,
\begin{equation}
\phi_{L(R)}(t)=\pm\varphi/2\pm\int V(t)dt\;.\label{eq:phase}
\end{equation}

Thus, by applying a bias voltage the phase becomes time dependent. 

Previously we had Green functions in superconductors, $\hat{g}^{R(A)}$, now we must add them to the calculations with the help of this phase factors,
\begin{equation}
\check{G_{j}}(t,t^\prime)=e^{i\phi_{j}(t)\hat{\tau_{3}}/2} \check{g_{j}}(t-t^\prime)e^{-i\phi_{j}(t)\hat{\tau_{3}}/2} \; .
\label{eq:G}
\end{equation}
Were $e^{i\phi_{j}(t)\hat{\tau_{3}}/2}$ corresponds to 
\begin{equation}
\left(\begin{array}{cc}
e^{i\phi_{j}(t)/2} & 0\\
0 & e^{-i\phi_{j}(t)/2}\end{array}\right)
\end{equation}
in Nambu space. The time dependent Green functions in the stationary case will be given by

\begin{equation}
\check{g}(t-t^\prime)=\int\frac{dE}{2\pi}e^{-i E t}\check{g}(E) \; .
\end{equation}

\bibliographystyle{unsrtnat}
\renewcommand{\bibname}{Bibliography of the Appendix} % changes default name Bibliography to References

%\end{document}

%% file: Thesis_AOzaeta.bbl
\begin{thebibliography}{10}
\bibitem{footnote1}
We neglect here the contribution to $I_A$  from partial Andreev reflection processes
at energies above the  gap. In the case of tunneling barriers, this contribution leads to
small corrections and can be neglected \cite{VBCH}.
\end{thebibliography}
\begin{thebibliography}{10}
\bibitem{note-IC}
In all the numerical calculations within the two-level model shown in this work
we have assumed that the system is initially in its ground state, i.e.,
$\Psi_0^\dagger=(0,1)$.
\end{thebibliography}
